\documentclass[longauth]{aa}
\usepackage{graphicx}
\usepackage{txfonts}
\usepackage{gensymb}
\usepackage[]{hyperref}
\usepackage{multirow}
\usepackage{graphicx}
\usepackage{times}
\usepackage{float}
\usepackage{pdflscape}
\usepackage{threeparttable}
\usepackage{amsmath}
\usepackage{amssymb}
\usepackage{xspace}
\usepackage{caption}
\usepackage{subcaption}
\usepackage{newtxtext}
\usepackage{color}
\usepackage{lscape}

		\newcommand{\fekab}{Fe\,K$\alpha$[$\beta$]\xspace}

		\newcommand{\feka}{Fe\,K$\alpha$\xspace}		
		 
		\newcommand{\fexxvxxvi}{Fe\,\textsc{xxv--xxvi}\xspace}
		\newcommand{\fcov}{f_{\rm cov}\xspace}
		
		\newcommand{\fexxv}{Fe\,\textsc{xxv}\xspace}
		\newcommand{\fexxvabs}{Fe\,\textsc{xxv}\,He$\alpha$\xspace}
		\newcommand{\fexxvi}{Fe\,\textsc{xxvi}\xspace}
		\newcommand{\fexxviabs}{Fe\textsc{xxvi}\,Ly$\alpha$\xspace}
		
		\newcommand{\pmc}{P_{ \mathcal{MC}}\xspace}
		\newcommand{\nhm}{\mathcal{NHM}\xspace}
		\newcommand{\mc}{$\mathcal{MC}$\xspace}

		\newcommand{\vout}{v_{\rm out}\xspace}
		\newcommand{\mout}{\dot M_{\rm out}}
					
		\newcommand{\vturb}{v_{\rm turb}\xspace}

		\newcommand{\chisq}{\chi^{2}}
		\newcommand{\nhgal}{N_{\rm H}^{\rm Gal}}
		\newcommand{\nh}{N_{\rm H}}
		\newcommand{\lognh}{\log(N_{\rm H}/\rm{cm}^{-2})}
		\newcommand{\logxi}{\log(\xi/\rm{erg\,cm\,s}^{-1})}

		\newcommand{\lbol}{L_{\rm bol}}

		\newcommand{\mbh}{M_{\rm BH}}
		\newcommand{\pout}{\dot p_{\rm out}}

		\newcommand{\ks}{\,\rm ks}

		\newcommand{\dC}{\Delta \mathcal{C} /\Delta\nu}
		\newcommand{\dc}{\Delta \mathcal{C}}
		\newcommand{\cstat}{\mathcal{C-}\rm stat\xspace}
		\newcommand{\leddratio}{L_{\rm bol}/L_{\rm Edd}}
		\newcommand{\sn}{SNR\xspace}

		\newcommand{\los}{line-of-sight\xspace}

		\newcommand{\fe}{Fe\,K\xspace}
        \newcommand{\iron}{iron\,K\xspace}
		\newcommand{\ergs}{\,\rm erg\,s^{-1}\xspace}
		\newcommand{\cmsq}{{\,\rm cm^{-2}}\xspace}
		
		\newcommand{\flux}{{\,\rm erg\,cm^{-2}\,s^{-1}}\xspace}

		\newcommand{\ev}{\,\rm eV}		
		\newcommand{\kev}{\,\rm keV}
				
		\newcommand{\kms}{\,\rm km\,s^{-1}\xspace}
		\newcommand{\ew}{equivalent width\xspace}

		\newcommand{\kb}{$\mathcal{KB}$\xspace}

		\newcommand{\athenaifu}{\emph{Athena}/X-IFU\xspace}
		\newcommand{\xrism}{\emph{XRISM}/Resolve\xspace}
		\newcommand{\suzaku}{\emph{Suzaku}\xspace} 
		\newcommand{\nustar}{\emph{NuSTAR}\xspace}		
		\newcommand{\xmm}{\emph{XMM-Newton}\xspace}

		\newcommand{\xmmnu}{\textit{XMM-Newton} and \textit{NuSTAR}\xspace}
		
		\newcommand{\sub}{SUBWAYS\xspace}

		\newcommand{\pds}{PDS\,456\xspace}		
		\newcommand{\mcg}{MCG--03--58--007\xspace}

		\newcommand{\tmosf}{2MASS\,J165315$+$2349\xspace}
		\newcommand{\lbqs}{LBQS\,1338$-$0038\xspace}
		\newcommand{\izw}{I\,Zwicky\,1\xspace}

		\newcommand{\xrade}{\textsc{xrade}\xspace}

		\newcommand{\xillver}{\texttt{xillver}\xspace}
		\newcommand{\relxill}{\texttt{relxill}\xspace}

		\newcommand{\spex}{\textsc{spex}\xspace}
		\newcommand{\xabs}{\textsc{xabs}\xspace}
		\newcommand{\heasoft}{\textsc{HEAsoft}\xspace}
		\newcommand{\xstar}{\textsc{xstar}\xspace}
		\newcommand{\xspec}{\textsc{xspec}\xspace}

		\newcommand{\sas}{\textsc{sas}\xspace}

        \newcommand{\ftgroup}{\textsc{ftgrouppha}\xspace}
        \newcommand{\specgroup}{\textsc{specgroup}\xspace}

\hypersetup{
    colorlinks=true,
    linkcolor=red,
    filecolor=magenta,      
    urlcolor=green,
    citecolor=blue
}

\defcitealias{Tombesi10}{T10}
\defcitealias{Tombesi11}{T11}
\defcitealias{Gofford13}{G13}
\defcitealias{Igo20}{Igo20}
\defcitealias{Chartas21}{C21}

\begin{document} 

\title{Supermassive Black Hole Winds in X-rays -- SUBWAYS. I. Ultra-fast outflows in QSOs beyond the local Universe}
\titlerunning{The SUBWAYS sample}
\authorrunning{G. A. Matzeu et al.,}

\author{G. A. Matzeu
\inst{1,2,3} \fnmsep\thanks{gabriele.matzeu@unibo.it},
M. Brusa$^{1,2}$
\and
G. Lanzuisi$^{2}$
\and
M. Dadina$^{2}$
\and
S. Bianchi$^{4}$
\and
G. Kriss$^{5}$
\and
M. Mehdipour$^5$
\and
E. Nardini$^6$
\and
G. Chartas$^7$
\and
R. Middei$^{8}$
\and
E. Piconcelli$^{9}$
\and
V. Gianolli$^{4,10}$
\and
A. Comastri$^{2}$
\and
A. L. Longinotti$^{11}$
\and
Y. Krongold$^{11}$
\and
F. Ricci$^{4}$
\and
P. O. Petrucci$^{10}$
\and
F. Tombesi$^{12,9,13,14}$
\and
A. Luminari$^{15,9}$
\and
L. Zappacosta$^{9}$
\and
G. Miniutti$^{16}$
\and
M. Gaspari$^{17}$
\and
E. Behar$^{18}$
\and
M. Bischetti$^{19}$
\and
S. Mathur$^{20,21}$
\and
M. Perna$^{16}$
\and
M. Giustini$^{16}$
\and
P. Grandi$^{2}$
\and
E. Torresi$^{2}$
\and
C. Vignali$^{1,2}$
\and
G. Bruni$^{15}$
\and
M. Cappi$^{2}$
\and
E. Costantini$^{22}$
\and
G. Cresci$^{6}$
\and
B. De Marco$^{23}$
\and
A. De Rosa$^{15}$
\and
R. Gilli$^{2}$
\and
M. Guainazzi$^{24}$
\and
J. Kaastra$^{22,25}$
\and
S. Kraemer$^{26}$
\and
F. La Franca$^{4}$
\and
A. Marconi$^{27}$
\and
F. Panessa$^{15}$
\and
G. Ponti$^{28}$
\and
D. Proga$^{29}$
\and
F. Ursini$^{4}$
\and
F. Fiore$^{19}$
\and
A. R. King$^{30}$
\and
R. Maiolino$^{31}$
\and
G. Matt$^{4}$
\and 
A. Merloni$^{32}$
}

\institute{
$^{1}$Department of Physics and Astronomy (DIFA), University of Bologna, Via Gobetti, 93/2, I-40129 Bologna, Italy\\
$^{2}$INAF-Osservatorio di Astrofisica e Scienza dello Spazio di Bologna, Via Gobetti, 93/3, I-40129 Bologna, Italy\\
$^{3}$European Space Agency (ESA), European Space Astronomy Centre (ESAC), E-28691 Villanueva de la Ca\~{n}ada, Madrid, Spain\\
$^4$Dipartimento di Matematica e Fisica, Universit\'{a} degli Studi Roma Tre, Via della Vasca Navale 84, I-00146, Roma, Italy\\
$^5$Space Telescope Science Institute, 3700 San Martin Drive, Baltimore, MD 21218, USA\\
$^{6}$INAF -- Osservatorio Astrofisico di Arcetri, Largo Enrico Fermi 5, I-50125 Firenze, Italy\\
$^7$  Department of Physics and Astronomy, College of Charleston, Charleston, SC, 29424, USA \\
$^{8}$Space Science Data Center - ASI, Via del Politecnico s.n.c., 00133 Roma, Italy\\
$^{9}$INAF - Osservatorio Astronomico di Roma, Via Frascati 33, 00078, Monte Porzio Catone (Roma), Italy\\
$^{10}$Univ. Grenoble Alpes, CNRS, IPAG, 38000, Grenoble, France\\
$^11$Instituto de Astronom\'{i}a, Universidad Nacional Aut\'{o}noma de M\'{e}xico, Circuito Exterior, Ciudad Universitaria, Ciudad de M\'{e}xico
04510, M\'{e}xico\\
$^{12}$ Department of Physics, University of Rome ‘Tor Vergata’, Via della Ricerca Scientifica 1, I-00133 Rome, Italy \\
$^{13}$ Department of Astronomy, University of Maryland, College Park, MD 20742, USA \\
$^{14}$ NASA/Goddard Space Flight Center, Code 662, Greenbelt, MD 20771, USA\\
$^{15}$ INAF — Istituto di Astrofisica e Planetologia Spaziali, Via Fosso del Cavaliere, I-00133 Roma, Italy \\
$^{16}$Centro de Astrobiolog\'ia (CSIC-INTA), Camino Bajo del Castillo s/n, Villanueva de la Ca\~nada, E-28692 Madrid, Spain\\
$^{17}$ Department of Astrophysical Sciences, Princeton University, 4 Ivy Lane, Princeton, NJ 08544-1001, USA\\ 
$^{18}$Physics Department, The Technion, 32000, Haifa, Israel\\
$^{19}$INAF - Osservatorio Astronomico di Trieste, Via G. B. Tiepolo 11, 34143, Trieste, Italy
$^{20}$Department of Astronomy, The Ohio State University, 140 West 18th Avenue, Columbus, OH 43210, USA \\
$^{21}$Center for Cosmology and Astroparticle Physics, 191 West Woodruff Avenue, Columbus, OH 43210, USA
$^{22}$SRON Netherlands Institute for Space Research, Niels Bohrweg 4, 2333 CA Leiden, The Netherlands\\
$^{23}$ Departament de Física, EEBE, Universitat Politècnica de Catalunya, Av. Eduard Maristany 16, 08019 Barcelona, Spain \\
$^{24}$ESA - European Space Research and Technology Centre (ESTEC), Keplerlaan 1, 2201 AZ, Noordwijk, The Netherlands
$^{25}$Leiden Observatory, P.O. Box 9513, 2300 RA Leiden, The Netherlands\\
$^{26}$Department of Physics, Institute for Astrophysics and Computational Sciences, The Catholic University of America, Washington, DC, 20064, USA\\
$^{27}$ Dipartimento di Fisica e Astronomia, Universit{\` a} di Firenze, via G. Sansone 1, 50019 Sesto Fiorentino, Firenze, Italy \\
$^{28}$ INAF -- Osservatorio Astronomico di Brera, Via Bianchi 46, I-23807 Merate (LC), Italy \\
$^{29}$Department of Physics \& Astronomy, University of Nevada, Las Vegas, USA \\
$^{30}$ Department of Physics \& Astronomy, University of Leicester, Leicester LE1 7RH, UK\\
$^{31}$Kavli Institute for Cosmology, University of Cambridge, Madingley Road, Cambridge, CB3 0HA, UK; Cavendish Laboratory, University of Cambridge, 19 J. J. Thomson Avenue, Cambridge, CB3 0HE, UK\\
$^{32}$Max-Planck-Institut f\"ur extraterrestrische Physik, Giessenbachstra{\ss}e 1, D-85748 Garching bei M\"unchen, Germany
}

 
  \abstract
{We present a new X-ray spectroscopic study of $22$ luminous ($2\times10^{45}\lesssim L_{\rm bol}/\ergs \lesssim 2\times10^{46}$) active galactic nuclei (AGNs) at intermediate-redshift ($0.1 \lesssim z \lesssim 0.4$), as part of the SUpermassive Black hole Winds in the x-rAYS (SUBWAYS) sample, mostly composed of quasars (QSOs) and type\,1 AGN. Here, 17 targets were observed with \textit{XMM-Newton} between 2019--2020 and the remaining 5 are from previous observations. The aim of this large campaign ($1.45\,\rm Ms$ duration) is to characterise the various manifestations of winds in the X-rays driven from supermassive black holes in AGN. In this paper we focus on the search and characterization of ultra-fast outflows (UFOs), which are typically detected through blueshifted absorption troughs in the Fe\,K band ($E>7\,\rm keV$). By following Monte Carlo procedures, we confirm the detection of absorption lines corresponding to highly ionised iron (e.g., Fe\,\textsc{xxv}\,H$\alpha$, Fe\,\textsc{xxvi}\,Ly$\alpha$) in 7/22 sources at the $\gtrsim95\%$ confidence level (for each individual line). The global combined probability of such absorption features in the sample is $>99.9\%$. The SUBWAYS campaign extends at higher luminosity and redshifts than previous local studies on Seyferts, obtained using \xmm and \suzaku observations. We find a UFO detection fraction of $\sim30\%$ on the total sample that is in agreement with the previous findings. This work independently provides further support for the existence of highly-ionised matter propagating at mildly relativistic speed ($\gtrsim0.1c$) in a considerable fraction of AGN over a broad range of luminosities, which is expected to play a key role in the self-regulated AGN feeding-feedback cycle, as also supported by hydrodynamical multiphase simulations. }

   \keywords{galaxies: active – galaxies: nuclei – X-rays: galaxies }

   \maketitle
%

\section{Introduction}

 It is widely accepted that supermassive black holes (SMBHs, 10$^{6}$-- 10$^{10}$ M$_\odot$ e.g., \citealt{Salpeter64,Magorrian98}) are hosted at the centre of virtually every known massive galaxy. The observed tight correlations between the host galaxy and the SMBH properties \citep[see][for a review]{Kormendy13} strongly suggest that their formation and evolution are profoundly coupled with each other. Some physical mechanisms must have therefore linked the regions where the SMBH gravitational field dominates to the larger scales, where its direct influence is negligible. At this stage, the key underlying ingredients at play in the co-evolution paradigms of Active Galactic Nuclei (AGN) and galaxies still need to be understood. It has been proposed that highly ionised gas outflows could play a pivotal role in this process \citep[e.g.,][]{King03,King05,Gaspari17_uni}. The presence of such powerful winds is expected to regulate accretion of material onto (and ejection from) compact objects. 

Through their mechanical power, ultra-fast outflows (UFOs) are accelerated at velocities larger than 10,000\,km\,s$^{-1}$ and up to a few tens of the speed of light. For these reasons, UFOs are also able to inject momentum and energy over wide spatial scales via the interaction with the inter-stellar medium (ISM) in the host galaxy. This process is expected to promote an efficient feedback mechanism \citep[e.g.,][]{Murray05,DiMatteo05,Ostriker10,Torrey20}, that is needed to reproduce the observed properties in galaxies, e.g. the scaling relations \citep{King11}, and to regulate their overall mass-size ecosystem \citep[e.g.,][]{Fabian12,KingPounds15,Heckman14}.

UFOs are routinely detected in the X-ray spectra of $30$--$40\%$ of local ($z\lesssim0.1$) AGN \citep[][hereafter \citetalias{Tombesi10,Gofford13,Igo20}]{Tombesi10,Gofford13,Igo20}, and in a handful of sources at intermediate to high 
redshift \citep[up to $z\sim3$; e.g.,][hereafter \citetalias{Chartas21}]{Chartas02,Lanzuisi12,Chartas21}. UFOs manifest themselves as absorption troughs associated with 
\fexxvabs and \fexxviabs transitions ($E_{\rm rest}=6.7$--$6.97\kev$) blueshifted 
at energies $E_{\rm rest}>7\kev$ (all the line energies will be given in the source rest-frame throughout this work). 
The degree of blueshift translates into the range of extreme outflow velocities observed between $\vout\sim-0.03c$ up to $-0.5c$, for gas column densities and ionisations of $\nh\sim10^{23-24}\cmsq$ and $\logxi\gtrsim3$, respectively \citep[e.g.,][]{Reeves03,Pounds09,Tombesi11,Matzeu17b,Reeves18PDS,Parker18iras,Braito18}. The frequent detection of these features, supported by a detailed modelling of the high energy spectra of the most powerful local QSO hosting X-ray winds, \pds, indicates that UFOs arise in wide angle outflows, 
implying that a significant amount of kinetic power is involved \citep{Nardini15,Luminari18}.

Evidence for low-ionisation UFO components 
have been also reported in the soft X-ray spectra in the $E_{\rm rest}=0.3$--$2\kev$ band \citep[e.g.,][]{Braito14,Longinotti15,Reeves16,Reeves18PG1211,Serafinelli19,Reeves20rgs,Krongold21}, usually observed as blueshifted oxygen and neon ions. Similar high-velocity outflows, arising directly from the accretion disc region, have also been found in the UV spectra via prominent blueshifted, ionised absorption and emission features in broad absorption line quasars (BAL QSOs), typically between $\lambda_{\rm rest}=50$--$2000$\,\AA~ \citep[e.g.,][]{Gaskell82,Wilkes87,Richards11}. These are associated with lower ionisation metal ions (C\,\textsc{iv}, Al\,\textsc{ii}, Fe\,\textsc{ii}, etc; e.g. \citealt{Crenshaw03,Green12,Hamann18,Kriss18,Kriss19,Mehdipour22,Vietri22arXiv}). It has been shown that at least some of the UV absorbing outflows in sub-Eddington systems can be driven by radiation pressure on spectral lines \citep[e.g.,][]{Murray95,Proga00}. 
We refer to the recent review by \citet{Giustini19} for more details.

Outflowing material at considerably lower velocity (typically within a  $\vout$ of $-5000\kms$) and less ionised than UFOs at $\logxi\sim1-3$ \citep[e.g.,][]{Sako00,Parker19Mrk335}, known as a warm absorber (WA), is also detected through absorption features and edges from He- or/and H-like ions of C, O, N, Ne, Mg, Al, Si and S in the X-rays  
\citep{Halpern84,Mathur97,Mathur98,Blustin05,Reeves13,Kaastra14,Laha14,Laha16}. WAs are detected in a substantial fraction of AGN, i.e. $\sim65\%$ \citep{Reynold97,Piconcelli05,McKernan07}. It was suggested by \citet{Tombesi13} that despite their physical distinction, UFOs and WAs might be connected somehow as part of the same wind, but originating from different locations (see \citealt[][]{Laha21NatAs} for a comprehensive review of ionised outflows). 


Finally, outflowing gas is also routinely observed at host-galaxy scales, in the ionised, neutral/atomic and molecular phases. These outflowing components observed at kpc scale or beyond are now traced with modern sensitive optical/far-IR/mm/radio facilities \citep[e.g.,][]{Morganti05,Feruglio10,Harrison14,Brusa15,Maiolino12,Cresci15,Feruglio17,Brusa18,Bischetti19} and show lower velocities with respect to the accretion disc winds ($ \vout \sim -500$ to $-2000\kms$, depending on the phase), and considerably higher mass outflow rates up to $100$--$1000$\,M$_\odot$/yr \citep[see][]{Cicone18b}. 

Some models predict that the fast outflowing gas is accelerated by the radiation pressure caused by highly accreting black holes approaching the Eddington limit \citep[e.g,][]{Zubova12}. Subsequently, the energy deposited via shocks by the UFO into the galaxy ISM generates the galaxy-wide outflows observed in lower-ionisation gas \citep[see][]{Fabian12,KingPounds15}. Alternatively, massive sub-relativistic outflows are expected also in systems accreting at lower Eddington ratio, due to magnetic \citep[e.g.,][]{Fukumura10,Fukumura17,Kraemer18} and/or thermal driving \citep[e.g.,][]{Woods96,Mizumoto19thermal_driving,Waters21}. 
The global AGN feeding-feedback self-regulated framework has been supported by three-dimensional hydrodynamical simulations unifying the micro and macro properties of 
the AGN environment \citep[e.g.,][]{Gaspari13_rev,Gaspari20,Sadowski17,Yang19,Wittor20}, 
which, in turn, have been corroborated by several multiwavelength observations \citep[e.g.,][]{Maccagni21,Eckert21,McKinley22,Temi22}.

Multi-phase tracers would therefore allow us to probe galactic outflows in their full extent, that is, from the nuclear ($<1$\,pc) to the largest scales ($>10$\,kpc), and at the same time to have a comprehensive view of their driving mechanism. 
\citet{Fiore17} reported a correlation between the velocity of the wind (for both UFOs and large-scale components) and the bolometric luminosity, ($L_{\rm bol} \propto v_{out}^{3.9}$), in agreement with that predicted by \citet{Costa14} for energy conserving outflows. However, the statistics of this work for the UFO sample is still limited ($\sim20$ AGN with UFOs), with only $50\%$ at $L_{\rm bol}\gtrsim10^{45}\ergs$. 

 A comparison between the momentum rates (i.e., $\pout=\mout\vout$) observed over a range of spatial scales can be used to disentangle the wind propagation mechanisms, i.e. energy (large-scale) vs. momentum (small-scale: UFO) conserving. 
 The first reported cases of molecular outflows in systems hosting an UFO are: IRAS\,F11119$+$3257 \citep{Tombesi15}, Mrk\,231 \citep{Feruglio15}, IRAS\,17020$+$4544 \citep{Longinotti15,Longinotti18} and APM\,08279$+$5255 \citep{Feruglio17}, supporting the energy-driven wind scenario, deemed as the smoking gun for a large scale feedback. Indeed, conserving energy is crucial to achieve an effective macro-scale feedback to quench cooling flows and star formation \citep[e.g.,][]{Gaspari19}. However, by analysing more sources, it emerged that not all the outflows supported this scenario. Further molecular outflows observed in UFO hosting sources, such as \pds \citep{Bischetti19PDS}, \mcg \citep{Sirressi19}, \izw \citep{Cicone14,ReevesBraito19} and Mrk\,509 \citep{Zanchettin21} revealed momentum rates at least two orders of magnitude below the expected value for an energy conserving wind (see also \citealt{Marasco20,Tozzi21}). These results are suggesting a more complex physical mechanism and range of efficiencies in transferring the nuclear wind out to the large-scale galaxy structure, or significant AGN variability over the lifetime of the flow \citep{NardiniZubovas,ZubovasNardini}, as supported by Chaotic Cold Accretion (CCA) simulations \citep{Gaspari17_cca}.
Related to this, an important parameter that is needed to constrain wind models is the UFO duty cycle, i.e. how persistent accretion disc winds are. Indeed, the derivation of the energy injection rate by UFOs into the galaxy ISM must take into acount this factor, with implications for the timescale and efficiency of propagation through the host galaxy \citep{ZubovasKing16}. The UFO duty cycle can be inferred from the fraction of AGN in which they are observed, but it is highly degenerate with the opening angle, hence, as of today, it is virtually unconstrained for sources above $L_{\rm bol}\sim10^{45}\ergs$. Observing large samples of sources at luminosities above the break $L \gtrsim L^{\star}$ of the AGN luminosity function \citep[e.g.,][]{Aird15}, with a range of Eddington ratios and enough statistics to constrain the wind duty cycle, has become crucial to overcome the limitations described above \citep[e.g.,][]{Bertola20}.  

This work is the first of a series of \sub publications, with the dedicated goals of investigating the various manifestations of UFOs emanating from the environments of supermassive black holes in AGNs. These include: (i) gaining significant advances in our understanding of the detection rate and physical properties of UFOs and (ii) their connection with WA features, as well as (iii) their role in providing a macro-scale feedback and finally (iv) mapping the physical properties of the outflows across different galaxy scales and gas phases at different wavelengths/ionisation states. In this paper we present our first results of the \sub campaign, specifically designed to provide a solid detection of blueshifted absorption features in the \fe band in context of a robust statistical grounds for high signal-to-noise ratio (\sn hereafter) sources at $L\gtrsim10^{45}\ergs$. 

The paper is organised as follows: Section\,\ref{sec:the subways campaign} presents the SUBWAYS sample and the target selection. In Section\,\ref{sec:Data reduction} we describe the reduction of the {\it  XMM-Newton} data. In Section\,\ref{sec:spectral analysis} we present all the details of the spectral analysis of the EPIC data, including the continuum modeling, the procedure we adopted to search for \fe emission and absorption features, the modeling of the \fe band and the Monte Carlo simulations we used to assign a robust significance level to the detections. In Section\,\ref{sec:results} we present our main results, i.e. the line detection rate as inferred from our spectral modeling, and in Section\,\ref{sec:Discussion} we discuss our findings in the light of recent results at both lower and higher redshift; Section\,\ref{sec:summary and conclusion} summarises our results. 
 Cosmological values of $H_0=70$\,km\,s$^{-1}$\,Mpc$^{-1}$, $\Omega_{\Lambda_{0}}=0.73$ and $\Omega_{\rm M}=0.27$ are assumed throughout this paper, and errors are quoted at the $90$ percent confidence level or a difference in C-statistic \citep[][i.e., $\dc=2.71$]{Cash79} for one parameter of interest, unless otherwise stated. The cosmic abundances are set to solar throughout the paper.

\section{The \sub campaign}
\label{sec:the subways campaign}

\begin{figure}
\begin{center}
\includegraphics[width=\linewidth]{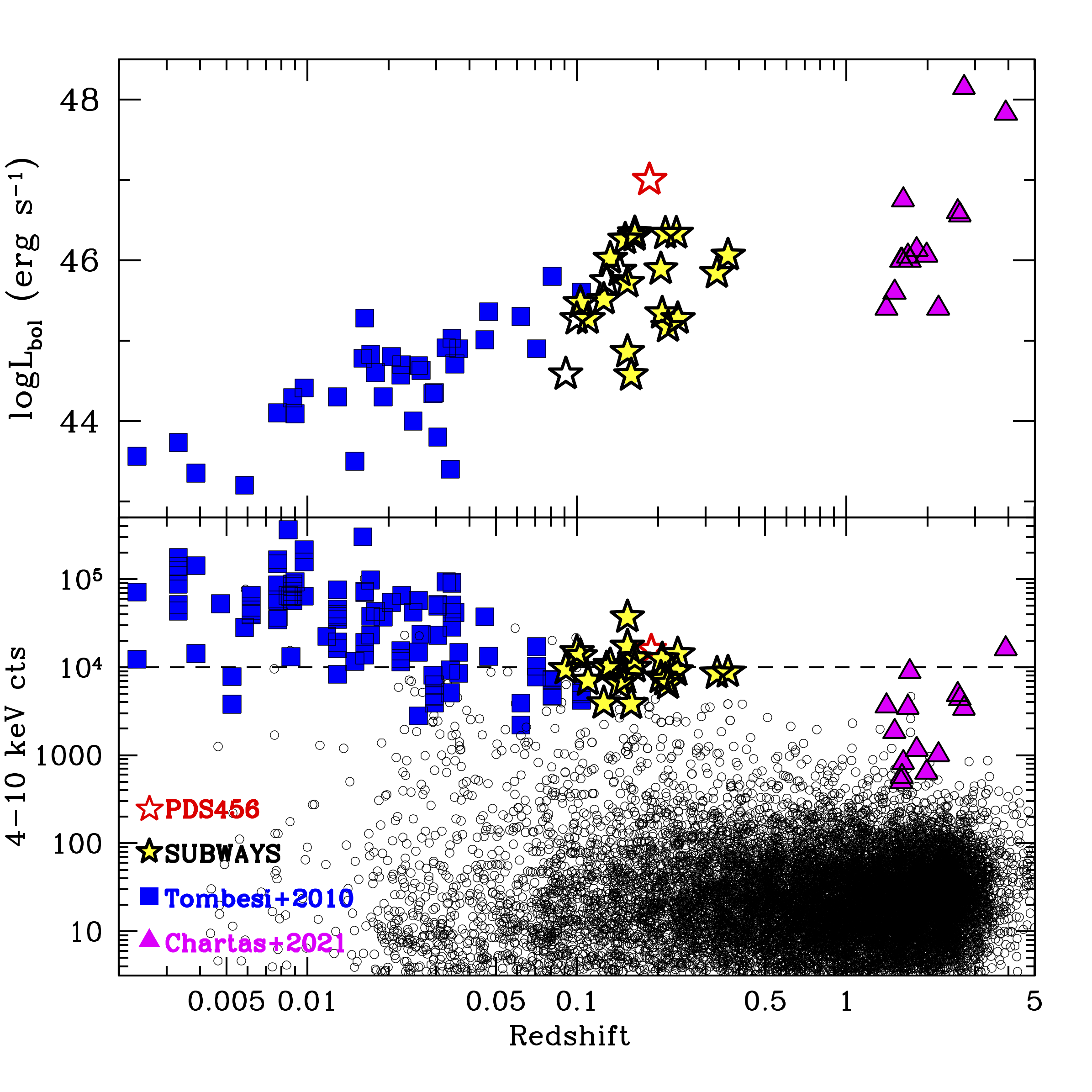}
\end{center}
\vspace{-0.6cm}
\caption{Luminosity (upper panel) and rest frame 4--10\,keV counts (lower panel) plotted against redshift for the objects in the SUBWAYS sample and the comparison samples \citepalias{Tombesi10,Chartas21}, as labelled. In the lower panel we also mark as small empty circles the sources of the 3XMM sample, used to select the SUBWAYS targets.}
\label{fig:sample}
\end{figure}

So far the characterisation of the fastest components of accretion disc winds has been mainly carried out through studies based on inhomogeneous archival data, restricted to two distinct cosmic epochs and luminosity regimes, merely for practical reasons: 
(i) at $z<0.1$, objects are close enough that it is relatively easy to collect $>5-10\times10^3$ counts in the $4$--$10\kev$ band in large samples ($\sim50$ objects), but mainly limited to Seyfert luminosities ($\lbol \lesssim 10^{45}\ergs$, e.g., \citetalias{Tombesi10,Gofford13,Igo20});
(ii) at $z \gtrsim 1.5$, on small and sparse samples ($<10$ objects) at $\lbol \gtrsim10^{46}\ergs$, 
which are mostly composed of gravitationally lensed objects (e.g., \citealt{Chartas09NewAR,Vignali15,Dadina16}, \citetalias{Chartas21}). 
The distribution of these two samples in the $L$--$z$ plane is shown in the upper panel of \autoref{fig:sample} (blue and magenta points, respectively). 

In order to gain significant advances in our understanding of the physical properties of UFOs in the quasi stellar object (QSO)-like regime, a systematic approach is needed.
The \sub sample consists of a total of 22 radio-quiet X-ray AGN, mostly Type\,1 and QSOs, where 17 sources have been observed with \xmm \citep{Jansen01} between May 2019 and June 2020 (see \autoref{Table:SUBWAYS_TOTAL}) as part of a large program (1.45 Ms, PI: M. Brusa) awarded in 2018 (cycle AO18 LP). In addition, the sample includes the data of 5 sources that meet the $L$--$z$--counts selection criteria (see below) already available in the archive\footnote{In this analysis we discarded all observations with a net count threshold of $\lesssim 1500$ in the $4$--$10\kev$ band.}. A companion \sub Paper II \citep{mehdipourpress} is primarily focused on the UV outflow spectroscopic analysis of Cosmic Origins Spectrograph \citep[COS,][]{Green12} data as part of a large complementary \sub observational campaign carried out with the Hubble Space Telescope (HST).  

The \sub selection criteria are based on the following requirements: 

\begin{enumerate}
\item presence of the source in the 3XMM-DR7 catalog\footnote{http://xmmssc.irap.omp.eu/Catalogue/3XMM-DR7/3XMM\_DR7.html}, matched to the SDSS-DR14 catalog\footnote{http://www.sdss.org/dr14/}, or to the Palomar-Green Bright QSO catalog \citep[PG QSO;][]{SchmidtGreen83};  

\item intermediate redshift in the range, $z=0.1$--$0.4$. This condition ensures that both WAs and UFOs can be studied at the same time, and provides the possibility to characterise the continuum up to $10\kev$. Indeed, in order to recognise faint absorption features, it is key to achieve a good handling of the continuum in spectra with high counting statistics up to $10\kev$;

\item count rate larger than $\sim$0.12\,cts/s, in order to ensure counts of the order of $\sim10^{4}$ in the $4$--$10\kev$ band in the EPIC-pn spectra, obtained within a single \xmm orbit. A by-product of this requirement also implies that our targets are QSOs ($\lbol\gtrsim10^{45}\ergs$; star points in \autoref{fig:sample}), complementing the data already available in the archives for this kind of studies; 

\item we discarded Narrow Line Seyfert\,1 (NLSy1s) due to the highly variable EPIC count rate, and QSOs in clusters/radio loud systems, in order to avoid contamination by processes other than AGN accretion and UFOs.

\end{enumerate}
The lower panel of \autoref{fig:sample} shows the currently available rest-frame $4$--$10\kev$ counts for the \sub sample (large stars), compared to those in 3XMM-SDSS, 3XMM-PGQSO and local and high-$z$ QSOs UFO samples (see labels and caption for details). In this paper, we will focus specifically on the detection and characterization of blueshifted absorption profiles in the \fe band in the 17 newly observed sources plus the additional 5 from previous observations (up to AO18 cycle), for a total of 22 targets. The properties of the targets are listed in \autoref{Table:SUBWAYS_TOTAL}.

\begin{figure}
\begin{center}
\includegraphics[height=7.3cm,width=7.3cm]{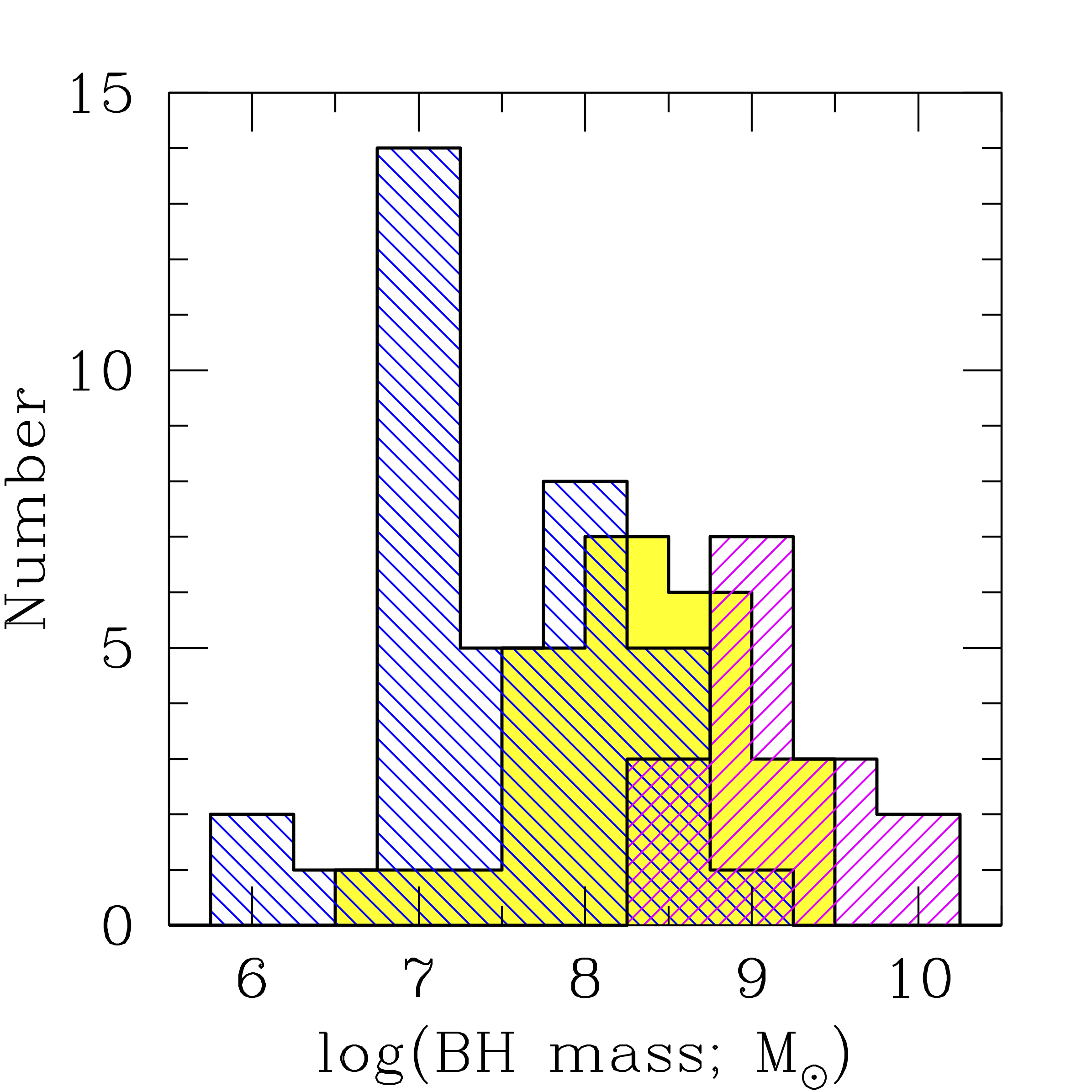}
\includegraphics[height=7.3cm,width=7.3cm]{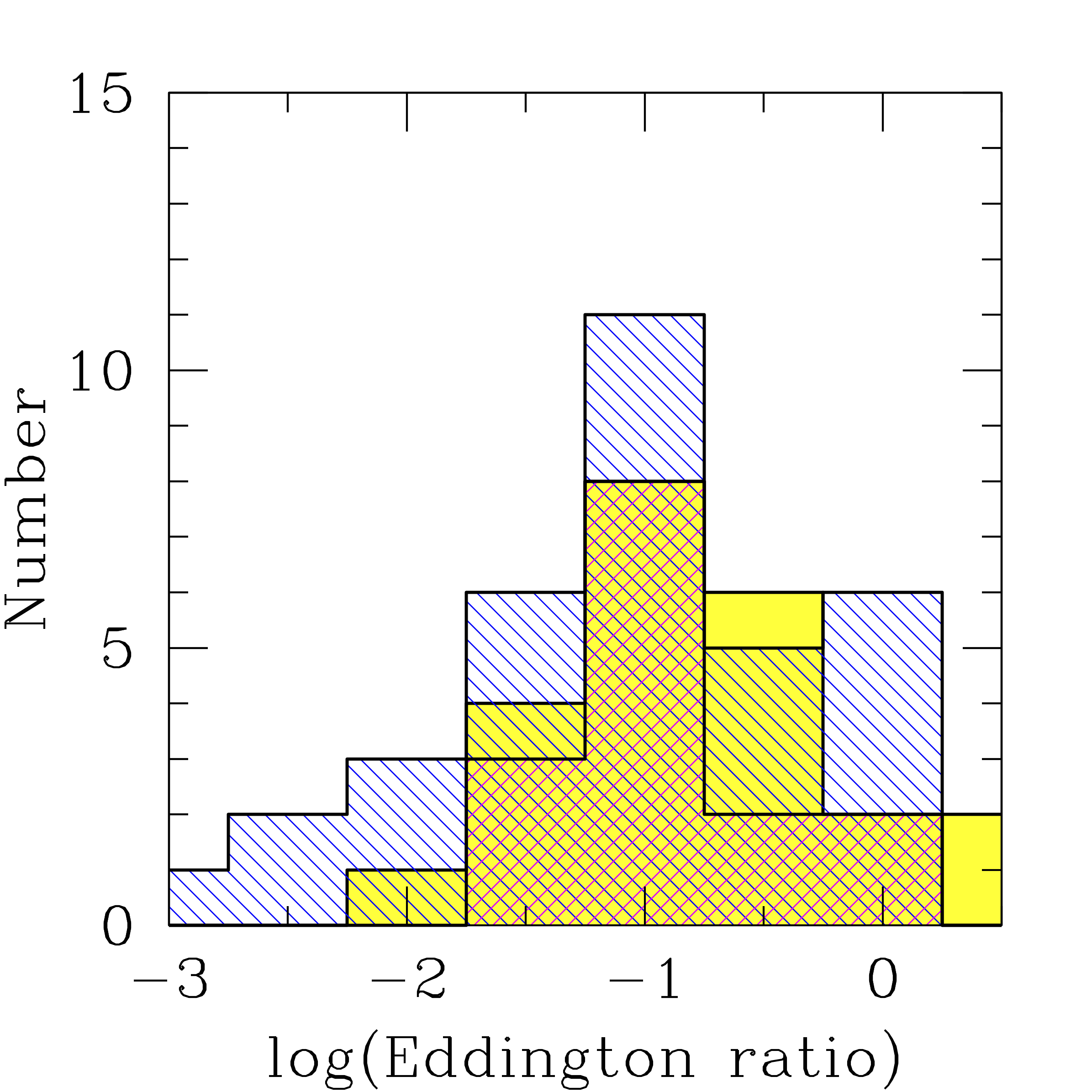}
\end{center}
\caption{The BH mass and Eddington ratio distributions of our targets (see \autoref{Table:SUBWAYS_TOTAL}) and the comparison samples. The color scheme for the samples is the same as in \autoref{fig:sample}. 
}
\label{fig:bh_histo}
\end{figure}

\begin{table*}
\footnotesize
\caption{Target properties of the large \sub campaign. 
Columns:
(1) Target name; (2) Redshift; (3) Galactic absorption measured by \citet{HI4PI16} or \citet{Murphy96NHGAL}$^\dagger$; (4) $2$--$10\kev$ intrinsic luminosity measured in this work by using the \texttt{clumin} convolution model in \xspec; (5) bolometric luminosity derived from the X-ray luminosity assuming the bolometric correction of \citet{Duras20}; (6) log of black hole mass from the measurements carried out in \citet{Bianchi09caixa1}$^a$, \citet{Xie17}$^b$, \citet{Perna17a}$^c$, \citet{Kaspi00revmap}$^d$ and  this work$^e$; (7) Eddington ratio; (8) $2$--$10\kev$ fluxes from the available spectra measured with \texttt{cflux} in \xspec. The $L_{\rm X}$ and $\lbol$ measured in this work are consistent with the ones that were used during the \sub selection process.}

\begin{center}
\begin{tabular}{c c c c c c c c}
\hline
Name\,(1) & $z$\,(2)  & $\nhgal$\,(3) & $\log L_{\rm X}$\,(4) &  $\log \lbol$\,(5) & $\log \mbh\,(6) $ & $\log \leddratio$\,(7) &  $\log F_{2-10\kev}$\,(8)  \\

\\

     &      & $10^{20}\cmsq$         & $\ergs$      & $\ergs$  & M$_\odot$    &         & $\flux$\\

\hline


\\
    PG0052+251              &$0.15445$  & $3.96$   &$44.609\pm0.003$   &$46.00$  &$8.41^a$              &$-0.51$  &$-11.204\pm0.003$ \\ 

    PG0953+414              &$0.23410$  & $1.09$   &$44.595\pm0.004$   &$46.08$  &$8.24^a$              &$-0.26$  &$-11.664\pm0.004$ \\  

    PG1626+554              &$0.13170$  &  $1.55^{\dagger}$  
                                                   &$44.076\pm0.004$   &$45.42$  &$8.54^a$      &$-1.22$ &$-11.600\pm0.004$ \\

    PG1202+281              &$0.16501$  &  $1.74$  &$44.400\pm0.004$           &$45.77$  &$8.61^a$      &$-0.94$  &$-11.469\pm0.004$ \\ 

    PG1435$-$067            &$0.12900$  &  $4.84$  &$43.684\pm0.007$           &$44.98$  &$7.77^b$      &$-0.89$  &$-11.962\pm0.007$ \\  
                                                                                                
    SDSSJ144414+0633       &$0.20768$  &  $2.57$   &$44.467\pm0.004$           &$45.84$  &$8.10^c$      &$-0.36$  &$-11.634\pm0.004$ \\ 

    2MASXJ165315+2349      &$0.10300$  &  $4.15$   &$43.790\pm0.006$           &$45.37$  &$6.98^e$        &$0.29$   &$-11.529\pm0.007$ \\ 

    PG1216+069             &$0.33130$  &  $1.51$   &$44.769\pm0.004$           &$46.24$  &$9.20^a$      &$-1.06$  &$-11.810\pm0.005$ \\

    PG0947+396\,(Obs1)     &$0.20553$  &  $1.91^{\dagger}$  
                                                   &$44.208\pm0.007$           &$45.55$  &$8.68^a$      &$-1.24$  &$-11.887\pm0.008$ \\ 

    PG0947+396\,(Obs2)     &$0.20553$  &  $1.91^{\dagger}$  
                                                   &$44.146\pm0.006$           &$45.55$  &$8.68^a$      &$-1.24$  &$-11.954\pm0.007$ \\ 

    WISEJ053756$-$0245    &$0.11000$  &  $15.0$    &$43.688\pm0.005$           &$44.86$  &$7.73^e$         &$-0.97$  &$-11.777\pm0.006$ \\ 

    HB891529+050          &$0.21817$  &  $3.93$    &$44.219\pm0.005$           &$45.52$  &$8.75^c$       &$-1.33$  &$-11.918\pm0.005$ \\

    PG1307+085            &$0.15384$  &  $2.10$    &$44.312\pm0.003$           &$45.70$  &$7.90^a$       &$-0.30$  &$-11.512\pm0.003$ \\ 

    PG1425+267            &$0.36361$  &  $1.57$    &$44.823\pm0.004$           &$46.31$  &$9.22^c$       &$-1.01$  &$-11.823\pm0.005$ \\

    PG1352+183            &$0.15147$  &  $1.55$    &$43.889\pm0.004$           &$45.21$  &$8.42^a$       &$-1.31$  &$-11.921\pm0.005$ \\ 

    2MASXJ105144+3539$^e$     &$0.15900$  &  $2.20$    &$43.692\pm0.008$           &$44.88$  &$8.40^c$       &$-1.62$  &$-12.068\pm0.009$ \\ 

    2MASXJ0220$-$0728     &$0.21343$  &  $2.42$    &$44.213\pm0.005$           &$45.51$  &$8.42^e$         &$-1.01$  &$-11.888\pm0.006$ \\ 

    LBQS1338$-$0038       &$0.23745$  &  $1.68$    &$44.520\pm0.003$           &$45.91$  &$7.77^c$       &$0.04$   &$-11.707\pm0.004$ \\

\hline

\multicolumn{8}{c}{Archival targets}\\

    PG0804$+$761$^d$     &$0.10000$  &  $7.09$  &$44.330\pm0.002$  &$45.71$ &$8.31^{d}$     &$-0.33$   &$-11.087\pm0.003$ \\ 

    PG1416$-$129         &$0.12900$  &  $3.34$  &$44.170\pm0.005$  &$45.58$ &$9.05^a$       &$-1.57$   &$-11.447\pm0.005$ \\

    PG1402+261           &$0.16400$  &  $1.22$  &$44.030\pm0.004$  &$45.39$ &$7.94^a$       &$-0.65$   &$-11.874\pm0.004$ \\ 

    HB89\,1257$+$286     &$0.09100$  &  $1.04$  &$43.551\pm0.002$  &$45.15$ &$7.46^c$       &$-0.41$   &$-11.831\pm0.002$ \\ 

    PG1114+445           &$0.14400$  &  $1.93^{\dagger}$  
                                                &$44.088\pm0.003$  &$45.53$ &$8.59^a$       &$-1.16$   &$-11.602\pm0.003$ \\ 

\hline
\end{tabular}
\end{center}
\smallskip
\label{Table:SUBWAYS_TOTAL}
 \end{table*}

\section{Data reduction}
\label{sec:Data reduction}
In this work we focus on the  EPIC-pn \citep{Struder01}, EPIC-MOS\,1 and MOS\,2 \citep{Turner2001} data. They were processed and cleaned by adopting the Science Analysis System \sas\,v18 \citep{Gabriel04} and the up-to-date calibration files. We initially checked for Cu instrumental emission in the EPIC-pn CCDs, between $7$--$8.5\kev$ and $7.8$--$8.2\kev$, for the source extraction and subsequent high-background screening. We followed the \citet{Piconcelli04} optimised procedure aimed at maximizing the \sn in the $4$--$10\kev$ band (in the EPIC-pn), rather than using the conservative criterion based on the fiducial rejection of time-intervals of high-background count rates (i.e., between $10$--$12\kev$). The \sn optimization procedure is necessary to identify any absorption feature that would otherwise be diluted \citep[e.g.,][]{Nardini19}, but insufficient if this does not also correspond to the optimal compromise between \sn and number of counts.

Given the relatively small EPIC-MOS collecting area at $E_{\rm obs}\geq4\kev$, a $4$--$10\kev$ band optimization would remove too many counts; we then optimised the filtering on the entire $0.3$--$10\kev$ band. Apart from the different reference bands, the applied method is the same for the 
pn and MOS instruments. We selected a background region free of instrumental features. These regions have dimensions of $40$ or $50$ arcsec depending on the possibility, for each observation, to find source-free regions on the detectors. In order to define the source regions different extraction radii were tested and for each radius we calculated the maximum level of background 
that can be tolerated in order to find the optimal \sn. Following \citet{Piconcelli04}, we define this level of background as Max\,Background (see Appendix\,\ref{appsec:SNR Optimization} for more details). The EPIC source spectra were individually inspected for the possible presence of photon pile-up by using the \sas task \textsc{epatplot}. The ratios of singles to double pixel events were found to be within 1\% of the expected nominal values, and thus no significant pile-up is present. The response files were subsequently generated with the \sas tasks \textsc{rmfgen} and \textsc{arfgen} with the calibration EPIC files version v3.12. In \autoref{Table:Summary_SUBWAYS} we show a summary of the individual observations of the 22 \sub targets that were selected adopting a threshold of $\geq1500$ EPIC-pn net counts in the 4--10\,keV band.


\section{Spectral Analysis}
\label{sec:spectral analysis}

The pioneering UFO studies were conducted on large archival samples of AGN. More specifically, \citetalias{Tombesi10} carried out a systematic hard-band (i.e., $3.5$--$10.5\kev$) analysis on a sample of 42 sources (for a total of 101 observations), drawn from the archival \xmm EPIC data, to carry out a blind search of \fexxvabs and \fexxviabs absorption lines. By analyzing the data of 51 AGN, obtained with the \suzaku observatory, \citetalias{Gofford13} constructed broadband spectral models over the entire band-pass, i.e. $0.6$--$10\kev$. 

For a robust analysis, we choose, as per \citetalias{Gofford13}, the entire EPIC band-pass, where additional spectral complexities like warm absorbers and/or strong soft excesses can also be taken into account. In this way we ensure that all our models accurately describe the overall continuum. We focus on the \xmm EPIC-pn, MOS\,1 and MOS\,2 data in the $0.3$--$10\kev$ range. 
We applied a blind-search procedure in each of the 41 observations by adopting four spectral binning methods (for a total of 164 blind-searches; see Appendix\,\ref{subapp:Spectral binning} for details). 
These binnings are: \texttt{grpmin1}, \texttt{SN5}, \texttt{OS3grp20}, by using the \sas routine \textsc{specgroup}, and the optimal binning of \citet[][\kb hereafter]{KaastraBleeker16}, by using the \heasoft routine \ftgroup.

We find that the spectral resolution delivered by the \texttt{SN5} and \texttt{OS3grp20} binning methods is too degraded compared to the \texttt{grpmin1} and \kb ones, and therefore not suitable for the detection of faint, narrow absorption and emission features. The \texttt{grpmin1} and \kb criteria produce nearly identical results, in terms of detection rate and statistical significance of the features. While \texttt{grpmin1} would certainly be a more conservative option, we finally chose the \kb binning, which is specifically developed to optimise the \sn in narrow, unresolved spectral features whilst maintaining the necessary spectral resolution. Such a choice provided the right compromise between these binning methods. In other words, \citet{KaastraBleeker16} worked out a binning scheme based on the resolution of the detector and the available number of photons at the energy of interest. Such a method allowed us to maximise the information provided by the \fe absorption lines, and in this framework we adopted a maximum likelihood statistic as $\cstat$.


\subsection{Continuum Modelling}
\label{sub:Continuum Modelling}
All the spectra were initially fitted with a power law and their corresponding Galactic absorption, modelled with \texttt{Tbabs} \citep{Wilms00}, with column densities obtained from the \citet{HI4PI16} survey. In order to accurately parameterise the properties of the underlying continuum, additional model components were required in the process such as warm or neutral absorption, or a soft excess, as outlined in Sections\,\ref{subsub:warm absorption} and \ref{subsub:soft excess}. 

We note that 
statistically identical results could have been achieved by modelling the continuum with distant and/or ionised Compton reflection models such as \xillver \citep{Garcia13} or \relxill \citep{Dauser14,Garcia14}. Since the \xmm bandpass is $0.3$--$10\kev$, the contribution from the Compton reflection continuum is not well constrained, therefore, for simplicity, here we adopt a power-law plus blackbody (when required) parameterization of the continuum. A similar approach was also adopted in the \textit{CAIXA} sample by \citet{Bianchi09caixa1}. Nonetheless, a thorough investigation using Compton and relativistic reflection models is addressed in a forthcoming companion paper, where we take advantage of the \nustar follow-up obtained in 2020 (PI: Bianchi). 

Our baseline phenomenological model can be described as:
\begin{equation}
  \texttt{Tbabs}\times\texttt{XABS}\times(\texttt{zbbody}_{1}+\texttt{zbbody}_{2}+\texttt{powerlaw}),
\label{eq:baseline}
\end{equation}
\noindent where \texttt{Tbabs} represents the absorption due to our Galaxy, \texttt{powerlaw} 
accounts for the primary emission component, and \texttt{zbbody}$_{1,2}$ are two layers of blackbody emission to account for the soft X-ray excess\footnote{In some \sub targets only one blackbody component is required (see Section\,\ref{subsub:soft excess}), while no blackbody component is needed for the absorbed sources 2MASS\,J105144$+$3539 and \tmosf.}. This parameterization of the soft excess is only phenomenological and hence the corresponding temperatures are not meaningful. \texttt{XABS} (when required) corresponds to a mildly ionised warm absorption component (see Section\,\ref{subsub:warm absorption}).

Once the best-fit of the $0.3$--$10\kev$ continuum spectra of each of the $41$ observations was reached, we performed a systematic search for \iron emission and absorption profiles between $4$--$10\kev$ through the following two methods: (i) blind-line search via energy--intensity plane contours plots (Section\,\ref{subsec:Fe searches}) and (ii) extensive Monte Carlo (e.g., \citealt[][]{Protassov02}; $\mathcal{MC}$ hereafter) simulations (Section\,\ref{sub:Monte Carlo simulations}). The modelling approach of each individual spectral component is described below and the detailed continuum and absorption parameters are tabulated in \autoref{Table:basecontSUBWAYS}

\subsubsection{Intrinsic Absorption}
\label{subsub:warm absorption}

Neutral or lowly ionised absorption is typically constituted by a distant ($\gtrsim$ pc scales), less ionised and denser circumnuclear material compared to UFOs, generally outflowing at velocities 
in the range of $\sim -100$ to $-1000\kms$ \citep[e.g.,][]{Kaastra00,Kaspi00,Blustin05}. These absorbers are detected in the soft X-ray part of the spectrum at energies $\lesssim2$--$3\kev$ and, depending on their properties, they can add a significant curvature to the spectra below $10\kev$ \citep[e.g.,][]{Matzeu16,Boller21}, and hence affect the overall continuum and line parameters in our broadband models. In the literature, the fraction of sources with reported warm absorbers is $>60\%$ (\citealt{Tombesi13}, \citetalias{Gofford13}).

In 2MASS\,J105144$+$3539, an intrinsic neutral absorption column of $\nh=6.7\pm0.4\times10^{21}\cmsq$ and two emission lines in the soft band, likely associated to collisionally ionised gas, are required. \tmosf is classified as a Seyfert\,2 and so requires a different model construction in the soft X-rays, with the presence of an intrinsic neutral absorber and emission arising from a distant scattered component. The continuum model differs from \autoref{eq:baseline} as:

\begin{equation}
    \texttt{Tbabs}\times(\texttt{powerlaw}_{\rm scatt}+\texttt{apec}_{1}+\texttt{apec}_{2}+\texttt{phabs}\times\texttt{powerlaw}_{\rm intr}),
\label{eq:baseline_absorbed}
\end{equation}
\noindent where \texttt{apec}$_{1,2}$ \citep{Smith01apec} are  thermal models accounting for two regions of emitting collisionally ionised plasma and \texttt{powerlaw}$_{\rm scatt}$ reproduces the distant scattered component. \texttt{powerlaw}$_{\rm intr}$ accounts for the primary continuum, which is absorbed by fully covering neutral material (\texttt{phabs}) with a column density of $\nh=2.2 \pm 0.9\times10^{23}\cmsq$. 

In this work we model the presence of fully covering mildly ionised absorption with a specifically generated \xabs model from \spex \citep{Kaastra96spex,Steenbrugge03xabs} converted to an \xspec \citep{Arnaud96} table\footnote{\url{https://www.michaelparker.space/xspec-models}} \citep[see appendix of][]{Parker19xba}. More specifically the \xabs table covers the following ranges in the parameter space: column density in the range $10^{20} \le \nh/\cmsq \le 10^{22}$ with $10$ logarithmic steps, and ionisation in the range $-4 \le  \logxi \le 6$ with $20$ linear steps, making this table well suited for investigating a wide range of absorbers in AGN such as warm absorbers. Although the presence of warm absorbers could have a minimal effect on the \fe region, we find it essential to include them as the most reliable continuum level must be determined. 

The \xabs table was generated by assuming the default \spex setting where the spectral energy distribution (SED) of NGC\,5548 was used as representative of a standard AGN input spectrum\footnote{We are aware that 
our \xabs table is built based on the ion balance calculated for a typical Seyfert, while our sample consists of QSOs of higher luminosity. By testing the same data with an \xstar grid with turbulent velocity of $\vturb=300\kms$ and a power-law SED input with a 
photon index of $\Gamma=2$, we get slightly higher ionisation parameters but consistent within the errors. For this reason, we allowed $\xi$ to explore the wide range of ionisation reported above.} \citep[see][for more details]{Steenbrugge05sed}. Another parameter in the model is the 2D root mean square velocity ($v_{\rm rms}$), which gives a measure of the velocity dispersion of the line profile.\footnote{\url{https://personal.sron.nl/~jellep/spex/manual.pdf}} At the energy resolution of the EPIC CCDs (see Appendix\,\ref{subapp:Spectral binning}), the individual soft X-ray absorption lines are indeed unresolved, so in the fitting procedure the RMS velocity broadening was freezed to its default value, i.e., $v_{\rm rms}=100\kms$ (unless specified otherwise) and the systemic velocities are set to $0\kms$. 

Here we also consider the possibility of partial covering along the \los. In this regime a fraction corresponding to $\fcov<1$ of the total flux is indeed absorbed, while a portion $1-\fcov$ leaks through the absorbing layer. This can also have a dramatic effect on the emerging continuum by imprinting a prominent spectral curvature at energies $<10\kev$ \citep[e.g.,][]{Matzeu16,Boller21}. On this basis a partial covering fraction was also added to the list of free parameters in our \xabs table, i.e. $0 \le \fcov \le 1$. A more comprehensive physical analysis of low- and high-ionisation outflows is presented in a companion paper.

\subsubsection{The Soft Excess}
\label{subsub:soft excess}

The soft X-ray excess is described as a strong featureless emission component that is often observed in unabsorbed AGN below $\sim2\kev$. The physical mechanism responsible for this emission is still the subject of active debates, i.e. a dual-coronal system \citep[e.g.,][]{Done12,Petrucci13,Rosanska15,Middei18,Petrucci18,Ursini20,Ballantyne20,Porquet21} or relativistic blurred reflection \citep[e.g.,][]{Ross05,Nardini11,Wilkins12,Walton13,Garcia19,Jiang19,Xu21,Mallick22}  

Regardless of the physical origin of the soft excess, in this work we take a completely empirical approach by fitting its profile with one/two layers of blackbody emission,
when required at the $\dC\gtrsim9.3/2$ threshold (i.e., $\gtrsim99\%$ confidence level for each blackbody component) in \xspec (e.g., \citealt[][]{Porquet04,Piconcelli05,Bianchi09caixa1}, \citetalias{Gofford13}). Although our phenomenological model largely ignores the detailed physics involved in the system, it allows us to fit and compare uniformly the underlying continua in our sample so that we can concentrate on the \fe band. There might be some degeneracies between the soft excess and partial covering components during fitting, however this is not an issue for our absorption line detections. We could have modelled equally well the soft excess with a reflection component and the final result would not change, as discussed above in Section\,\ref{sub:Continuum Modelling}.

\begin{figure}
\includegraphics[width=\linewidth]{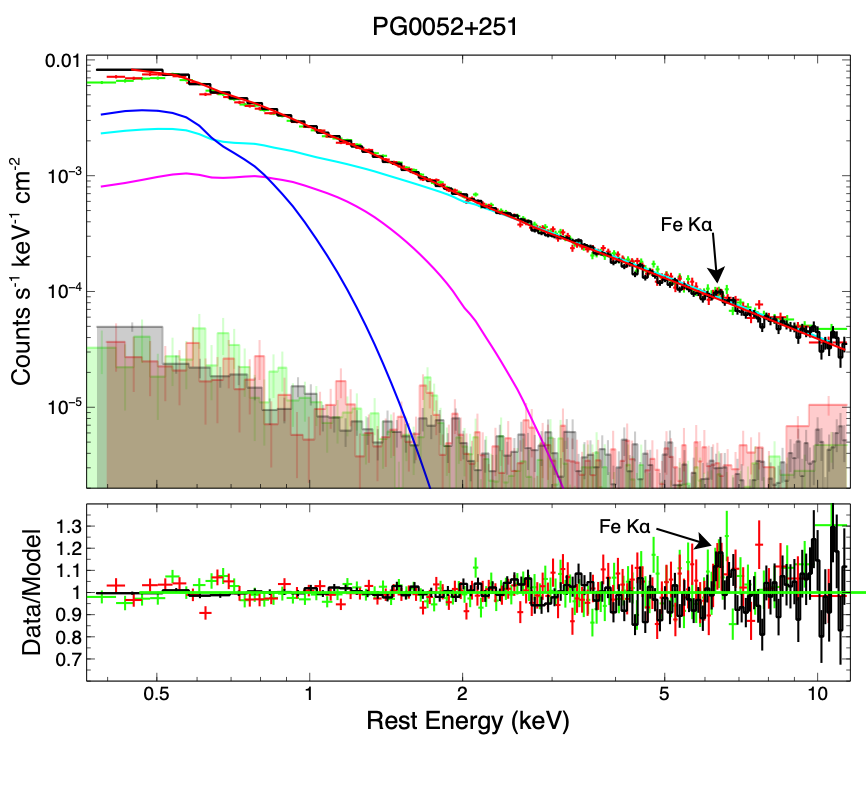}
\caption{Broadband EPIC data and best-fit continuum of PG\,0052$+$251 between the rest-frame energy of $0.3$--$10\kev$. Top: background subtracted \xmm spectra (EPIC-pn in black, EPIC-MOS\,1 in red, EPIC-MOS\,2 in green) and their corresponding background levels (shaded areas). The broadband best-fit continuum model (solid red) is shown alongside the individual model components: absorbed power-law (cyan), low- (blue) and high-temperature blackbody (magenta). Bottom: corresponding residuals of the data compared to the best-fitting model.}
\label{fig:SUB_LDATA}
\end{figure}

\subsection{Search for Fe\,K emission and absorption}
\label{subsec:Fe searches}

In \autoref{fig:SUB_LDATA}\,(top) we report, as an example, the background subtracted \xmm spectra (EPIC-pn, black; EPIC-MOS\,1, red; EPIC-MOS\,2, green) and their corresponding background spectra of PG\,0052$+$251. This source is one of the most luminous in the sample and is on the bright side of the luminosity/counts distribution in Type\,1 AGN (see \autoref{figapp:SUB_LDATA}). 

The $0.3$--$10\kev$ (rest-frame) best-fitting broadband continuum model (excluding \iron emission and/or absorption lines) is indicated in solid-red. The residuals are shown in the bottom panel. Our baseline continuum models include the absorbed power-law (cyan) and the two blackbody components, with high and low temperature ($kT$), respectively in magenta and blue. The continua are well reproduced with our baseline model (see the model statistics reported in \autoref{Table:basecontSUBWAYS}). Few exceptions were found in the sample, when the soft X-ray emission was parameterised with one or two regions of collisionally ionised plasma modelled with \texttt{apec} in 2MASS\,J105144$+$3539 and \tmosf, respectively. The presence of strong residuals from neutral \feka core emission at $E_{\rm rest}=6.4\kev$ from distant material is ubiquitous in the \sub sample (e.g., see \autoref{fig:SUB_LDATA}). 

In some observations (see \autoref{figapp:SUB_LDATA}) we observed strong absorption residuals (also in the \fe band) likely associated with \fexxv/\fexxvi transitions. Therefore we ran a blind search, simultaneously in both EPIC-pn and EPIC-MOS spectra for every observation in the sample, in order to have a first assessment of energy, strength, shape and significance of any absorption or emission line relative to the underlying continuum model. We performed an inspection of the deviation in $\vert \Delta \mathcal{C} \vert $ from the best-fitting continuum model by generating two-dimensional energy--intensity contours plots. This method was adopted by \citetalias{Tombesi10} and \citetalias{Gofford13}, and extra details are described in \citet{Miniutti06}.


The search was performed with our baseline continuum model (\autoref{eq:baseline} or \autoref{eq:baseline_absorbed}) plus a narrow Gaussian line (with the velocity width fixed at $\sigma_{\rm width}=10\ev$). We also let the power-law photon index and normalization 
free to vary during the search. In adopting our broadband `multi-component' continuum model above, we ensure a better reproduction accuracy of the continuum level compared to a simpler two-component power law plus Gaussian line model restricted on the \fe band. For this routine we freeze all the soft X-rays parameters from the broadband continuum to their best-fit values, re-fit, and run the scan along with the $5$--$10\kev$, rest frame, energy band.

The blind search method adopted here is carried out based on the following steps:

\begin{enumerate}

    \item[(i)] We have a baseline continuum model between $0.3$--$10\kev$ (described above) plus the unresolved Gaussian line required for the scan. The \feka emission line at $6.4\kev$ was not included in the baseline model. For each run, the energy of the Gaussian is scanning the three EPIC spectra simultaneously between $5$--$10\kev$ in intervals of $\Delta E=50\ev$. The normalization of the Gaussian component probes the intensity of the spectral line and is free to vary between negative and positive values in $250$ steps.

    \item[(ii)] Each individual step in the energy--intensity plane with respect to the baseline model was recorded into a file including the corresponding $\Delta\mathcal{C}$.

    \item[(iii)] The resulting confidence contours are plotted according to a mapped $\dc$ deviation of $-2.3$, $-4.61$, $-9.21$, $-13.82$ and $-18.52$ for 2 parameters of interest corresponding to the nominal $68\%$, $90\%$, $99\%$, $99.9\%$ and $99.99\%$ confidence levels.

    \item[(iv)] We inspect the contour plots to check whether there is any evidence of  emission and/or absorption in the spectrum (see text below).

\end{enumerate}

\begin{figure}
\includegraphics[width=\linewidth]{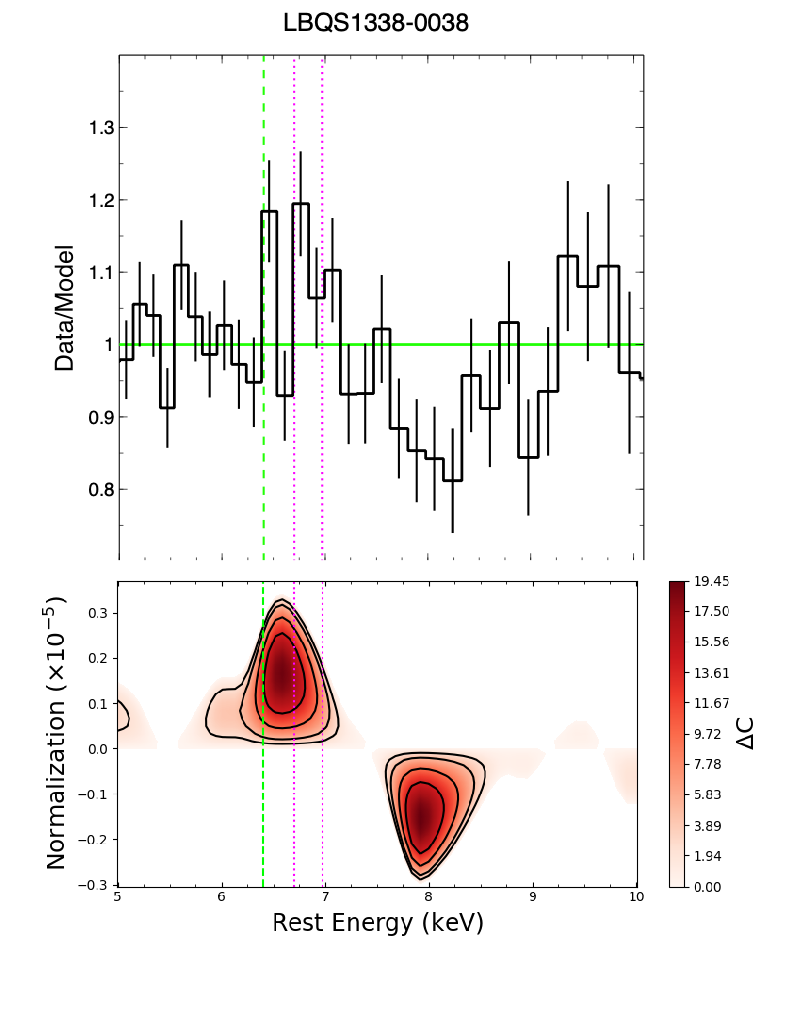}
\caption{Top: data/model ratio plot for the EPIC-pn spectrum (MOS\,1 and MOS\,2 are not included for clarity) showing the residuals in the $5$--$10\kev$ band of to the type\,1 AGN LBQS\,1338$-$0038, showing a broad emission and absorption profiles respectively at $\sim6.6\kev$ and $\sim8\kev$ suggesting a P-Cygni-like profile. The vertical lines indicate the \feka laboratory transition at $E_{\rm lab}=6.40\kev$ (lime green) and \fexxv and \fexxvi at $E_{\rm lab}=6.70\kev$ and $6.97\kev$ respectively (magenta). The outer to inner closed contours correspond to a $\dc$ significance of ($-2.3=68\%$), ($-4.61=90\%$), ($-9.21=99\%$) and ($-13.82=99.9\%$) relative to the best-fitting continuum. Bottom: Corresponding blind-search contours showing the evidence of strong emission profiles at high confidence level (see colour bar on the right).}
\label{fig:2massj165_lbqs}
\end{figure}

If an emission or absorption line is detected, it is parameterised by using a Gaussian profile. All the key Gaussian absorption and emission parameters accounting for the detected lines, using the \kb binning, are tabulated in \autoref{Table:basegaussSUBWAYS}. The $\dc$ mapping provided by the energy--intensity contours is a powerful tool to detect emission or absorption profiles by visually assessing the location and strength of the line relative to the underlying continuum model. The spectral complexity within the $5$--$10\kev$ band can be enhanced by a number of atomic features such as ionised emission lines from \fexxv and \fexxvi, at $6.7\kev$ and $6.97\kev$ respectively. As we have learned from previous work, in ultra-fast wind systems the ionised emission due to scattered photons from the outflowing material can be as important as the absorption \citep[e.g.,][]{Sim08,Nardini15,Luminari18,Matzeu22}. In several \sub spectra, the shape of the emission and/or absorption profiles is indeed complex/broad, which suggests a superposition/blending of multiple ionised lines.
Steps\,(i)--(iv) were carried out in each fitted EPIC spectrum of the sample.

Examples of this method are shown in \autoref{fig:2massj165_lbqs}. In the top panel we show the corresponding residual of EPIC-pn spectrum (MOS\,1 and MOS\,2 are not included for clarity) without the emission and absorption components. The vertical dashed lines denote the position of the laboratory energy transition of \feka\,(lime green), \fexxv\,(magenta) and \fexxvi\,(magenta).

In \autoref{fig:2massj165_lbqs} we show the result of the same blind-search procedure applied to the type\,1 AGN LBQS\,1338$-$0038. The energy--intensity contour map is showing the presence of highly significant emission and absorption profiles at centroid rest-frame energies of $E_{\rm em}\sim6.6\kev$ and $E_{\rm abs}\sim8\kev$, respectively. What differs from \tmosf is that both the emission and absorption have comparable width and the emission (as well as the absorption) seems to be originating from ionised material. Such features are highly reminiscent of the well-established P-Cygni-like profile detected in \pds \citep{Nardini15}, where the emission component arises from photons scattered back into our \los from the same outflowing ionised outflow, averaged over all the viewing angles. The complete set of blind-search line contours of all the remaining observations in the sample are plotted in Appendix\,\ref{appsec: blind line search}, and  \autoref{figapp:sub_ALLOBS_SCAN_RA}. The visual inspection of the residuals makes the \fe emission/absorption profiles detections largely qualitative at this stage. Nonetheless, we have now a strong basis to carry out a systematic identification of \iron absorption features that might be arising from an ultra-fast outflow.

\subsection{Modelling the Fe\,K band}
\label{sub:modelling the fek band}

In this paper, we only adopt phenomenological models that are homogeneously fitted across the whole sample. All the \fe emission and absorption residuals arising from the line blind-search (see \autoref{figapp:sub_ALLOBS_SCAN_RA}) are fitted with a cosmologically redshifted Gaussian model (\texttt{zgauss} in \xspec) added to equations\,\ref{eq:baseline} and/or \ref{eq:baseline_absorbed}. 
Modelling this way the \fe features allows us to characterise the significance and intrinsic properties of the lines, i.e. centroid rest energy, line-width and the overall strength with respect to the underlying continuum. In some cases, when the feature is broader, we fix their line-widths (for simplicity) between $\sigma_{\rm width}=10\ev$--$400\ev$ (depending on which produces a larger improvement to the fit). When the lines are broad/resolved, the line-width is left as a free parameter (see \autoref{Table:basegaussSUBWAYS}). A strong emission profile corresponding to the neutral \feka core at $\sim6.4\kev$ is present in almost every target. In some observations, of PG\,0953$+$414, PG\,1425$+$267, PG\,1626$+$554, PG\,1352$+$183, PG\,1216$+$069, PG\,0947$+$396\textunderscore\,obs1, 2MASS\,J105144$+$3539 and PG\,0804$+$761, the \fe emission lines are more complex (see \autoref{figapp:sub_ALLOBS_SCAN_RA}). Their \fe emissions are broader and the centroid energies correspond to highly ionised iron emission in the \fexxv and \fexxvi domain. In the context of UFOs, broader \iron emission can arise from scattered photons from outflowing ionised material and can have crucial implications in determining the covering fraction of the outflowing gas \citep{Sim08,Sim10,Nardini15,Matzeu16,ReevesBraito19}.

For the absorption features we estimated the outflow velocity by assuming that \iron absorption observed at energy $E_{\rm abs}$ is associated with H-like iron (\fexxviabs) gas, with laboratory rest-frame energy of $E_{\rm 0}=6.97\kev$. Assuming $E_{\rm 0}=6.7\kev$ (i.e. \fexxvabs) would correspond in the above calculations to a faster $v_{\rm out}$, given the larger degree of blueshift. Our choice of $E_{\rm 0}$ can be considered as a conservative choice for the outflow velocity estimate, based on the reference energy assumed. For a more appropriate identification of line transitions, i.e. either \fexxvabs, \fexxviabs or \fe-shell edges, we will need to carry out a photoionisation modelling of the \fe features (with \xabs and \xstar models), which can yield an accurate description of the physical conditions of the gas, e.g. its ionisation state. A comprehensive physically motivated analysis of the emission/absorption profiles of the entire sample is the subject of a forthcoming companion paper.

For quantifying the statistical significance of the Gaussian lines in our modelling, we first adopt the $\dc$ improvement as used for our blind-search method. More specifically, we compute the $P_{\rm F}$ significance derived by first obtaining the $\dC$ between the best-fit with and without a specific Gaussian line and subsequently compute the F-test probability. We detected blueshifted \fe absorption lines in 11 out of 22 sources 
at $P_{\rm F}\gtrsim95\%$, where 9/22 have $P_{\rm F}\gtrsim99\%$.

Numerous authors in the literature (e.g., \citealt{Vaughan03,Porquet04monte,Markowitz06}, \citetalias{Tombesi10,Gofford13,Igo20}, \citealt{Parker20}) have established that to obtain an adequate statistical test when determining the significance of a detection of atomic lines in rather complex spectra, an extensive $\mathcal{MC}$ approach is required. In an absorption-line search framework we might detect unexpected lines at a specific energy, without any prior justification, \citep[e.g.,][]{Protassov02} over an arbitrary energy range. Indeed, as discussed in the next section, we find that our $P_{\rm F}$ improvements over-predict the detection probabilities, as opposed to a more robust $\mathcal{MC}$ simulation approach.

\subsection{Monte Carlo approach}
\label{sub:Monte Carlo simulations}

The $\mathcal{MC}$ simulation method has now been extensively adopted in the literature \citep[e.g.][]{Porquet04monte,Miniutti06,Tombesi10,Gofford13,Nardini19,Parker20,Middei20leda} in order to achieve a robust determination of the significance of a spectral line 
independently from the spectral noise and the quality of the detector. The $\mathcal{MC}$ approach overcomes the limitations of the often used F-test, which can sometimes over-predict the statistical significance of the line detection when compared to extensive $\mathcal{MC}$ simulations. In this paper the $\mathcal{MC}$ approach is focused on the \fe absorption lines detected in 11 sources on the basis of $P_{\rm F}\gtrsim95\%$. We report the results on the detection probability based on $\mathcal{MC}$ in \autoref{Table:basegaussSUBWAYS}. This process was carried out by following these steps.

\begin{enumerate}

    \item The continuum baseline null-hypothesis model ($\nhm$ hereafter) is our final best-fitting $0.3$--$10\kev$ model (see \autoref{Table:basecontSUBWAYS} in Appendix\,\ref{appsec:Broadband continuum modelling}) re-adjusted after removing the Gaussian absorption component. For each test, we simulated $1000$ EPIC-pn, $1000$ EPIC-MOS\,1 and $1000$ EPIC-MOS\,2 source and background spectra, by using the \texttt{fakeit} command in \xspec. The simulated spectra were generated with the same exposure times and response files from the original data and grouped accordingly. We adopt the \kb binning (as described in Appendix\,\ref{subapp:Spectral binning}) with \ftgroup.

	\item Our $0.3$--$10\kev$ simulated spectra are then fitted with the $\nhm$, which takes into account the associated uncertainties from the $\nhm$ itself. During this procedure, we fixed the line-width of any broad Gaussian emission features present in the spectra (both in the soft and hard X-ray band) at their corresponding best-fit energy values from the real data and we let their intensities free to vary. The dual blackbody temperature, normalizations and any Galactic, intrinsic neutral/warm absorptions were frozen to their best-fit values of the real data reported in \autoref{Table:basecontSUBWAYS}. 

	\item A narrow Gaussian profile, with line width fixed at zero, was then added to the $\nhm$, with normalization also set to zero, but free to fluctuate between negative and positive values, 
	in order to probe both absorption and emission features. The rest-energy centroid of the Gaussian line was stepped between 5 and 10 kev in $\Delta E=25\ev$ increments with the \textsc{steppar} command in \xspec. Additionally, to prevent a local minimum during fitting we also enable the \textsc{shakefit} procedure developed by Simon Vaughan \citep[see Section 3.2.2 in][]{Hurkett08}. This process maps the $\dc$ variations relative to $\mathcal{C}_{\rm null}$, which are recorded after each step as $|\dc_{\rm noise}|$. The degrees of freedom corresponding to both models are also recorded.

	\item The above steps were repeated through $S=1000$ iterations for each test, which produced a $|\dc_{\rm noise}|$ distribution under the null-hypothesis by mapping the statistical significance of any deviations from $\nhm$ due to random photon noise in the spectra.

    \item The initial significance of the line derived from the real data $|\dc_{\rm line}|$ was compared to the $|\dc_{\rm noise}|$ distribution so that the number $N$ of simulated spectra with a random noise fluctuation larger than the observed one can be evaluated. In case when the $N$ simulated spectra have $|\dc_{\rm noise}| \geq |\dc_{\rm line}|$, the $\mathcal{MC}$ statistical significance ($\pmc$) of the absorption line detection can be calculated as $\pmc=1-\left(\frac{N}{S}\right)$ and reported in \autoref{Table:basegaussSUBWAYS}.

In order to compare with literature results, and following e.g. \citetalias{Tombesi10}, we consider as robust detections only absorption lines with $P_{\rm F}\gtrsim99\%$ and $P_{ \mathcal{MC}}\gtrsim95\%$.
The discussion of the \fe emission/absorption detection rate in our sample is presented in Section\,\ref{sub:line detection rate}. 
\end{enumerate}

\begin{landscape}
\begin{table}
\setlength{\tabcolsep}{0.4pt}
\centering
\caption{\small{Gaussian emission and absorption parameters in the \fe band corresponding to a total of $41$ observations. Notes: (1) Source name; (2) observation ID; (3) measured Gaussian emission line energy in the rest-frame; (4) emission line energy width. The absorption lines were fitted with widths ranging between $\sigma_{\rm width}10\ev$--$400\ev$; (5) emission line intensity in units of $\rm ph\,cm^{-2}\,s^{-1}$; (6) equivalent width; (7) change in $\mathcal{C}$-stat fit statistic, degrees of freedom and the corresponding significance (in per cent) when the emission line is removed from the best fit. (8)--(11) same as (3)--(6) but for the Gaussian absorption line; (12) corresponding outflow velocity inferred from the rest-frame energy of the absorption line with $E_{\rm rest}=6.97\kev$ as a reference energy; (13) same as (7); (14) $\mathcal{MC}$ significance of the absorption line where the $\gtrsim95\%$ values are highlighted in bold. $\dagger$ In this observation the absorption profile is likely arising from a blend of a \fexxvxxvi pair, which could be resolved by binning the spectra with \texttt{grpmin1} (see Appendix\,\ref{subapp:Spectral binning}). 
$^{\ddagger}$ as the two absorption lines are consistent with a \fexxvxxvi pair, we inferred the null probability of both profiles simultaneously as a false detection by multiplying the probabilities of each line.
$\star$ In this P-Cygni-like profile the line-width of the emission is tied to that of the absorption 
as $\sigma_{\rm em}=(\sigma_{\rm abs}/E_{\rm abs})\times E_{\rm em}$. $^f$ denotes a fixed parameter during fitting. 
$^{\blacklozenge}$ A Monte Carlo test was not applied for this feature as it is likely associated with a neutral \fe edge.
}}

\begin{tabular}{l c c c c c c c c c c c c c c c c}

\hline
&              &\multicolumn{6}{c}{\texttt{Emission Lines}} &\multicolumn{7}{c}{\texttt{Absorption Lines}}\\

Source\,(1)&\textit{XMM}\,(2)  &&$E_{\rm em}$\,(3) &$\sigma_{\rm width}$\,(4) &Int\,(5) &$EW$\,(6) &$\dC\,(P_{\rm F})$\,(7) &$E_{\rm abs}$\,(8) &$\sigma_{\rm width}$\,(9) &Int\,(10)   &$EW$\,(11)  &$v_{\rm Gau}$\,(12)  &$\dC\,(P_{\rm F})$\,(13) & $P_{ \mathcal{MC}}$\,(14)\\

   &        ObsID   &&keV          &eV      &$10^{-6}$  &eV   &                   &keV           &eV        &$10^{-6}$  &eV     &$c$            &                  &\\

\hline
\\
PG0052$+$251

&0841480101     &&$6.40_{+0.06}^{-0.06}$ &$10^{f}$  &$3.3_{+1.5}^{-1.5}$ &$33_{+15}^{-15}$   &$16.3/2\,(>99.9\%)$&&&&&&\\

 \\
\multirow{2}{*}{PG0953$+$414} 

&\multirow{2}{*}{0841480201}     
&&$6.42_{-0.06}^{+0.06}$ &$10^{f}$     &$1.9_{-1.0}^{+1.0}$  &$36_{-20}^{+20}$   &$11.95/2\,(99.75\%)$           													  

&$7.82_{-0.10}^{+0.10}$ &$10^{f}$     &$-1.5_{-0.8}^{+0.9}$ &$-43_{-24}^{+24}$  & $-0.116_{-0.012}^{+0.012}$ &$8.56/2\,(98.62\%)$   &$82.1\%$\\
								
&&&$6.92_{-0.08}^{+0.07}$ &$100^{\rm t}$ &$3.2_{-1.0}^{+1.1}$ &$73_{-23}^{+26}$  &$24.87/2\,(>99.99\%)$&&&&&\\

 \\
\multirow{2}{*}{PG1626$+$554}

&\multirow{2}{*}{0841480401} 

&&$6.40_{-0.07}^{+0.08}$ &$100^{f}$ &$2.6_{-0.9}^{+0.9}$  &$63_{-22}^{+22}$   &$21.02/2\,(>99.99\%)$&&&&&&\\

&&&$7.01_{-0.15}^{+0.12}$ &$100^{f}$ &$2.1_{-1.1}^{+1.1}$  &$60_{-36}^{+36}$   &$9.87/2\,(99.28\%)$&&&&&&\\

 \\
\multirow{2}{*}{PG1202$+$281}

&\multirow{2}{*}{0841480501} 
&&$6.45_{-0.05}^{+0.05}$  &$10^{f}$   &$2.1_{-1.0}^{+1.0}$     &$37_{-18}^{+18}$    &$13.7/2\,(99.89\%)$        															     

&$7.78_{-0.06}^{+0.06}$   &$10^{f}$   &$-1.9_{-0.9}^{+0.9}$    &$-45_{-21}^{+21}$   &$-0.113_{-0.012}^{+0.012}$ &$10.3/2\,(99.42\%)$ &$\mathbf{95.0\%}$\\

&&&$7.09_{-0.08}^{+0.09}$ &$10^{f}$   &$1.67_{-1.0}^{+1.0}$     &$34_{+21}^{+21}$    &$8.0/2\,(98.17\%)$ &&&&&     &\\

 \\
PG1435$-$067

&0841480601 
&&$6.37_{-0.06}^{+0.06}$  &$10^{f} $   &$1.8_{-0.74}^{+0.78}$   &$100_{-36}^{+39}$  &$19.0/2\,(>99.99\%)$        															                     
&$8.28_{-0.45}^{+0.70}$   &$400^{f}$   &$-2.2_{-1.4}^{+1.4}$    &$-184_{-117}^{+117}$ &$-0.166_{-0.053}^{+0.079}$ &$9.52/2\,(99.14\%)$  &$90.9\%$\\

SDSS\,J144414$+$0633

&0841480701 
&&$6.43_{-0.03}^{+0.03}$  &$10^{f}$   &$2.8_{-0.8}^{+0.8}$  &$64_{-18}^{+18}$   &$37.48/2\,(>99.99\%)$    &&&&&&\\

\\
\multirow{3}{*}{2MASS\,J165315$+$2349$^{\dag}$}

&\multirow{3}{*}{0841480801} 
&&$6.37_{-0.03}^{+0.03}$  &$10^{f}$  &$3.4_{-0.9}^{+0.9}$  &$73_{-20}^{+20}$   &\multirow{2}{*}{$49.09/2\,(>99.99\%)$}                       										

&$7.42_{-0.06}^{+0.06}$  &$10^{f}$ &$-1.8_{-0.8}^{+0.8}$ &$-48_{-21}^{+21}$ &$-0.082_{-0.012}^{+0.012}$  &$14.39/2\,(>99.9\%)$  &\multirow{2}{*}{$\mathbf{>99.9\%}^{\ddagger}$}\\

&&&&&&&&$7.77_{-0.07}^{+0.06}$  &$10^{f}$ &$-1.8_{-0.8}^{+0.9}$ &$-56_{-25}^{+28}$ &$-0.108_{-0.010}^{+0.009}$  &$14.97/2\,(>99.9\%)$  &\\

&&&$8.87_{-0.17}^{+0.13}$ &$100^{f}$ &$2.5_{-1.3}^{+1.3}$ &$87_{-39}^{+43}$ &$10.48/2\,(99.64\%)$ &&&&&&\\

 \\

PG1216$+$069  

&0841480901  
&&$6.47_{-0.13}^{+0.12}$ &$182_{-129}^{+181}$ &$2.8_{-1.2}^{+1.5}$ &$68_{-29}^{+36}$ &$21.28/3\,(>99.99\%)$ &&&&&\\

\multirow{2}{*}{PG0947$+$396\,(Obs\,1)}

&\multirow{2}{*}{0841481001}  
&&$6.43_{-0.06}^{+0.06}$ &$10^{f}$ &$2.2_{-1.0}^{+1.0}$ &$88_{-31}^{+31}$     &$23.8/2\,(>99.99\%)$  																                                    		&$9.51_{-0.11}^{+0.10}$ &$10^{f}$ &$-2.2_{-0.9}^{+1.0}$ &$-169_{-69}^{+77}$ & $-0.300_{-0.010}^{+0.010}$  &$11.65/2\,(99.7\%)$  &$\mathbf{96.9\%}$\\	 
		          		           
&&&$6.96_{-0.11}^{+0.12}$ &$10^{f} $&$1.4_{-0.9}^{+1.0}$     &$64_{-45}^{+45}$     &$9.47/2\,(99.12\%)$ &&&&&&\\

PG0947$+$396\,(Obs\,2)

&0841482301 
&&$6.35_{-0.06}^{+0.06}$ &$100^{f}$ &$3.0_{-0.9}^{+0.9}$ &$132_{-40}^{+40} $&$35.02/2\,(>99.99\%)$&&&&&&\\

WISE\,J053756$-$0245

&0841481101 
&&$6.40_{-0.06}^{+0.06}$ &$10^{f}$ &$1.5_{-0.6}^{+0.6}$ &$81_{-32}^{+32}$ &$21.0/2\,(>99.99\%)$&&&&&&\\

HB\,891529$+$050 

&0841481301 
&&$6.47_{-0.04}^{+0.04}$ &$10^{f}$ &$2.0_{-0.5}^{+0.5}$ &$103_{-26}^{+26}$   &$38.88/2\,(>99.99\%)$&&&&&&\\

PG1307$+$085

&0841481401 
&&$6.35_{-0.06}^{+0.06}$ &$100^{f}$ &$3.5_{-1.0}^{+1.0}$ &$64_{-18}^{+18}$    &$35.74/2\,(>99.99\%)$&&&&&&\\

\\
\multirow{2}{*}{PG1425$+$267} 

&\multirow{2}{*}{0841481501} 

&&$6.40_{-0.03}^{+0.03}$ &$10^{f}$ &$3.0_{-0.7}^{+0.7}$ &$73_{-17}^{+17}$ &$48.12/2\,(>99.99\%)$               													                               		&\multirow{2}{*}{$8.24_{-0.06}^{+0.06}$} &$10^{f}$ &$-1.2_{-0.6}^{+0.6}$ &$-46_{-24}^{+23}$ & $-0.166_{-0.009}^{+0.009}$  &$10.09/2\,(99.36\%)$  &$93.2\%$\\

&&&$6.88_{-0.06}^{+0.06}$ &$10^{f}$ &$1.4_{-0.7}^{+0.7}$ &$38_{-20}^{+20}$ &$11.91/2\,(98.74\%)$ &&&&&  &\\

\\		

PG1352$+$183

&0841481601 
&&$6.53_{-0.12}^{+0.13}$ &$468_{-113}^{+138}$ &$6.0_{-1.6}^{+1.8}$ &$319_{-85}^{+96}$ &$74.69/3\,(>99.99\%)$&&&&&&\\

2MASS\,J105144$+$3539 

&0841481701 
&&$6.37_{-0.08}^{+0.08}$ &$100^{f}$  &$1.7_{-0.5}^{+0.5}$ &$156_{-56}^{+75}$ &$35.66/3\,(>99.99\%)$  																                    		&$8.51_{-0.07}^{+0.08}$ &$100^{f}$ &$-0.78_{-0.33}^{+0.38}$&$-108_{-46}^{+52}$ & $-0.197_{-0.009}^{+0.009}$ &$10.56/2\,(99.49\%)$&$\mathbf{95.9\%}$\\


2MASS\,J0220$-$0728

&0841481901 
&&$6.39_{-0.06}^{+0.05}$ &$10^{f}$ &$1.3_{-0.5}^{+0.5}$ &$60_{-23}^{+23}$ &$15.58/2\,(>99.9\%)$ &&&&&&\\

\multirow{2}{*}{LBQS\,1338$-$0038}

&\multirow{2}{*}{0841482101} 
&&\multirow{2}{*}{$6.62_{-0.21}^{+0.94}$} &\multirow{2}{*}{$311_{-124}^{+432}$} &\multirow{2}{*}{$2.5_{-1.3}^{+1.1}$}  &\multirow{2}{*}{$67_{-35}^{+/}$}    &\multirow{2}{*}{$15.09/2\,(>99.99\%)$} 

&$8.03_{-1.06}^{+0.20}$    &$378^{\star}$  &$-3.6_{-1.1}^{+1.1}$ &$-130_{-41}^{+41}$ &  $-0.139_{-0.02}^{+0.02}$ &$24.14/3\,(>99.9\%)$ &$\mathbf{>99.9\%}$\\

&&&&&&&&$11.06_{-0.29}^{+0.29}$    &$378^{f}$  &$-3.01_{-1.3}^{+1.3}$ &$-187_{-81}^{+81}$ &  $-0.432_{-0.041}^{+0.041}$ &$10.00/2\,(99.33\%)$ &$87.2\%$\\

\\

\hline
\label{Table:basegaussSUBWAYS}
\end{tabular}
\end{table}
\end{landscape}

\begin{landscape}
\begin{table}
\setlength{\tabcolsep}{0.4pt}
\centering
\begin{tabular}{l c c c c c c c c c c c c c c c c c}
\hline
&              &\multicolumn{6}{c}{\texttt{Emission Lines}} &\multicolumn{7}{c}{\texttt{Absorption Lines}}\\
Source\,(1)&\textit{XMM}\,(2)  &&$E_{\rm em}$\,(3) &$\sigma_{\rm width}$\,(4) &Int\,(5) &$EW$\,(6) &$\dC\,(P_{\rm F})$\,(7) &$E_{\rm abs}$\,(8) &$\sigma_{\rm width}$\,(9) &Int\,(10)   &$EW$\,(11)  &$v_{\rm Gau}$\,(12)  &$\dC\,(P_{\rm F})$\,(13) & $P_{ \mathcal{MC}}$\,(14)\\

   &       ObsID    &&keV          &eV      &$10^{-6}$  &eV   &                   &keV           &eV        &$10^{-6}$  &eV     &$c$            &                  &\\
\hline

\\

\multirow{5}{*}{PG0804$+$761}
&\multirow{2}{*}{0102040401}

&&$6.33_{-0.07}^{+0.08}$  &$10^{f}$   &$8.2_{-5.1}^{+5.4}$     &$60_{-37}^{+40}$    &$7.43/2\,(97.56\%)$        															                         	    &$7.93_{-0.06}^{+0.06}$   &$10^{f}$   &$-11.7_{-4.2}^{+4.6}$    &$-133_{-48}^{+52}$   &$-0.128_{-0.006}^{+0.007}$ &$15.99/2\,(>99.99\%)$ &$\mathbf{99.8\%}$\\

&&&$6.96_{-0.17}^{+0.16}$ &$10^{f}$   &$8.4_{-5.3}^{+5.7}$     &$74_{+47}^{+50}$    &$7.50/2\,(97.65\%)$ &&&&&     &\\

&\multirow{2}{*}{0605110101}

&&$6.41_{-0.07}^{+0.08}$  &$100^{f}$   &$8.0_{-2.8}^{+2.8}$     &$69_{-24}^{+24}$ &$25.77/2\,(>99.99\%)$\\        															                         	    
&&&$6.81_{-0.17}^{+0.16}$ &$10^{f}$   &$4.7_{-2.2}^{+2.3}$     &$44_{+21}^{+22}$  &$13.89/2\,(>99.99\%)$ &&&&&     &\\

&0605110201
&&$6.58_{-0.09}^{+0.09}$  &$388_{-80}^{+104}$   &$26.4_{-5.7}^{+6.5}$     &$328_{-71}^{+81}$    &$115.0/3\,(>99.99\%)$\\

\\
PG1416$-$129
&0203770201 
&&$6.36_{-0.05}^{+0.04}$ &$10^{f}$ &$3.0_{-1.1}^{+1.1}$ &$61_{-22}^{+22}$ &$22.13/2\,(>99.99\%)$&&&&&&\\       

\\

\multirow{4}{*}{PG1402$+$261}

&\multirow{2}{*}{0400200101}
&&$6.37_{-0.07}^{+0.07}$ &$10^{f}$ &$1.9_{-1.0}^{+1.0}$ &$82_{-49}^{+55}$ &$10.88/2\,(99.57\%)$\\               													                               		
&&&$7.40_{-0.10}^{+0.09}$ &$10^{f}$ &$1.5_{-0.9}^{+1.0}$ &$76_{-40}^{+40}$ &$7.42/2\,(97.55\%)$ &&&&&&\\

&\multirow{2}{*}{0830470101}
&&$6.15_{-0.16}^{+0.15}$ &$100^{f}$ &$1.1_{-0.7}^{+0.7}$ &$42_{-27}^{+27}$ &$11.69/2\,(99.71\%)$\\               													                               		
&&&$8.36_{-0.06}^{+0.06}$ &$10^{f}$ &$0.86_{-0.72}^{+0.74}$ &$67_{-56}^{+58}$ &$7.74/2\,(97.91\%)$ &&&&&  &\\

\\
\multirow{7}{*}{HB89\,1257$+$286~~~~} 
&0204040101
&&$6.34_{-0.05}^{+0.04}$ &$100^{f}$ &$2.2_{-0.6}^{+0.6}$ &$85_{-23}^{+23}$ &$38.3/2\,(>99.99\%)$               										
&$8.78_{-0.16}^{+0.15}$ &$10^{f}$ &$-1.1_{-0.7}^{+0.7}$ &$-81_{-54}^{+54}$ & $-0.227_{-0.018}^{+0.016}$&$6.33/2\,(95.78\%)$&$65.8\%$\\

&0204040201
&&$6.37_{-0.05}^{+0.04}$ &$10^{f}$ &$1.5_{-0.5}^{+0.5}$ &$93_{-31}^{+31}$ &$27.2/2\,(>99.99\%)$&&&&&&\\               										
&\multirow{2}{*}{0204040301}
&&$6.35_{-0.08}^{+0.09}$ &$100^{f}$ &$1.7_{-0.7}^{+0.7}$ &$82_{-34}^{+34}$ &$18.9/2\,(>99.99\%)$\\               										

&&&$6.93_{-0.09}^{+0.09}$ &$100^{f}$ &$1.5_{-0.7}^{+0.7}$ &$84_{-38}^{+38}$ &$14.5/2\,(98.74\%)$ &&&&&&\\

&0304320201
&&$6.46_{-0.13}^{+0.15}$ &$284_{-200}^{+213}$ &$4.7_{-1.5}^{+1.8}$ &$156_{-50}^{+60}$ &$27.2/3\,(>99.99\%)$&&&&&&\\

&0304320301
&&$6.49_{-0.08}^{+0.08}$ &$10^{f}$ &$1.2_{-0.7}^{+0.7}$ &$66_{-39}^{+39}$ &$9.7/2\,(99.23\%)$&&&&&&\\               										

&0304320801
&&$6.41_{-0.13}^{+0.13}$ &$10^{f}$ &$1.1_{-0.8}^{+0.8}$ &$44_{-32}^{+32}$ &$6.36/2\,(95.84\%)$&&&&&&\\               		

\\
\multirow{12}{*}{PG\,1114$+$445~~~~}

&0109080801 
&&$6.56_{-0.07}^{+0.07}$ &$200_{-47}^{+57}$ &$11.2_{-2.2}^{+2.3}$  &$279_{-55}^{+57}$    &$103.9/2\,(>99.99\%)$\\ 

&0651330101 
&&$6.48_{-0.06}^{+0.06}$ &$143_{-56}^{+74}$ &$8.3_{-2.2}^{+2.4}$  &$230_{-61}^{+67}$    &$54.9/2\,(>99.99\%)$ 

&$7.49_{-0.28}^{+0.29}$    &$300^{f}$  &$-4.2_{-1.9}^{+2.0}$ &$-158_{-71}^{+75}$ &  $-0.072_{-0.042}^{+0.042}$ &$11.71/2\,(99.7\%)$ &$\mathbf{95.5\%}$\\

&0651330301 
&&$6.42_{-0.05}^{+0.05}$ &$50^{f}$ &$4.4_{-1.4}^{+1.6}$  &$133_{-40}^{+43}$    &$29.2/2\,(>99.99\%)$ 

&$7.35_{-0.19}^{+0.12}$    &$100^{f}$  &$-2.6_{-1.1}^{+1.2}$ &$-94_{-35}^{+38}$ &  $-0.053_{-0.027}^{+0.017}$ &$11.9/2\,(99.7\%)$ &$\mathbf{97.2\%}$\\

&0651330401 
&&$6.48_{-0.07}^{+0.06}$ &$170_{-55}^{+75}$ &$10.6_{-3.2}^{+4.4}$  &$261_{-79}^{+108}$    &$56.9/3\,(>99.99\%)$ 
			
&$7.21_{-0.14}^{+0.11}$    &$10^{f}$  &$-2.1_{-1.3}^{+1.4}$ &$-68_{-42}^{+45}$ &  $-0.034_{-0.015}^{+0.020}$ &$8.3/2\,(98.4\%)$ &$90.5\%$\\

&0651330501

&&$6.42_{-0.04}^{+0.05}$ &$10^{f}$ &$6.9_{-1.9}^{+2.1}$  &$178_{-49}^{+54}$    &$49.6/2\,(>99.99\%)$ 

&$7.04_{-0.08}^{+0.12}$  &$10^{f}$  &$-2.7_{-1.3}^{+1.4}$ &$-79_{-38}^{+41}$ & --  &$10.02/2\,(99.33\%)$ & (N/A)$^{\blacklozenge}$\\

&0651330601
&&$6.51_{-0.09}^{+0.05}$ &$100^{f}$ &$8.5_{-4.2}^{+4.7}$  &$121_{-60}^{+67}$    &$32.3/2\,(>99.99\%)$\\

&0651330701
&&$6.49_{-0.04}^{+0.04}$ &$50^{f}$ &$6.6_{-2.3}^{+2.6}$  &$117_{-41}^{+46}$    &$29.4/2\,(>99.99\%)$\\ 

&\multirow{2}{*}{0651330801}
&&$6.42_{-0.09}^{+0.09}$ &$100^{f}$ &$5.3_{-2.1}^{+1.2}$  &$126_{-50}^{+29}$    &$21.6/2\,(>99.99\%)$\\ 

&&&$9.02_{-0.17}^{+0.16}$ &$227_{-156}^{+177}$ &$5.7_{-3.0}^{+3.6}$  &$266_{-140}^{+168}$&$12.9/3\,(99.51\%)$ &&&&&&\\

&0651330901
&&$6.44_{-0.07}^{+0.06}$ &$50^{f}$ &$4.4_{-1.7}^{+2.2}$  &$96_{-37}^{+46}$    &$21.5/2\,(>99.99\%)$\\

&0651331001
&&$6.50_{-0.06}^{+0.05}$ &$134_{-64}^{+67}$ &$8.6_{-2.4}^{+2.8}$  &$214_{-60}^{+70}$    &$48.8/3\,(>99.99\%)$\\

&0651331101
&&$6.42_{-0.09}^{+0.09}$ &$50^{f}$ &$4.3_{-2.0}^{+2.6}$  &$92_{-43}^{+66}$    &$13.1/2\,(99.86\%)$\\

\\
\hline

\end{tabular}
\end{table}
\end{landscape}

\begin{figure*}
\centering
\includegraphics[scale=0.33]{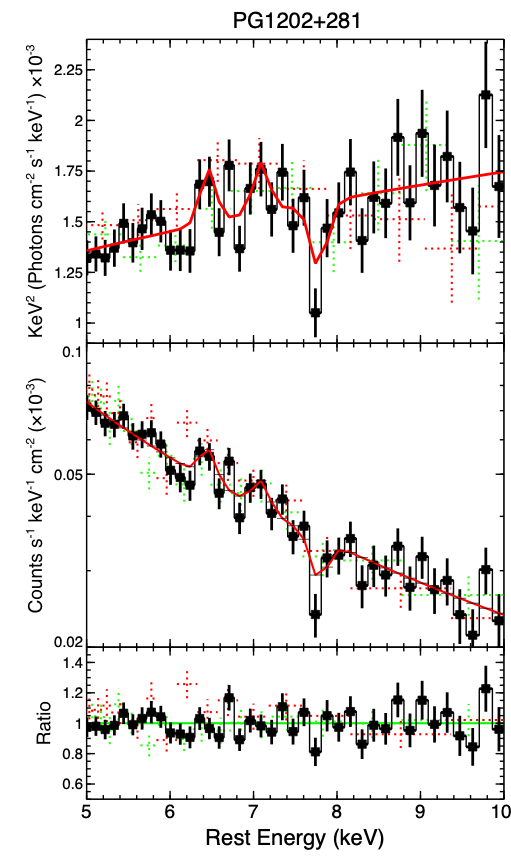}
\includegraphics[scale=0.33]{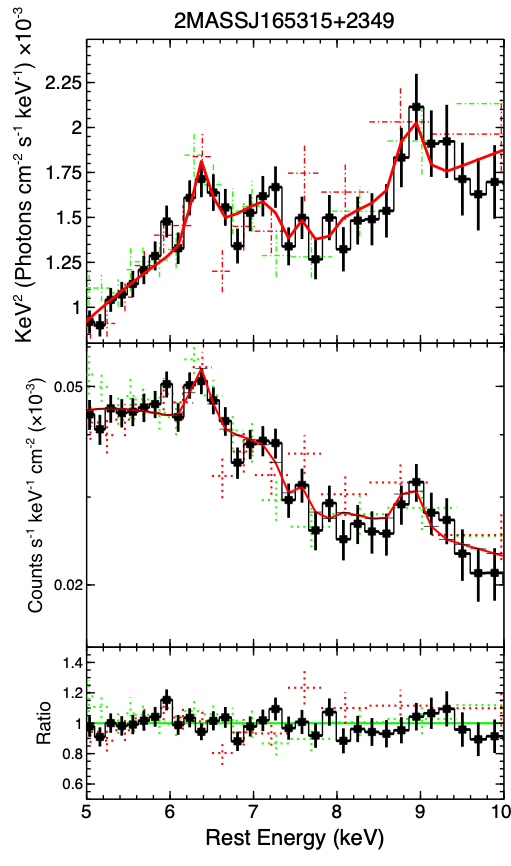}
\includegraphics[scale=0.33]{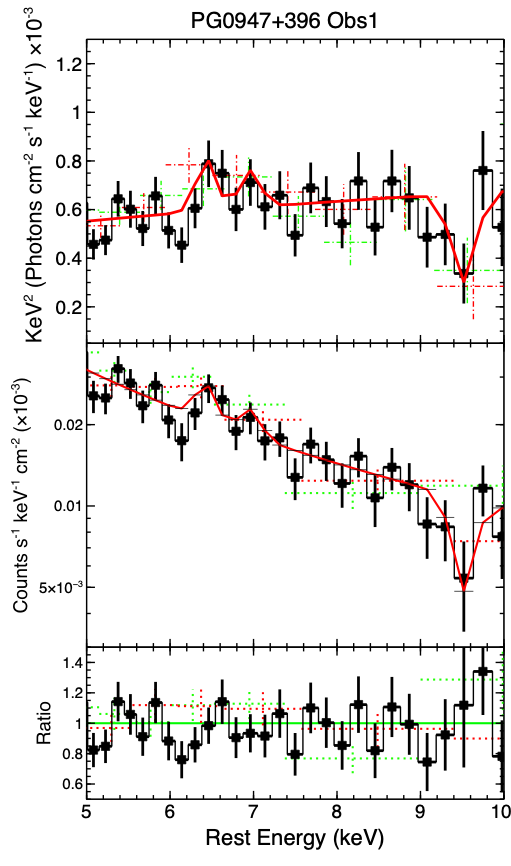}
\includegraphics[scale=0.33]{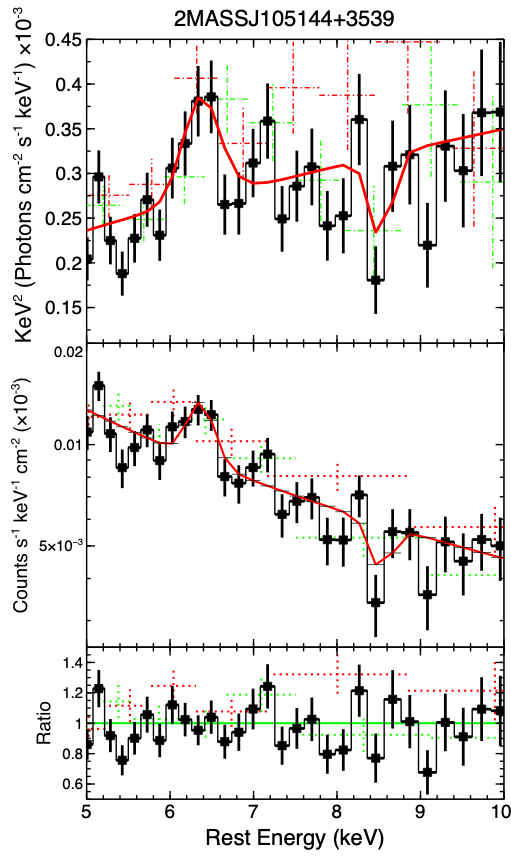}
\includegraphics[scale=0.33]{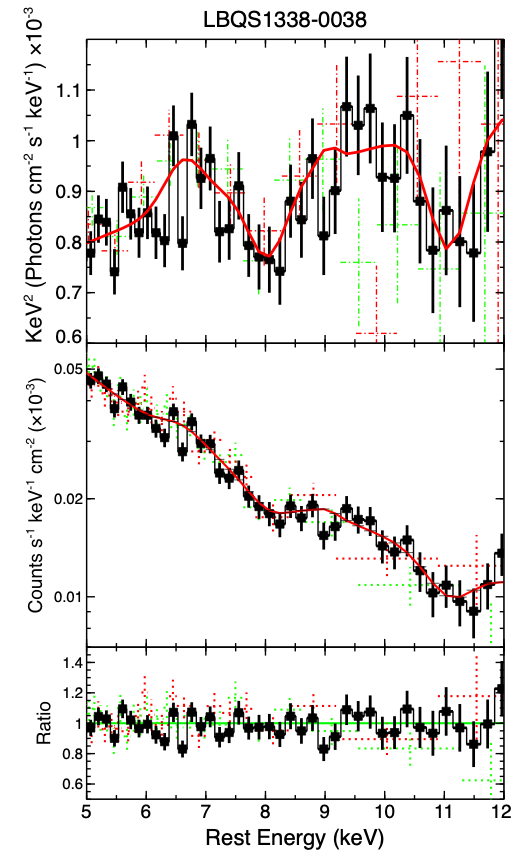}
\includegraphics[scale=0.33]{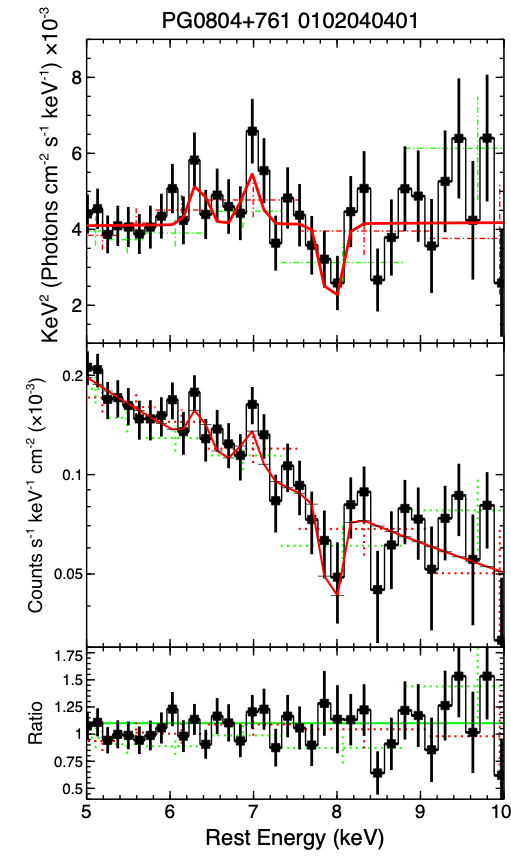}
\includegraphics[scale=0.33]{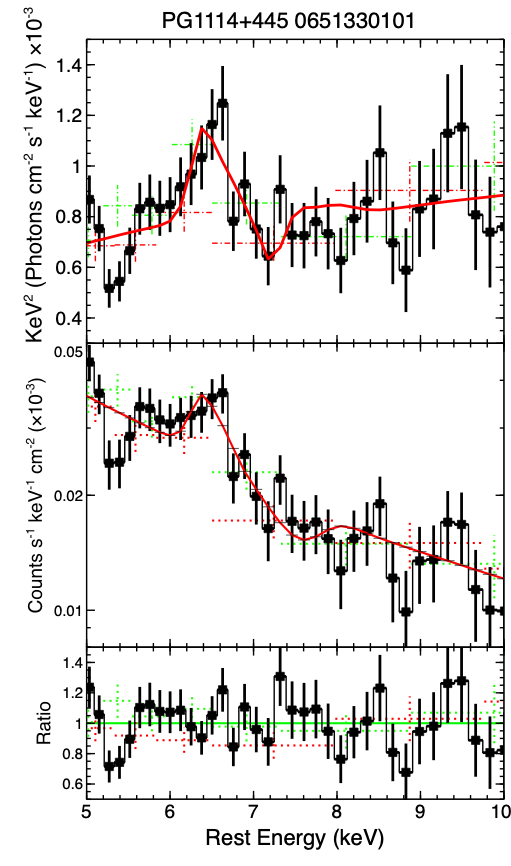}
\includegraphics[scale=0.33]{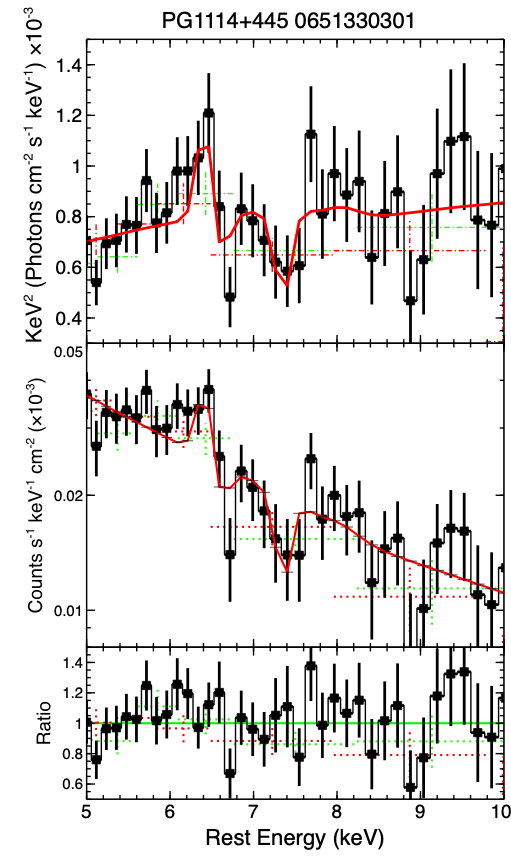}
\caption{Top: Unfolded EPIC-pn (black), EPIC-MOS\,1 (dashed red) and EPIC-MOS\,2 (dashed green) spectra, between $5$--$10\kev$, of the 8 observations where the \fe absorption line was detected at $\pmc\gtrsim95\%$. The spectra were firstly unfolded against a 
power law of $\Gamma=2$ and their corresponding best-fitting model (solid red) overlaid on top afterwards. Middle: Corresponding EPIC data counts and best-fitting model. Bottom: Data/Model ratio. The presumed absorption features at $\sim9.2\kev$ and $\sim9.5\kev$ present in e.g., 2MASSJ105144$+$3539 and \tmosf, respectively, are simply not significant enough to be considered as detections (see \autoref{fig:2massj165_lbqs} and \autoref{figapp:sub_ALLOBS_SCAN_RA}), and hence they were not included in the best-fitting models.}
\label{fig:fek_gamod}
\end{figure*}

\begin{table*}
\centering
\caption{\xstar parameters for the \fe absorption features detected in the \sub sample with $\pmc \gtrsim 95\%$. Notes: (1) Source name; (2) observation ID; (3) intrinsic velocity broadening of the \xstar grid; (4) column density and (5) gas ionisation state both in log scale; (6) change in $\cstat$ fit statistic over degrees of freedoms when \xstar is removed; (7) corresponding significance in per cent.}

\begin{tabular}{l c c c c c c c}

\hline



\multicolumn{8}{c}{\xstar parameters}\\

\\
Source\,(1)   &ObsID\,(2)     &$\sigma_{\rm turb}$\,(3)   &$\log(\nh)$\,(4)         &$\log(\xi)$\,(5)       &$v_{\rm out}$\,(6)        &$\dC$\,(7)       &$(P_{\rm F})$\,(8)\\

\\
        &     &$\kms$   &$\cmsq$   &$\rm erg\,cm\,s^{-1}$  &$c$& & \\ 

\hline
\\
PG1202$+$281

&0841480501     &$1000$   & > 23.84        & > 5.02         & $-0.108^{+0.010}_{-0.08}$ &$8.60/3$          &$96.49\%$\\

\\
2MASS\,J165315$+$2349$^{\dag}$

&0841480801  &$5000$   & 23.76$^{+ 0.39}_{- 0.15}$   & 4.76$^{+ 0.59}_{- 0.42}$  & $-0.110^{+ 0.008}_{- 0.015}$ &$-22.98/3$          &$>99.99\%$\\

\\
PG0947$+$396\,(Obs\,1)

&0841481001   &$5000$ & > 23.68     & 5.38$^{+ 0.44}_{- 1.27}$  & $-0.305^{+ 0.037}_{- 0.019}$ &$10.8/3$          &$98.71\%$\\

\\
2MASS\,J105144$+$3539 

&0841481701  &$1000$      & 22.78$^{+ 0.67}_{- 0.33}$   & 4.10$^{+ 0.35}_{- 0.38}$  & $-0.237^{+ 0.011}_{- 0.010}$ &$10.9/3$          &$98.77\%$\\

\\

LBQS\,1338$-$0038

&0841482101   &$10000$
& 23.16$^{+ 1.01}_{-0.22}$ & $5.01_{-0.38}^{+0.64}$  &$-0.152_{-0.023}^{+0.020}$ &23.27/3& $>99.99\%$\\

\\
PG0804$+$761 

&0102040401  &$5000$  &$23.93^{+ 0.33}_{- 0.27}$ & > 4.49    & $-0.130^{+ 0.011}_{- 0.011}$ &$11.1/3$   &$98.88\%$\\

\\
\multirow{3}{*}{PG1114$+$445}  

&0651330101  &$5000$  &$23.90_{-0.16}^{+0.25}$  &$4.65_{-0.44}^{+0.77}$  &$-0.072_{-0.016}^{+0.017}$ &$11.91/3$    &$99.23\%$\\

\\
 
&0651330301  &$5000$  &$23.61_{-0.15}^{+0.10}$  &$3.56_{-0.13}^{+0.10}$  &$-0.070_{-0.015}^{+0.017}$ &$62.08/3$          &$>99.99\%$\\

\\
\hline
\label{Table:XSTAR_TABLE}
\end{tabular}
\end{table*}

\section{Results}
\label{sec:results}

A total of 14 absorption features with energies $E_{\rm rest}\gtrsim7.1\kev$ and $P_{\rm F}\gtrsim99\%$ are found. Of these, 8 are robustly detected with $P_{ \mathcal{MC}}\gtrsim95\%$ while 6 have $\pmc<95\%$ and are therefore considered non-detections. In PG\,1114$+$445 (ObsID 0651330501) an absorption line at $E_{\rm rest}=7.04_{-0.08}^{+0.12}\kev$ was detected at the $P_{\rm F}>99\%$ confidence level. However, such a feature is likely consistent with a neutral iron\,K edge so no Monte Carlo test was applied here. 

In \autoref{fig:fek_gamod}\,(top), we plot the unfolded EPIC-pn (black), MOS\,1\,(red) and 2\,(green) data showing the 8 \fe absorption lines detections with $P_{ \mathcal{MC}}\gtrsim95\%$. To avoid model and data convolution issues, the spectra in each panel are initially unfolded against a simple $\Gamma=2$ power law (with normalisation of 1) and subsequently their corresponding best-fitting model are superimposed (solid red). In \autoref{fig:fek_gamod}\,(middle) the plot is in terms of the EPIC data counts normalised by the effective areas and \autoref{fig:fek_gamod}\,(bottom) are their corresponding residuals. 

We conservatively identify these \fe absorption lines as highly ionised iron, specifically \fexxviabs K-shell transitions, all blueshifted with respect to their laboratory rest energies. Some of these lines can be a blend of both \fexxvabs and \fexxviabs resonant transitions and might be indistinguishable with the current EPIC energy resolution. In a forthcoming paper (Matzeu et al., in preparation), we will carry out a comprehensive physical modelling of these features where an accurate measurement of the ionisation balance, as well as density and velocity, of the outflowing gas can be achieved. A photoionisation analysis of the \fe lines will also help towards a quantitative identification of the absorption/emission features. As presented in Section\,\ref{sub:modelling the fek band}, the corresponding outflow velocities were conservatively estimated by choosing $E_{\rm rest}=6.97\kev$ as a reference energy (see \autoref{Table:basegaussSUBWAYS}).

\subsection{Line Detection Rate}
\label{sub:line detection rate}

Here, we quantify the probability of whether or not the detected absorption lines are caused by statistical fluctuation (`shot noise'). This can be done by using the binomial distribution \citepalias[e.g.,][]{Tombesi10,Gofford13}. For an event with a null-probability $p$, the likelihood of $n$ detections after $N$ trials is given by the expression:  

\begin{equation}
P(n; N, p) = \frac{N!}{n!(N-n)!} p^{n}(1-p)^{N-n}.
\end{equation}
In this context, $n$ is the number of absorption lines detected in $N$ systems and depending on the latter quantity we investigate two different cases where we take into account: 
case\,(i) all the individual targets, or $N_{\rm (i)}=22$; and 
case\,(ii) all the individual observations (with total net counts of $\geq1500\,\rm cts$ in the $4$--$10\kev$ band), or $N_{\rm (ii)}=41$. 

In case\,(i) we have $n_{95}=7$ 
\fe absorption line systems detected in $N_{\rm (i)}=22$ observations at a significance of $P_{ \mathcal{MC}}\geq95\%$. 
So the probability of one of these absorption profiles being due to fluctuating noise can be taken as $p<0.05$. The probability of all of the observed absorption systems being associated with noise is then reasonably low, with $P_{\rm 95,(ii)}<6.17\times10^{-5}\,(\lesssim0.006\%)$. This suggests that the observed lines are unlikely to be associated with simple statistical fluctuations in the spectra. 

In case\,(ii) we have a total of $n_{95}=8$ detections out of $N_{\rm (ii)}=41$ individual observations. Here we have $P_{\rm 95, (ii)}<6.89\times10^{-4}\,(\lesssim0.07\%)$.

\subsection{Photoionisation modelling of Fe\,K features: initial results}
\label{sub:Photoionization modelling with xabs}

Although this paper is solely focused on UFO detection, we present a preliminary photoionization analysis of the \fe features and we provide first-order physical measurements of their properties. This analysis is carried out so that our \sub results can be compared with those previously obtained in \citet[][T11 hereafter]{Tombesi11} \citetalias{Tombesi11}, \citetalias{Gofford13} and \citetalias{Igo20}. The search for absorption features and the Gaussian modelling of the absorption profiles in the \fe band described in Section\,\ref{sub:modelling the fek band} suggest they can be ascribed to outflowing and highly ionised material likely associated with \fexxvabs--\fexxviabs transitions. 

In contrast with the phenomenological models used before, modelling the absorption features with \xstar allows us to probe the physical properties of the absorbing medium. More specifically we are able to quantify the ionization state $(\xi)$, the column density $(\nh)$ and the systemic redshift of the material relative to the observed one, which translates into the outflow velocity ($v_{\rm out}$; see below for more details). Through the photoionization modelling approach it is also possible to infer the geometric properties, such as the radial distance from the ionizing source, the covering factor and the resulting overall kinematics \citep[e.g.,][]{Gofford15,Matzeu17b}. 

We replaced the Gaussian absorption profiles, detected at the  $\pmc\gtrsim95\%$ confidence level, with \xstar photoionization models, generated with a power-law SED input spectrum of $\Gamma=2$, by using the \xstar suite v2.54a \citep{BautistaKallman01,Kallman04}. We adopted various \xstar grids with different turbulent velocity, defined as $\sigma_{\rm turb}=\sqrt{2}\sigma$, so that an accurate description of the width of the \fe absorption lines could be provided. Choosing a grid with a smaller $\sigma_{\rm turb}$ results in a smaller \ew of the profile in the data, and the absorption would saturate too quickly at lower $\nh$. So for each source we adopted grids with $\sigma_{\rm turb}$ ranging between $1000$--$10000\kms$ (see \autoref{Table:XSTAR_TABLE}).

The measured column densities are ranging between $10^{23} \lesssim \nh/\rm cm^{-2} \lesssim 10^{24}$ with a mean value of $\log(\overline{N}_{\rm H}/\rm cm^{-2})\sim23.6$ and a median of $\log(\Tilde{N}_{\rm H}/\rm cm^{-2})\sim23.8$. We also report the measured ionization distribution, which is found to extend between $  3.5 \lesssim \logxi \lesssim 5.5 $, with mean/median values of $\log(\overline{\xi}/\rm{erg\,cm\,s}^{-1})\sim4.7$ and $\log(\Tilde{\xi}/\rm{erg\,cm\,s}^{-1})\sim4.8$, respectively.

The outflow velocity distribution measured with \xstar
\footnote{The systemic redshift of the absorber obtained from fitting with \xstar is given in the observer's rest-frame $z_{\rm abs}$ and related to $v_{z_{\rm abs}}=\left[(1+z_{\rm abs})^{2} - 1 \right]/\left[(1+z_{\rm abs})^{2}+1 \right]$, and correcting for the systemic velocities of the sources $u$ we obtain $v_{\rm out}/c = \left( u-v_{z_{\rm abs}} \right) / \left[1-(uv_{z_{\rm abs}}) \right]$.} ranges between $-0.3 \lesssim v_{\rm out}/c \lesssim -0.1$ (see \autoref{Table:XSTAR_TABLE}). The mean/median values are $\overline{v}_{\rm out}\sim-0.144c$ and $\Tilde{v}_{\rm out}\sim-0.110c$ respectively. We also compare the outflow velocities measured with \xstar and with the one measured from the Gaussian fitting. For the latter, the measured velocities are found to lie between $-0.3 \lesssim v_{\rm out}/c \lesssim -0.05$  (see \autoref{Table:basegaussSUBWAYS}) with mean/median values of $\overline{v}_{\rm out,Gauss}\sim-0.131c$ and $\Tilde{v}_{\rm out,Gauss}\sim-0.113c$ respectively. We find that both the phenomenological and the physical modelling of the \fe absorption features have the same distribution, as confirmed at the $99.7\%$ confidence level by a Kologoromv-Smirnov test. The \xstar-based approach returns a $\sim10\%$ higher mean velocity but a comparable median, within $\sim3\%$.

In our sample a considerable fraction of the \fe absorbers are characterised by material with high column density and highly ionised material, likely H-like iron. Such result does not come as surprise when considering the hard average photon-index ($\overline{\Gamma}\sim1.8$) measured on the entire sample (see \autoref{fig:sub_histo_comparison}), which would overionise the outflowing material \citep[see][Figs. 2 and 3 ]{Matzeu22}. A comprehensive photoionisation analysis with \xstar (and other physically motivated wind models) will be presented in a companion paper, where customised photoionization tables will be generated with more realistic optical/UV/X-ray SED inputs for each individual \sub source \citep[e.g.,][]{Nardini15,Matzeu16}.

\section{Discussion}
\label{sec:Discussion}

In this work, we searched for \fe absorption features in a sample of 22 targets (41 observations), of which 17 were observed as part of the \xmm large program carried out in AO18. Through a systematic blind line scan performed in all the observations and supported by a \mc procedure, we detected \iron absorption lines in 7/22 sources (i.e., $\sim30\%$) at the $\pmc\gtrsim95\%$ confidence level. Through our statistical approach, we have found 2 robust \fe absorption line detections at $\pmc\gtrsim99\%$ in \tmosf and \lbqs. The remaining 5 detections are still significant but with $ 95\% \lesssim P_{ \mathcal{MC}} \lesssim  99\% $, in PG\,1202$+$281, 2MASS\,J105144+3539, PG\,1114$+$445, PG\,0804$+761$ and PG\,0947$+$396 (see \autoref{Table:basegaussSUBWAYS}).

Such absorption (and sometimes emission) profiles are associated with highly ionised He- and/or H-like iron, arising from the outflowing material, as their centroid energy is blueshifted with respect to the QSO systemic redshift. In this paper we only focused on the search and phenomenological analysis of such features, thus for the estimate of the outflow velocities we assumed \fexxviabs at $E_{\rm rest}=6.97\kev$ as a reference energy for a conservative result, which correspond to the lowest possible outflow velocity. Accordingly, we found that the average outflow velocity measured in our sample is $\overline{v}_{\rm out}=-0.133c$, as shown in \autoref{fig:histo_vout_compare}. 

For our search of \fe absorption and emission features, an accurate parameterization of the underlying broadband continuum (i.e., $0.3$--$10\kev$) in each observation was required. We therefore summarise the phenomenological continuum findings of \sub.
We found that 27 out of the 
41 ($\sim65\%$) \sub observations are characterised by intrinsic soft X-ray absorption. More specifically, 20/27 systems 
can be identified as fully covering, mildly ionised (warm) absorbers, while 5/27 
are partially covering the \los. We note that 2/27 spectra are modified by a fully covering neutral absorber, where in \tmosf the spectral curvature at energies $<2\kev$ is caused by a column density $\nh\sim10^{23}\cmsq$, consistent with the values measured in Seyfert\,2 galaxies. We find that a prominent soft excess, at a 
$>99\%$ significance, is present in the majority of the spectra in our sample, i.e. 39/41,
where one \texttt{zbbody} component is required in 13/39 
and two components in 26/39 sources. 

\begin{figure}

\includegraphics[width=1.1\linewidth]{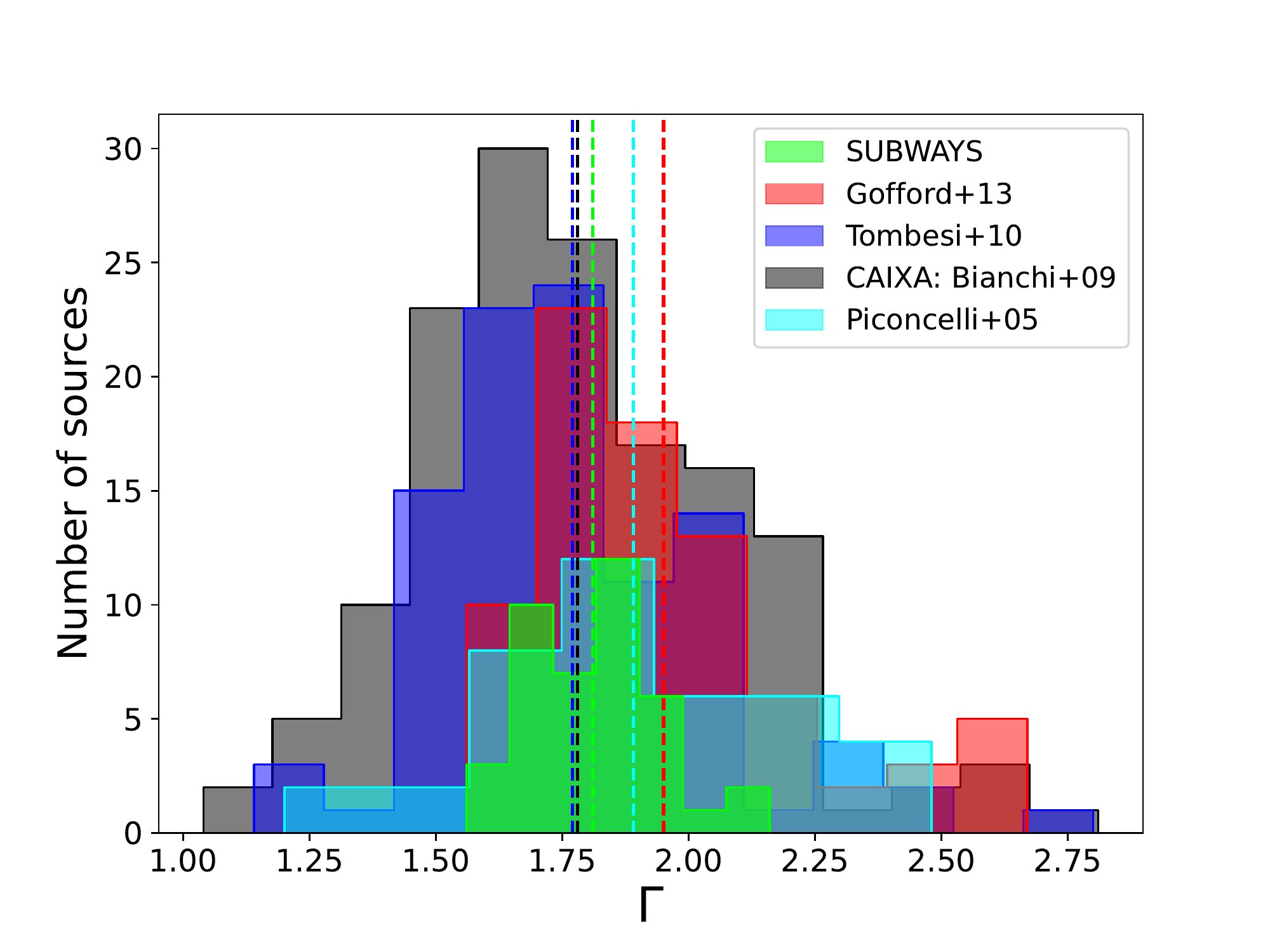}
  \caption{Histogram comparing the photon-index ($\Gamma$) measured in \sub (green), this work, with mean value of $\overline{\Gamma}=1.81$ (vertical line), with those measured in an X-ray selected sample (CAIXA sample, 150 sources with $\overline{\Gamma}=1.78$; \citealt{Bianchi09caixa1}); PG QSO sample (\citealt{Piconcelli05}, 40 sources $\overline{\Gamma}=1.89$); local 
  AGN (\citetalias{Tombesi10,Gofford13} with $\overline{\Gamma}=1.77$ and $\overline{\Gamma}=1.95$ respectively). } 
\label{fig:sub_histo_comparison}
\end{figure}

In \autoref{fig:sub_histo_comparison} we compare the primary continuum photon indices ($\Gamma$) with those measured in previous works in the literature: a purely X-ray selected sample, CAIXA \citep{Bianchi09caixa1}, an optically selected PG QSO sample \citep{Piconcelli05}, low-$z$ 
AGN analyzed with \xmm \citepalias{Tombesi10} and \suzaku \citepalias{Gofford13}. We find that the \sub sample tends to be characterised by $\Gamma < 2$ ($\overline{\Gamma}=1.81$), only slightly harder compared to the optically selected PG QSO sample ($\overline{\Gamma}=1.89$) and largely consistent with the CAIXA \citep{Bianchi09caixa1} ($\overline{\Gamma}=1.78$) and \citetalias{Tombesi10} ($\overline{\Gamma}=1.77$) samples. The mean photon index of the \suzaku sample is softer, with ($\overline{\Gamma}=1.95$).

\subsection{Sample comparisons}
\label{sub:sample comparisons}

The \sub sample size is indeed small compared to those in \citetalias{Tombesi10,Gofford13} and \citetalias{Igo20}. With this in mind this selection of targets must be taken as an initial exploration of the intermediate-$z$ population that is bridging the gap of UFOs studies between low- and high-$z$ sources.

We find that our overall measurements seem to be skewed to higher values of $\nh$ and $\vout$ compared to T10 and G13, whilst the ionization state of the absorber is on the same order of magnitude (slightly lower). The latter parameter can be influenced by the SED input assumed when generating the photoionization grids. We also find that the outflow velocity measured in PG0947$+$396\,(Obs1), being the highest in the sample, is unlikely to be associated with outflowing, highly ionised material, but rather the result of an artefact of the EPIC CCD or/and some background issue. Although the absorption line is significantly detected at the $P_{\rm F}>99\%$ (Gaussian modelling) and $P_{\rm MC}\sim97\%$ (Monte Carlo approach) confidence level, it is weakly detected with \xstar at $\sim90\%$. Furthermore, an outflow velocity of $\vout>0.30c$ is generally considered on the high end of the scale of ultra fast winds and can carry a huge amount of kinetic power \citep[e.g.,][]{Matzeu17b,Reeves18PDS}. These events are more likely to be present in highly accreting sources where $\leddratio \rightarrow 1$ (or above) and Eddington fractions of $\leddratio \sim10\%$ \citep{Bianchi09caixa1} might be not enough to drive such strong outflows, although it cannot be ruled out as magneto-hydrodynamic (MHD) driving  mechanisms could come into play \citep[e.g,][]{Fukumura10,Kraemer18,Luminari21,Fukumura22}, especially in low-Eddington regimes. 

\begin{figure} 
    \centering
    \includegraphics[width=1.1\linewidth]{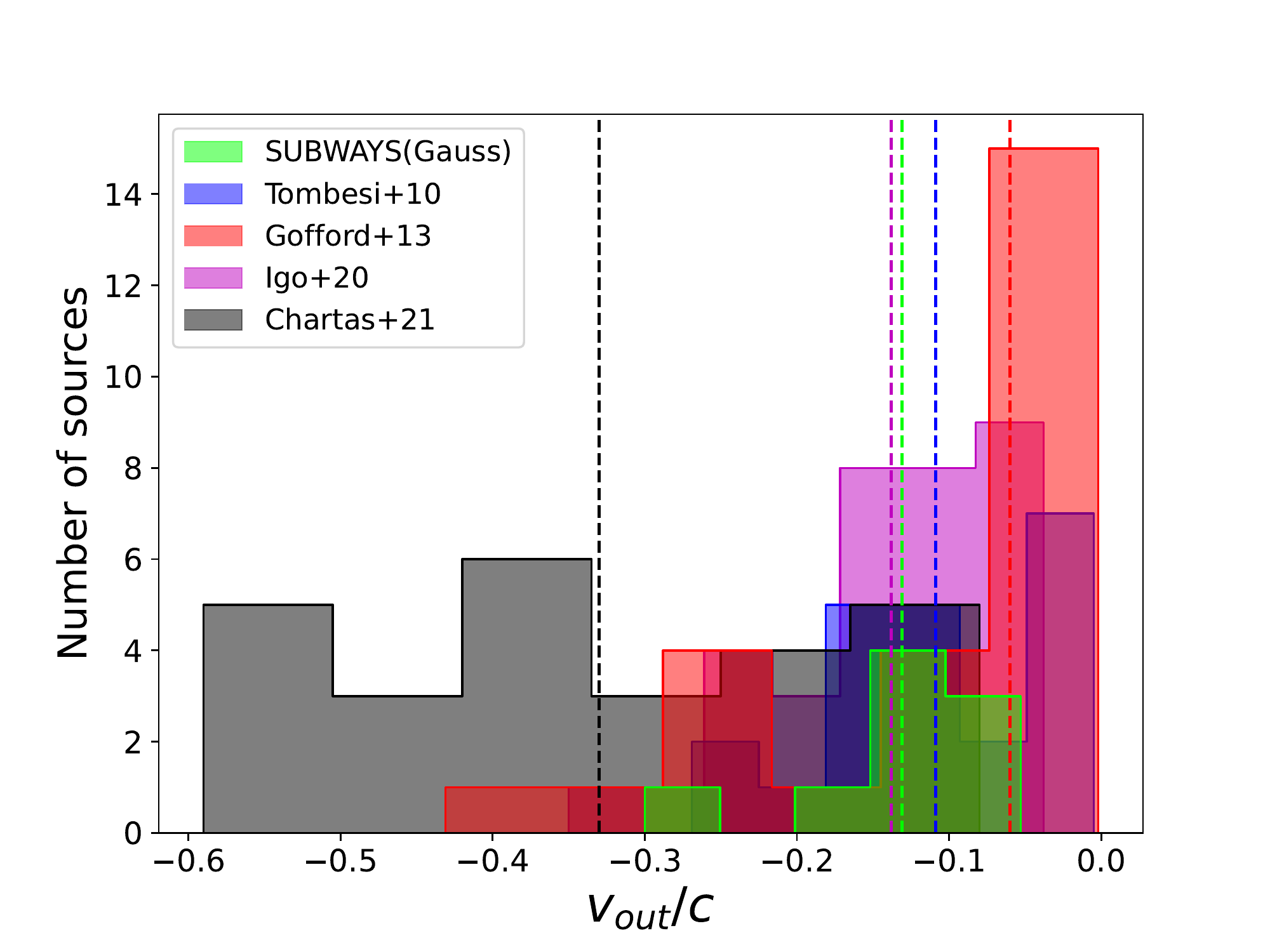}
    \caption{Distributions and mean outflow velocity values, indicated by the vertical dashed lines, as measured in this work (\sub) $\overline{v}_{\rm out}=-0.133c$; \citetalias{Tombesi10} $\overline{v}_{\rm out}=-0.109c$; \citetalias{Gofford13} $\overline{v}_{\rm out}=-0.056c$; \citetalias{Igo20} $\overline{v}_{\rm out}=-0.138c$, \citetalias{Chartas21} $\overline{v}_{\rm out}=-0.330c$. 
    }
    \label{fig:histo_vout_compare}
\end{figure}

\begin{figure}
\begin{center}
\includegraphics[width=\linewidth]{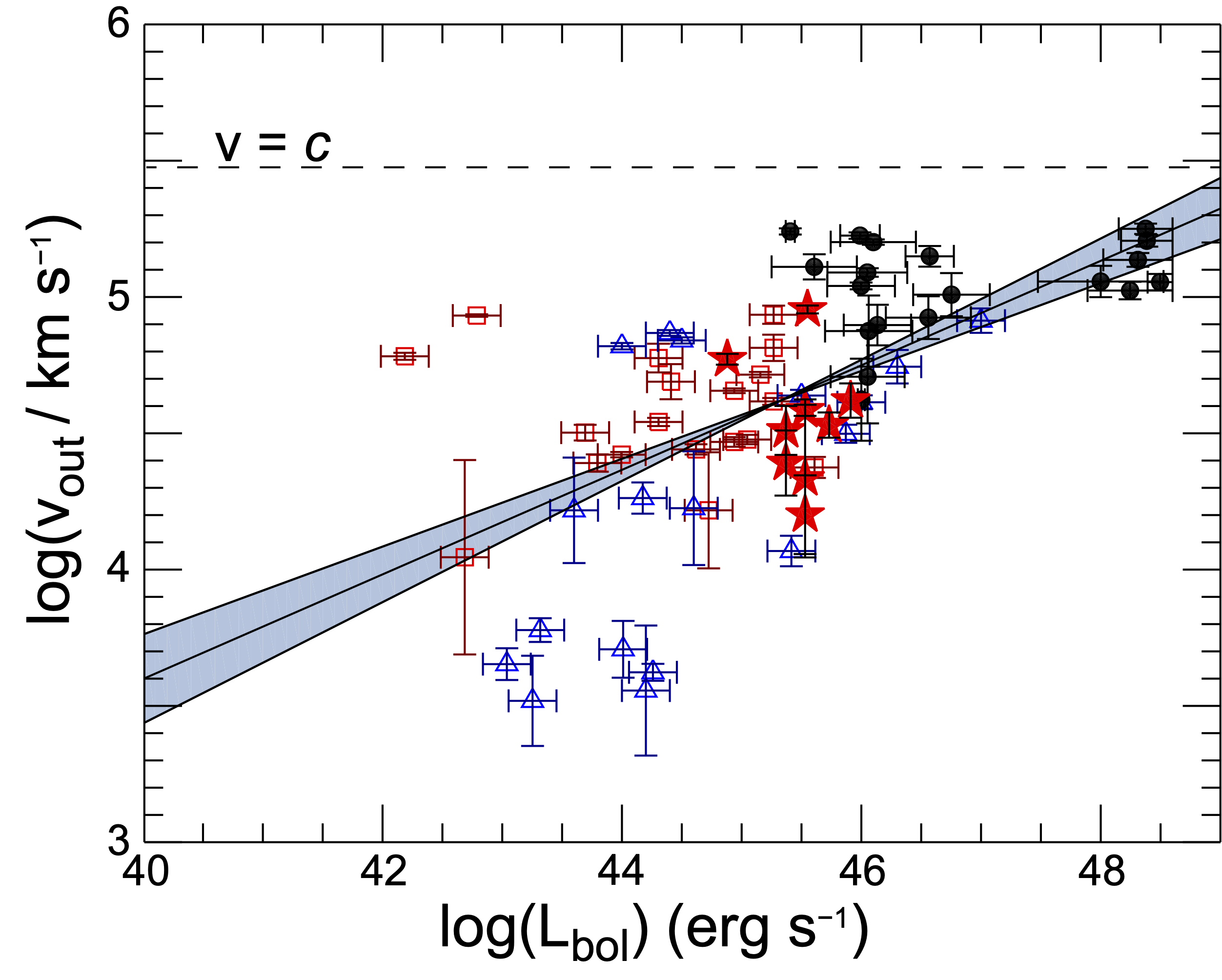}
\end{center}
\caption{
Outflow velocity of the ionised absorber for our \sub targets (red filled stars) plotted against their bolometric luminosity. Also plotted are the same quantities as derived in the analysis of the low-$z$ AGN samples of \citet[][red empty squares]{Tombesi12}, \citetalias[][(blue empty triangles)]{Gofford13} and in the high-$z$ AGN sample of \citetalias[][(black filled circles)]{Chartas21}. The power-law least-squares fit 
to the combined samples (a total of 49 individual objects and 63 observations) is shown with a solid line. The shaded area represents the uncertainty of the slopes of our fits to the data. Here the outflow velocity is simply described as the absolute value of $\vout$.}
\label{fig:v_lbol_plus_subways}
\end{figure}

In \autoref{fig:histo_vout_compare}, we show the $\vout$ distributions, and their mean values, measured with \xmm in previous works in the literature, such as in \citetalias[][$\overline{v}_{\rm out}=-0.109c$]{Tombesi10}; \citetalias[][$\overline{v}_{\rm out}=-0.138c$]{Igo20}; \citetalias[][$\overline{v}_{\rm out}=-0.330c$]{Chartas21}. An interesting trend is shown in \autoref{fig:histo_vout_compare}. By looking at all the measurements, the UFO outflow velocities seem to increase with redshift. Although the statistical footing of this trend is beyond the scope of this paper, we can recognise that such behaviour does arise from a high-$\lbol$ selection bias 
expected in sources at progressively higher redshift as the feeding becomes stronger \citep[e.g.,][]{DiMatteo05}, in particular due to the larger inflow of cold gas mass triggering CCA and boosting accretion rates by a few orders of magnitude compared with quiescent hot modes \citep[e.g.,][]{Gaspari17_cca}. Another way to interpret this trend is simply realise that the outflow velocity seems to increase with the luminosity \citep[e.g.,][\citetalias{Chartas21}]{Saez11,Matzeu17b,Chartas18IRAS13224}, as shown below in \autoref{fig:v_lbol_plus_subways}, and on the other hand the most luminous sources are observed at higher $z$. Additionally, another possible bias that is involved at higher $z$ is that higher velocity shifts become more detectable with increasing redshift. 


In \autoref{fig:v_lbol_plus_subways} we show the wind velocities, as measured from our X-ray spectral fits with Gaussian lines and tabulated in \autoref{Table:basegaussSUBWAYS}, plotted against the bolometric luminosity of the \sub targets (see \autoref{Table:SUBWAYS_TOTAL}) shown as red stars. 
We added and recomputed the fit of the correlation of v$_{\rm wind}$ versus L$_{\rm bol}$ of the low-$z$ \citepalias[][red squares and blue triangles respectively]{Tombesi10,Gofford13} and high-$z$ samples already presented in \citetalias{Chartas21}, including also our \sub measurements. For all samples, bolometric luminosities are consistenly computed from the 2-10 keV luminosities assuming a luminosity-dependent bolometric correction \citep{Duras20}. The best fit parameters of a linear relation in log space of $\log(v_{\rm out})$ = $A(L_{\rm bol})^{B}$ are $A=-3.88\pm1.40$ and $B=0.19\pm0.03$. Overall, the \sub data fit right between the low- and the high-$z$ data. We find a Kendall's (rank) correlation coefficient of $\tau=0.45$ with a null probability of $p_{\rm null}=1.8\times10^{-7}$. The strength and slope of this correlation is partially driven by the 6 data-points in \citetalias{Chartas21}, with $\log(\lbol/\ergs)\gtrsim48$, that may be affected by additional uncertainties associated with the magnification factor due to lensing. 

The low- and high-$z$ fit correlation in \citetalias{Chartas21} (see their Table\,10) returned a slope of $B=0.20\pm0.03$, a correlation coefficient of $\tau=0.51$ with a null probability $p_{\rm null}=6.0\times10^{-8}$. Their slope is consistent with our measurement, whereas their coefficient is about $10\%$ higher, which is suggesting that our correlation, with all the 4 samples included, is slightly weaker than in \citetalias{Chartas21}.  A similar correlation was also observed by \citet{Matzeu17b} in the luminous QSO \pds between the \xmm, \nustar and \suzaku observation from 2001--2014. The slope of the correlation in \citet[][i.e., $0.22\pm0.04$]{Matzeu17b} is largely consistent with what we have found here.

Overall, with our result we can conclude that there is a correlation between the outflow velocities and bolometric luminosities within the overall low- intermediate- high-$z$ samples. 
A positive correlation with a slope of 0.5 between outflow velocities and the luminosities of the AGN is what it would be expected in a radiatively-driven wind scenario as the radiation pressure plays a key role in driving the outflows.

The fact that the observed slope is lower than the expected value can be explained in several ways. 
One possible explanation, already suggested in \citetalias{Chartas21}, is that we did not include outflows with velocities $\la10000\kms$, as in \citetalias{Tombesi10}.

Another plausible explanation is that, as the luminosity keep increasing, the inner part of the UFOs detected in \sub might be over-ionised, with weaker absorption features \citep[e.g.,][]{Parker17,Pinto18} leading to their observability being pushed to the outer streamlines. Within this regime the observed velocities, due to their radial dependence, would appear slightly slower and such a physical condition leads to an overall flattening of the slope \citep{Matzeu17b} from the nominal value of $0.5$. The outflows shown in \autoref{fig:v_lbol_plus_subways} have a range of mass outflow rates and black hole masses, so it is not clear that a simple scaling of velocity with luminosity is likely. If instead we assume that all the systems are close to their Eddington luminosities $L_{\rm Edd}$ and the outflows have the Eddington momenta $L_{\rm Edd}/c$ \citep[see e.g. ][and references therein]{King03,KingPounds15} one finds that the velocities should be of order $0.1c$, as observed (see also \autoref{fig:histo_vout_compare}).

In reality, the driving (and launching) mechanism responsible for the observed UFOs are likely the result of a complex interaction between radiation pressure and MHD driving. Indeed, it was previously found, in \citet{Matzeu16} that in the powerful disk-wind observed in \pds in 2013, the radiation pressure alone, imparted from a strong flare, could have not deposit enough kinetic power on the outflowing material and hence suggesting that an additional launching mechanism, such as MHD, was also involved. Decoupling and assessment of each individual contribution remains a challenging subject in disk-wind physics with the current CCD detectors.

\subsection{Strong features around rest-energies of 9\,keV}
\label{sub:Strong features around 9 keV}
In our \sub sample, the \fe absorption line with the highest degree of blueshift was detected in PG\,0947$+396$ (Obs1) at $E_{\rm rest}=9.5\pm0.1\kev$ ($E_{\rm obs}\sim7.9\kev$).

Despite its reasonable significance (see \autoref{Table:basegaussSUBWAYS}), there are a few caveats that could rule out its UFO identification. The energy shift from the H-like iron rest-energy is rather large (even larger if He-like Fe is considered), and corresponds to an outflow velocity of $\vout=-0.30\pm0.01c$, which is, by far, the fastest of the sample. Such result could be (in principle) at odds considering the relatively low-Eddington fraction of this source i.e., $\leddratio\sim0.06$. These kind of outflow velocities are more common in sources that are accreting near or above their Eddington limit, e.g. \pds. In this campaign, a second $\sim60\ks$ observation (PG\,0947$+396$ Obs2) was carried out about 5 months later and no \fe absorption line was detected in the spectra (see \autoref{figapp:sub_ALLOBS_SCAN_RA}). The same applies for the first $20\ks$ \xmm observation in 2001. Having said that, it is not impossible to have such powerful UFO in a low-Eddington regime as other driving mechanisms, such as MHD, could play a key role \citep[e.g., ][]{Fukumura17}. Indeed, further monitoring of this source will shed some light on the presence of an UFO.

Strong residuals in emission at $E_{\rm rest} \sim 9\kev$ are detected in \tmosf and PG1114$+$445\,(0651330801) at the $P_{\rm F}>99\%$ confidence level, which might be associated with a possible high-order \iron transition or perhaps associated with an instrumental calibration artifact. The origin of this feature could be associated with 
Ni\,\textsc{xxvii}\,He$\beta$ $1s \rightarrow 3p$, or a blend thereof, however no lower transitions are observed in the spectra. Such a feature is also observed in PG 1114$+$445, as blueshifted \fexxvi\,Ly$\beta$ ($8.25\kev$). The dominant emission line in the pn background is the Cu\,K$\alpha$ line at 8.04 keV. Weaker surrounding lines include Ni\,K$\alpha$ (7.47 keV), Zn\,K$\alpha$ (8.63 keV) and Cu\,K$\beta$ (8.90\,keV). 

So another possible origin of these spurious features might arise from background subtraction issues during the data reduction process, or even from the detector itself such as in PG 1114$+$445. After a careful check we confirm that the line detections are genuine as no such issues were found. A thorough characterization of the physical properties of the winds responsible for the detected \fe lines will be carried out by using physically motivated models such as \xstar, \xabs, \textsc{wine} \citep{Luminari21} and \xrade \citep[][]{Matzeu22}, and will be presented in a forthcoming \sub paper. 


A second absorption line at $E_{\rm rest}\sim11\kev$ ($E_{\rm obs}\sim9\kev$) is detected in \lbqs at the $P_{\rm F}=99.3\%$ confidence level however, following a $\mathcal{MC}$ procedure the significance of the line drops considerably to $P_{ \mathcal{MC}}=87.2\%$. This discrepancy in the detection can be likely attributed to the lack of data at energies $E_{\rm obs}>10\kev$ being at the edge of the \xmm bandpass, and hence extra care is needed. Nonetheless, this high energy absorption feature is also present during a $\sim50\ks$ \nustar exposure one year later in 2020 (PI Bianchi). A joint analysis focused on of the \xmmnu data of \lbqs will be presented in a companion letter (Matzeu et al., in prep.). 


\section{Summary and Conclusions}
\label{sec:summary and conclusion}


We carried out a systematic search focused on absorption features 
in the \fe band in a sample of 22 (41 observations) luminous ($2\times10^{45}\lesssim L_{\rm bol}/\ergs \lesssim 2\times10^{46}$) active galactic nuclei (AGN) at intermediate redshift ($0.1 \lesssim z \lesssim 0.4$), as part of the large \xmm program \sub. For each \xmm observation, the data reduction was performed by optimising the level of background in order to increase of the signal-to-noise (SNR) between the $4$--$10\kev$ band. Afterwards, an additional and crucial step was the appropriate choice of an optimal spectral binning in order to avoid loss of information for any search of weak features like UFOs. We applied the four spectral binning that are most used in the literature and subsequently after cross-checking them all our results are based on the \kb binning.

The main results are summarised below: 

\begin{itemize}

\item 
We carried out an \xmm broadband analysis between $0.3$--$10\kev$. We find that in 27 out of 41 observations ($\sim65\%$) our targets have intrinsic absorption in the soft X-rays, of which ($\sim70\%$) can be identified as fully covering, mildly ionised (warm) absorbers and 5/27 ($\sim30\%$) are partially covering the \los. \\

\item We then analysed the EPIC spectra of each of the 41 observations by first performing a series of blind-search line scans, in both emission and absorption, focused on the $5$--$10\kev$ band. For the overall \fe emissions, the energies range between $6.2 \la E_{\rm rest}/\rm keV \la 9 $ where $\sim90\%$ are consistent with the \feka core (detected in 20/22 targets) and $\sim36\%$ consistent with \fexxvxxvi (detected in 8/22 targets). Their equivalent width ranges between 10 $\la  EW /\rm eV \la 468$, where the high-end values are due to complex emission features such as in PG 1352$+$183, likely arising from a blend between \fekab and \fexxv. \\

\item For the \fe absorption features we detected 14 absorption lines at the $P_{\rm F} \ga 99\%$ confidence level with energies ranging between $7 \la E_{\rm rest}/\rm keV \la 11 $, where $\sim85\%$ are consistent with \fexxvxxvi and one detection is consistent with an \iron edge. Their equivalent widths are ranging between $-200 \la EW/\rm eV \la -40$, which are in line with what is expected for relatively narrow \fe absorption profiles. \\

\item Thanks to extensive Monte Carlo simulations, we confirmed absorption lines corresponding to highly ionised iron in 7/22 sources. These findings yield an UFO detection fraction of $\sim30\%$ on the total sample, at a $\pmc\gtrsim95\%$ significance level. These features likely correspond to either \fexxvabs and/or \fexxviabs. By using the \fexxvi lab transition as reference energy, we measures outflow velocities ranging between $-0.3 \lesssim v_{\rm out}/c \lesssim -0.05$ with average and median velocities of $\overline{v}_{\rm out}=-0.133c$ and $\Tilde{v}_{\rm out}=-0.110c$.\\

\item In this work we also presented preliminary results of photoionisation modelling of the \iron features detected at the $\pmc\gtrsim95\%$ confidence level, with \xstar. We find median values of   $\log(\Tilde{N}_{\rm H}/\cmsq)\sim23.8$ and $\log(\Tilde{\xi}/{\rm erg\,cm\,s}^{-1})\sim4.8$ for the
column densities and ionization parameter, respectively. \\

\item The measured outflow velocities with \xstar are ranging between $-0.3 \lesssim v_{\rm out}/c \lesssim -0.1$, where the mean/median values are $\overline{v}_{\rm out}\sim-0.144$ and $\Tilde{v}_{\rm out}\sim-0.110$, respectively. Such distribution is largely comparable with the outflow velocities measured with the phenomenological (Gaussian) modelling. Such results confirm that the absorption detected in the \fe band arise from fast highly ionised material with high column density as typically observed in UFOs. \\

\item By comparing our results with previous work, we computed a power-law least-squares fit to the low-$z$ (\citetalias{Tombesi10,Gofford13}), intermediate-$z$ (\sub) and high-$z$ (\citetalias{Chartas21}) data, which show a positive correlation between outflow velocity and bolometric luminosity within the overall low-intermediate-high $z$ samples, with slope $0.19\pm0.03$. Such $v_{\rm out}$--$L_{\rm bol}$ correlation is also observed in \citet{Matzeu17b} with a slope of $0.22\pm0.04$. \\

\end{itemize}

The outcome of this work independently provides further support for the existence of highly ionised matter propagating at mildly relativistic speed, which are expected to play a key role in the self-regulated AGN feeding-feedback loop that shapes galaxies, as shown by hydrodynamical multi-phase simulations \citep[][for a review]{Gaspari20}. These results suggest that the likely dominant driving mechanisms of UFOs is radiation pressure arising in high-accretion regimes. It is important to note that MHD also play a key role in the driving and launching mechanism of disk-winds and future observations at micro-calorimeter resolution will contribute towards distinguishing each component. 

An alternative scenario that has been put forward is that the origin of \fe absorption features can be attributed to a layer of hot gas located at the surface of the accretion disk rather than from an outflowing wind \citep{Gallo13}. Thus the prominent and blueshifted absorption lines are the result of strong relativistic reflection component that dominate the hard X-ray continuum rather than the primary emission. Such model was successfully applied to NLSy1 IRAS13224$-$3809 by \citet{Fabian20}. The unprecedented spectral micro-calorimeter resolution from future UFOs observations, such as \xrism (and \athenaifu), will greatly contribute towards disentangling each of these scenarios including the disk-wind's launching/driving physical mechanism (e.g., \citealt{Giustini12,Fukumura22,Dadinainprep,matzeuinprep}.

\begin{acknowledgements}
GAM and all the italian co-authors acknowledge support and fundings from Accordo Attuativo ASI-INAF n. 2017-14-H.0. MB is supported by the European Union’s Horizon 2020 research and innovation programme Marie Skłodowska-Curie grant No 860744 (BID4BEST). MG acknowledges partial support by HST GO-15890.020/023-A, the {\it BlackHoleWeather} program, and NASA HEC Pleiades (SMD-1726). BDM acknowledges support via Ramón y Cajal Fellowship RYC2018-025950-I. SM is grateful for the NASA ADAP grant 80NSSC20K0438. AL acknowledges support from the HORIZON-2020 grant “Integrated Activities for the High Energy Astrophysics Domain" (AHEAD-2020), G.A. 871158. SRON is supported financially by NWO, the Netherlands Organization for Scientific Research. M.Gi. is supported by the ``Programa de Atracci\'on de Talento'' of the Comunidad de Madrid, grant number 2018-T1/TIC-11733.
We warmly thank Katia Gkimisi and Raffaella Morganti for useful discussions.   

\end{acknowledgements}

\thispagestyle{empty}
\bibliographystyle{aa}
\bibliography{matzeu_references}

\appendix

\section{Signal to noise optimization method}
\label{appsec:SNR Optimization}

By following the \citet{Piconcelli04} optimization method, we maximised the level of background that can be tolerated resulting into an increase of the SNR. This is carried out by testing, through an iterative process, different extraction source radii and for each radius the level of the background, defined as Max\,Background, is derived together with the corresponding SNR. 

For observations that are not affected by background flares, the background level remains stable along the exposure and, consequently, we have no necessity to filter-out any particular time intervals. In this case the Max\,Background level is stable independently from the extraction region radius and the highest SNR is achieved at the maximum allowed extraction radius, that, in our case is $r=40\,\rm arcsec$. Differently, for observations affected by strong background flares, the dimension of the extraction regions regulates the dominance of the source signal over the background and thus, the Max\,Background level changes as a function of the extraction radius itself. For those cases, the SNR depends on the interplay between the relative source dominance that tends to diminish as the source extraction radius increases leading to lower acceptable Max\,Background, and the overall source counts that tend to increase as the extraction radius increases. This interplay, at the end, defines a couple of values for the extraction radius and the Max\,Background for which we have the maximum SNR for our data. Between the extreme cases, we have observations for which there are short and/or weak background flares. In these cases we can appreciate some changes of the Max\,Background, but the best extraction region has the dimensions of maximum allowed radius. It is finally worth noting how the different effective areas among MOS and pn drive different instrument responses to background flares and, thus, different SNR and Max\,Background curves. 


\section{Summary of \sub observations}

In \autoref{Table:Summary_SUBWAYS} we summarise the individual observation details from the 22 \sub targets such as their ID, net exposure time and the corresponding total number of counts for each EPIC detector. We also tabulated the maximum signal-to-noise ratio obtained from our optimization procedure. We discarded all the observations with a total count $\lesssim1500$ in the EPIC-pn.

\begin{table*}
\setlength{\tabcolsep}{0.4pt}
\centering
\caption{(1) Source name; (2) Starting date of observation; (3) Net exposure after background optimization; (4) EPIC optimum SNR in the $4$--$10\kev$ (pn) and $0.3$--$10\kev$ (MOS) band and it is not reported in coadded spectra; (5) EPIC total net count in the $4$--$10\kev$ band.}
\label{Table:Summary_SUBWAYS}
\begin{tabular}{l c c c c c c c c c c c}
		\hline
		\\
\multicolumn{12}{c}{Optimised EPIC Data Reduction:$4$--$10\kev$ Band}\\

\\
Source$^1$  &XMM ObsID &Date$^2$ &\multicolumn{3}{c}{Net\,Exposure$^3$\,(ksec)}   &\multicolumn{3}{c}{maxSNR$^4$} &\multicolumn{3}{c}{Total counts$^5$}\\

        &      &         &     &&&                                                                                             &&&\\


 &            &yyyy-mm-dd          &pn      &MOS\,1 &MOS\,2    &pn        &MOS\,1     &MOS\,2   &pn       &MOS\,1  &MOS\,2\\
\\
\hline
\\
\multicolumn{12}{c}{\sub targets observed during AO18 cycle}\\
\\
PG0052$+$251          &0841480101 &2019-07-15  &$34.4$ &$50.1$ &$50.1$ &$92.3~~$  &$236.2~~$  &$238.9$  &$~~9119$ &$3771$  &$4229$\\
PG0953$+$414          &0841480201 &2020-04-14  &$45.5$ &$50.9$ &$50.9$ &$69.2~~$  &$208.8~~$  &$204.9$  &$~~5244$ &$1869$  &$1897$\\
PG1626$+$554          &0841480401 &2019-05-27  &$51.4$ &$60.9$ &$60.9$ &$72.2~~$  &$205.4~~$  &$206.4$  &$~~5902$ &$2300$  &$2375$\\
PG1202$+$281          &0841480501 &2020-06-29  &$46.8$ &$54.9$ &$57.6$ &$78.6~~$  &$179.1~~$  &$181.8$  &$~~7141$ &$2671$  &$2746$\\
PG1435$-$067          &0841480601 &2019-07-24  &$42.3$ &$59.4$ &$64.3$ &$39.8~~$  &$108.4~~$  &$112.7$  &$~~2022$ &$899$   &$973$\\
SDSS\,J144414$+$0633  &0841480701 &2019-07-28  &$66.3$ &$80.0$ &$82.7$ &$78.1~~$  &$188.1~~$  &$193.9$  &$~~7067$ &$2480$  &$2774$\\
2MASS\,J165315$+$2349 &0841480801 &2020-02-11  &$59.8$ &$78.0$ &$78.6$ &$82.3~~$  &$57.9~~$   &$63.1$   &$~~7652$ &$2641$  &$3039$\\
PG1216$+$069          &0841480901 &2020-06-05  &$63.4$ &$78.6$ &$81.2$ & $65.2~~$ &$164.7~~$  &$167.9$  &$~~4792$ &$1807$  &$1902$\\
\\

\multirow{2}{*}{PG0947$+$396}

&0841481001\,(Obs\,1)& 2019-11-25       &$31.1$&$43.9$&$44.8$   &$32.9~~$  &$100.8~~$  &$105.9$    &$~~1682$  &$789$  &$847$\\
     
&0841482301\,(Obs\,2) &2020-04-17       &$45.7$&$54.1$&$54.1$   &$45.6~~$  &$119.3~~$  &$117.5$    &$~~2436$  &$908$  &$947$\\
  
\\
WISE\,J053756$-$0245  &0841481101  &2020-03-02       &$69.8$&$90.3$&$91.4$  &$57.6~~$  &$112.4~~$  &$113.7$    &$~~3900$ &$1563$ &$1576$\\
HB\,891529$+$050      &0841481301  &2019-08-22       &$75.3$&$92.9$&$93.0$   &$56.3~~$ &$134.1~~$  &$134.6$   &$~~3592$ &$1407$ &$1439$\\

\multirow{1}{*}{PG1307$+$085}

&0841481401   &2020-01-22         &$73.1$&$86.9$&$86.8$    &   &    &                        &$10547$  &$3971$ &$3787$\\

PG1425$+$267      &0841481501       &2020-02-05       &$73.3$&$89.9$&$90.2$    &$63.7~~$  &$152.2~~$  &$152.0$    &$~~4708$ &$2013$ &$2004$\\

PG1352$+$183      &0841481601       &2020-01-26       &$74.4$&$93.1$&$93.0$    &$58.3~~$  &$173.6~~$   &$171.8$   &$~~3951$ &$1646$ &$1685$\\ 

2MASS\,J105144$+$3539 &0841481701  &2020-05-26       &$80.3$&$95.3$&$93.0$    &$41.6~~$  &$59.1~~$   &$60.5$     &$~~2139$ &$811$ &$859$\\
2MASS\,J0220$-$0728 &0841481901     &2020-02-03       &$77.1$&$101.9$&$102.2$  &$57.5~~$  &$128.1~~$  &$130.8$    &$~~3856$ &$1661$ &$1678$\\
LBQS\,1338$-$0038 &0841482101       &2020-02-01       &$91.6$&$114.3$&$114.4$  &$81.9~~$  &$202.9~~$  &$201.3$    &$~~7775$ &$3085$ &$3139$\\
\\
\hline

\\
\multicolumn{12}{c}{\sub targets observed prior AO18 cycle}\\
\\

\multirow{3}{*}{PG0804$+$761}      
						& 0102040401 &2000-11-04    &$2.33$ &$6.56$ &$6.56$  &        &$138.9$  &$141.0$  &$1823$ &$845$ &$889$ \\ 

						& 0605110101    						   &2010-03-10    &$29.2$ &$27.6$ &$27.6$  &$78.3$  &$270.2$  &$270.1$  &$8426$ &$3070$ &$3316$ \\ 

						& 0605110201                             &2010-03-12    &$25.2$ &$35.1$ &$35.1$  &$67.1$  &$270.2$  &$238.0$  &$5741$ &$3006$ &$2547$ \\

\\

PG1416$-$129        & 0203770201  &2004-07-14   &$44.3$ &$48.9$ &$49.0$   &$68.9$  &$133.8$  &$134.5$  &$5693$  &$1997$  &$2037$  \\

\\

\multirow{2}{*}{PG1402+261}        

						& 0400200101  &2006-12-21   &$28.0$ &$16.6$ &$16.6$   &$39.7$  &       &       &$1752$  &$674$   &$683$  \\


						& 0830470101  &2018-12-17   &$67.2$ &$75.5$ &$75.5$   &$59.7$  &$201.0$  &$197.8$  &$3894$  &$1347$  &$1531$    \\

\\

\multirow{6}{*}{HB89\,1257$+$286}  
						& 0204040101  &2004-06-06    &$72.8$ &$85.3$ &$85.4$  &$70.3$  &$185.2$  &$186.6$  &$5374$ &$2125$ &$2011$ \\

						& 0204040201  &2004-06-18    &$79.1$ &$91.9$ &$93.6$  &$57.7$  &$166.2$  &$166.6$  &$3880$ &$1690$ &$1613$  \\ 

						& 0204040301  &2004-07-12    &$77.4$ &$96.3$ &$96.4$  &$60.5$  &$162.4$  &$163.5$  &$4341$ &$1750$ &$1856$  \\ 

						& 0304320201  &2005-06-27    &$64.1$ &$76.7$ &$78.6$  &$61.4$  &$152.7$  &$155.8$  &$4211$ &$1605$ &$1551$ \\ 

						& 0304320301  &2005-06-28    &$46.2$ &$54.1$ &$54.3$  &$45.3$  &$119.5$  &$123.1$  &$2498$ &$953$ &$1001$ \\ 

						& 0304320801  &2006-06-06    &$47.5$ &$62.3$ &$62.5$  &$54.0$  &$154.1$  &$154.0$  &$3539$ &$1441$ &$1546$ \\ 
\\

\multirow{11}{*}{PG1114$+$445}        
    							                                          
 						 &0109080801 &2002-05-14   &$37.0$ &$42.2$ &$42.2$   &$59.2$   &$93.4$   &$92.4$    &$3787$  &$1351$  &$1293$ \\ 

						 &0651330101 &2010-05-19   &$24.5$ &$29.4$ &$30.0$   &$41.9$   &$54.9$   &$55.4$    &$1922$  &$682$   &$778$\\ 


						 &0651330301  &2010-05-23   &$30.3$ &$34.3$ &$34.3$   &$32.2$   &$50.2$   &$51.2$    &$1802$  &$626$   &$627$\\

 						 &0651330401  &2010-06-10   &$29.9$ &$38.0$ &$38.5$   &$36.2$   &$63.7$   &$65.5$    &$1861$  &$853$   &$937$\\

						 &0651330501  &2010-06-14   &$25.4$ &$29.8$ &$29.6$   &$33.1$   &$55.3$   &$55.1$    &$1843$  &$710$   &$715$\\

						 &0651330601  &2010-11-08   &$18.5$ &$30.1$ &$30.1$   &$51.6$   &$82.6$   &$83.4$    &$2838$  &$1289$  &$1248$\\

						 &0651330701  &2010-11-16   &$19.5$ &$26.1$ &$26.1$   &$43.6$   &$68.6$   &$68.6$    &$2137$  &$915$   &$937$  \\

						 &0651330801  &2010-11-18   &$23.6$ &$28.4$ &$29.6$   &$43.9$   &$63.8$   &$63.7$    &$2053$  &$794$   &$857$ \\

						 &0651330901  &2010-11-20   &$22.9$ &$26.9$ &$27.5$   &$45.6$   &$69.3$   &$69.0$    &$2261$  &$906$   &$1000$ \\
 
						 &0651331001  &2010-11-26   &$21.4$ &$25.1$ &$25.0$   &$41.0$   &$59.9$   &$59.7$    &$1914$  &$658$   &$739$ \\

						 &0651331101  &2010-12-12   &$17.1$ &$6.89$ &$6.72$   &$38.7$   &         &        &$1672$  &$553$   &$638$ \\

\hline
\end{tabular}
\end{table*}

\section{Spectral binning}
\label{subapp:Spectral binning}

In this analysis we we were interested in searching narrow and strong absorption features in the EPIC spectra. After maximizing the \sn, an additional and crucial step was to carefully include the appropriate choice of an optimal spectral binning in order to avoid the loss of information in the search of weak features like UFOs. We applied the four spectral binning methods most used in the literature (listed below) and subsequently cross-checked their results:

\begin{enumerate}

	\item \underline{\textit{\texttt{grpmin1}}} - we over-sampled the EPIC-pn and EPIC-MOS 
	resolution by imposing that each energy channel contains a minimum of 1 count through the \textsc{specgroup} task within \sas (statistics: $\mathcal{C}$stat). This resulted in a binning resolution of $\Delta E = 5\ev$ and $15\ev$ per energy bin throughout the $0.3$--$10\kev$ energy band for pn and MOS respectively for all the observations. 

	\item \underline{\textit{\texttt{SN5}}} - the data were binned to ensure a significance of at least $5\sigma$ per energy channel with \textsc{specgroup} (statistics: $\chisq$). For the pn, such binning resulted in a binning resolution of $\Delta E = 5\ev$ at $1\kev$ and $\Delta E\sim20\ev$ at $6.4\kev$, whilst $\Delta E=15\ev$ and $\Delta E\sim60\ev$ for MOS.

	\item \underline{\textit{\texttt{OS3grp20}}} - this binning (also obtained with \specgroup) is commonly adopted in the literature and corresponds to an oversampling approximately $3\times$ of the instrumental FWHM. Additionally, the data are grouped to a minimum of $20$ counts per bin in order to use $\chisq$ statistics. This binning prescription is a safe approach especially in cases where the level of significance of the measured signal is not known a priory. Such binning yielded a resolution of $\Delta E\sim40\ev $, $\sim45\ev$ at $1\kev$ and $\Delta E\sim75\ev$, $80\ev$ at $6.4\kev$ in EPIC-pn and -MOS respectively.

	\item \underline{\textit{Kaastra--Bleeker (\kb)}} - the \citet{KaastraBleeker16} optimal binning option in \ftgroup task within the \heasoft package (statistics: $\mathcal{C}$stat) in order to maximise the signal in the spectra. In this sophisticated method, a variable binning scheme is followed so that each bin resolution matches the CCDs FWHM energy resolution of the EPIC detectors (or at least is not smaller than $1/3$). Depending on the spectra in question, the \kb binning produced a EPIC-pn data resolution ranging between $\Delta E\sim50$--$70\ev$ and $\Delta E\sim100$--$150\ev$ at $1\kev$ and $6.4\kev$ and a EPIC-MOS resolution between $\Delta E\sim40$--$60\ev$ and $\Delta E\sim100$--$140\ev$ at $1$ and $6.4\kev$. 
\end{enumerate}

\section{Broadband continuum modelling}
\label{appsec:Broadband continuum modelling}

\begin{figure*} 
\centering
\caption{Broadband fitting and \fe residuals of the \xmm observations in the \sub sample. The EPIC-pn\,(black), MOS\,1\,(red) and MOS\,2\,(green) spectra the corresponding background spectra (shaded area) and the best-fit model (solid red) are plotted. The individual model components are: absorbed power-law (cyan), black-body low (blue) and high (magenta) $kT$, soft X-ray Gaussian lines (dark green) and scattered power-law (orange). EPIC-MOS are visually binned to $10\sigma$ for clarity.}
\label{figapp:SUB_LDATA}
\includegraphics[scale=0.25]{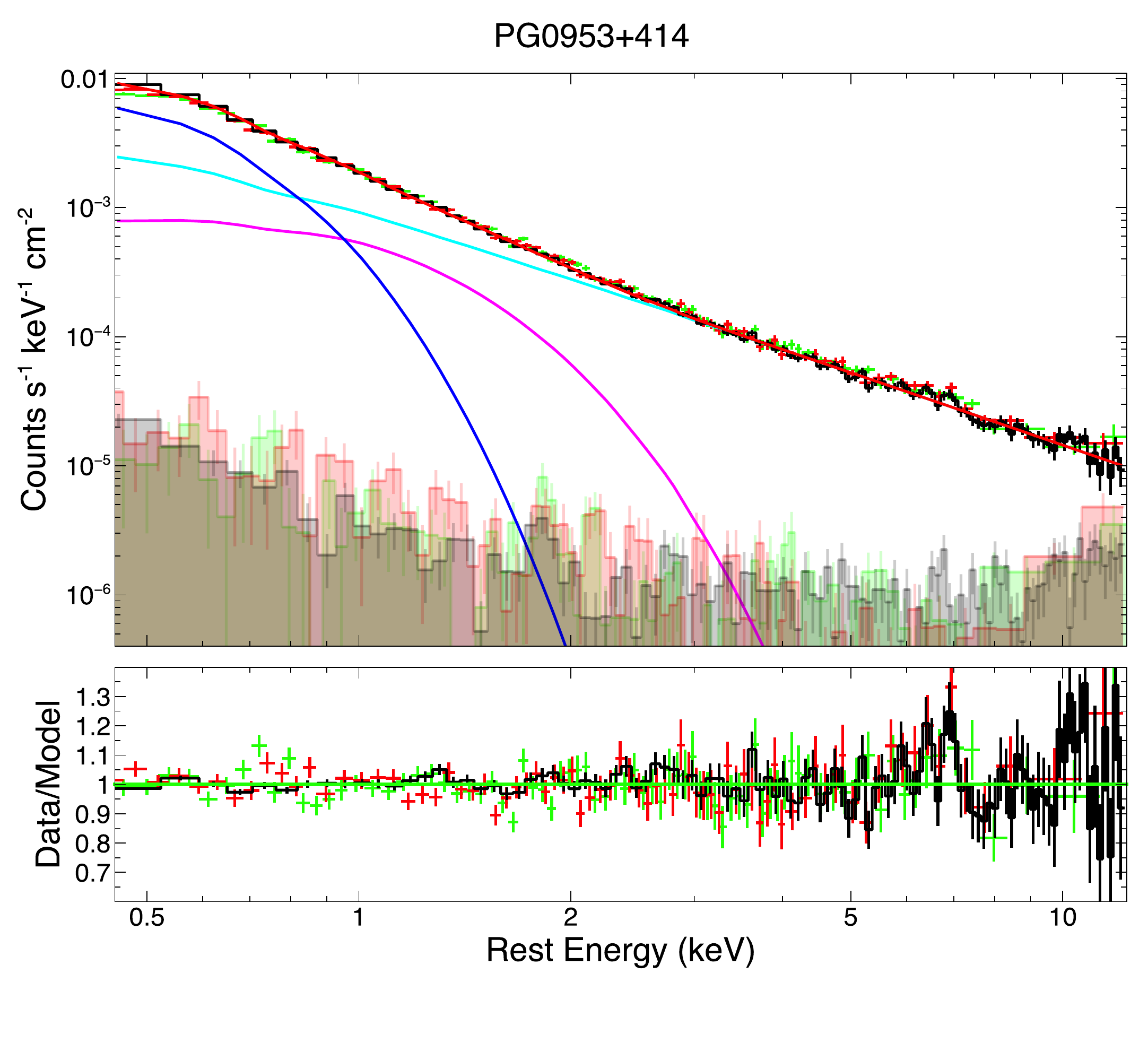}
\includegraphics[scale=0.25]{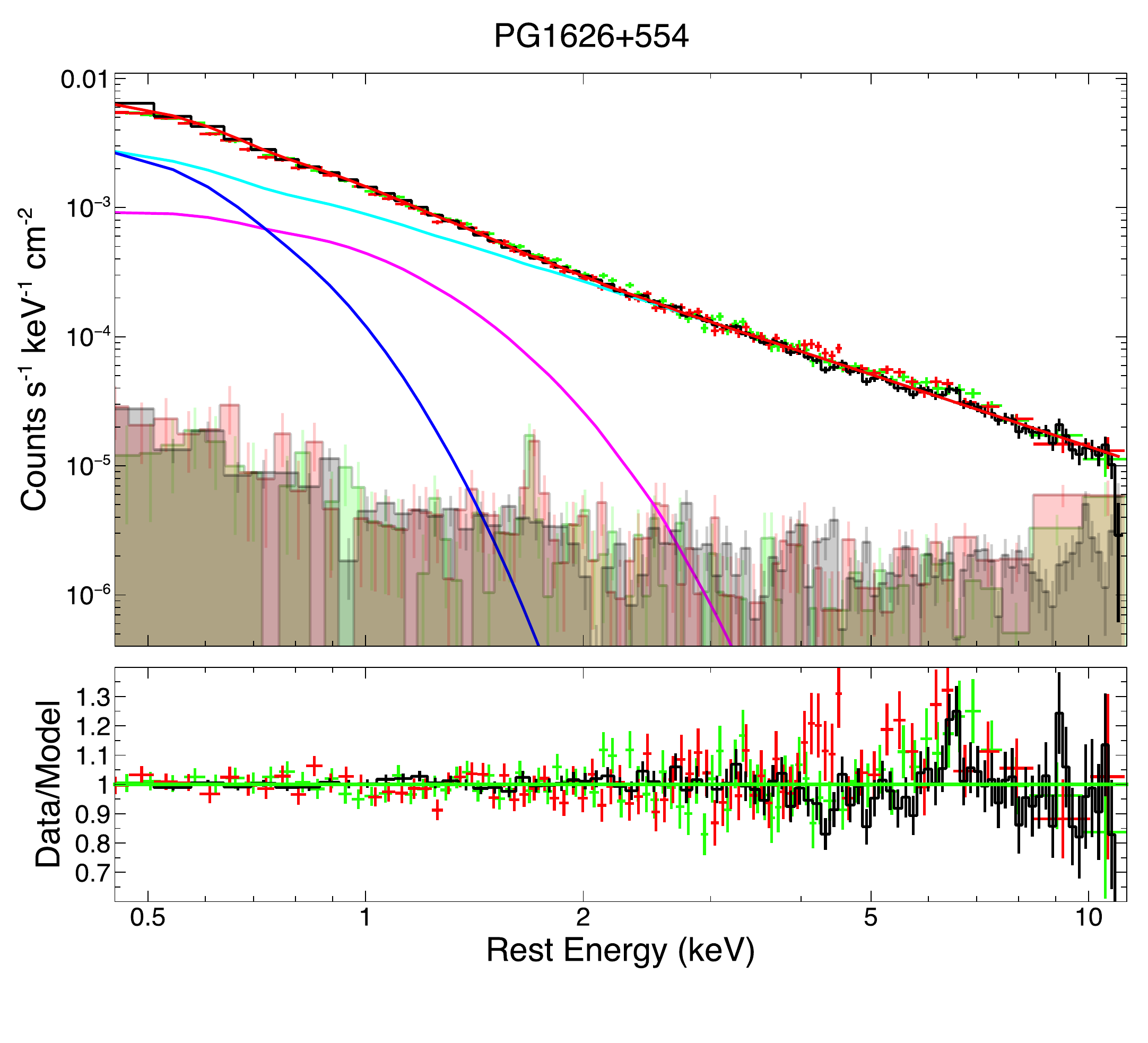}
\includegraphics[scale=0.25]{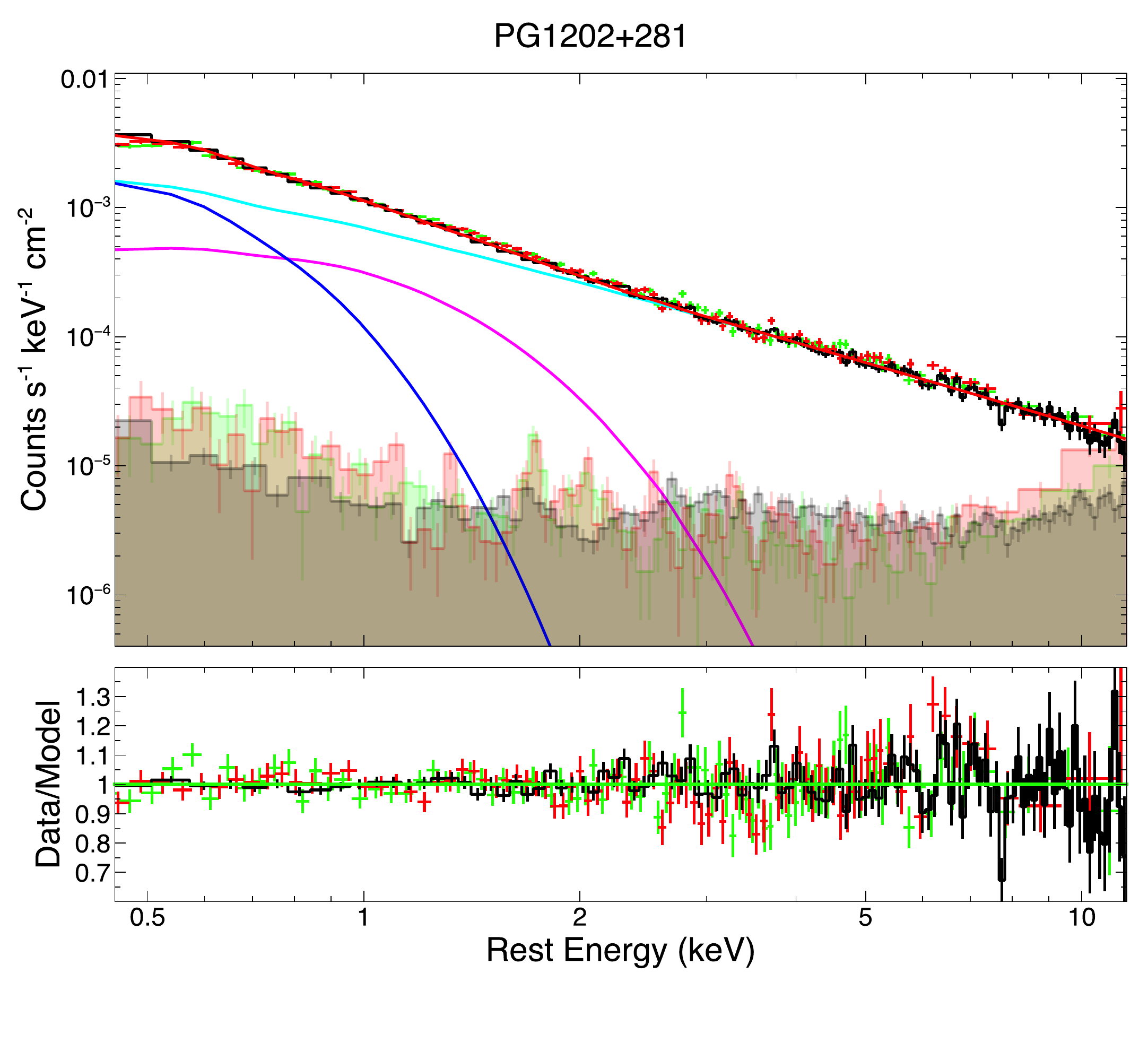}
\includegraphics[scale=0.25]{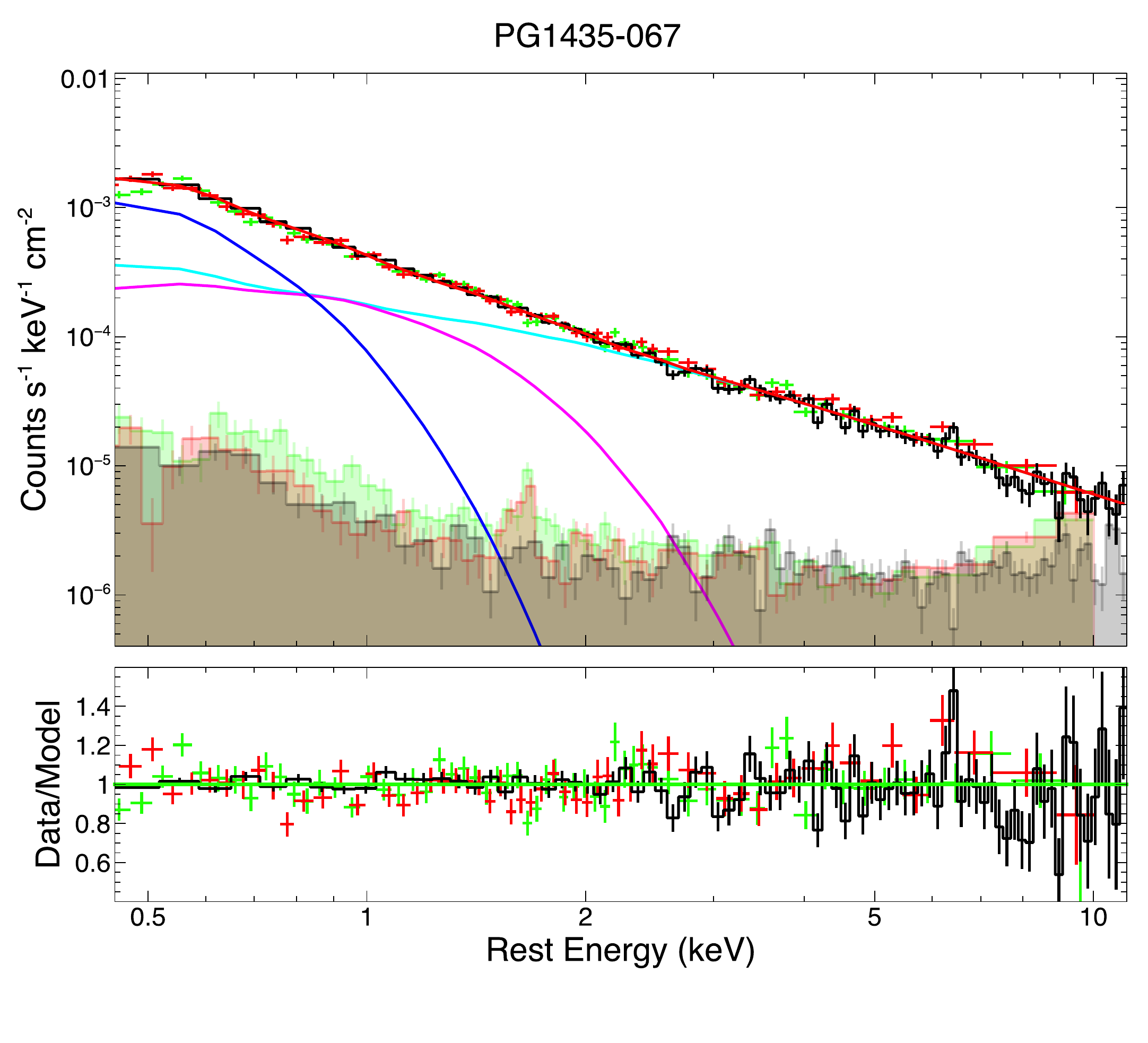}
\includegraphics[scale=0.25]{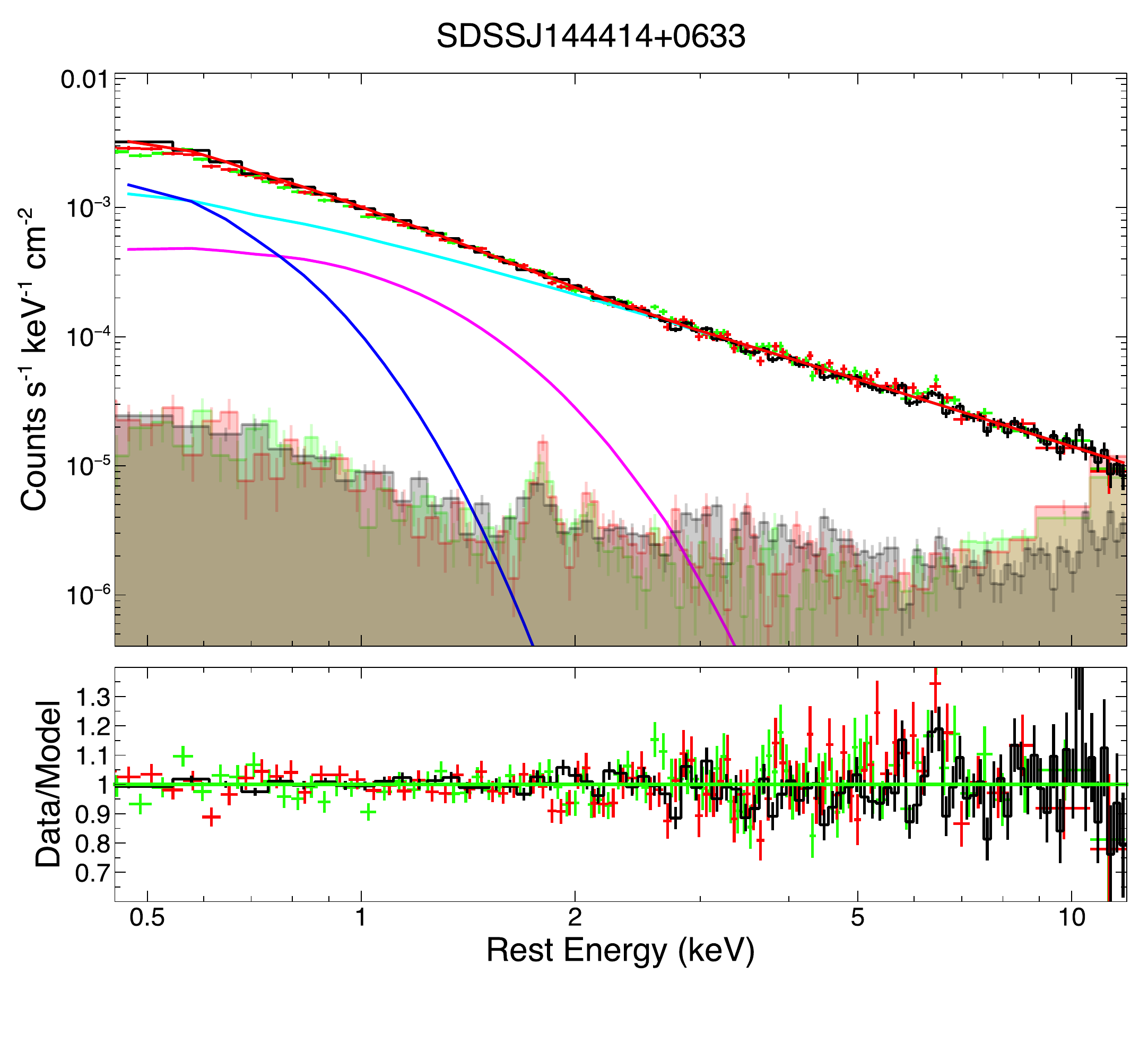}
\includegraphics[scale=0.25]{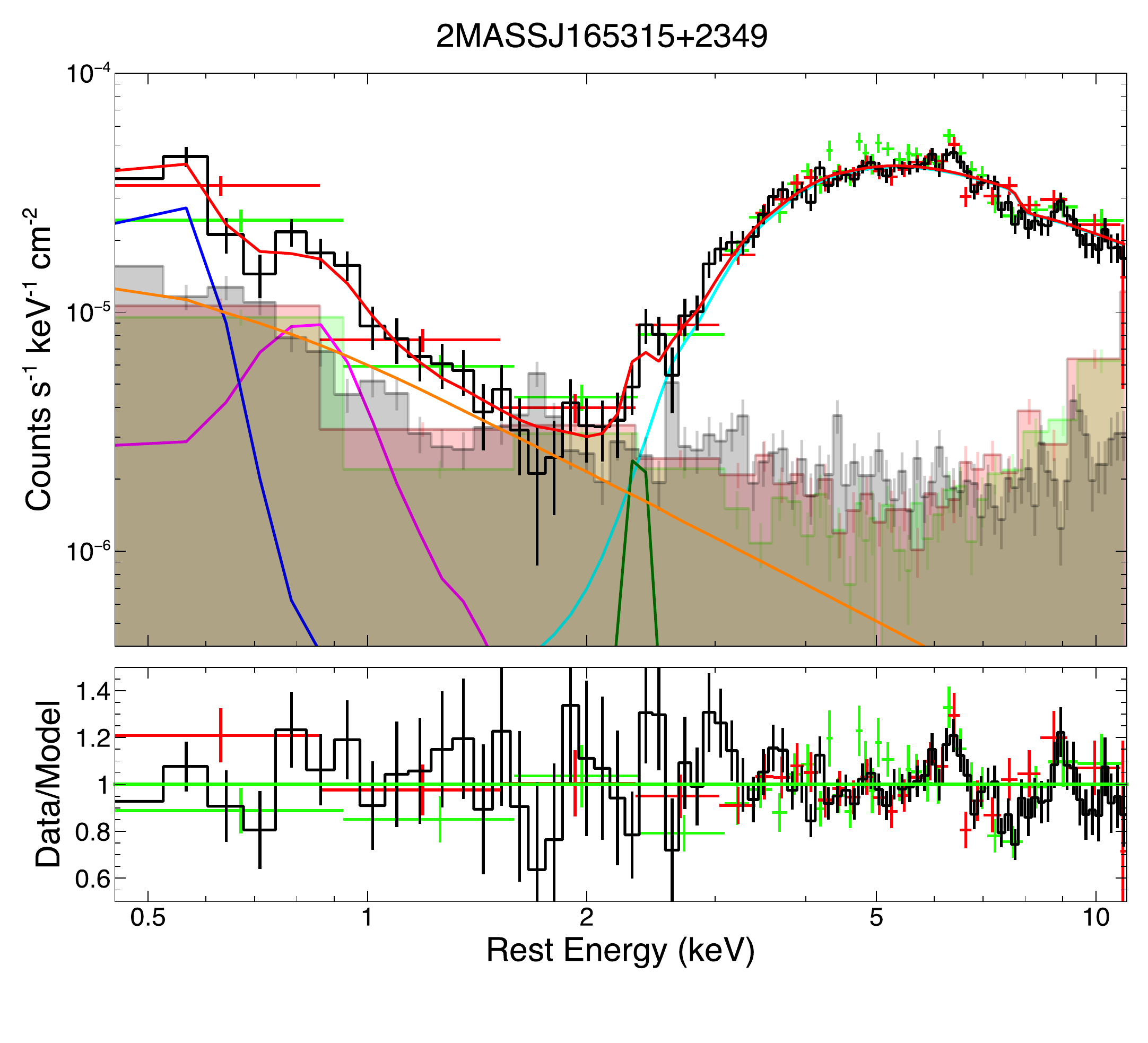}
\includegraphics[scale=0.25]{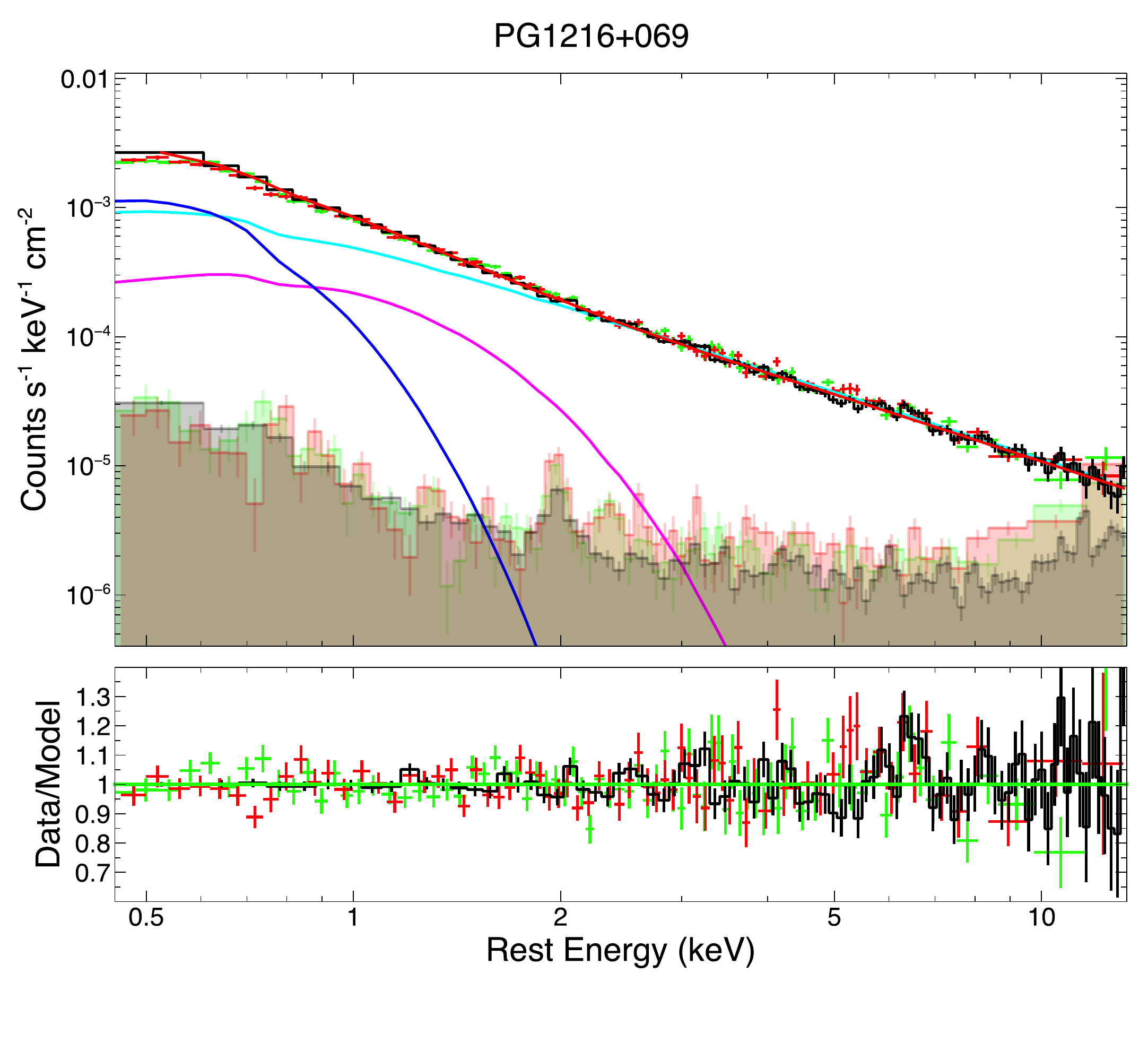}
\includegraphics[scale=0.25]{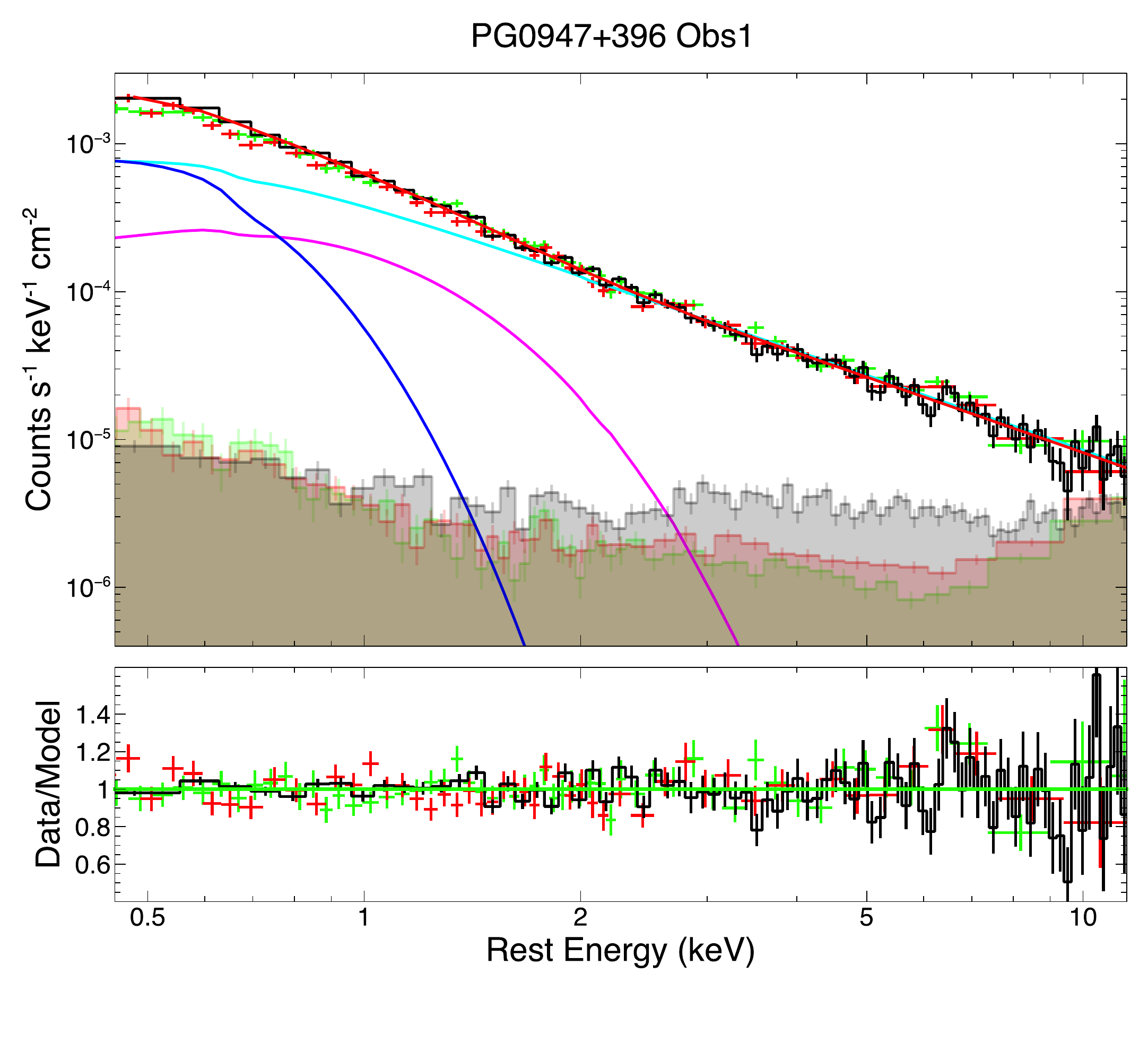}
\includegraphics[scale=0.25]{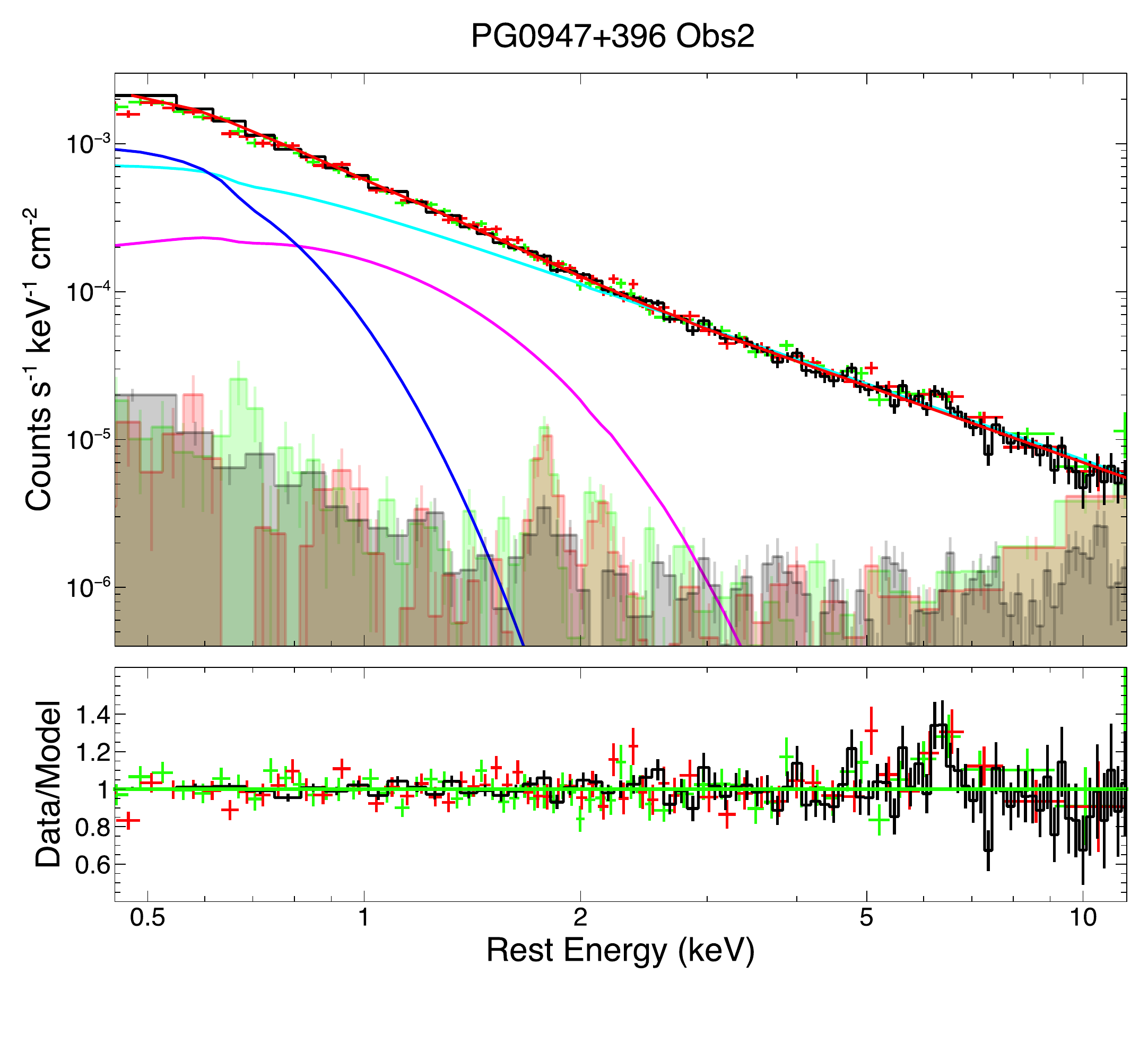}
\includegraphics[scale=0.25]{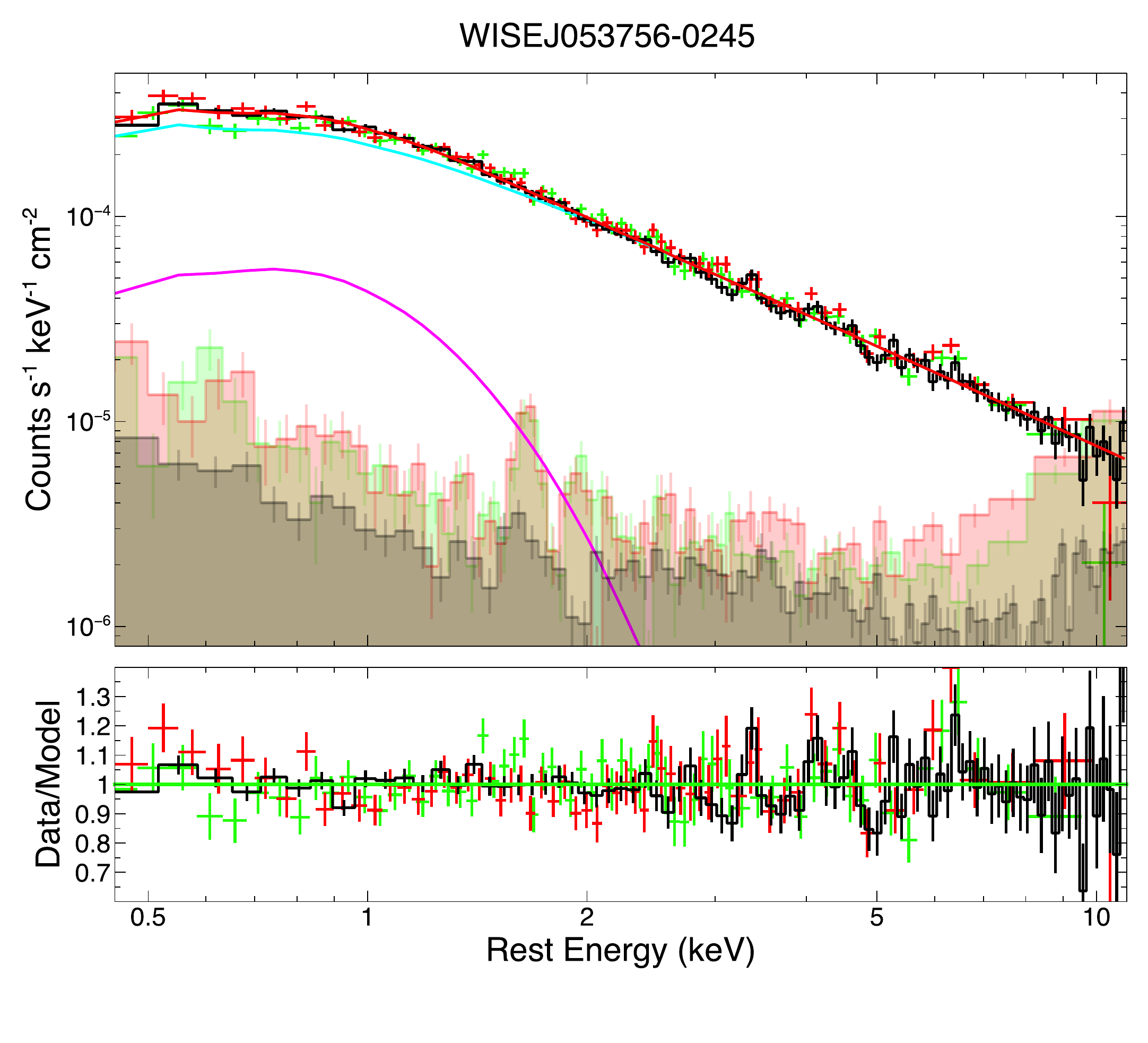}
\includegraphics[scale=0.25]{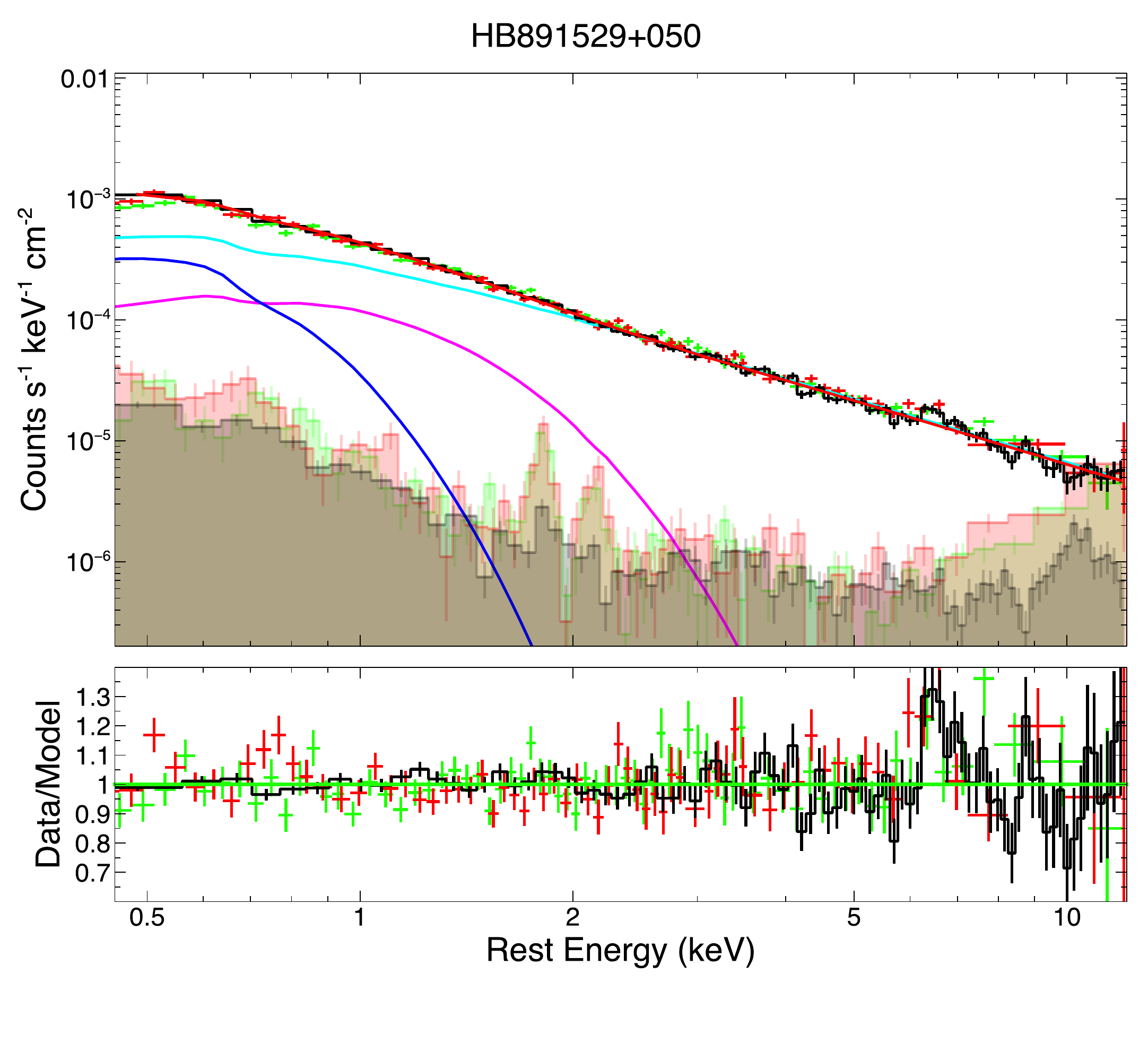}
\includegraphics[scale=0.25]{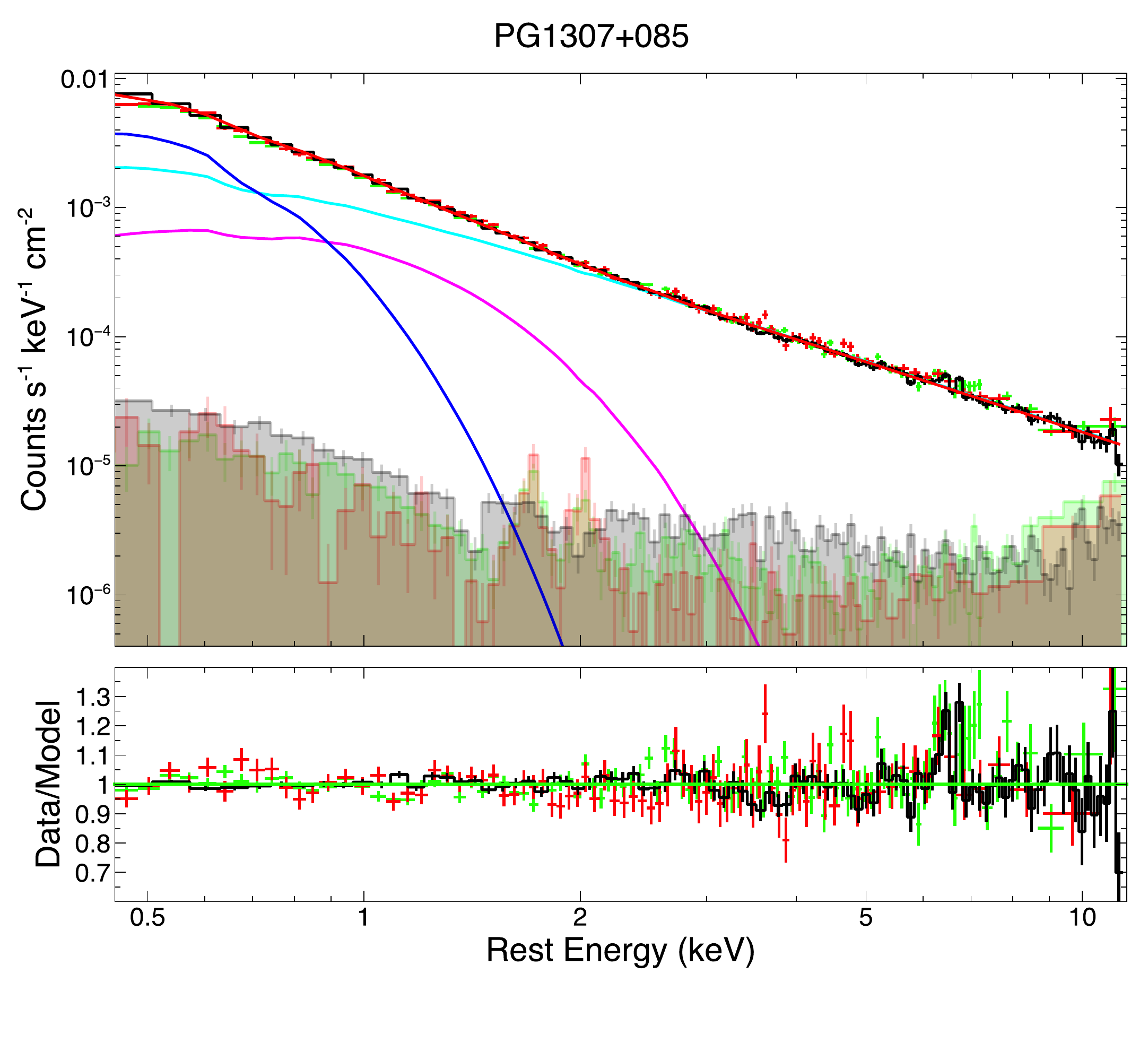}
\includegraphics[scale=0.25]{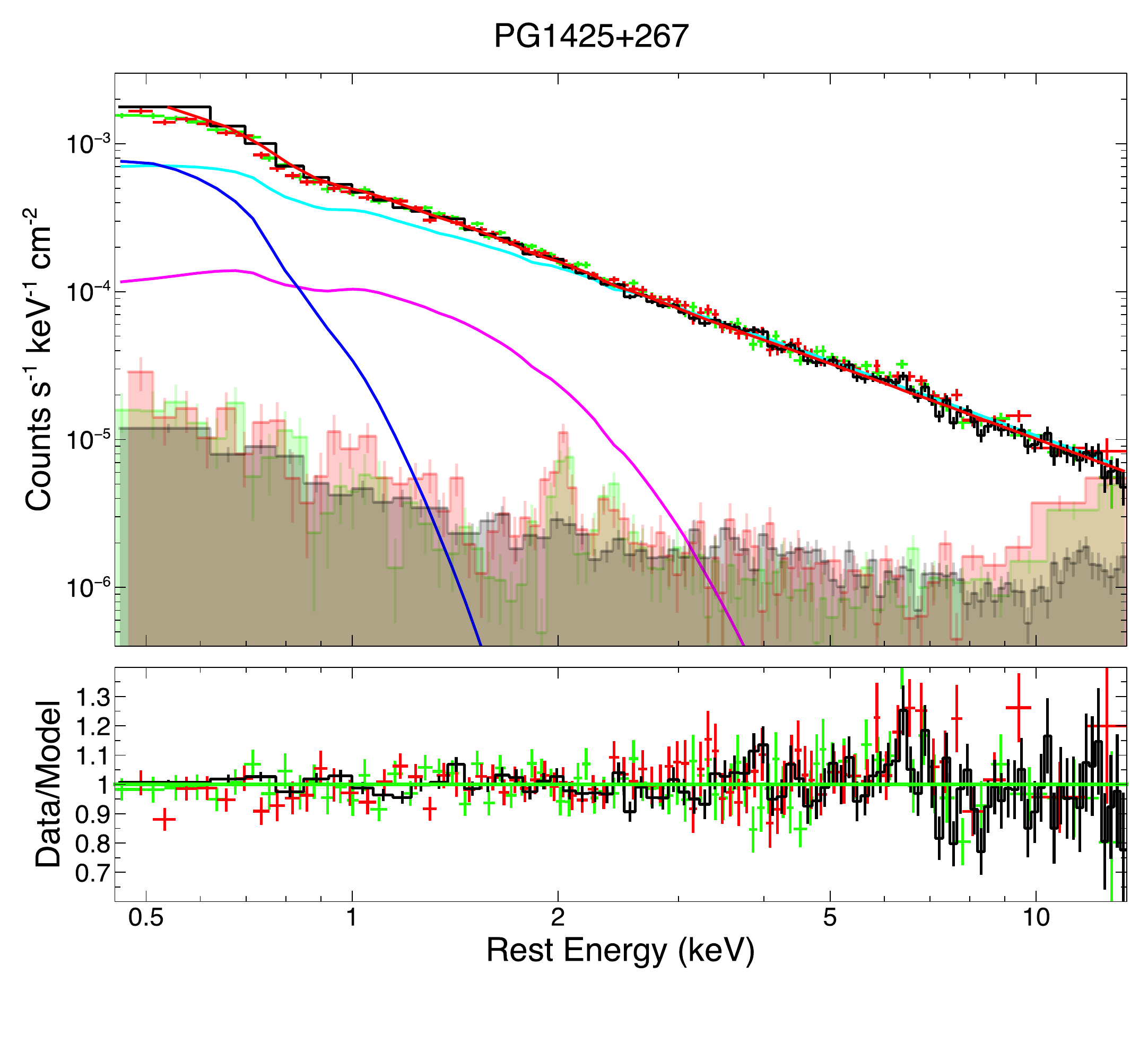}
\includegraphics[scale=0.25]{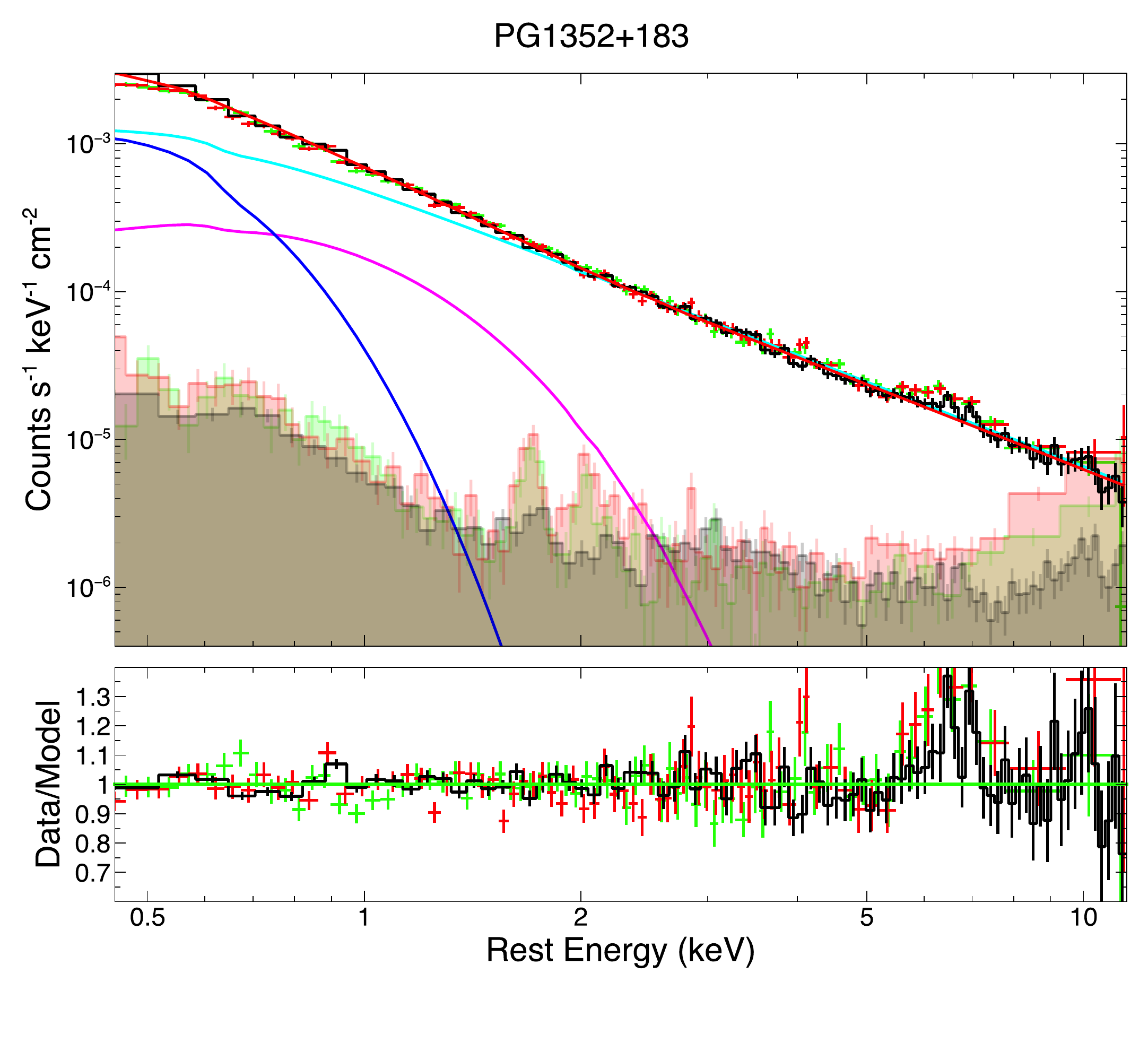}
\includegraphics[scale=0.25]{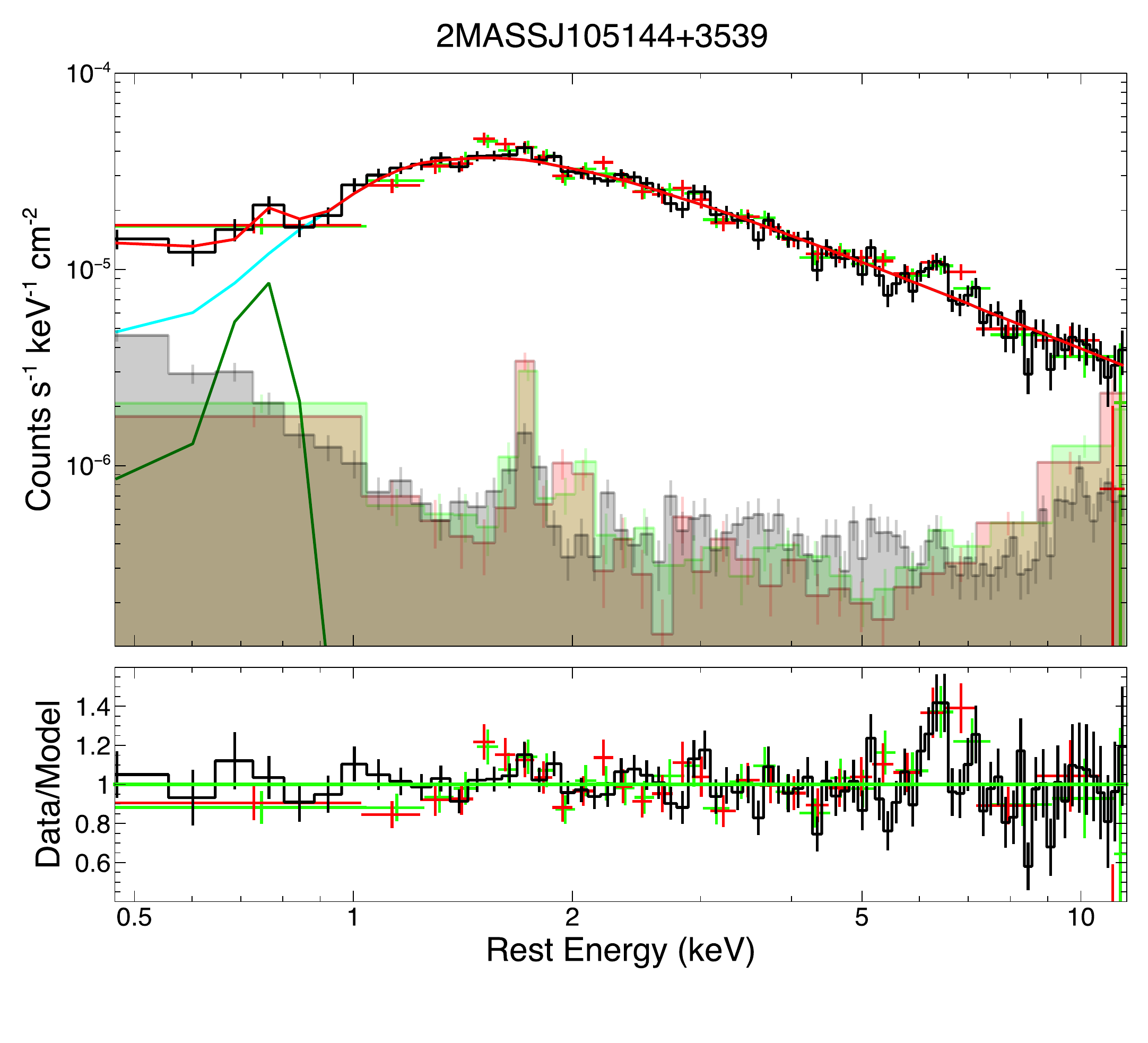}
\end{figure*}

In this work, despite being mainly focused on the \iron band, we carefully modelled the $0.3-10\kev$ continuum with the simplest phenomenological solution see \autoref{eq:baseline}. The summary of the best-fitting continuum parameters and the overall statistics, including for each EPIC detector, are tabulated in \autoref{Table:basecontSUBWAYS}. In \autoref{figapp:SUB_LDATA} we show the individual plots for each \sub observation as per in \autoref{fig:SUB_LDATA}. It is important to note that the final errors for the best-fit continuum models are recalculated and propagated to the final continuum plus \iron emission/absorption lines model.

\begin{landscape}
\begin{table}
\caption{\small{Summary of the phenomenological final best-fitting continuum model parameters of each observation of the \sub sample, including the \iron emission and absorption lines, with their Gaussians model parameters reported separately in \autoref{Table:basegaussSUBWAYS}. The $\nhm$ used in Section\,\ref{sub:Monte Carlo simulations} consist of the best-fitting models reported here minus the Gaussian absorption components. Flux in the $4$--$10\kev$ band in units of $\flux)$. Reduced $\mathcal{C}$stat and number of degrees of freedom ($\nu $) for the final best-fit model. All the warm absorber components are required at $P_{\rm F}>99\%$. $\cstat$ using the number of bins ($\zeta$) in each EPIC detector and the total $\cstat$ with the degree of freedoms ($\nu$) corresponding to the final best-fit model.} \\
\small{$\boxplus$ A narrow ($10\ev$) Gaussian absorption line of rest-frame energy of $E=3.51\pm0.06\kev$ was required at $3\sigma\,(99.7\%)$ confidence level. \\
$\dagger$ Seyfert\,2 target with an intrinsic neutral absorber with column density of $\lognh=23.25_{-0.02}^{+0.02}$ required at $P_{\rm F} VALUE$. The soft X-ray band $<2\kev$ was modelled with two thermal emission components with \texttt{apec} yielding $kT_{\rm low}=145_{-36}^{+34}\ev$, $\rm norm_{\rm low}=1.7_{-0.5}^{+2.3}\times10^{-5}$ and $kT_{\rm high}=732_{-20}^{+21}\ev$, $\rm norm_{\rm low}=2.3_{-0.3}^{+0.9}\times10^{-6}$ with normalizations in unit of $\frac{10^{-14}}{4\pi[D_{A}(1+z)]^{2}}\int n_{e} n_{\rm H} dV$ where $D_{A}$ is the angular diameter distance to the source in cm, $n_e$ and $n_H$ are the electron and H densities (cm$^{-3}$), respectively. A narrow/strong Gaussian emission line of $E=2.35\pm0.04\kev$ was required at $3.6\sigma\,(>99.9\%)$ confidence level.\\
$^{\ddagger}$ Intrinsic neutral absorber with $\lognh=21.83_{-0.01}^{+0.01}$ required at $P_{\rm F} >99.99\%$ and two narrow soft X-ray emission at $E_{\rm rest}=0.54_{-0.03}^{+0.03}\kev$ and $E_{\rm rest}=0.75_{-0.03}^{+0.02}\kev$ both at $P_{\rm F} >99.99\%$}}
\label{Table:basecontSUBWAYS}

\setlength{\tabcolsep}{1.1pt}
\small
\begin{tabular}{l c c c c c c c c c c c r r r c r r}

\hline

\hline
                   \multicolumn{18}{c}{Baseline Model: Continuum parameters}\\

& &\multicolumn{3}{c}{\texttt{power-law}} &\multicolumn{2}{c}{\texttt{bbody}$_{\rm low}$} &\multicolumn{2}{c}{\texttt{bbody}$_{\rm high}$} &\multicolumn{3}{c}{\texttt{XABS}}&\multicolumn{4}{c}{Stats}&\multicolumn{2}{c}{Cross-cal.}\\
\\
Source&XMM ObsID& $\Gamma$  &norm   &$F_{\rm 4-10\,keV,obs}$      &$kT_{\rm low}$  &norm$_{\rm low}$ &$kT_{\rm high}$ &norm$_{\rm high}$ &$\nh$  &$\log(\xi)$  &$f_{\rm cov}$ &$\mathcal{C}_{\rm pn}/\zeta$ &$\mathcal{C}_{\rm mos1}/\zeta$ &$\mathcal{C}_{\rm mos2}/\zeta$ &$\mathcal{C}_{\rm tot}/\nu$&MOS\,1 &MOS\,2  \\

 && &$10^{-3}$     &  $10^{-12}$  &eV               &$10^{-5}$   &eV  &$10^{-5}$   &$\rm 10^{21}\,cm^{-2}$    &erg\,cm\,s$^{-1}$  &$\%$   &&&\\
\hline

\\
PG0052$+$251&0841480101
&$1.76_{-0.03}^{+0.03}$&$1.71_{-0.49}^{+0.48}$ &$4.10_{-0.06}^{+0.06}$ 

&$110_{-4}^{+4}$ &$9.0_{-2.4}^{+4.3}$ &$248_{-4}^{+4}$ &$3.2_{-0.7}^{+0.8}$  

&$6.0_{-3.4}^{+1.4}$ &$-1.1_{-0.4}^{+0.5}$&$45_{-16}^{+17}$ 

&$125.7/102$ &$124.4/119$ &$127.0/127$ &~~$377.1/335$

&~~$1.02_{-0.01}^{+0.01}$ &$1.04_{-0.01}^{+0.01}$\\

\\
PG0953$+$414&0841480201 
&$1.87_{-0.04}^{+0.04}$ &$0.90_{-0.05}^{+0.05}$ &$1.81_{-0.03}^{+0.04}$    

&$115_{-3}^{+3}$ &$8.8_{+0.5}^{+0.6}$ &$280_{-14}^{+14}$  &$2.0_{-0.2}^{+0.2}$  

&$0.88_{-0.22}^{+0.20}$ &$1.1_{-0.2}^{+0.3}$ &$100^{f}$ 

&$109.7/106$ &$109.7/114$ &$152.5/120$ &~~$371.8/324$ 

&~~$1.06_{-0.01}^{+0.01}$ &$1.08_{-0.01}^{+0.01}$\\

\\
PG1626$+$554&0841480401 
&$1.90_{-0.03}^{+0.03}$ &$0.96_{-0.04}^{+0.04}$ &$1.80_{-0.04}^{+0.03}$   

&$102_{-6}^{+5}$ &$4.0_{-0.4}^{+0.4}$ &$225_{-14}^{+14}$  &$1.6_{-0.2}^{+0.2}$  

&$0.86_{-0.32}^{+0.23}$ &$0.84_{-0.23}^{+0.34}$ &$100^{f}$   

&$79.0/90$ &$125.2/115$ &$115.1/120$ &~~$320.1/311$ 

&~~$1.01_{-0.01}^{+0.01}$ &$1.05_{-0.01}^{+0.01}$\\

\\
PG1202$+$281
&0841480501 
&$1.64_{-0.01}^{+0.01}$ &$0.80_{-0.06}^{+0.06}$  &$2.43_{-0.05}^{+0.04}$  

&$115_{-8}^{+7}$ &$2.3_{-0.4}^{+0.9}$ &$265_{-20}^{+21}$  &$2.3_{-0.3}^{+0.9}$  

&$7.4_{-5.1}^{+11.1}$ &$0.57_{-0.53}^{+0.44}$ &$28_{-13}^{+17}$   

&$92.1/108 $ &$112.6/118$ &$127.0/124$ &~~$331.8/333$ 

&~~$1.07_{-0.01}^{+0.01}$ &$1.08_{-0.01}^{+0.01}$\\

\\
PG1435$-$067
&0841480601 

&$1.68_{-0.07}^{+0.06}$ &$0.30_{-0.03}^{+0.03}$  &$0.80_{-0.02}^{+0.03}$

&$103_{-4}^{+4}$ &$1.7_{-0.1}^{+0.1}$ &$283_{-28}^{+26}$  &$0.36_{-0.08}^{+0.08}$
&--  &--  &--    

&$86.8/83$ &$124.5/104$ &$159.4/112$ &~~$370.7/287$ 

&~~$1.02_{-0.02}^{+0.02}$ &$1.02_{-0.02}^{+0.02}$\\

\\
SDSS\,J144414$+$0633$^{\boxplus}$
&0841480701 

&$1.72_{-0.03}^{+0.03}$&$0.65_{-0.22}^{+0.23}$&$1.64_{-0.03}^{+0.03}$  

&$103_{-5}^{+5}$ &$2.3_{-0.1}^{+0.1}$ &$259_{-15}^{+15}$  &$0.96_{-0.10}^{+0.09}$  
&--  &--  &--    
&$117.3/91 $ &$117.4/118$ &$124.2/124$ &~~$358.9/321$ 

&~~$1.02_{-0.01}^{+0.01}$ &$1.02_{-0.01}^{+0.01}$\\

\\
2MASS\,J165315$+$2349$^{\dag}$
&0841480801 

&$1.65_{-0.08}^{+0.09}$ &$1.00_{-0.20}^{+0.20}$ &$2.17_{-0.07}^{+0.04}$ 

&--   &--   &--    &--    

&--   &--   &--     

&$74.2/82 $ &$83.2/100$ &$105.7/107$ &~~$263.1/273$ 

&~~$1.01_{-0.03}^{+0.03}$ &$1.07_{-0.03}^{+0.03}$\\

\\
PG1216$+$069  
&0841480901 
&$1.74_{-0.01}^{+0.01}$ &$0.48_{-0.02}^{+0.02}$ &$1.19_{-0.03}^{+0.02}$ 

&$117_{-7}^{+8}$ &$2.6_{-0.2}^{+0.3}$ &$276_{-20}^{+21}$  &$0.95_{-0.12}^{+0.12}$  

&$1.4_{-0.8}^{+1.1}$ &$2.0_{-0.4}^{+0.3}$ &$100^{f}$   

&$83.2/89 $ &$122.4/114$ &$141.8/121$ &~~$347.4/311$ 

&~~$1.03_{-0.01}^{+0.01}$ &$1.05_{-0.01}^{+0.01}$\\

\\
PG0947$+$396\,(Obs\,1) 
&0841481001&
$1.71_{-0.07}^{+0.06}$ &$0.36_{-0.03}^{+0.03}$  &$0.93_{-0.04}^{+0.03}$  

&$110_{-10}^{+9}$ &$1.3_{-0.2}^{+0.2}$ &$269_{-28}^{+29}$  &$0.64_{-0.11}^{+0.11}$  

&-- &-- & --   

&$96.8/82 $ &$102.0/100$ &$108.2/109$ &~~$307.0/277$ 

&~~$0.99_{-0.02}^{+0.02}$ &$1.03_{-0.02}^{+0.02}$\\

\\
PG0947$+$396\,(Obs\,2)
&0841482301 
&$1.73_{-0.05}^{+0.05}$ &$0.33_{-0.02}^{+0.02}$ &$0.84_{-0.03}^{+0.02}$  

&$108_{-6}^{+6}$ &$1.5_{-0.1}^{+0.1}$ &$274_{-22}^{+22}$  &$0.59_{-0.08}^{+0.08}$  

&-- &-- &--  

&$81.1/84 $ &$112.5/107$ &$99.7/112$ &~~$293.2/293$ 

&~~$1.04_{-0.02}^{+0.02}$ &$1.04_{-0.02}^{+0.02}$\\

\\
WISE\,J053756$-$0245
&0841481101 
&$1.65_{-0.06}^{+0.06}$ &$0.30_{-0.01}^{+0.01}$ &$0.91_{-0.02}^{+0.02}$  

&-- &--  &$228_{-20}^{+21}$  &$0.17_{-0.04}^{+0.04}$  

&-- &-- &-- 

&$96.1/86 $ &$121.8/109$ &$121.1/114$ &~~$339.0/301$ 

&~~$1.06_{-0.02}^{+0.02}$ &$1.04_{-0.02}^{+0.02}$\\

\\
HB\,891529$+$050 
&0841481301 
&$1.75_{-0.05}^{+0.05}$ &$0.31_{-0.02}^{+0.02}$   &$0.77_{-0.03}^{+0.02}$   

&$116_{-14}^{+13}$ &$0.74_{-0.15}^{+0.18}$ &$275_{-20}^{+21}$  &$0.42_{-0.07}^{+0.07}$  

&$0.68_{-0.45}^{+0.42}$ &$1.2_{-0.5}^{+0.7}$ &$100^{f}$   

&$84.3/86 $ &$112.1/113$ &$129.4/118$ &~~$325.7/304$ 

&~~$1.04_{-0.01}^{+0.01}$ &$1.04_{-0.01}^{+0.01}$\\

\\ 
PG1307$+$085
&0841481401 
&$1.82_{-0.02}^{+0.02}$ &$1.10_{-0.04}^{+0.04}$   &$2.28_{-0.03}^{+0.03}$   

&$110_{-3}^{+3}$ &$6.8_{-0.6}^{+0.6}$ &$260_{-11}^{+11}$  &$1.8_{-0.1}^{+0.1}$  

&$0.98_{-0.19}^{+0.20}$ &$0.36_{-0.21}^{+0.21}$ &$100^{f}$   

&$114.8/94 $ &$116.1/119$ &$186.5/126$ &~~$399.4/327$ 

&~~$1.05_{-0.01}^{+0.01}$ &$1.02_{-0.01}^{+0.01}$\\

\\
PG1425$+$267 
&0841481501 

&$1.71_{-0.11}^{+0.11}$ &$0.42_{-0.05}^{+0.05}$ &$1.09_{-0.02}^{+0.02}$ 

&$96_{-6}^{+6}$ &$2.7_{-0.3}^{+0.3}$ &$321_{-20}^{+21}$  &$0.60_{-0.3}^{+0.9}$  

&$7.6_{-1.8}^{+2.1}$ &$2.3_{-0.11}^{+0.11}$ &$100^{f}$   

&$97.8/88 $ &$102.4/115$ &$107.5/120$ &~~$307.6/307$ 

&~~$1.03_{-0.01}^{+0.01}$ &$1.05_{-0.01}^{+0.01}$\\

\\
PG1352$+$183
&0841481601 

&$1.92_{-0.04}^{+0.04}$  &$0.45_{-0.03}^{+0.03}$   &$0.85_{-0.03}^{+0.02}$  

&$98_{-5}^{+4}$ &$1.7_{-0.1}^{+0.1}$ &$244_{-17}^{+17}$  &$0.55_{-0.08}^{+0.08}$  

&-- &-- &-- 

&$119.9/88 $ &$128.8/115$ &$112.9/119$ &~~$361.7/311$ 

&~~$1.03_{-0.01}^{+0.01}$ &$1.04_{-0.01}^{+0.01}$\\

\\
2MASS\,J105144$+$3539$^{\ddagger}$ 
&0841481701
&$1.56_{-0.02}^{+0.02}$& $0.14_{-0.05}^{+0.10}$ &$0.46_{-0.02}^{+0.01}$ 

&-- &-- &--&--

&-- &-- &--   

&$94.7/80 $ &$87.2/94$ &$122.3/102$ &~~$304.3/263$ 

&~~$1.10_{-0.04}^{+0.04}$ &$1.02_{-0.03}^{+0.03}$\\

\\
2MASS\,J0220$-$0728
&0841481901 

&$1.67_{-0.06}^{+0.05}$ &$0.29_{-0.07}^{+0.07}$ &$0.83_{-0.01}^{+0.02}$ 

&-- &-- &$197_{-11}^{+11}$  &$0.21_{-0.04}^{+0.04}$  

&$21.9_{-0.5}^{+0.3}$ &$0.58_{-0.52}^{+0.44}$ &$29_{-13}^{+14}$   

&$79.7/78 $ &$149.7/114$ &$117.6/119$ &~~$347.0/311$ 

&~~$1.03_{-0.02}^{+0.02}$ &$1.05_{-0.02}^{+0.02}$\\

\\
LBQS\,1338$-$0038
&0841482101 
&$1.69_{-0.04}^{+0.03}$ &$0.26_{-0.02}^{+0.02}$ &$1.34_{-0.03}^{+0.03}$    

&$119_{-8}^{+7}$ &$1.7_{-0.2}^{+0.3}$ &$282_{-20}^{+20}$  &$0.71_{-0.09}^{+0.09}$  

&$0.72_{-0.32}^{+0.33}$ &$0.95_{-0.31}^{+0.35}$ &$100^{f}$   

&$101.5/92 $ &$128.8/119$ &$100.1/125$ &~~$330.4/320$ 

&~~$1.02_{-0.01}^{+0.01}$ &$1.03_{-0.01}^{+0.01}$\\

\\
\hline
\end{tabular}
\end{table}
\end{landscape}

\begin{landscape} 
\begin{table}
\setlength{\tabcolsep}{1.1pt}
\small
\begin{tabular}{l c c c c c c c c c c c r r r c r r}

\hline
                   \multicolumn{18}{c}{Baseline Model: Continuum parameters}\\

& &\multicolumn{3}{c}{\texttt{power-law}} &\multicolumn{2}{c}{\texttt{bbody}$_{\rm low}$} &\multicolumn{2}{c}{\texttt{bbody}$_{\rm high}$} &\multicolumn{3}{c}{\texttt{XABS}}&\multicolumn{4}{c}{Stats}&\multicolumn{2}{c}{Cross-cal.}\\
\\
Source&XMM ObsID& $\Gamma$  &norm   &$F_{\rm 4-10\,keV,obs}$     &$kT_{\rm low}$  &norm$_{\rm low}$ &$kT_{\rm high}$ &norm$_{\rm high}$ &$\nh$  &$\log(\xi)$  &$f_{\rm cov}$ &$\mathcal{C}_{\rm pn}/\zeta$ &$\mathcal{C}_{\rm mos1}/\zeta$ &$\mathcal{C}_{\rm mos2}/\zeta$ &$\mathcal{C}_{\rm tot}/\nu$&MOS\,1 &MOS\,2  \\

 && &$10^{-3}$     &$10^{-12}$    &eV &$10^{-5}$   &eV  &$10^{-5}$   &$\rm  	10^{21}\,cm^{-2}$    &erg\,cm\,s$^{-1}$  &$\%$   &&&\\
\hline

\\

\multirow{3}{*}{PG0804$+$761}

&0102040401
&$1.97_{-0.06}^{+0.05}$ &$3.93_{-0.29}^{+0.28}$ &$6.71_{-0.24}^{+0.21}$

&$93_{-7}^{+6}$ &$3.2_{-0.8}^{+0.8}$ &$239_{-29}^{+27}$ &$0.11_{-0.01}^{+0.01}$  

&-- &--  &--    

&$88.8/92 $ &$137.6/112$ &$131.1/112$ &~~$357.6/302$ 

&~~$1.14_{-0.02}^{+0.02}$ &$1.16_{-0.02}^{+0.02}$ \\

&0605110101
&$1.96_{-0.03}^{+0.03}$ &$3.35_{-0.15}^{+0.15}$ &$5.67_{-0.06}^{+0.07}$

&$92_{-2}^{+2}$ &$2.1_{-0.8}^{+0.7}$ &$252_{-12}^{+11}$ &$0.48_{-0.04}^{+0.04}$  

&-- &--  &--    

&$116.2/106 $ &$153.0/121$ &$131.9/126$ &~~$401.0/333$ 

&~~$1.07_{-0.01}^{+0.01}$ &$1.09_{-0.01}^{+0.01}$ \\

&0605110201
&$2.08_{-0.03}^{+0.03}$ &$3.03_{-0.11}^{+0.11}$ &$4.23_{-0.07}^{+0.08}$

&$85_{-3}^{+3}$ &$1.7_{-0.8}^{+0.7}$ &$200_{-12}^{+12}$ &$0.41_{-0.04}^{+0.04}$  

&-- &--  &--    

&$129.9/103 $ &$133.6/124$ &$139.6/120$ &~~$403.0/30$ 

&~~$1.04_{-0.01}^{+0.01}$ &$1.09_{-0.01}^{+0.01}$ \\

\\

PG1416$-$129$^{\star}$
&0203770201
&$1.74_{-0.02}^{+0.02}$ &$0.69_{-0.18}^{+0.17}$ &$2.28_{-0.04}^{+0.02}$ 

&$153_{-8}^{+7}$ &$1.0_{-0.3}^{+0.2}$ &--  &--  

&$1.20_{-0.05}^{+0.05}$ &$0.80_{-0.27}^{+0.51}$ &$100^{f}$   

&$132.5/109 $ &$120.5/118$ &$136.5/118$ &~~$389.6/335$
&~~$1.53_{-0.01}^{+0.02}$ &$1.06_{-0.01}^{+0.02}$ \\

\\

\multirow{2}{*}{PG1402+261} 
&0400200101
&$2.00_{-0.05}^{+0.05}$ &$0.63_{-0.04}^{+0.04}$ &$1.00_{-0.03}^{+0.03}$ 

&$90_{-5}^{+5}$ &$4.2_{-0.3}^{+0.3}$ &$213_{-18}^{+18}$  &$1.0_{-0.2}^{+0.1}$  

&-- &-- &--   

&$94.7/99 $ &$90.0/106$ &$117.2/108$ &~~$301.9/301$ 
&~~$1.04_{-0.02}^{+0.02}$ &$1.09_{-0.02}^{+0.02}$\\

&0830470101
&$2.16_{-0.05}^{+0.05}$ &$0.79_{-0.04}^{+0.04}$ &$0.90_{-0.02}^{+0.02}$ 

&$88_{-5}^{+5}$ &$3.9_{-0.3}^{+0.3}$ &$217_{-13}^{+13}$  &$1.1_{-0.1}^{+0.1}$  

&-- &-- &--   

&$122/104 $ &$136.4/116$ &$107.0/121$ &~~$365.3/329$ 
&~~$1.03_{-0.01}^{+0.01}$ &$1.05_{-0.01}^{+0.01}$\\

\\

\multirow{6}{*}{HB89\,1257$+$286~~~~} 
&0204040101
&$1.83_{-0.02}^{+0.02}$ &$0.59_{-0.17}^{+0.47}$ &$1.20_{-0.03}^{+0.02}$

&$102_{-4}^{+5}$ &$2.1_{-0.3}^{+0.5}$ &$229_{-15}^{+18}$ &$0.89_{-0.20}^{+0.39}$  

&$6.00_{-3.71}^{+4.33}$ &$-0.14_{-0.63}^{+0.40}$ &$35_{-6}^{+15}$   

&$117.2/98 $ &$121.7/119$ &$137.5/120$ &~~$376.4/322$ 

&~~$1.03_{-0.01}^{+0.01}$ &$1.05_{-0.01}^{+0.01}$ \\

&0204040201
&$1.83_{-0.04}^{+0.04}$ &$0.41_{-0.03}^{+0.02}$ &$0.87_{-0.03}^{+0.02}$

&$105_{-5}^{+5}$ &$2.0_{-0.3}^{+0.4}$ &$243_{-20}^{+20}$ &$0.56_{-0.09}^{+0.08}$  

&$1.00_{-0.35}^{+4.32}$ &$-0.27_{-0.45}^{+0.48}$ &$100^{f}$   

&$111.0/96 $ &$151.3/117$ &$157.8/117$ &~~$420.1/318$ 

&~~$1.03_{-0.01}^{+0.01}$ &$1.06_{-0.01}^{+0.01}$ \\

&0204040301
&$1.87_{-0.09}^{+0.08}$ &$0.52_{-0.08}^{+0.08}$ &$1.00_{-0.04}^{+0.02}$

&$100_{-7}^{+7}$ &$2.0_{-0.9}^{+0.9}$ &$222_{-19}^{+24}$ &$1.2_{-0.7}^{+1.1}$  

&$13.35_{-0.83}^{+0.70}$ &$-0.52_{-0.83}^{+0.67}$  &$52_{-27}^{+15}$   

&$85.4/96 $ &$165.7/119$ &$130.7/119$ &~~$381.8/319$ 

&~~$1.05_{-0.01}^{+0.01}$ &$1.05_{-0.01}^{+0.01}$ \\

&0304320201
&$1.78_{-0.05}^{+0.05}$ &$0.45_{-0.03}^{+0.03}$ &$1.05_{-0.03}^{+0.02}$

&$113_{-7}^{+6}$ &$1.3_{-0.2}^{+0.3}$ &$256_{-29}^{+27}$ &$0.47_{-0.09}^{+0.09}$  

&$0.88_{-0.35}^{+0.29}$ &$0.60_{-0.45}^{+0.31}$  &$100^{f}$   

&$93.9/94 $ &$104.5/115$ &$91.4/116$ &~~$289.9/312$ 

&~~$1.03_{-0.01}^{+0.01}$ &$1.05_{-0.01}^{+0.01}$ \\

&0304320301
&$1.75_{-0.06}^{+0.06}$ &$0.39_{-0.03}^{+0.07}$ &$0.93_{-0.03}^{+0.02}$

&$105_{-7}^{+6}$ &$0.96_{-0.11}^{+0.11}$ &$266_{-25}^{+24}$ &$0.43_{-0.08}^{+0.08   }$  

&-- &--  &--   

&$78.7/91$ &$107.4/112$ &$119.9/113$ &~~$305.1/306$ 

&~~$1.02_{-0.21}^{+0.02}$ &$1.05_{-0.02}^{+0.02}$ \\

&0304320801
&$1.82_{-0.05}^{+0.05}$ &$0.58_{-0.04}^{+0.04}$ &$1.25_{-0.03}^{+0.03}$

&$100_{-6}^{+5}$ &$0.51_{-0.3}^{+0.9}$ &$262_{-23}^{+22}$ &$0.34_{-0.18}^{+0.18}$  

&-- &--  &--    

&$101.2/93 $ &$100.0/112$ &$110.7/112$ &~~$311.9/307$ 

&~~$1.03_{-0.01}^{+0.01}$ &$1.03_{-0.01}^{+0.01}$ \\
\\

\multirow{22}{*}{PG1114$+$445~~~~} 

&\multirow{2}{*}{0109080801} 
&\multirow{2}{*}{$1.98_{-0.03}^{+0.03}$} &\multirow{2}{*}{$1.09_{-0.02}^{+0.02}$} &\multirow{2}{*}{$1.54_{-0.03}^{+0.04}$}    

&\multirow{2}{*}{$153_{-17}^{+12}$} &\multirow{2}{*}{$1.8_{-0.8}^{+0.5}$} &\multirow{2}{*}{--}  &\multirow{2}{*}{--} 

&$73.01_{-0.27}^{+0.20}$&$3.00_{-0.07}^{+0.03}$ &\multirow{2}{*}{$100^{f}$}

&\multirow{2}{*}{$118.7/104$} &\multirow{2}{*}{$110.5/108$} &\multirow{2}{*}{$122.0/99$} &\multirow{2}{*}{~~$351.1/297$} 
&\multirow{2}{*}{~~$1.09_{-0.02}^{+0.02}$} &\multirow{2}{*}{$1.04_{-0.02}^{+0.02}$}\\
&&&&&&&&&$16.69_{-0.04}^{+0.04}$&$1.03_{-0.08}^{+0.04}$ &\\


&\multirow{2}{*}{0651330101} 
&\multirow{2}{*}{$1.71_{-0.12}^{+0.15}$} &\multirow{2}{*}{$0.58_{-0.12}^{+0.20}$} &\multirow{2}{*}{$1.30_{-0.03}^{+0.04} $}    

&\multirow{2}{*}{$110_{-10}^{+11}$} &\multirow{2}{*}{$2.1_{-1.1}^{+1.5}$} &\multirow{2}{*}{--}  &\multirow{2}{*}{--} 

&$91.65_{-31.72}^{+50.22}$&$2.78_{-0.13}^{+0.27}$ &\multirow{2}{*}{$100^{f}$}

&\multirow{2}{*}{$88.0/95$} &\multirow{2}{*}{$92.8/97$} &\multirow{2}{*}{$107.5/101$} &\multirow{2}{*}{~~$288.3/278$} 
&\multirow{2}{*}{~~$1.11_{-0.04}^{+0.04}$} &\multirow{2}{*}{$1.11_{-0.04}^{+0.04}$}\\
&&&&&&&&&$16.85_{-0.66}^{+0.86}$&$0.70_{-0.11}^{+0.18}$ &\\

&\multirow{2}{*}{0651330301}
&\multirow{2}{*}{$1.84_{-0.11}^{+0.39}$} &\multirow{2}{*}{$0.76_{-0.15}^{+0.31}$} &\multirow{2}{*}{$1.23_{-0.11}^{+0.06}$}    

&\multirow{2}{*}{$122_{-10}^{+13}$} &\multirow{2}{*}{$3.2_{-0.1}^{+1.3}$} &\multirow{2}{*}{--}  &\multirow{2}{*}{--} 

&$124.97_{-3.73}^{+6.41}$&$3.00_{-0.09}^{+0.03}$
&\multirow{2}{*}{$100^{f}$}

&\multirow{2}{*}{$86.0/96$} &\multirow{2}{*}{$92.2/91$} &\multirow{2}{*}{$112.2/95$} &\multirow{2}{*}{~~$290.3/268$}
&\multirow{2}{*}{~~$1.12_{-0.04}^{+0.04}$} &\multirow{2}{*}{$1.07_{-0.04}^{+0.04}$}\\
&&&&&&&&&$27.96_{-0.34}^{+0.34}$&$0.85_{-0.14}^{+0.16}$ &\\

&\multirow{2}{*}{0651330401}
&\multirow{2}{*}{$1.91_{-0.14}^{+0.12}$} &\multirow{2}{*}{$0.90_{-0.21}^{+0.23}$} &\multirow{2}{*}{$1.42_{-0.06}^{+0.07}$}    

&\multirow{2}{*}{$119_{-12}^{+14}$} &\multirow{2}{*}{$2.3_{-0.1}^{+1.0}$} &\multirow{2}{*}{--}  &\multirow{2}{*}{--} 

&$79.90_{-3.08}^{+5.02}$&$3.01_{-0.10}^{+0.08}$&\multirow{2}{*}{$100^{f}$}

&\multirow{2}{*}{$113.1/96$} &\multirow{2}{*}{$103.1/96$} &\multirow{2}{*}{$106.4/106$} &\multirow{2}{*}{~~$323.1/288$} 

&\multirow{2}{*}{~~$1.09_{-0.03}^{+0.03}$} &\multirow{2}{*}{$1.07_{-0.03}^{+0.03}$}\\
&&&&&&&&&$19.07_{-0.32}^{+0.32}$&$0.87_{-0.13}^{+0.15}$ &\\

&\multirow{2}{*}{0651330501}
&\multirow{2}{*}{$1.90_{-0.09}^{+0.12}$} &\multirow{2}{*}{$0.97_{-0.17}^{+0.25}$} &\multirow{2}{*}{$1.48_{-0.11}^{+0.06}$}    

&\multirow{2}{*}{$89_{-5}^{+5}$} &\multirow{2}{*}{$1.9_{-0.6}^{+1.4}$} &\multirow{2}{*}{--}  &\multirow{2}{*}{--} 

&$98.80_{-33.58}^{+197.53}$&$3.00_{-0.15}^{+0.22}$
&\multirow{2}{*}{$100^{f}$}

&\multirow{2}{*}{$95.0/96$} &\multirow{2}{*}{$115.8/97$} &\multirow{2}{*}{$127.4/101$} &\multirow{2}{*}{~~$338.2/286$} 
&\multirow{2}{*}{~~$1.07_{-0.04}^{+0.04}$} &\multirow{2}{*}{$1.04_{-0.04}^{+0.04}$}\\
&&&&&&&&&$15.81_{-0.70}^{+0.53}$&$-0.17_{-0.20}^{+0.20}$ &\\

&\multirow{2}{*}{0651330601}
&\multirow{2}{*}{$1.88_{-0.07}^{+0.07}$} &\multirow{2}{*}{$1.38_{-0.43}^{+0.45}$} &\multirow{2}{*}{$2.31_{-0.09}^{+0.05}$}    

&\multirow{2}{*}{$139_{-14}^{+15}$} &\multirow{2}{*}{$3.29_{-1.1}^{+1.1}$} &\multirow{2}{*}{--}  &\multirow{2}{*}{--} 

&$66.08_{-28.45}^{+33.18}$&$2.95_{-0.09}^{+0.08}$
&\multirow{2}{*}{$100^{f}$}

&\multirow{2}{*}{$112.0/98$} &\multirow{2}{*}{$135.3/106$} &\multirow{2}{*}{$122.3/109$} &\multirow{2}{*}{~~$369.7/300$} 
&\multirow{2}{*}{~~$1.07_{-0.03}^{+0.03}$} &\multirow{2}{*}{$1.05_{-0.02}^{+0.03}$}\\
&&&&&&&&&$16.70_{-0.08}^{+0.21}$&$0.92_{-0.11}^{+0.12}$&\\

&\multirow{2}{*}{0651330701}
&\multirow{2}{*}{$1.90_{-0.10}^{+0.10}$} &\multirow{2}{*}{$1.12_{-0.20}^{+0.23}$} &\multirow{2}{*}{$1.75_{-0.07}^{+0.05}$}    

&\multirow{2}{*}{$112_{-13}^{+14}$} &\multirow{2}{*}{$2.6_{-1.0}^{+1.1}$} &\multirow{2}{*}{--}  &\multirow{2}{*}{--} 

&$91.24_{-25.29}^{+28.70}$&$2.86_{-0.11}^{+0.15}$
&\multirow{2}{*}{$100^{f}$}

&\multirow{2}{*}{$113.0/97$} &\multirow{2}{*}{$102.9/101$} &\multirow{2}{*}{$123.3/105$} &\multirow{2}{*}{~~$339.4/291$} 
&\multirow{2}{*}{~~$1.06_{-0.03}^{+0.03}$} &\multirow{2}{*}{$1.06_{-0.03}^{+0.03}$}\\
&&&&&&&&&$16.69_{-0.08}^{+0.41}$&$0.81_{-0.10}^{+0.12}$ &\\

&\multirow{2}{*}{0651330801}
&\multirow{2}{*}{$1.95_{-0.11}^{+0.11}$} &\multirow{2}{*}{$1.03_{-0.20}^{+0.26}$} &\multirow{2}{*}{$1.47_{-0.09}^{+0.04}$}    

&\multirow{2}{*}{$102_{-13}^{+14}$} &\multirow{2}{*}{$1.5_{-1.0}^{+1.1}$} &\multirow{2}{*}{--}  &\multirow{2}{*}{--} 

&$119.34_{-32.14}^{+31.03}$&$2.89_{-0.13}^{+0.14}$
&\multirow{2}{*}{$100^{f}$}

&\multirow{2}{*}{$85.6/96$} &\multirow{2}{*}{$83.8/100$} &\multirow{2}{*}{$134.6/104$} &\multirow{2}{*}{~~$304.0/288$} 
&\multirow{2}{*}{~~$1.04_{-0.03}^{+0.03}$} &\multirow{2}{*}{$1.03_{-0.03}^{+0.03}$}\\
&&&&&&&&&$18.16_{-0.21}^{+0.53}$&$0.89_{-0.13}^{+0.17}$ &\\

&\multirow{2}{*}{0651330901}
&\multirow{2}{*}{$1.80_{-0.09}^{+0.10}$} &\multirow{2}{*}{$0.88_{-0.14}^{+0.19}$} &\multirow{2}{*}{$1.72_{-0.08}^{+0.04}$}    

&\multirow{2}{*}{$108_{-13}^{+14}$} &\multirow{2}{*}{$3.3_{-1.0}^{+1.1}$} &\multirow{2}{*}{--}  &\multirow{2}{*}{--} 

&$85.35_{-27.47}^{+31.26}$&$2.93_{-0.13}^{+0.09}$
&\multirow{2}{*}{$100^{f}$}

&\multirow{2}{*}{$96.0/96$} &\multirow{2}{*}{$65.3/101$} &\multirow{2}{*}{$113.2/105$} &\multirow{2}{*}{~~$274.2/290$} 
&\multirow{2}{*}{~~$1.12_{-0.03}^{+0.03}$} &\multirow{2}{*}{$1.11_{-0.03}^{+0.03}$}\\
&&&&&&&&&$16.77_{-0.06}^{+0.38}$&$0.93_{-0.13}^{+0.10}$ &\\

&\multirow{2}{*}{0651331001}
&\multirow{2}{*}{$1.75_{-0.09}^{+0.10}$} &\multirow{2}{*}{$0.69_{-0.14}^{+0.19}$} &\multirow{2}{*}{$1.49_{-0.06}^{+0.05}$}    

&\multirow{2}{*}{$106_{-13}^{+14}$} &\multirow{2}{*}{$3.3_{-1.0}^{+1.1}$} &\multirow{2}{*}{--}  &\multirow{2}{*}{--} 

&$71.07_{-21.34}^{+34.65}$&$2.99_{-0.12}^{+0.07}$
&\multirow{2}{*}{$100^{f}$}

&\multirow{2}{*}{$92.4/95$} &\multirow{2}{*}{$91.1/98$} &\multirow{2}{*}{$78.7/103$} &\multirow{2}{*}{~~$262.2/284$} 
&\multirow{2}{*}{~~$1.08_{-0.03}^{+0.04}$} &\multirow{2}{*}{$1.07_{-0.03}^{+0.04}$}\\
&&&&&&&&&$16.78_{-0.06}^{+0.38}$&$1.00_{-0.13}^{+0.10}$ &\\

&\multirow{2}{*}{0651331101}
&\multirow{2}{*}{$1.84_{-0.10}^{+0.12}$} &\multirow{2}{*}{$0.94_{-0.19}^{+0.24}$} &\multirow{2}{*}{$1.60_{-0.08}^{+0.05}$}    

&\multirow{2}{*}{$109_{-13}^{+13}$} &\multirow{2}{*}{$2.1_{-1.0}^{+1.0}$} &\multirow{2}{*}{--}  &\multirow{2}{*}{--} 

&$92.09_{-31.35}^{+36.90}$&$2.90_{-0.15}^{+0.12}$
&\multirow{2}{*}{$100^{f}$}

&\multirow{2}{*}{$112.2/94$} &\multirow{2}{*}{$72.7/98$} &\multirow{2}{*}{$122.0/102$} &\multirow{2}{*}{~~$306.9/282$} 
&\multirow{2}{*}{~~$0.99_{-0.03}^{+0.04}$} &\multirow{2}{*}{$1.05_{-0.04}^{+0.04}$}\\
&&&&&&&&&$16.57_{-0.11}^{+0.31}$&$0.85_{-0.13}^{+0.17}$ &\\

\\
\hline

\hline
  \end{tabular}
 \end{table}
\end{landscape}
\begin{figure*}
\centering
\includegraphics[scale=0.25]{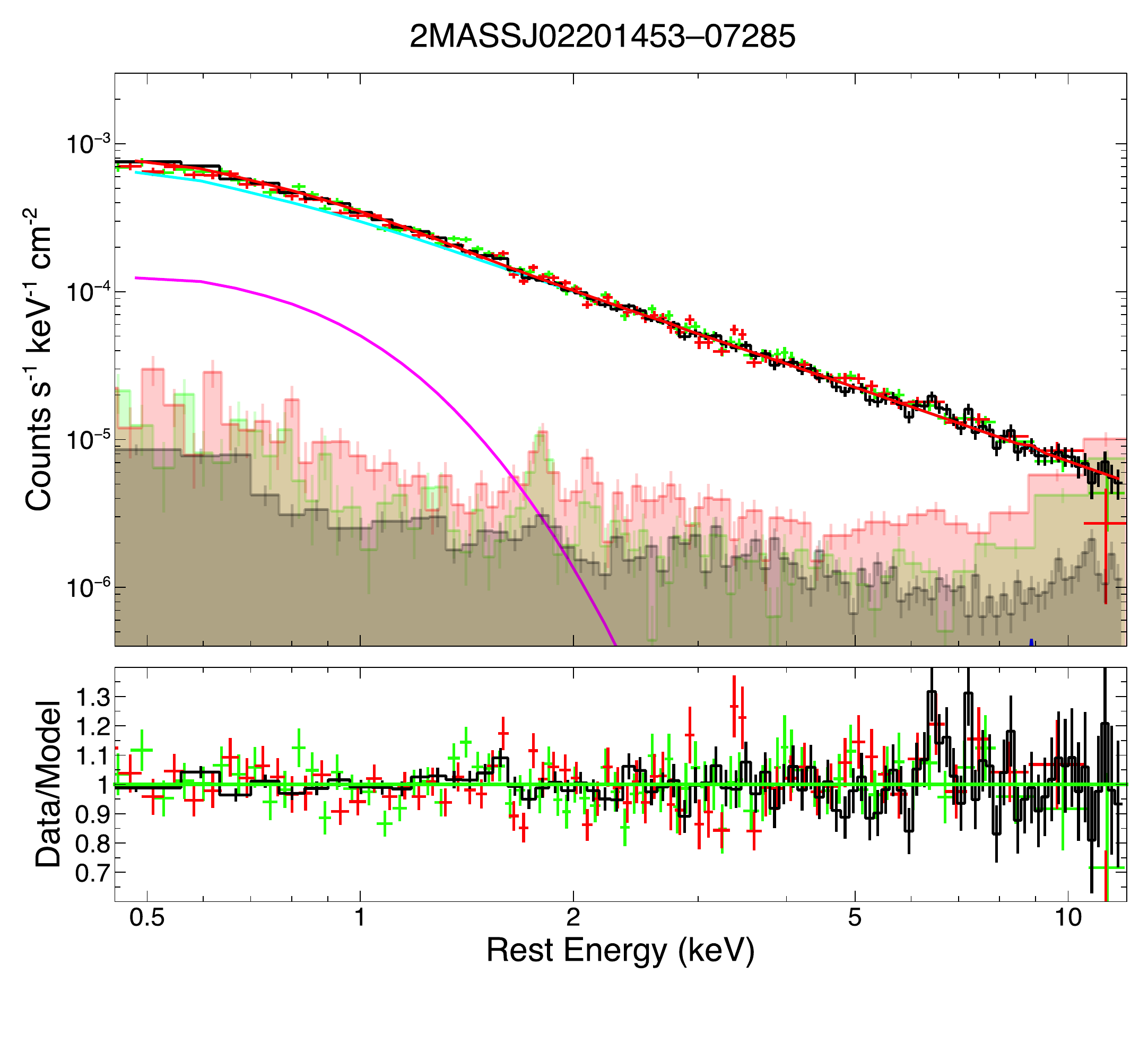}
\includegraphics[scale=0.25]{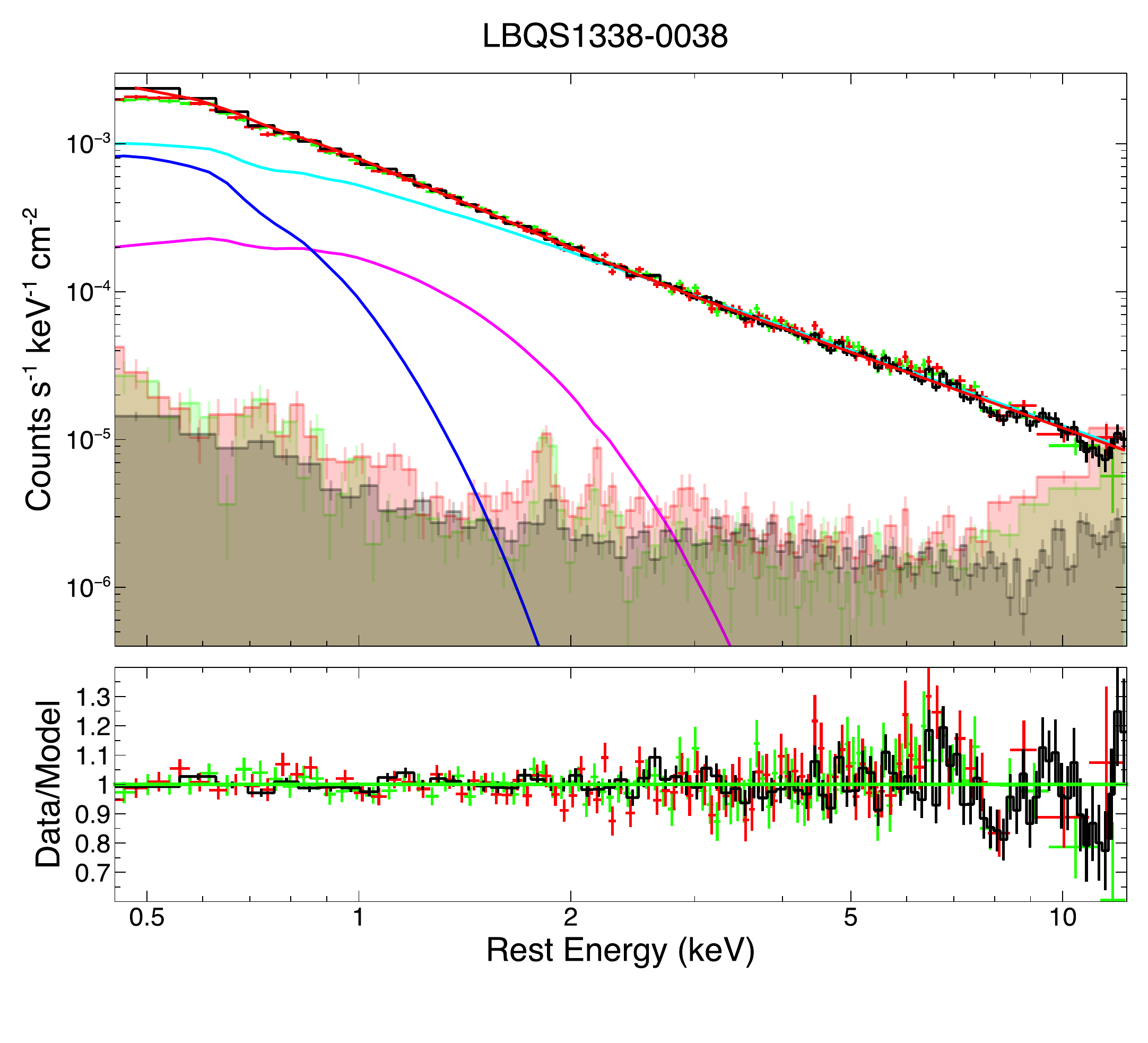}
\includegraphics[scale=0.25]{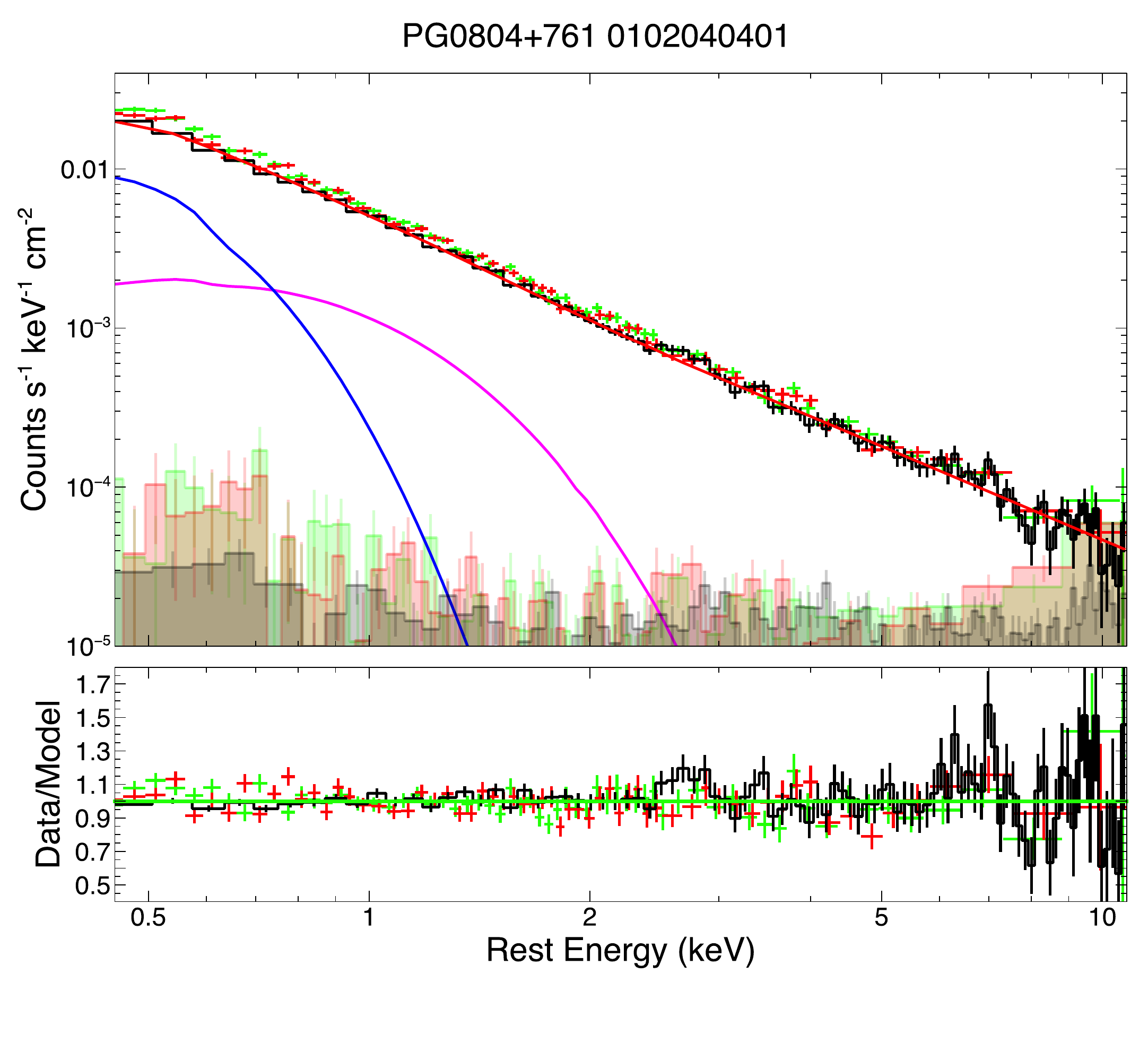}
\includegraphics[scale=0.25]{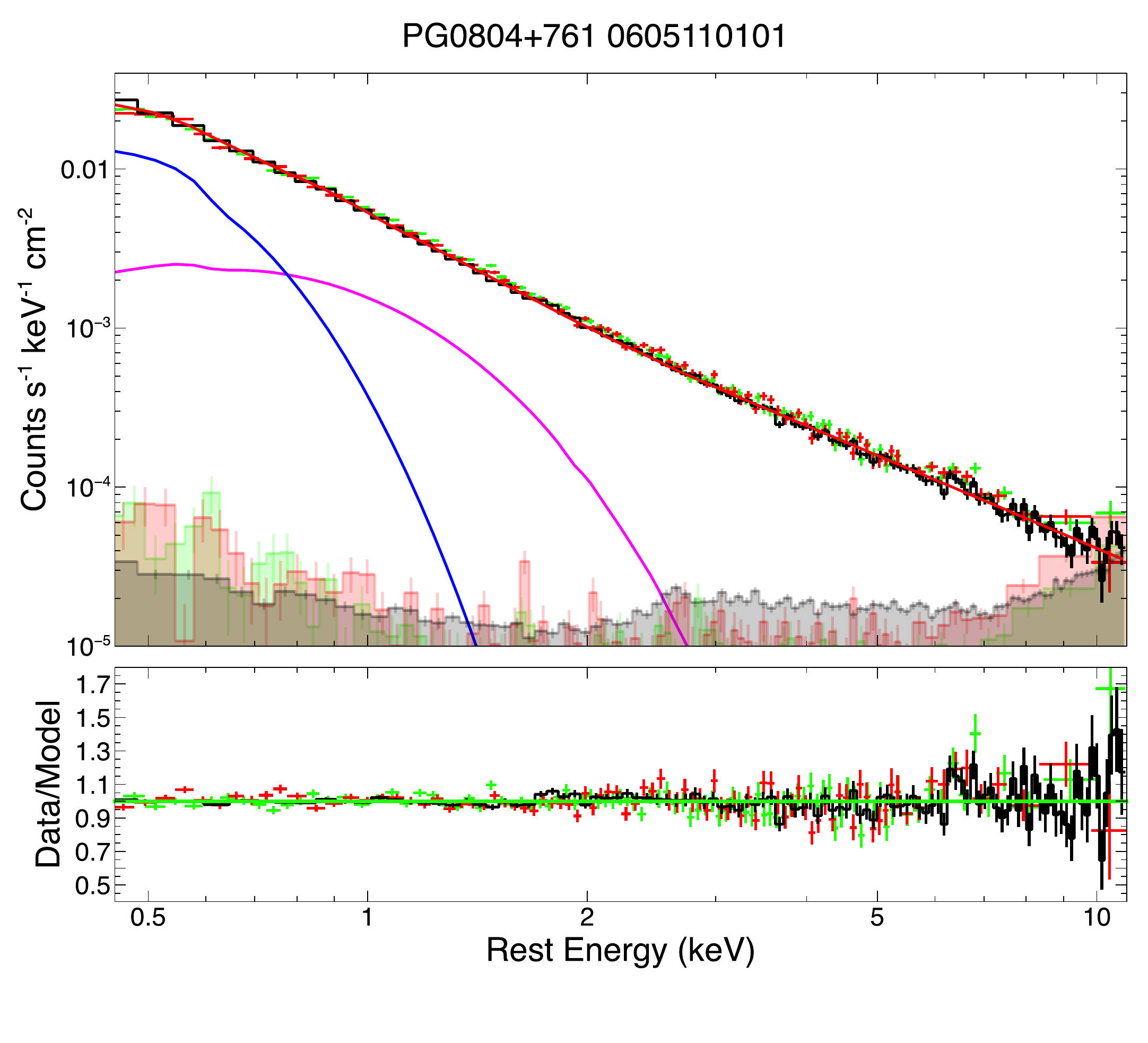}
\includegraphics[scale=0.25]{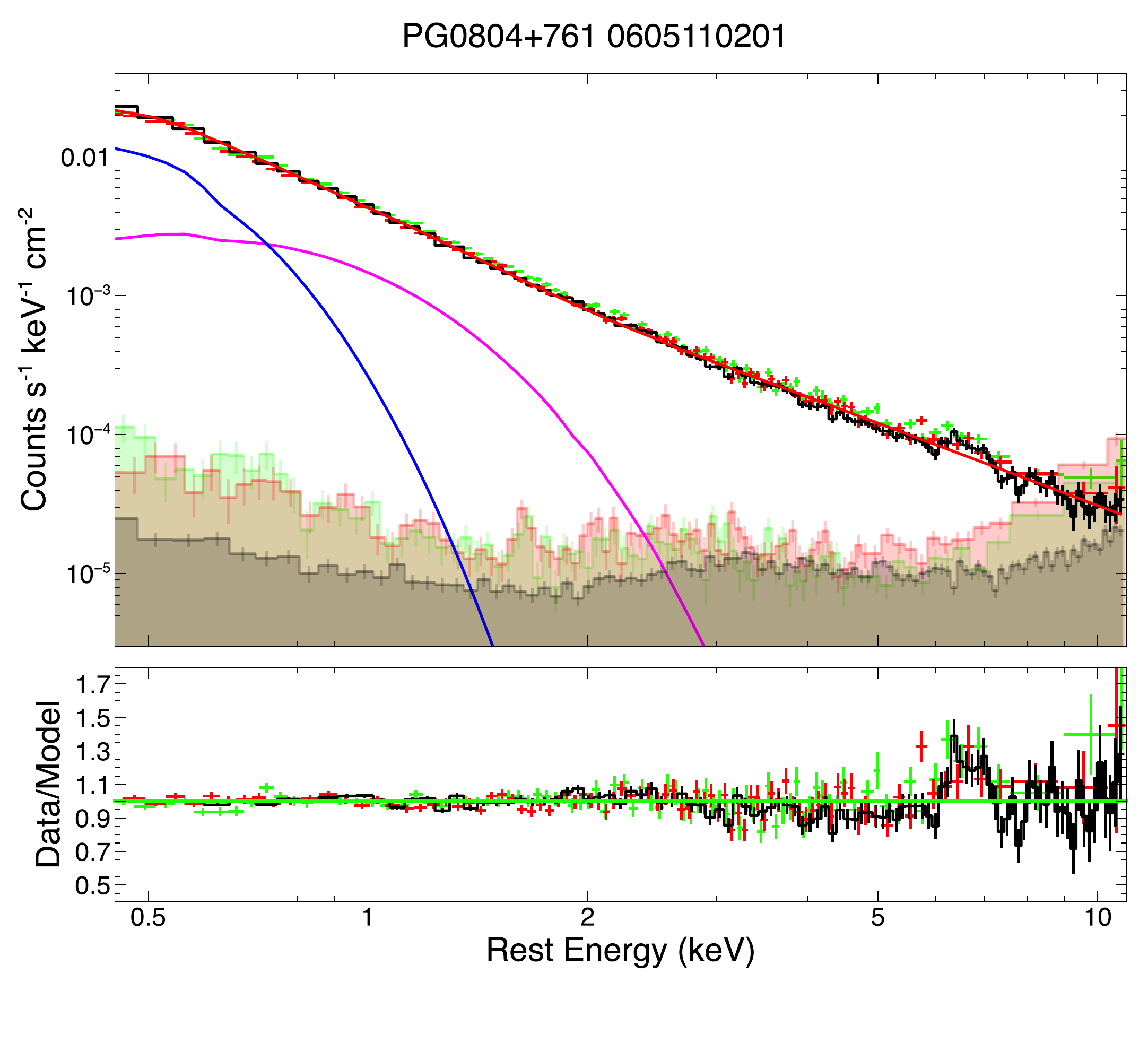}
\includegraphics[scale=0.25]{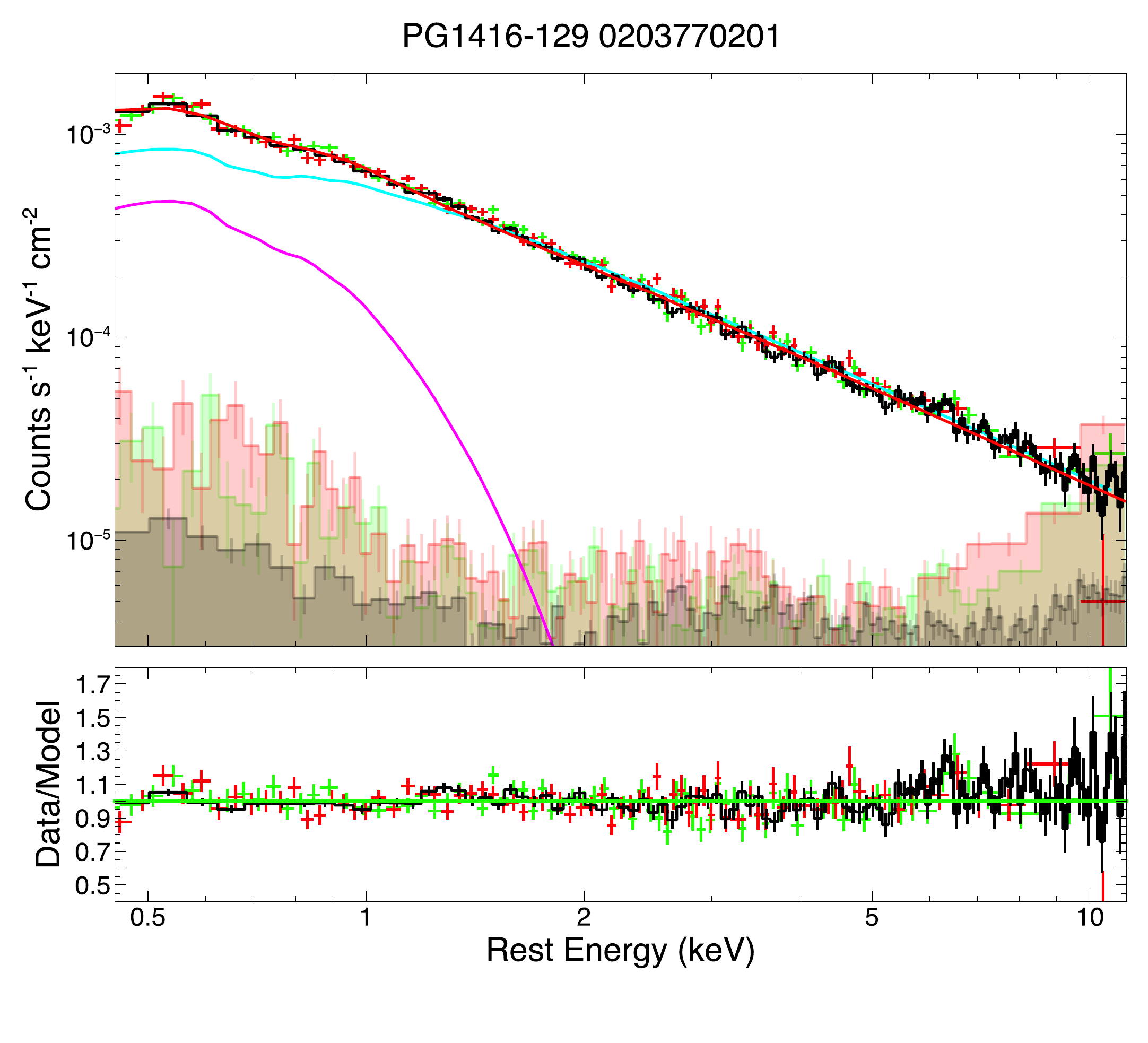}
\includegraphics[scale=0.25]{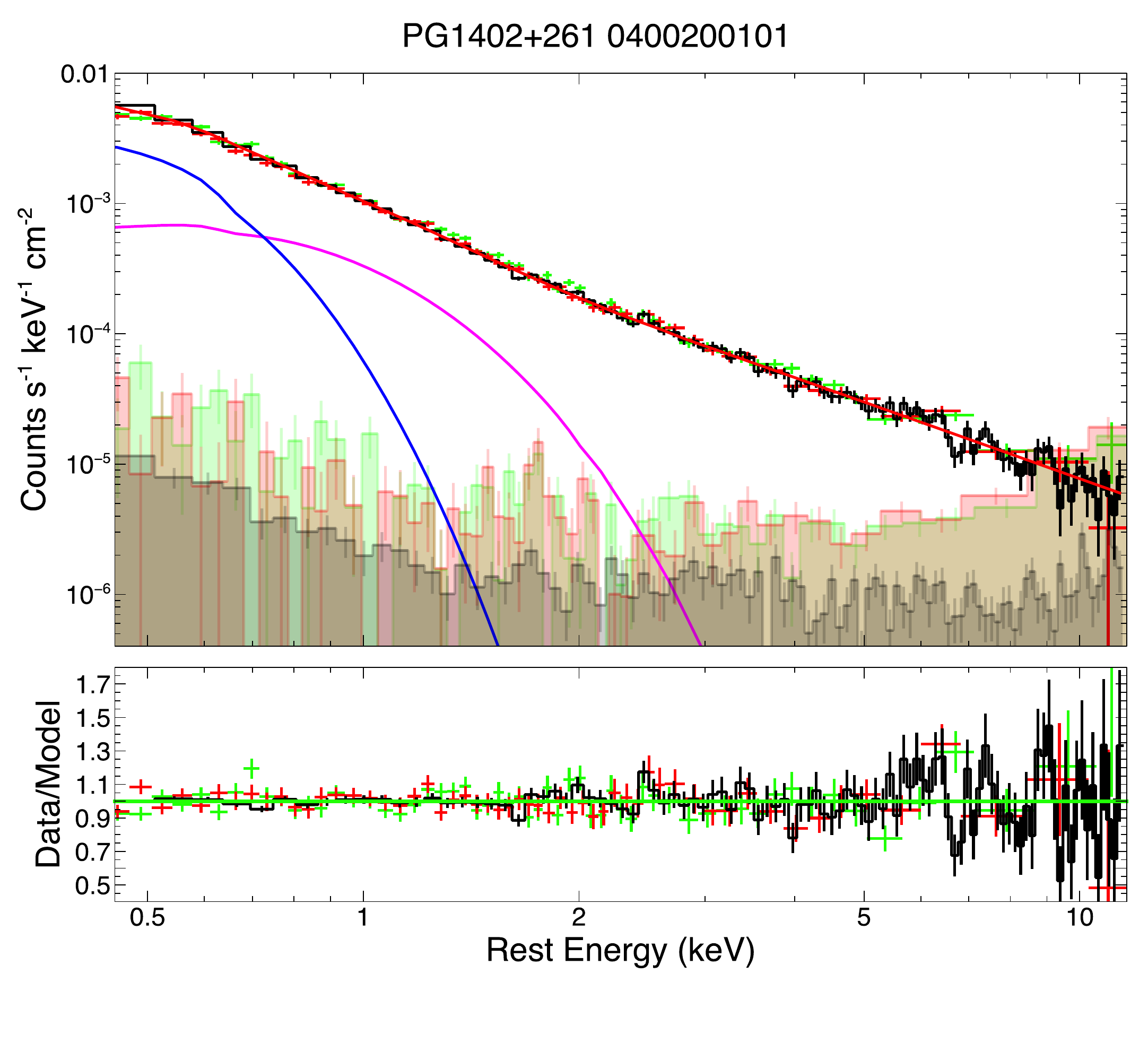}
\includegraphics[scale=0.25]{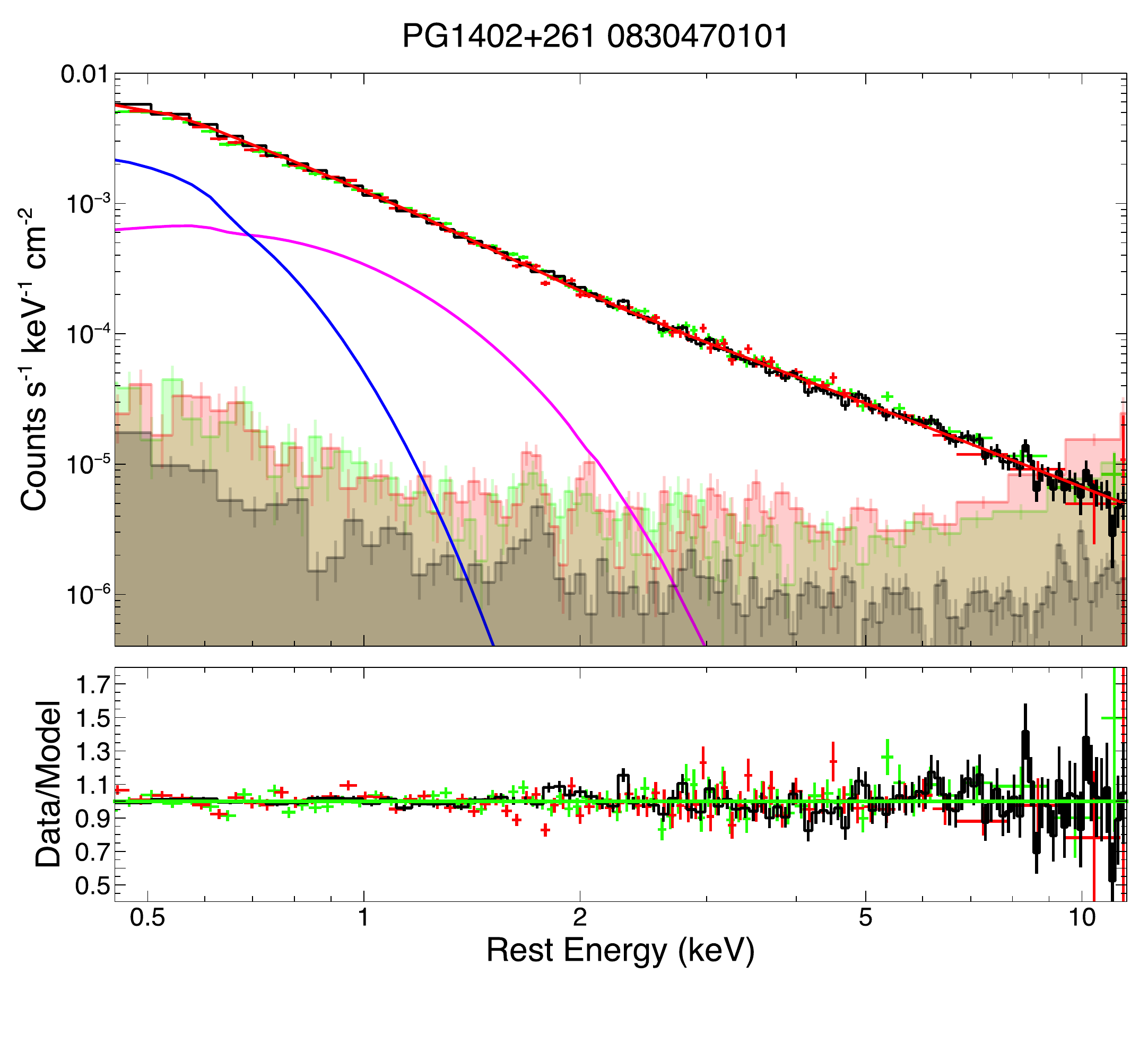}
\includegraphics[scale=0.25]{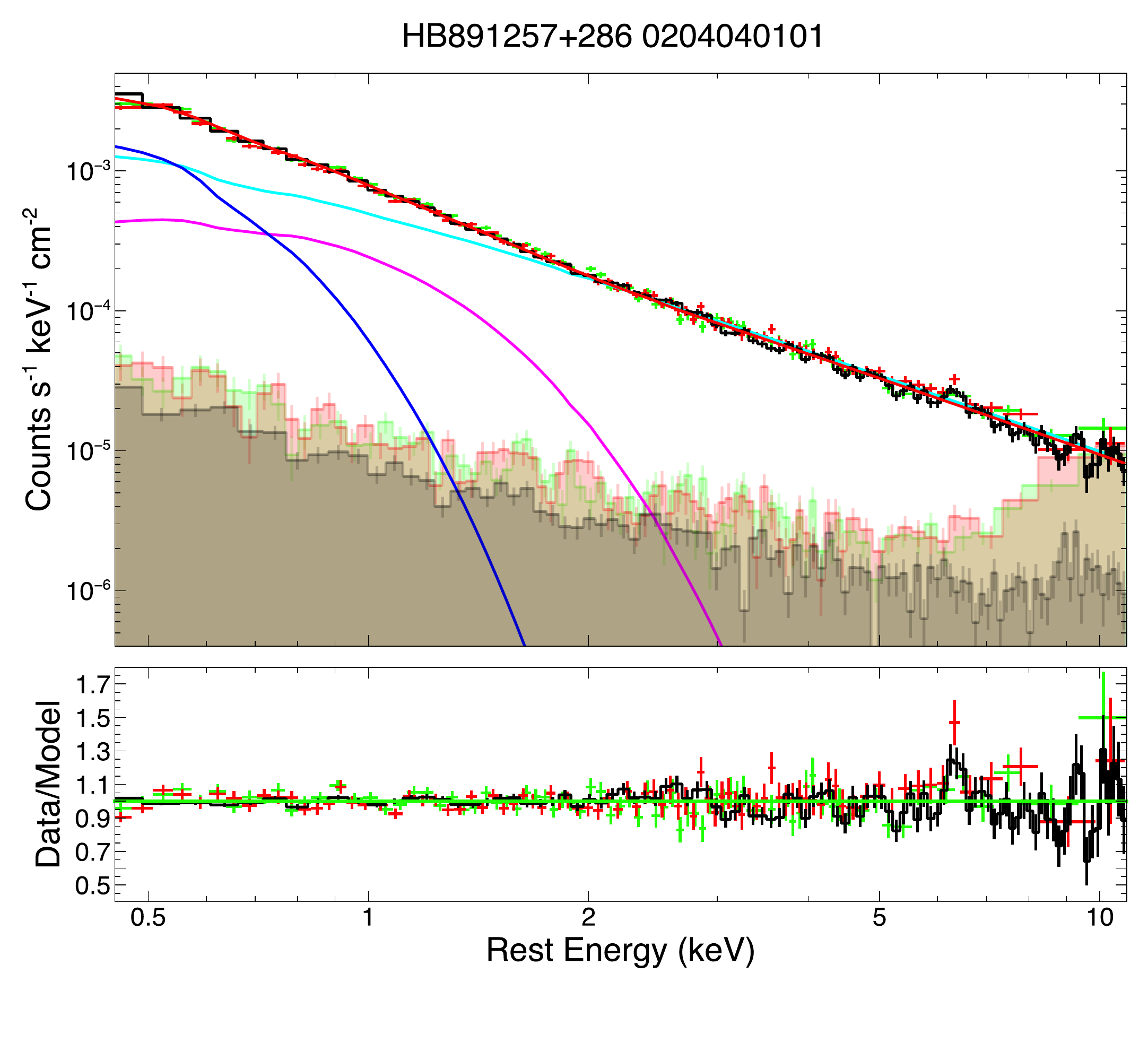}
\includegraphics[scale=0.25]{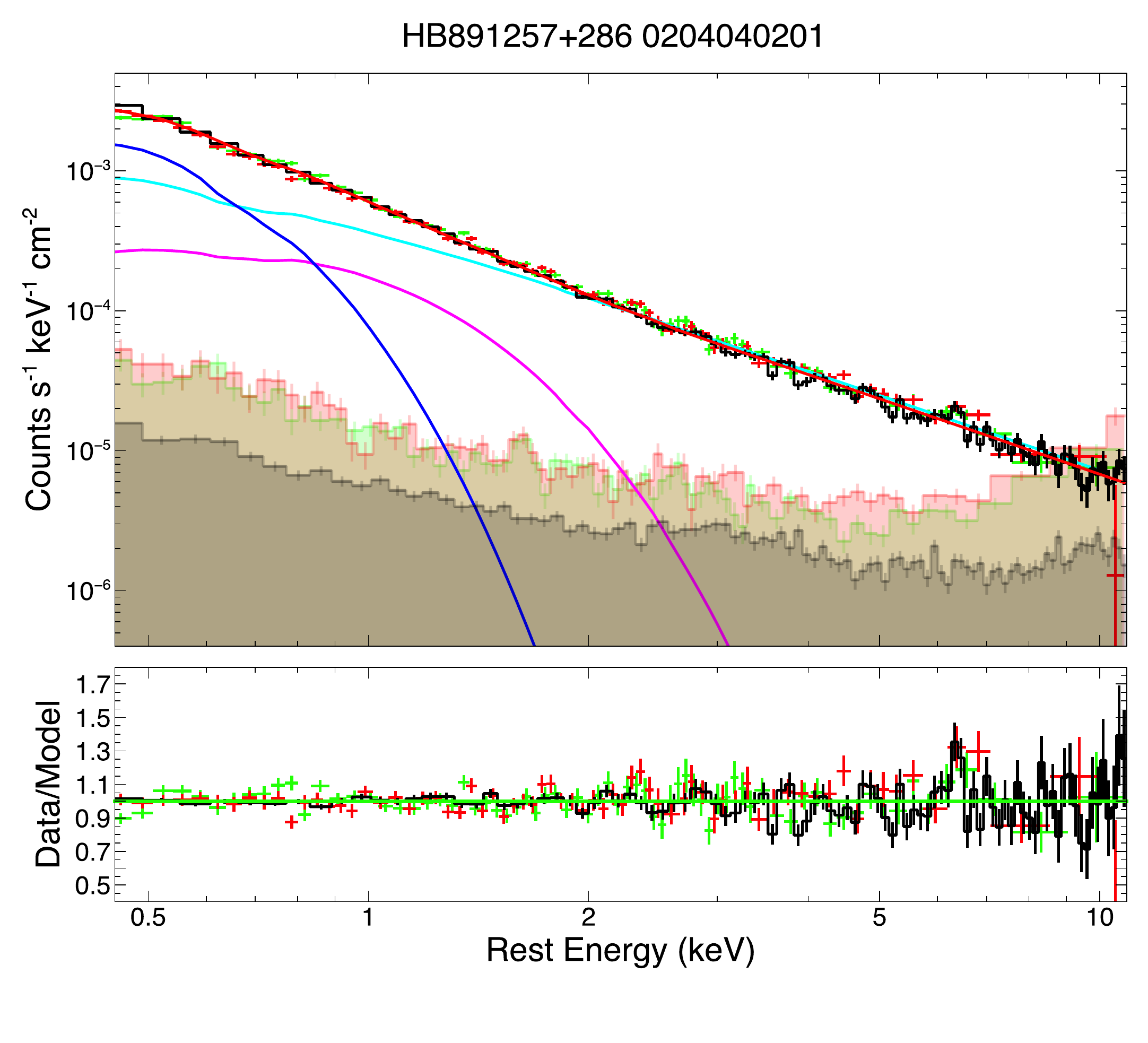}
\includegraphics[scale=0.25]{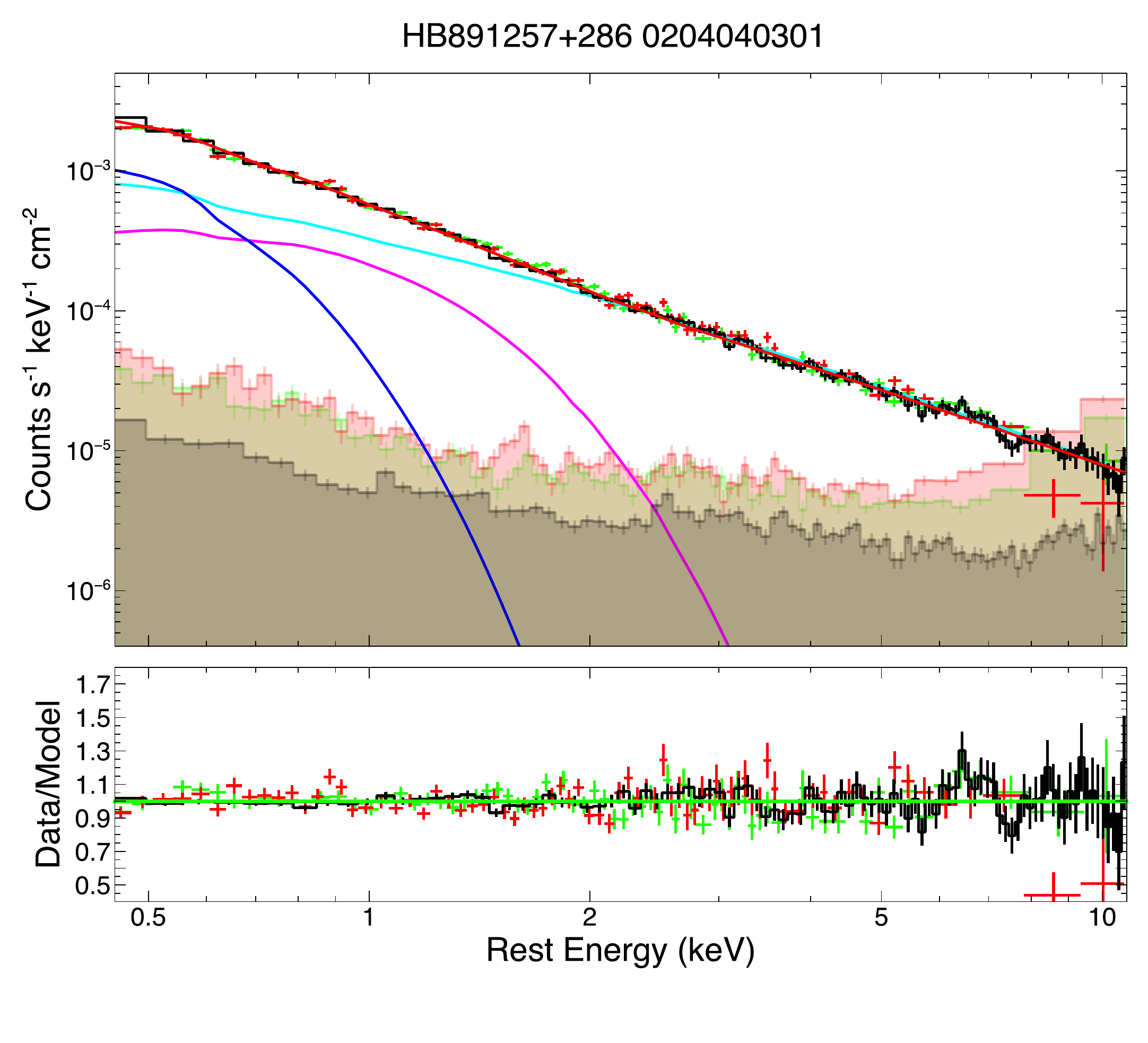}
\includegraphics[scale=0.25]{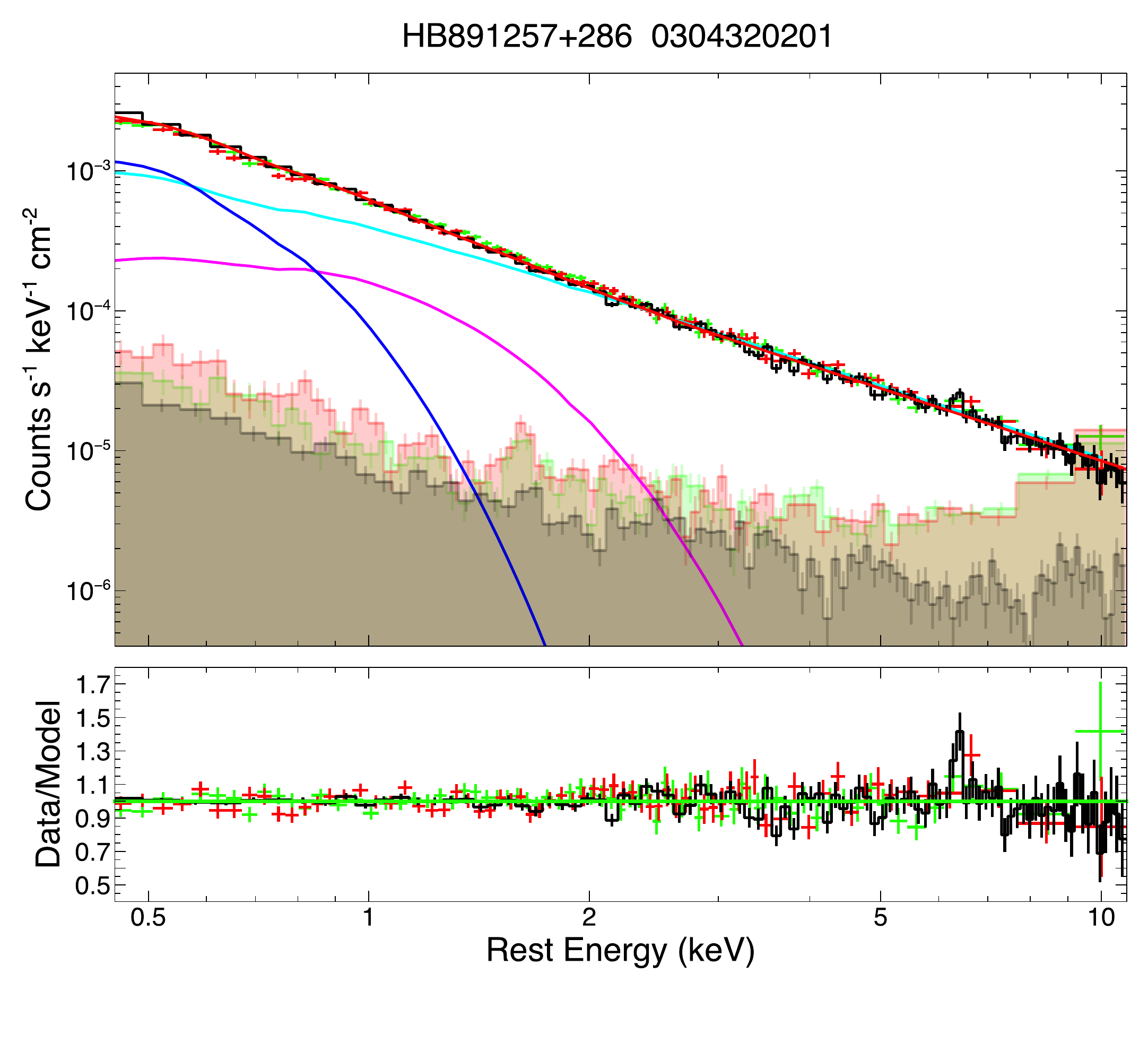}
\includegraphics[scale=0.25]{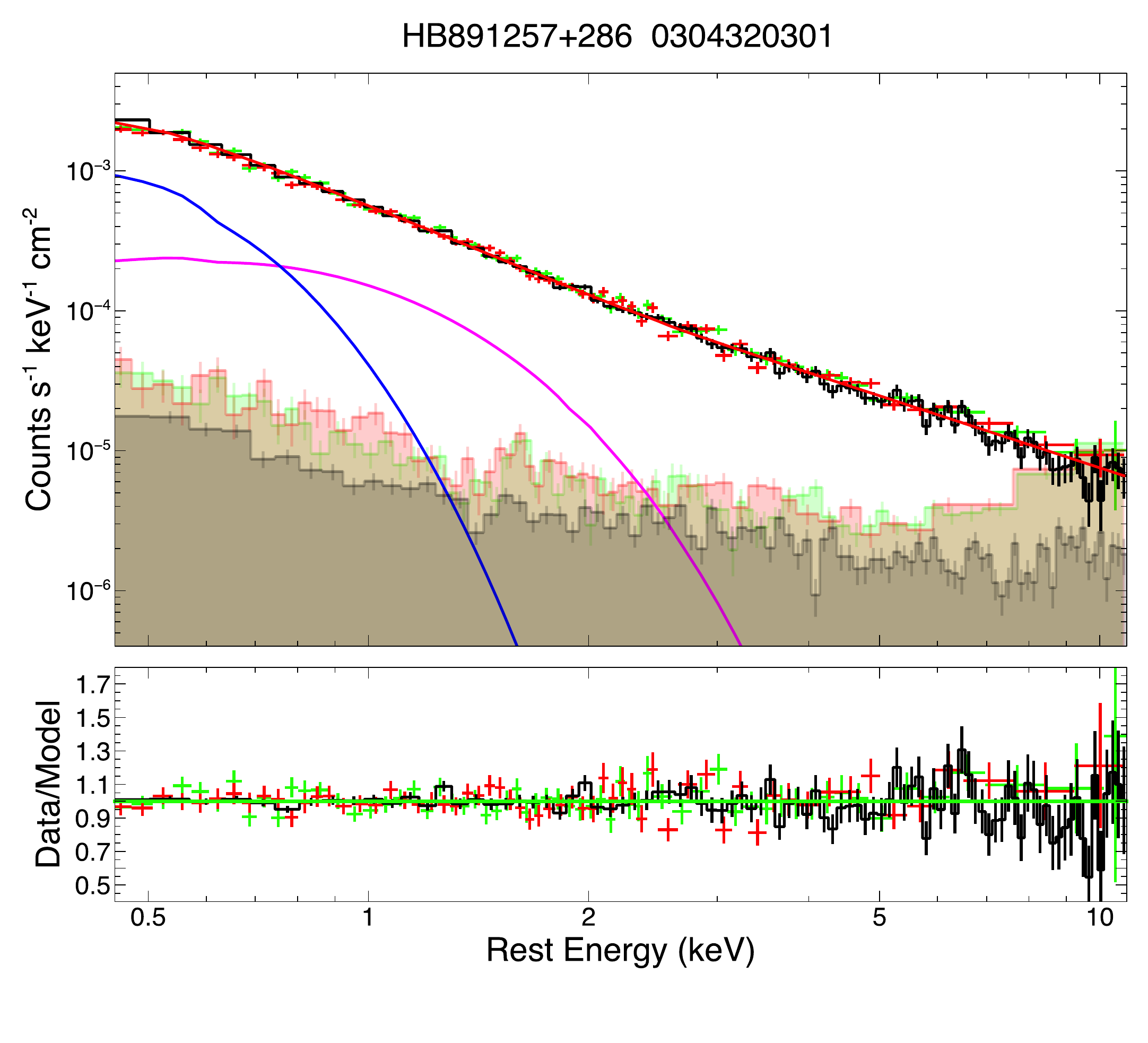}
\includegraphics[scale=0.25]{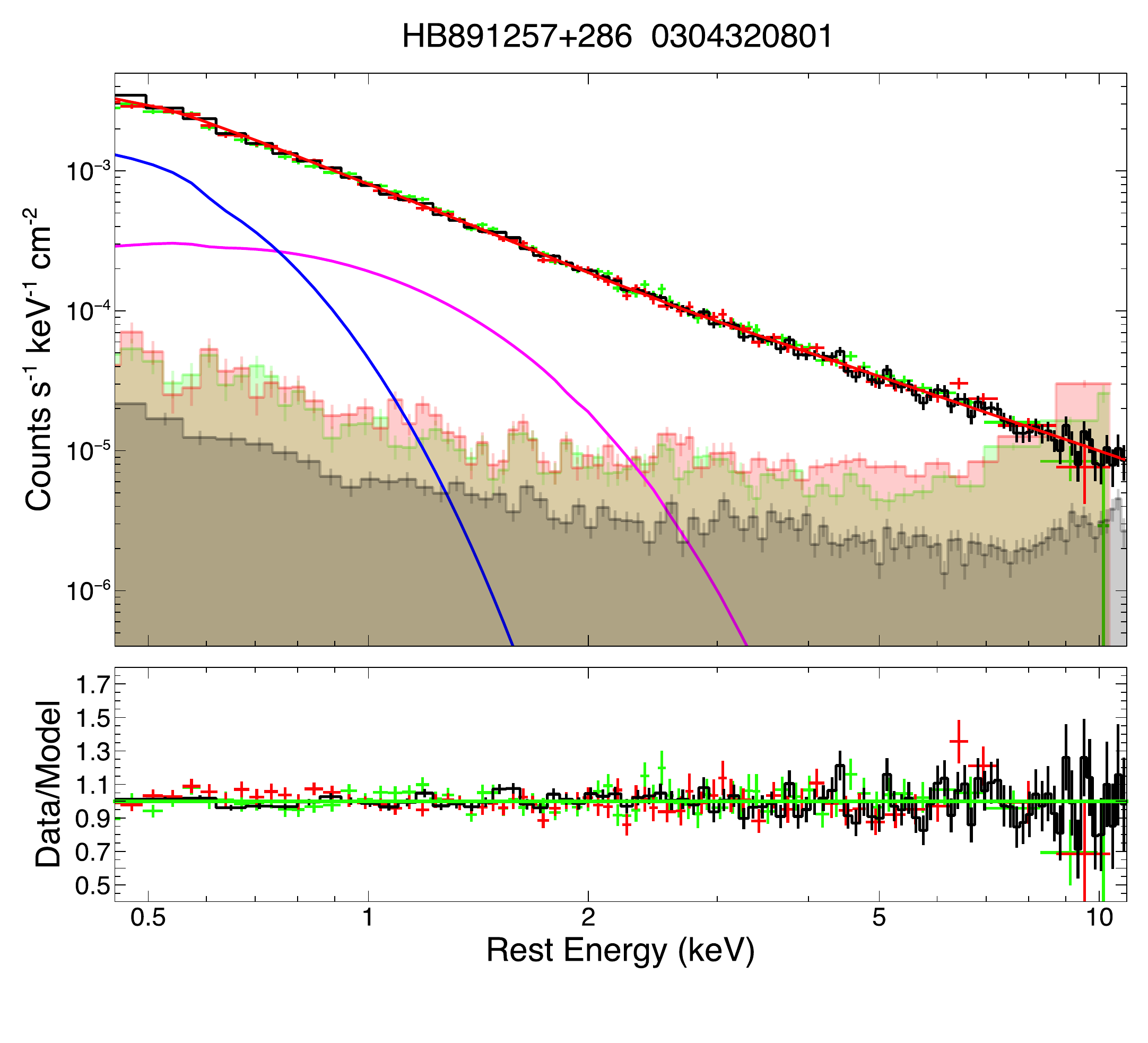}
\includegraphics[scale=0.25]{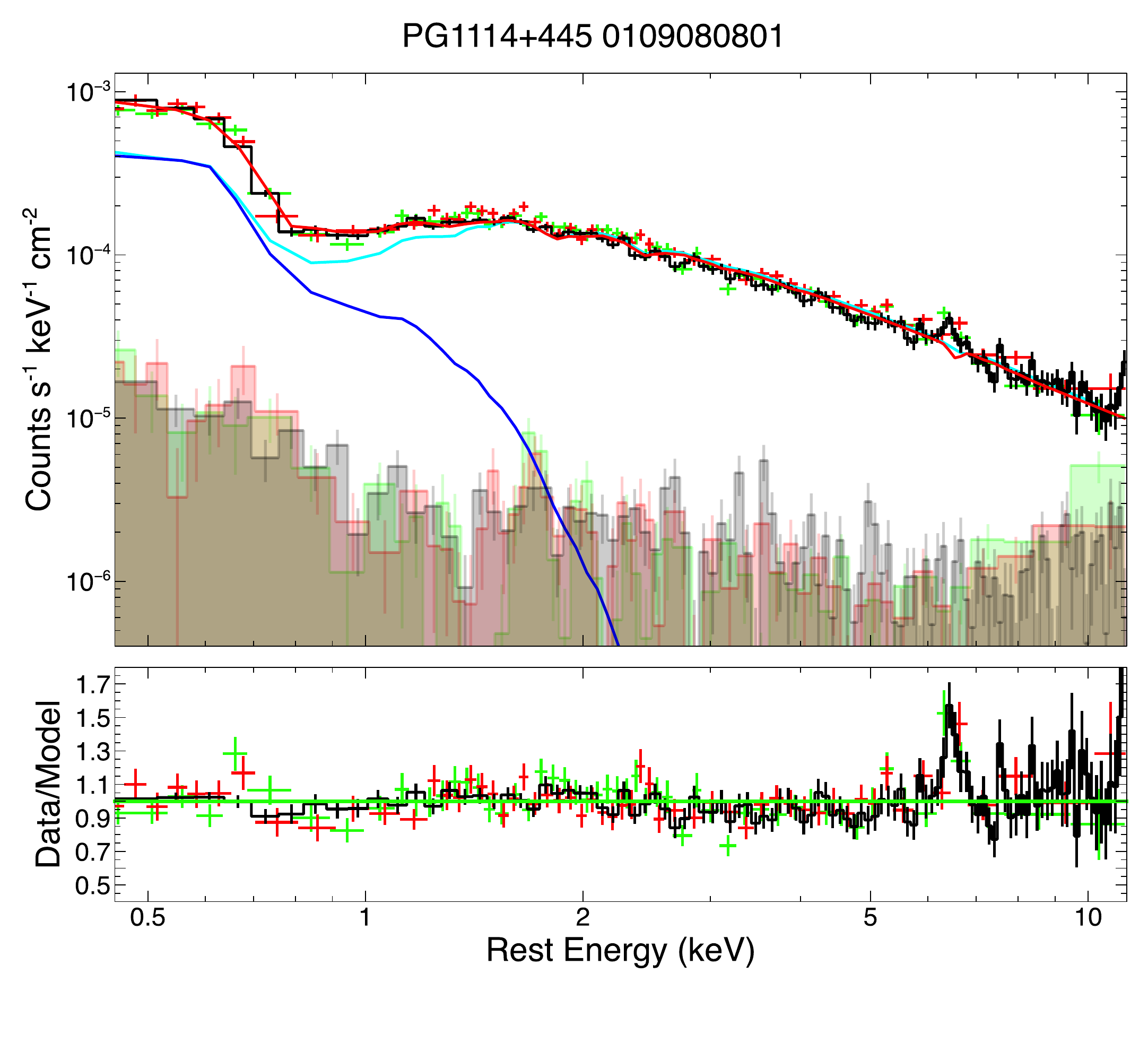}
\end{figure*}
\begin{figure*}
\centering
\includegraphics[scale=0.25]{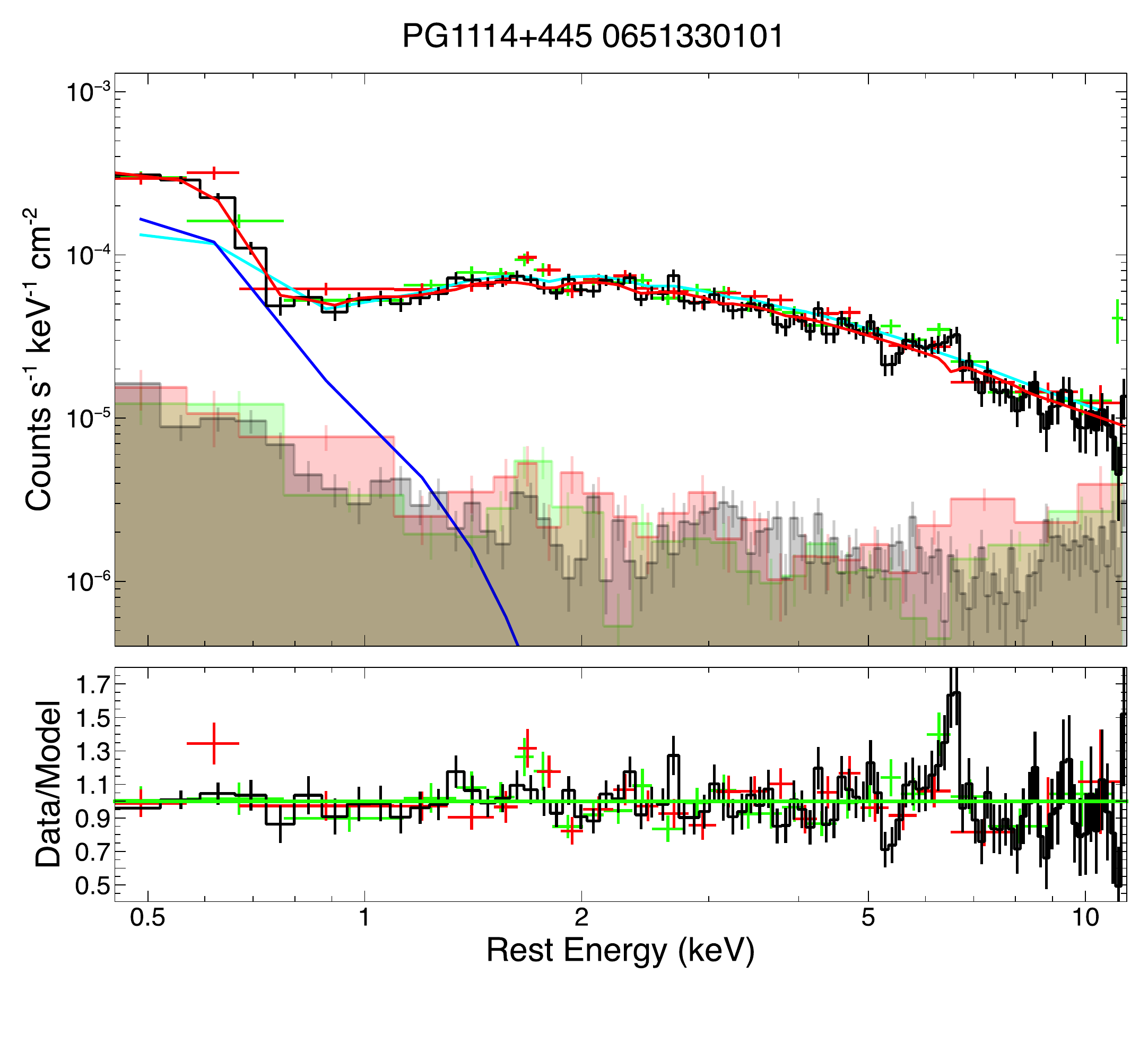}
\includegraphics[scale=0.25]{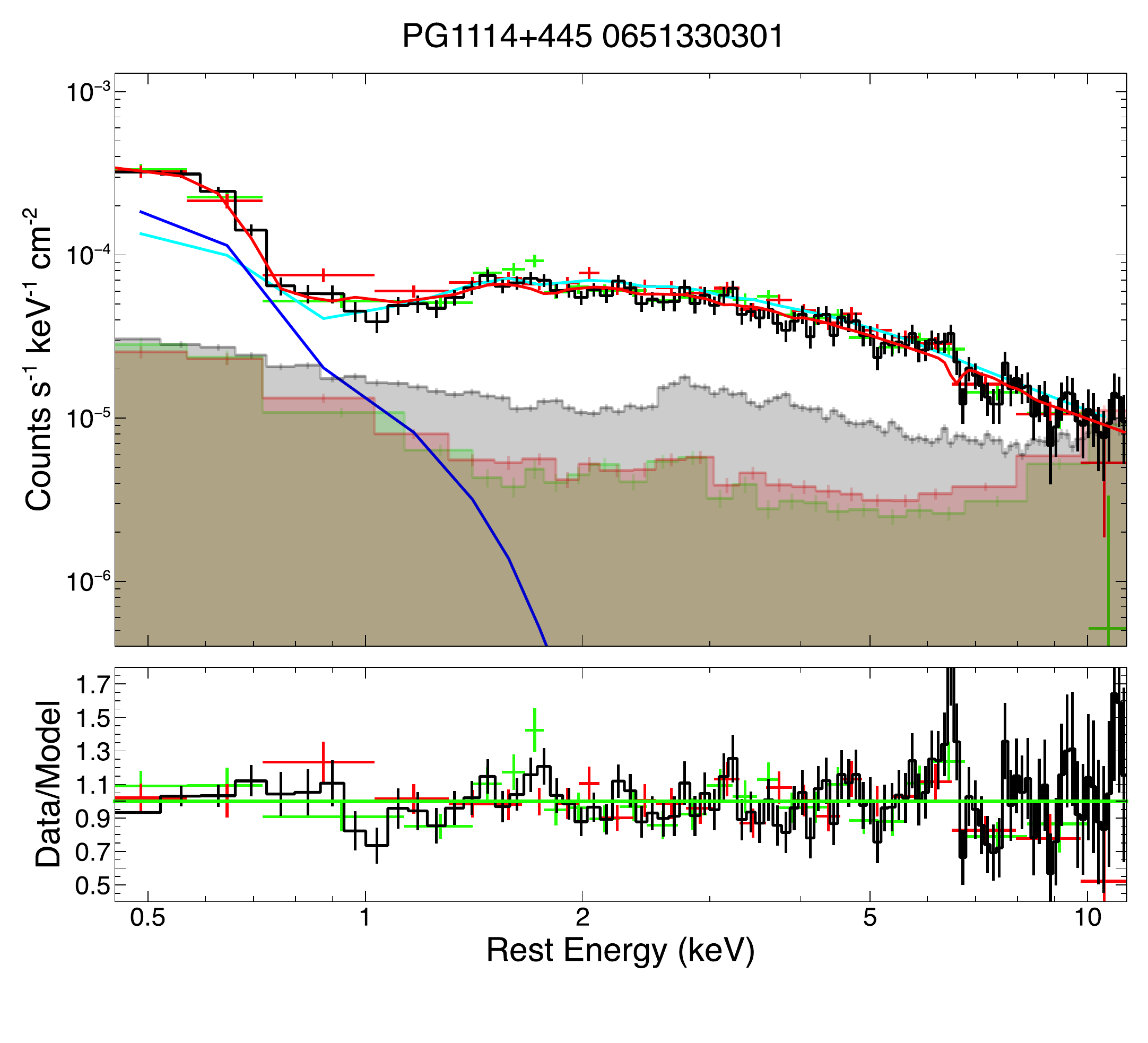}
\includegraphics[scale=0.25]{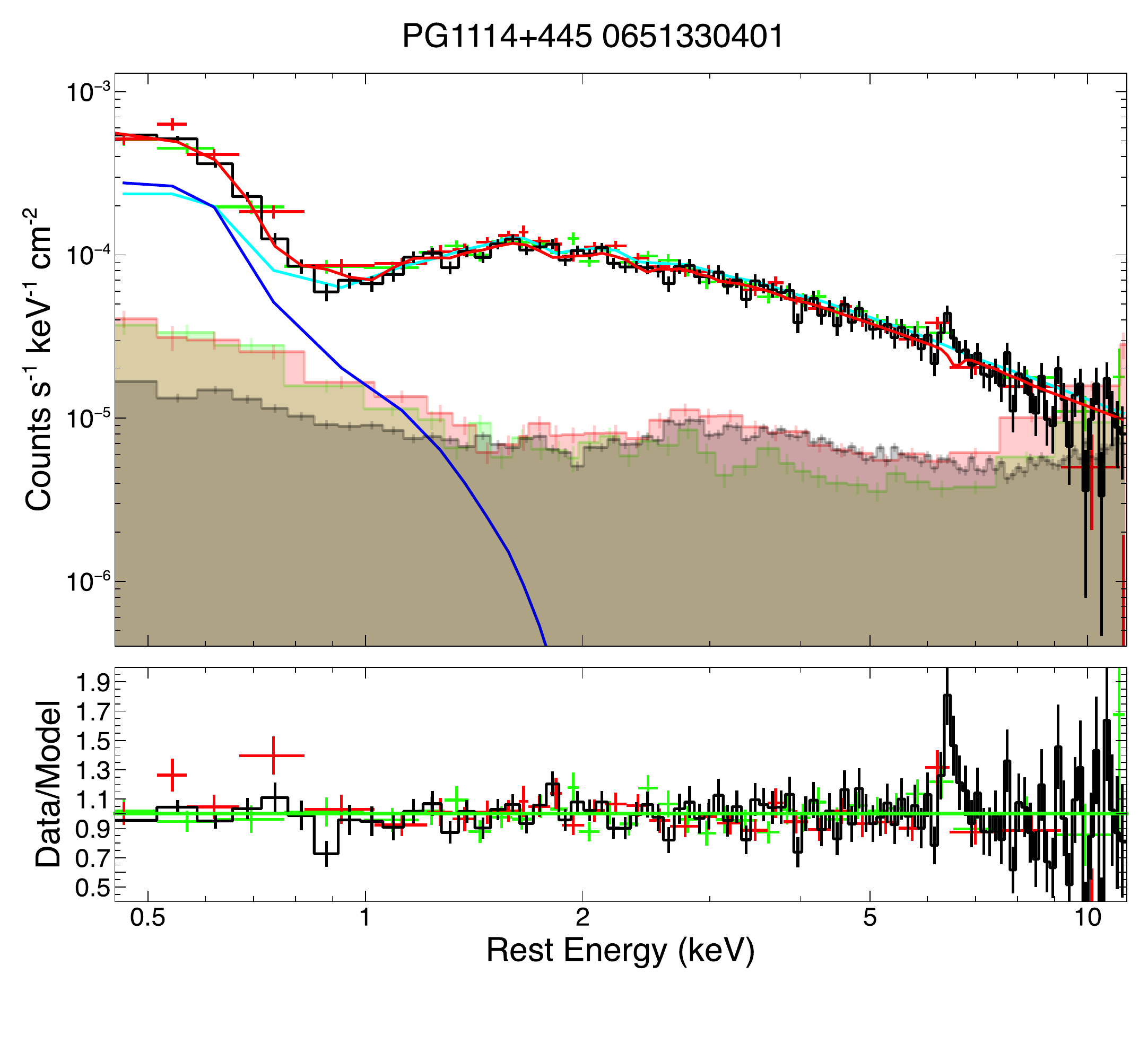}
\includegraphics[scale=0.25]{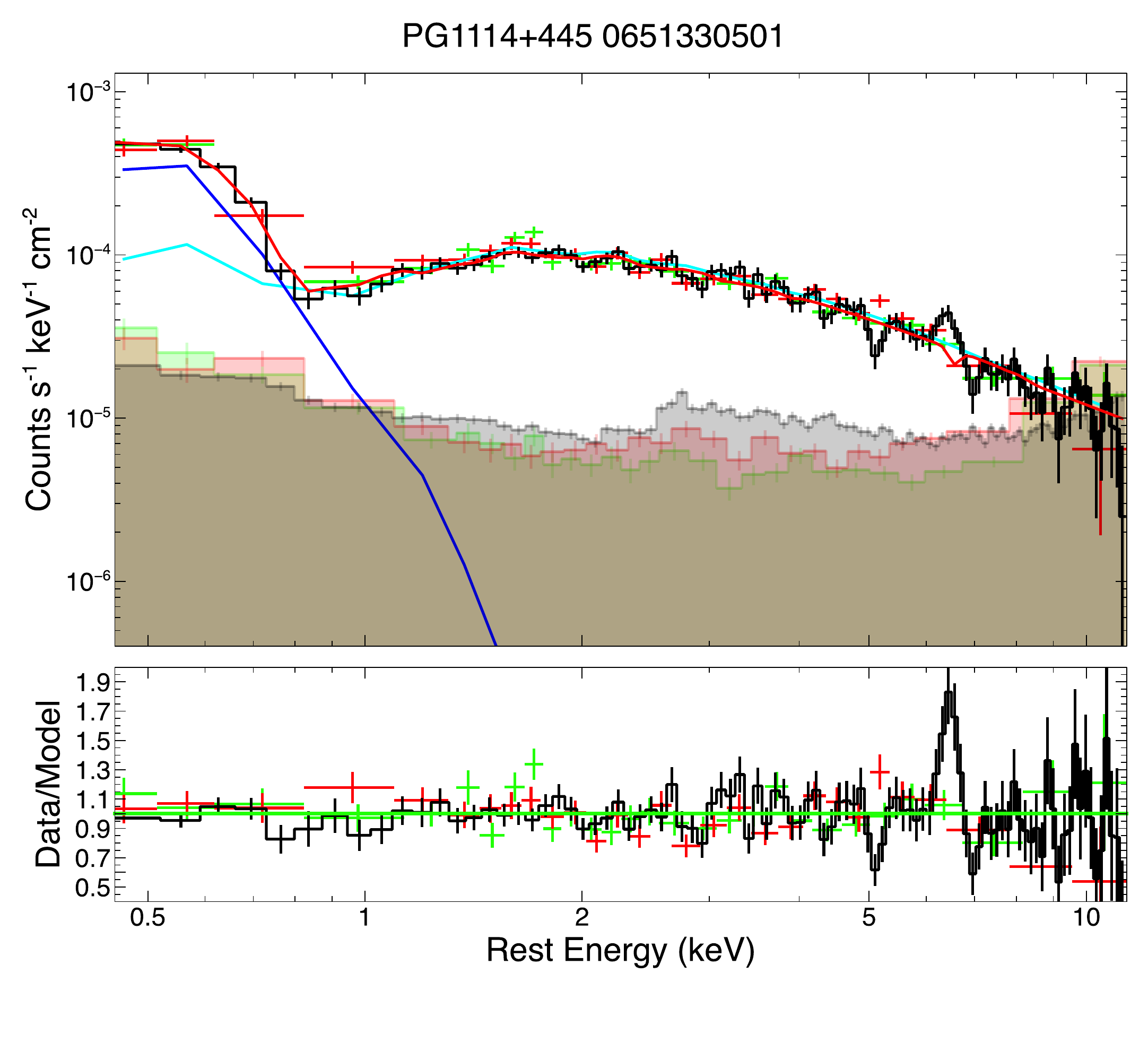}
\includegraphics[scale=0.25]{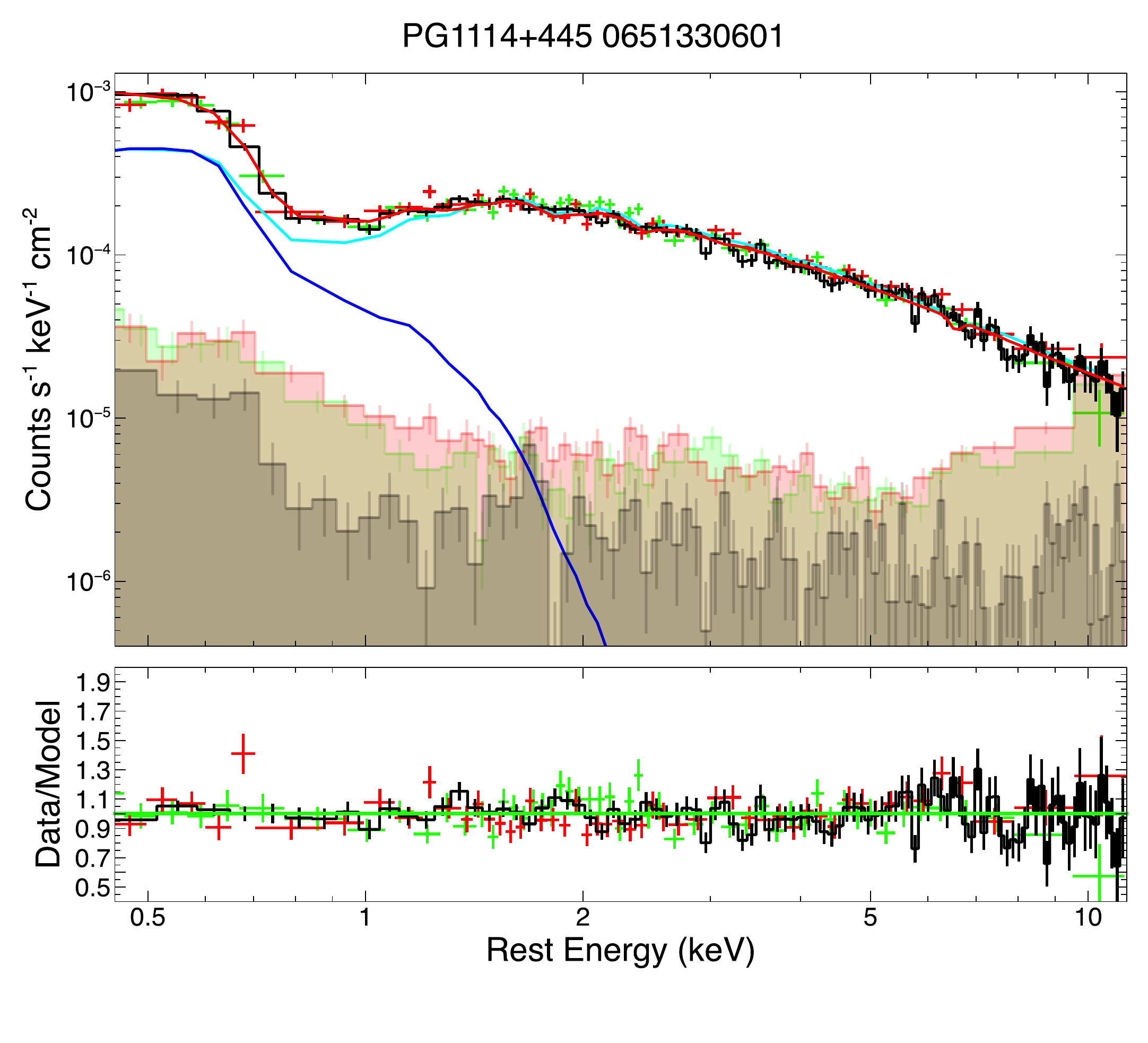}
\includegraphics[scale=0.25]{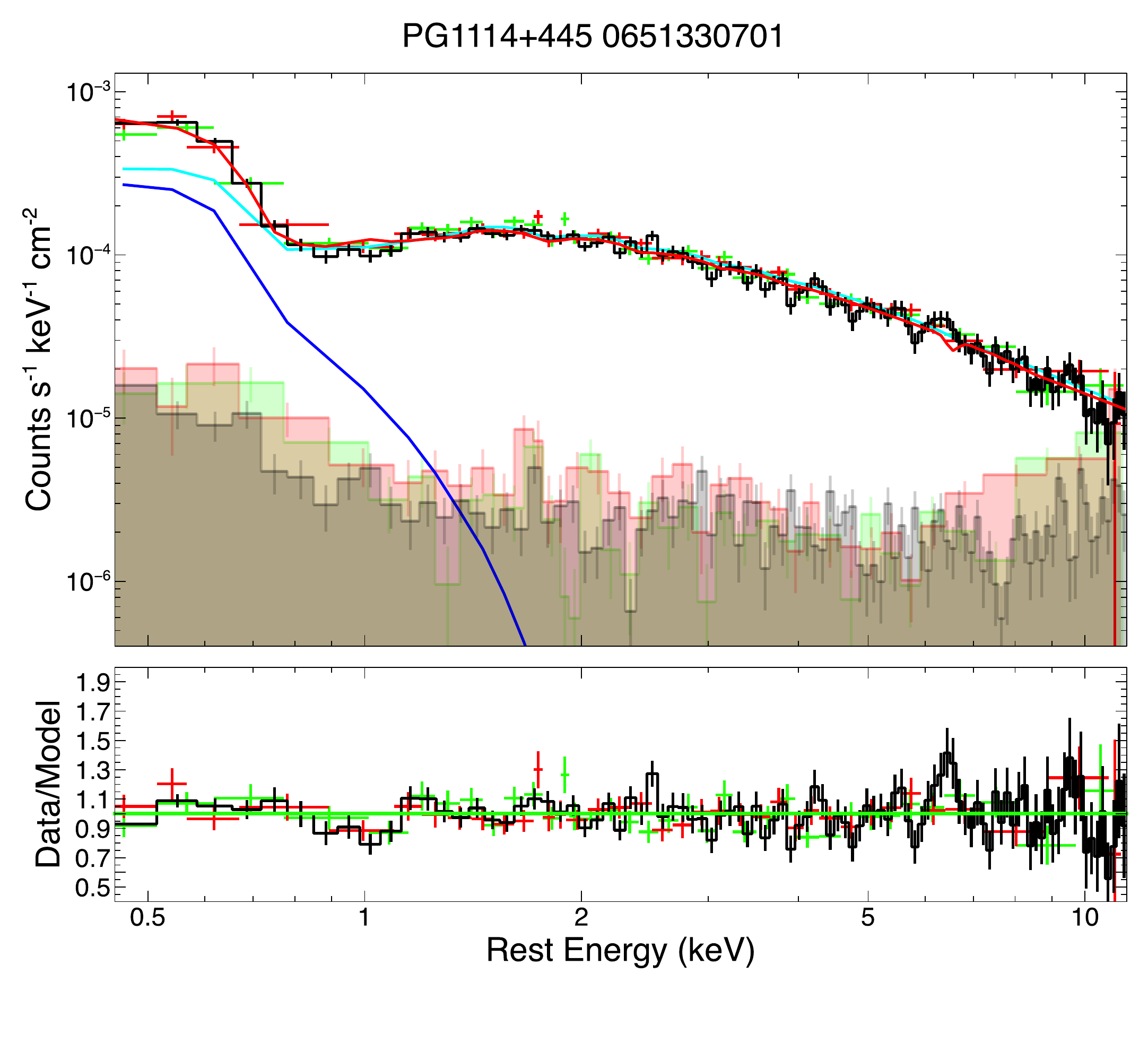}
\includegraphics[scale=0.25]{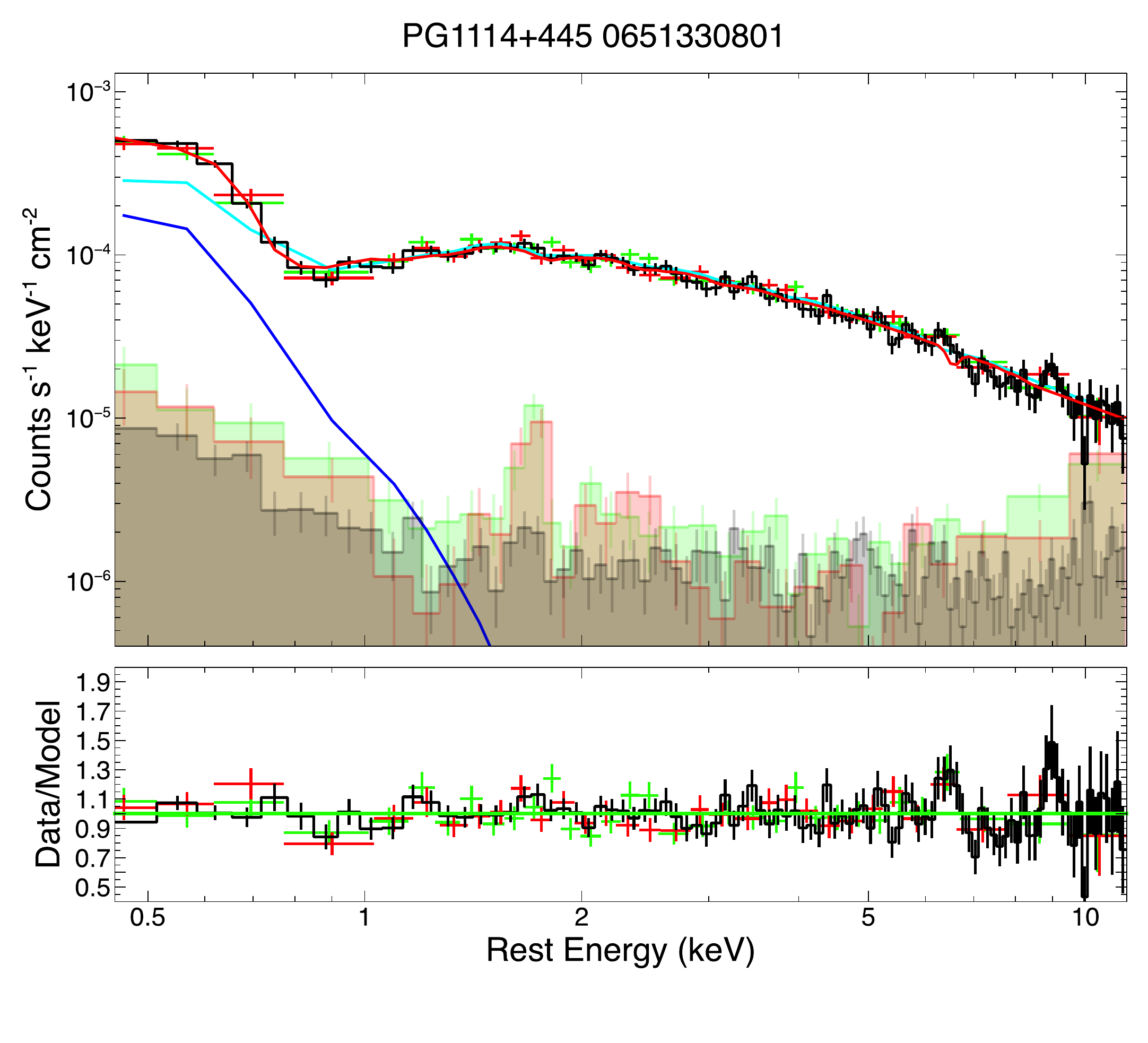}
\includegraphics[scale=0.25]{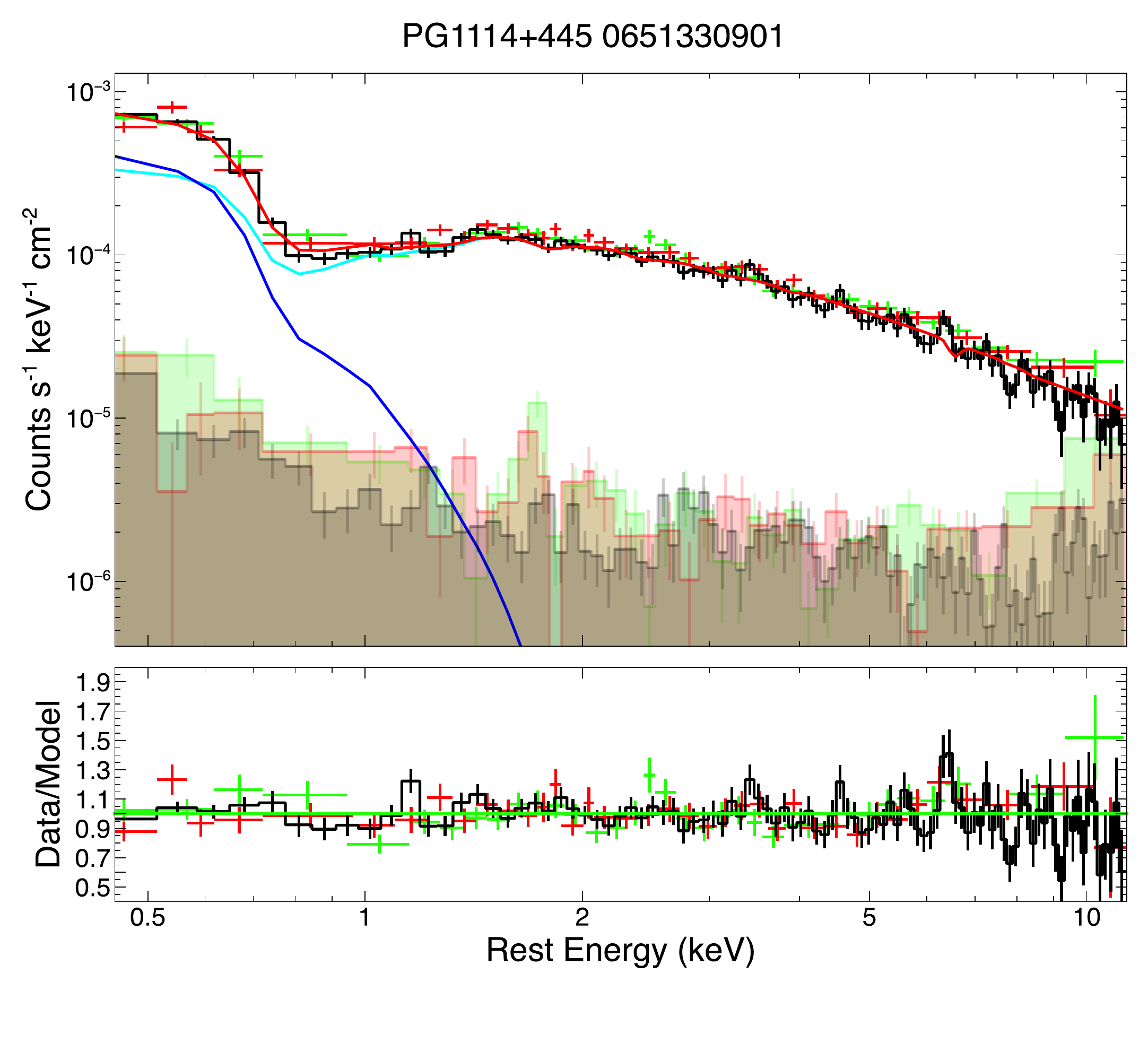}
\includegraphics[scale=0.25]{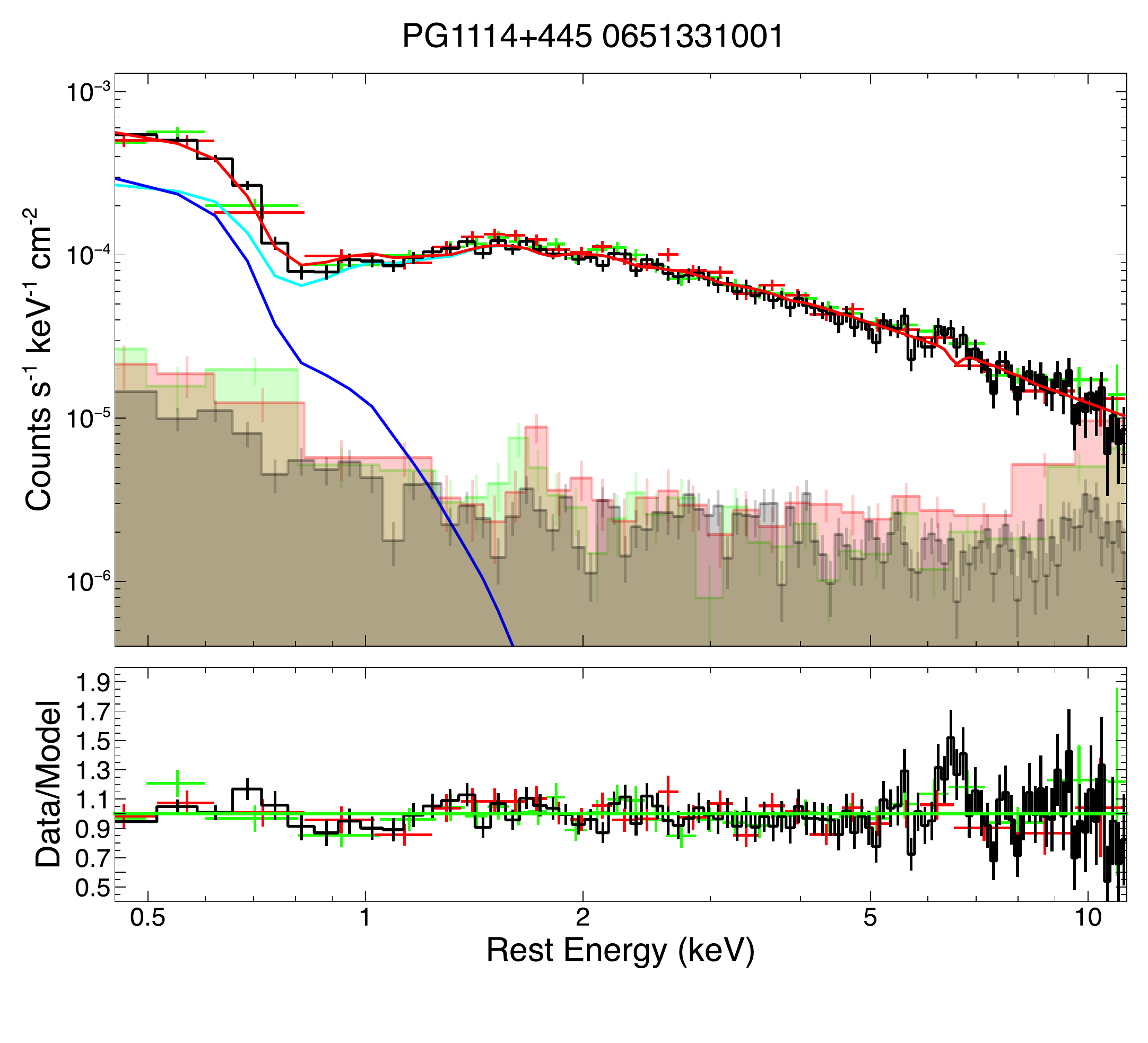}
\includegraphics[scale=0.25]{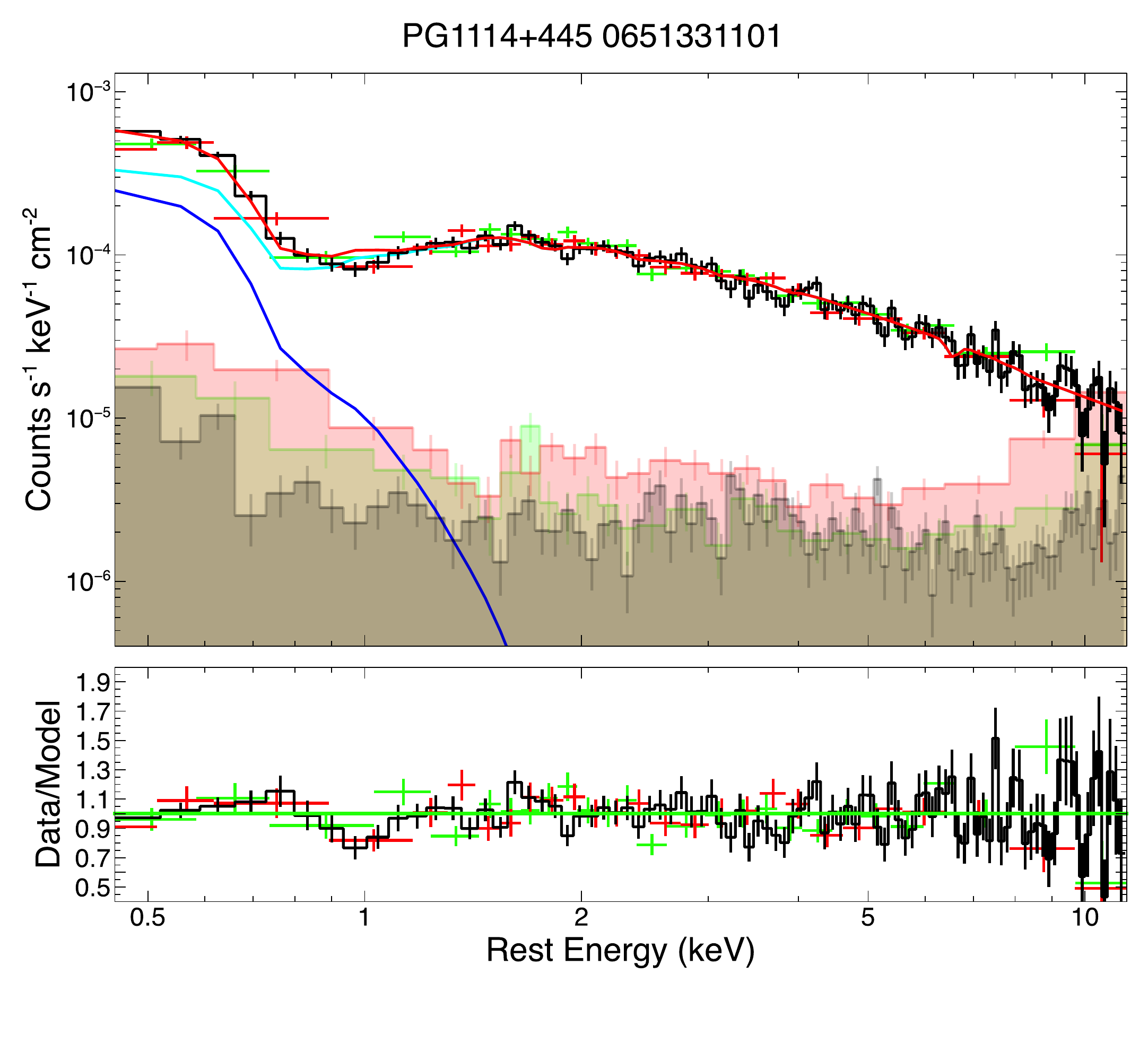}
\end{figure*}

\section{Blind-line search results}
\label{appsec: blind line search}

As in \autoref{fig:2massj165_lbqs}, we shown the residual (top) and blind-scan search contours (bottom) for the remaining 39 in Appendix\,\ref{figapp:sub_ALLOBS_SCAN_RA}.

\begin{figure*} 
\centering
\caption{Blind line search results, as in \autoref{fig:2massj165_lbqs}, for the \sub sample with the corresponding residuals of the EPIC-pn spectrum in the rest-frame energy between $5$--$10\kev$. }
\label{figapp:sub_ALLOBS_SCAN_RA}
\includegraphics[scale=0.21]{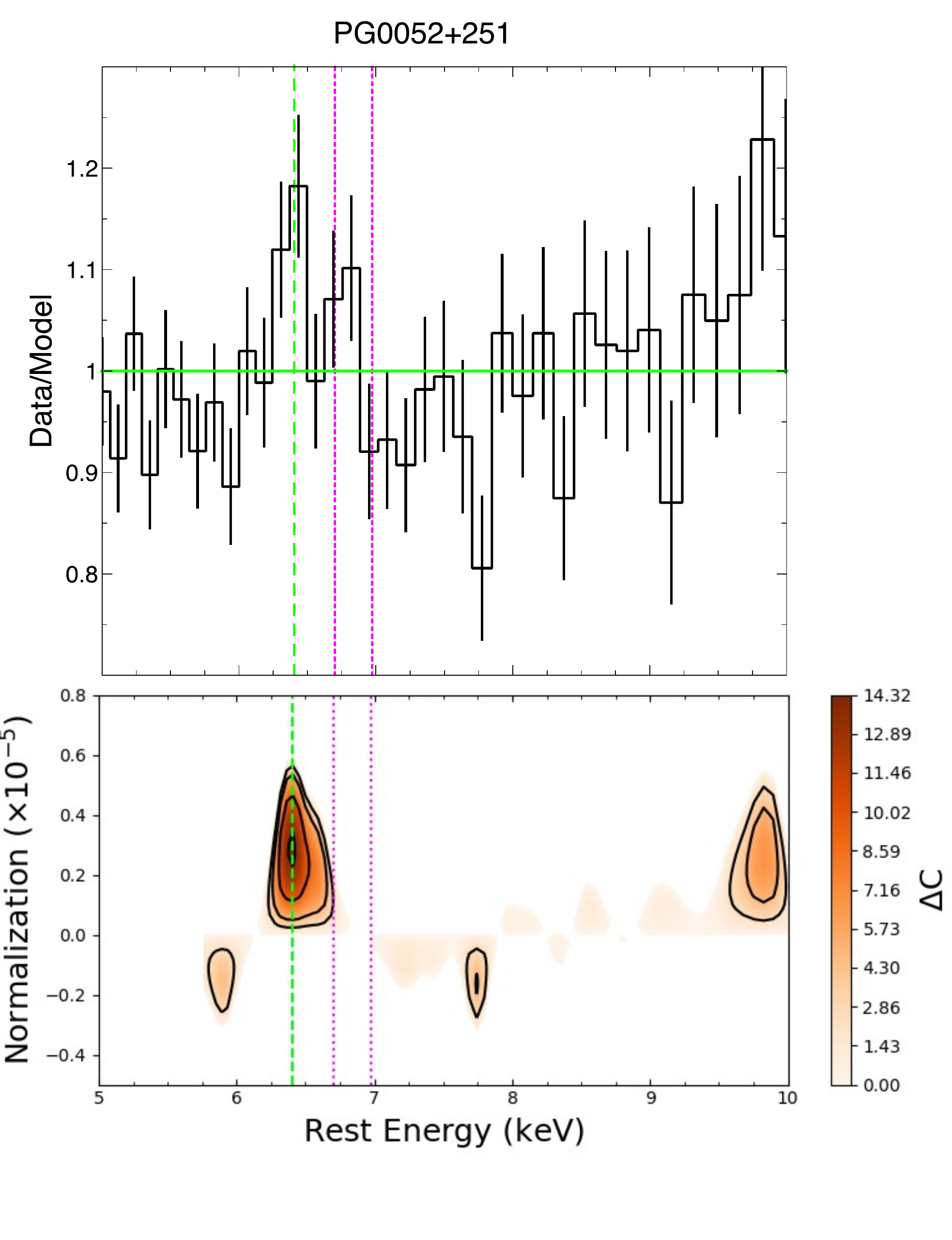}
\includegraphics[scale=0.21]{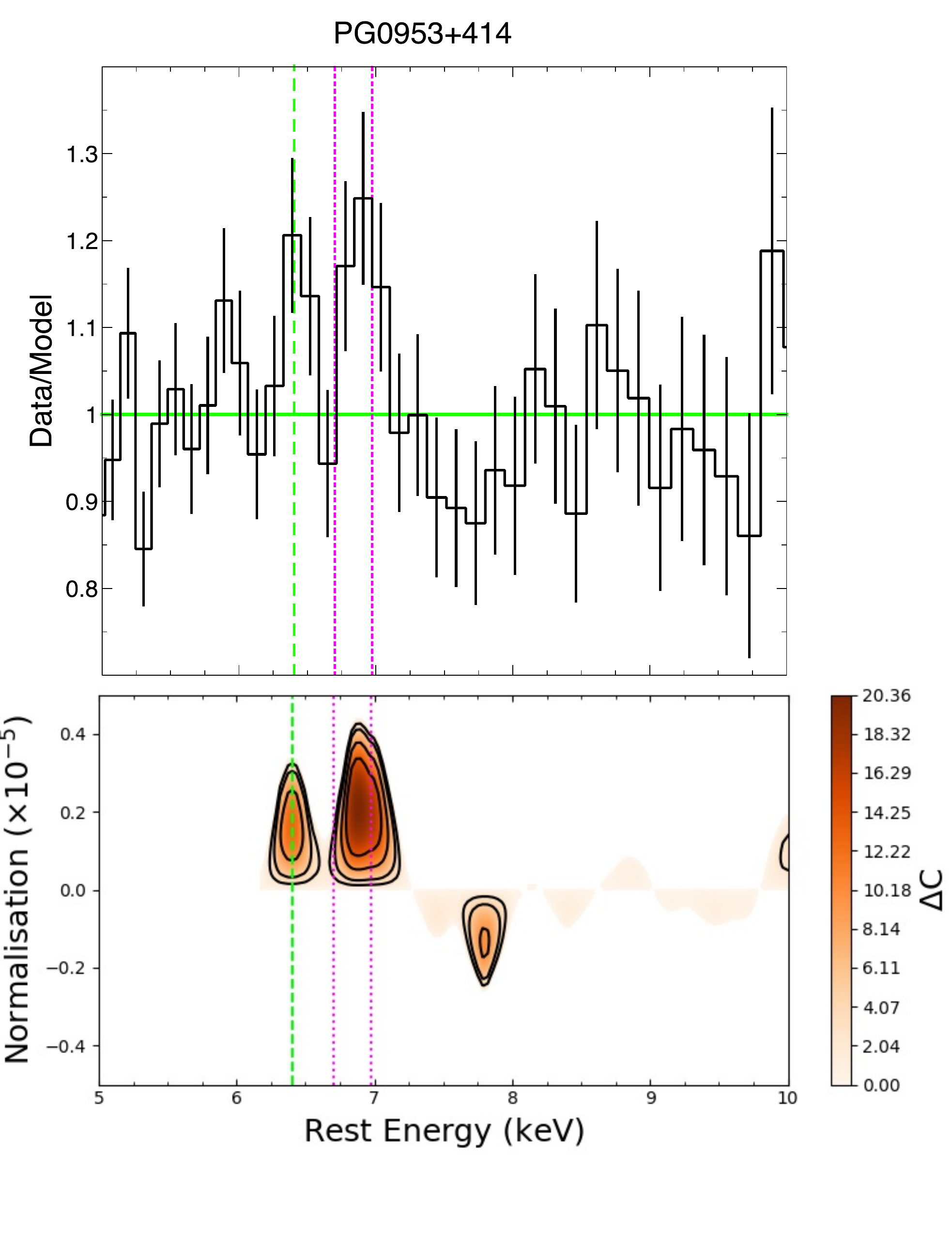}
\includegraphics[scale=0.21]{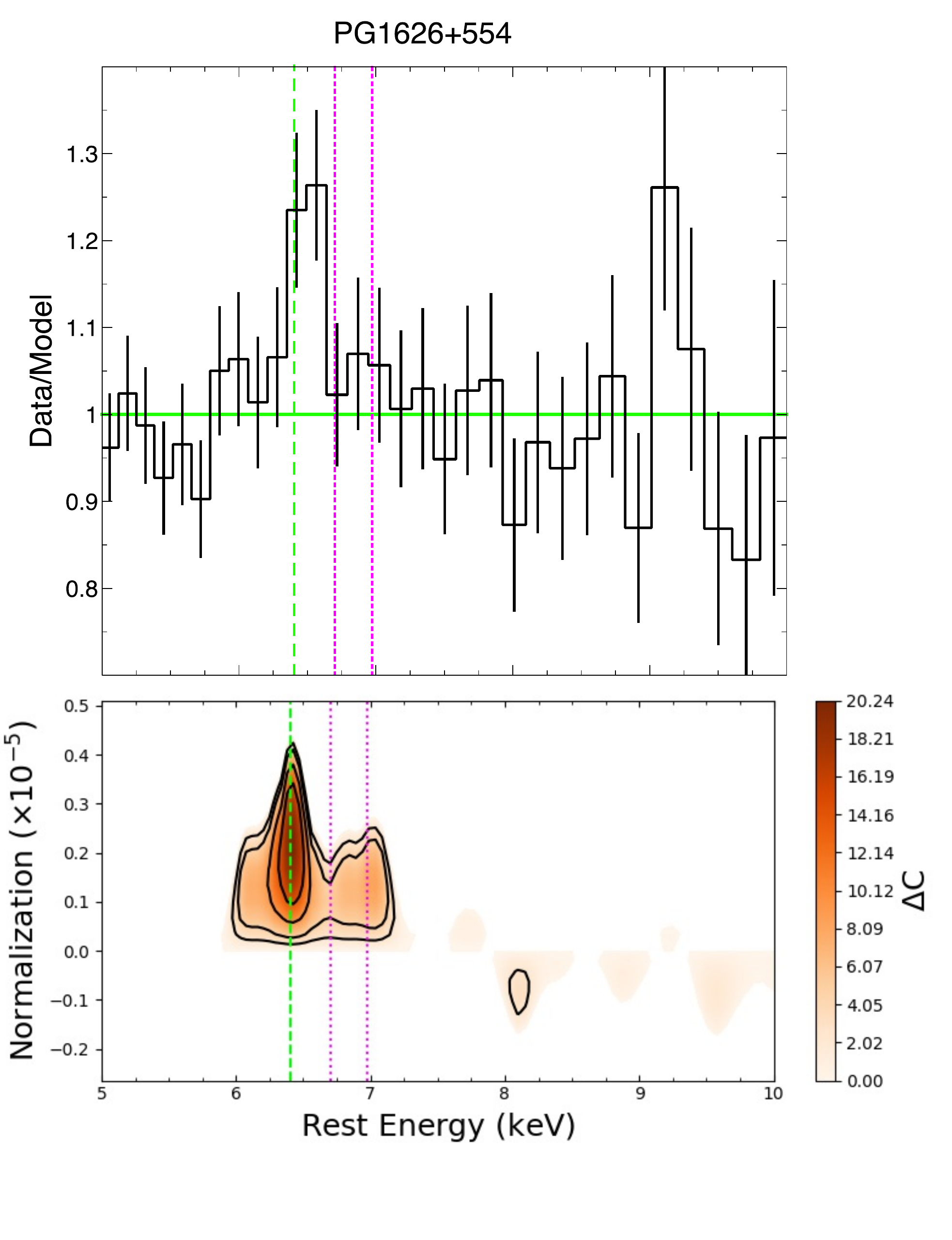}
\includegraphics[scale=0.21]{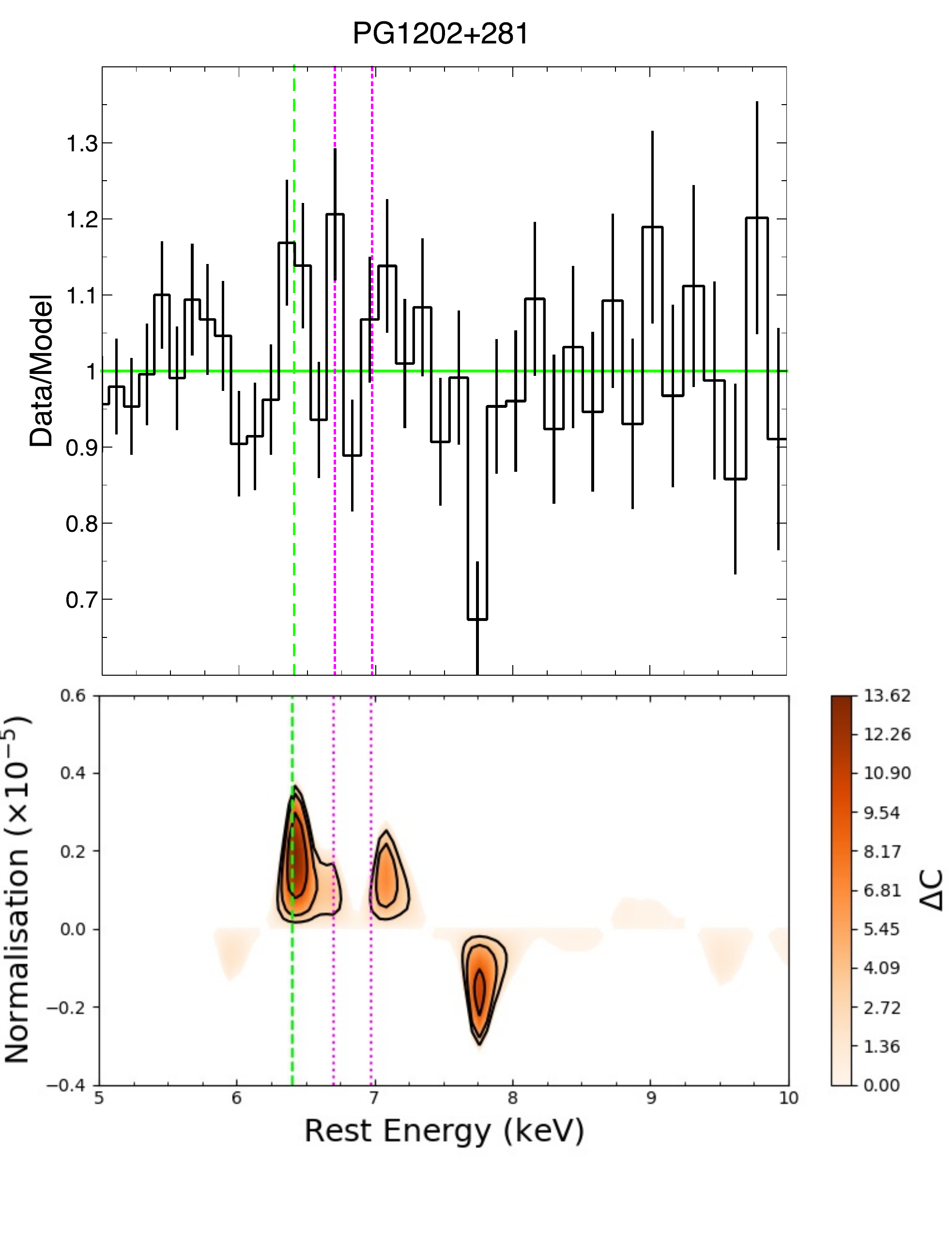}
\includegraphics[scale=0.21]{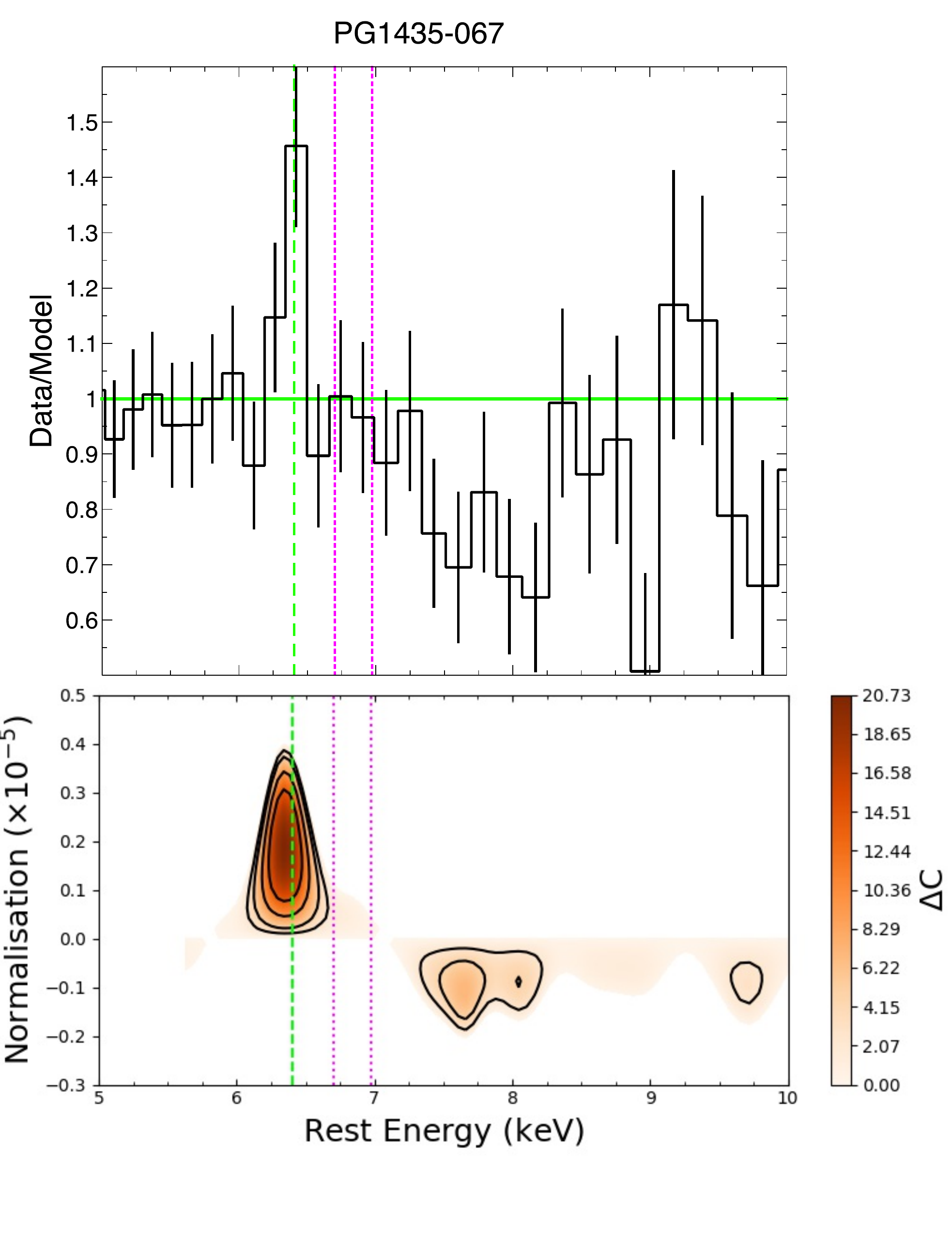}
\includegraphics[scale=0.21]{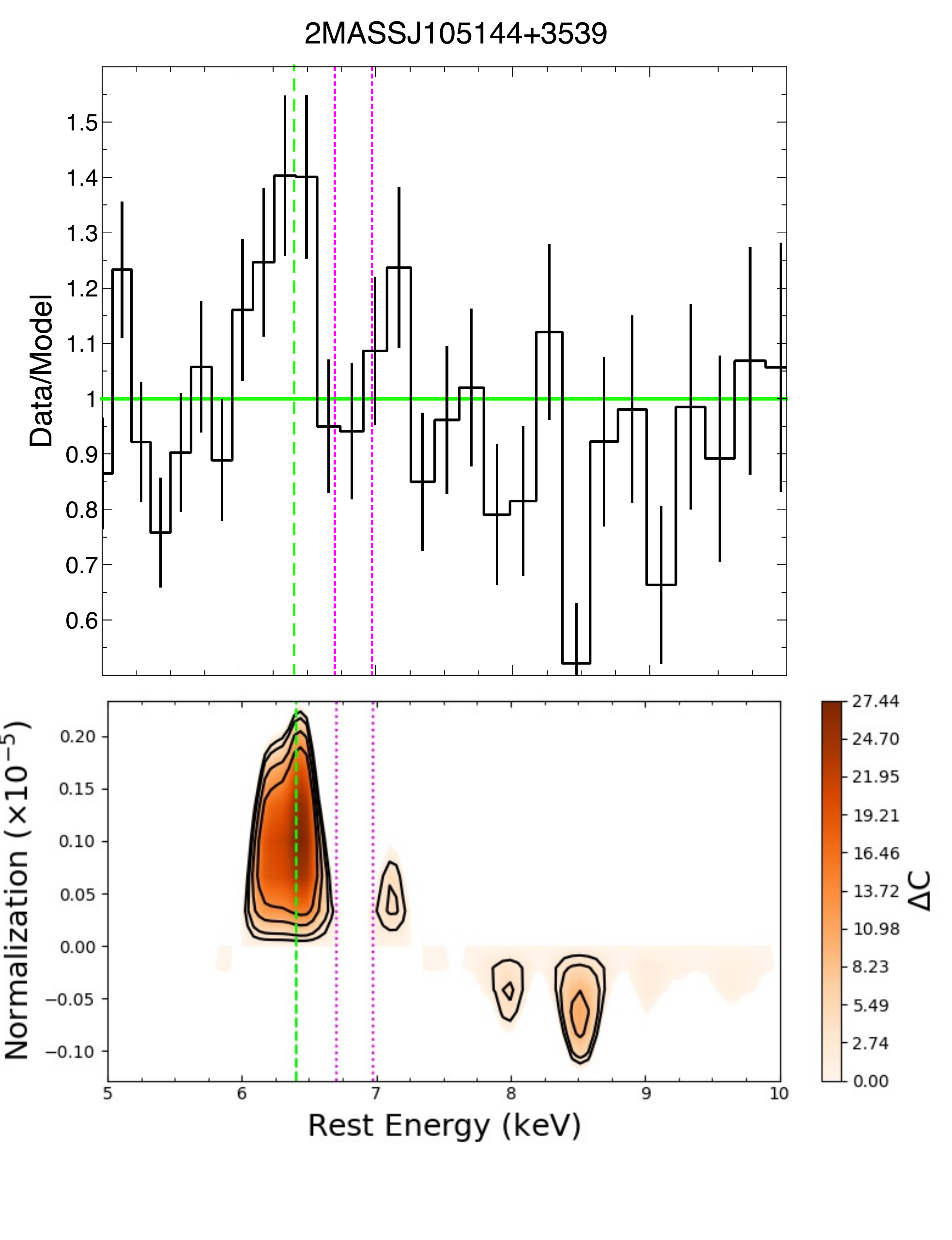}
\includegraphics[scale=0.21]{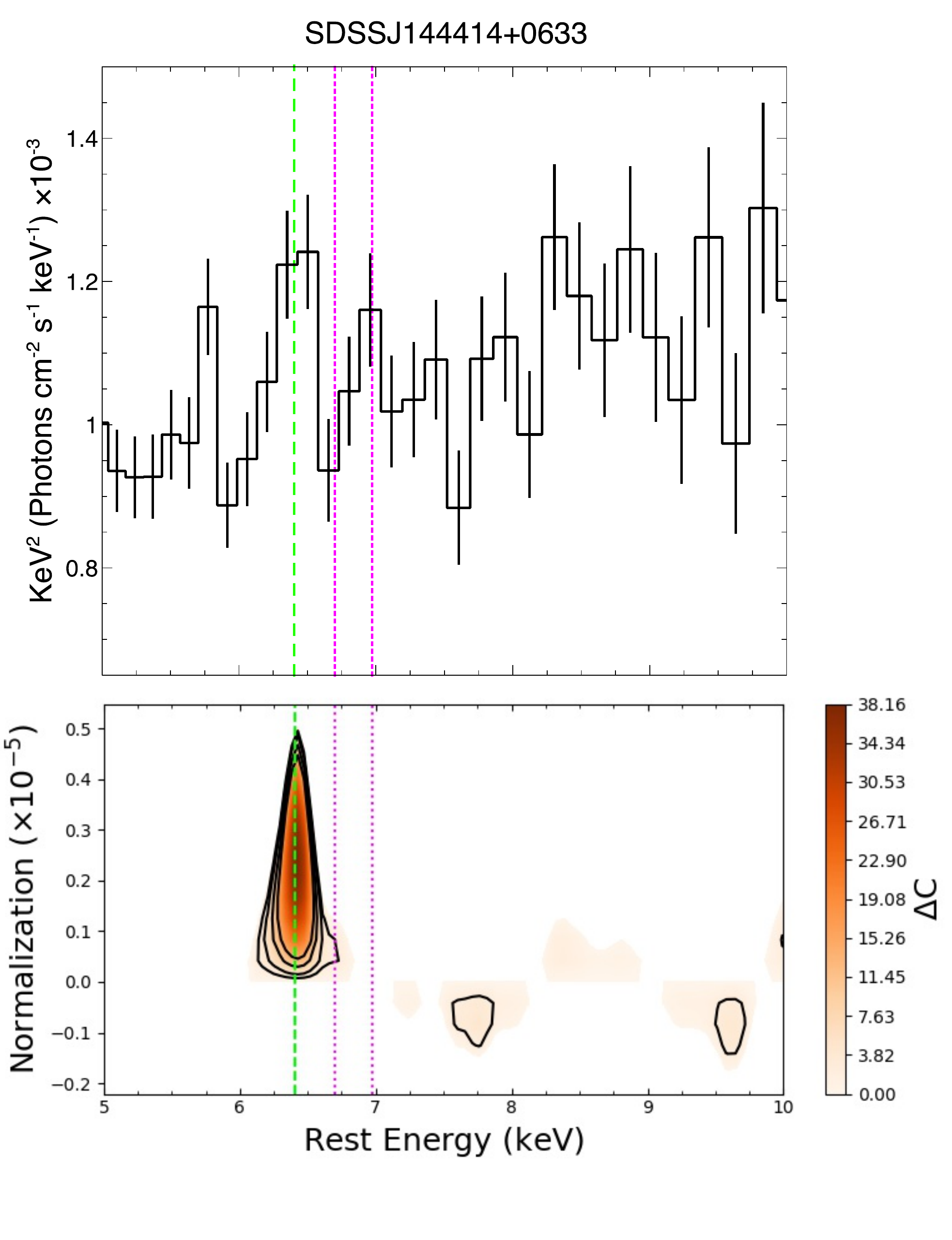}
\includegraphics[scale=0.21]{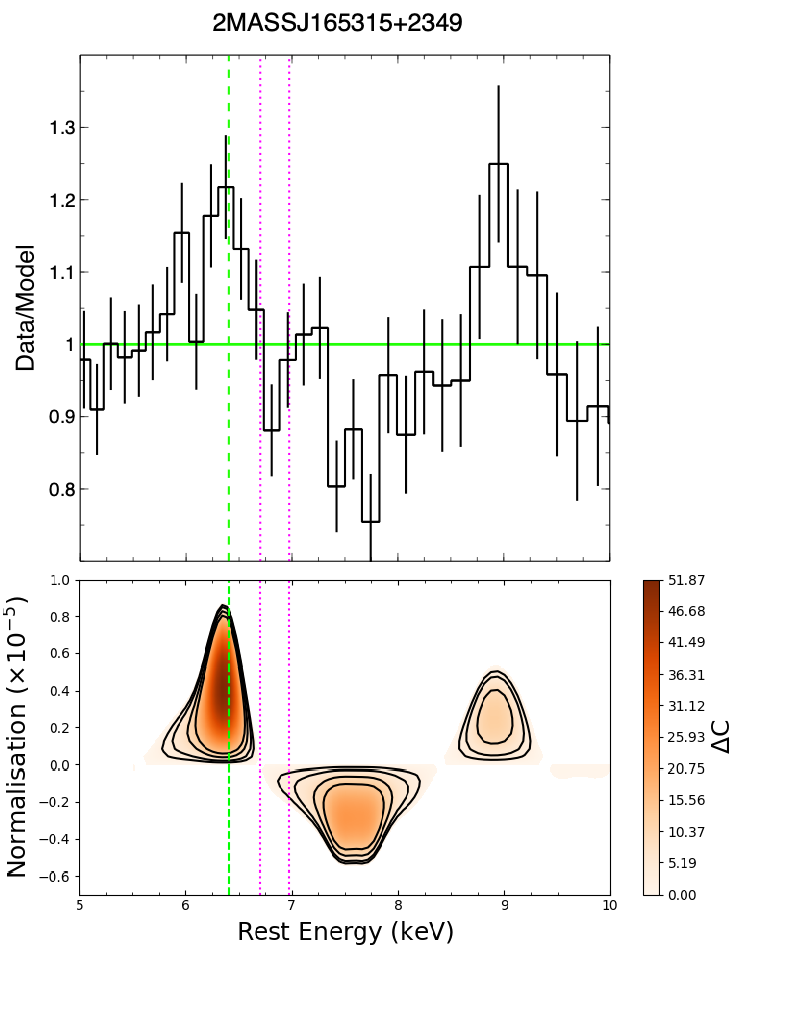}
\includegraphics[scale=0.21]{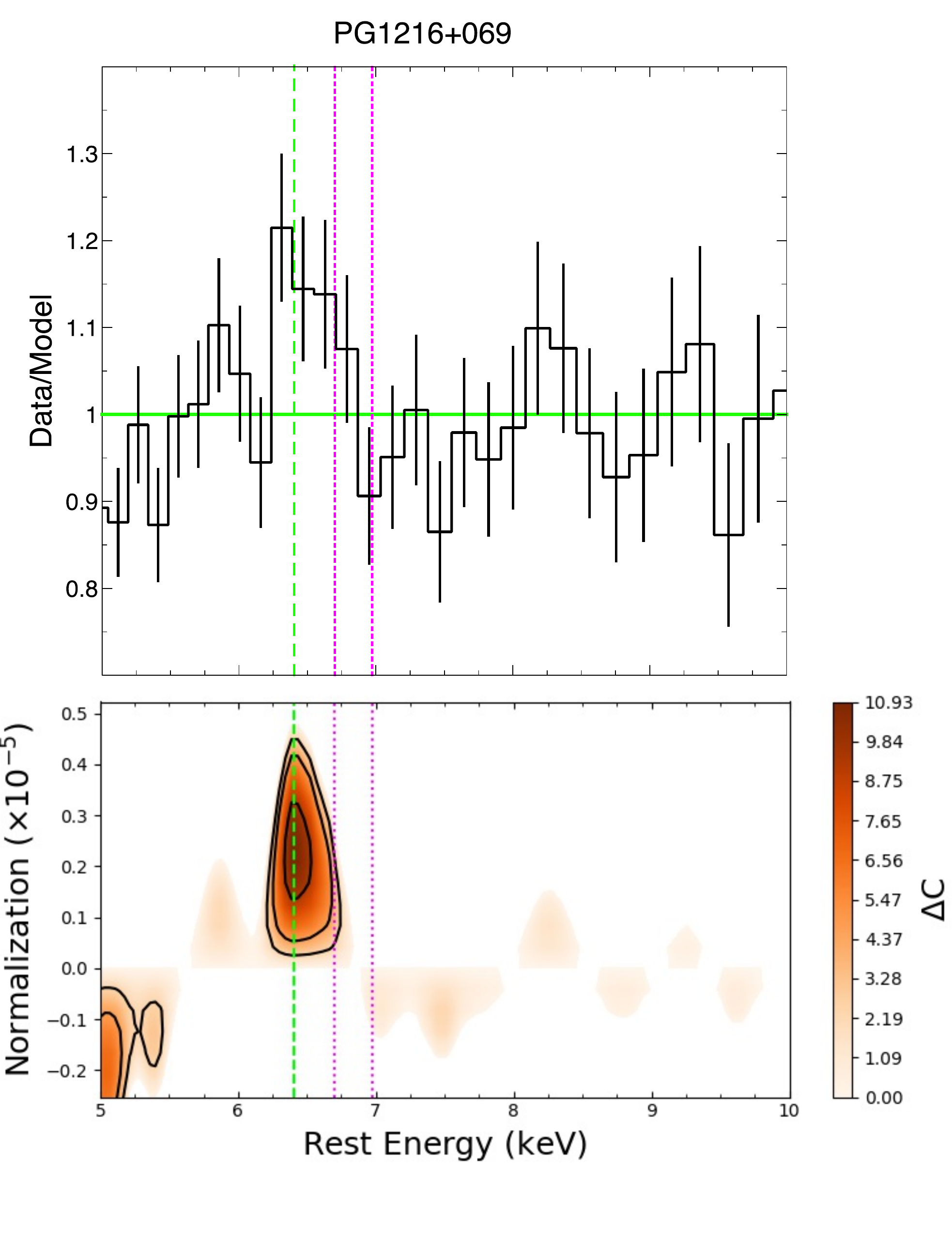}
\includegraphics[scale=0.21]{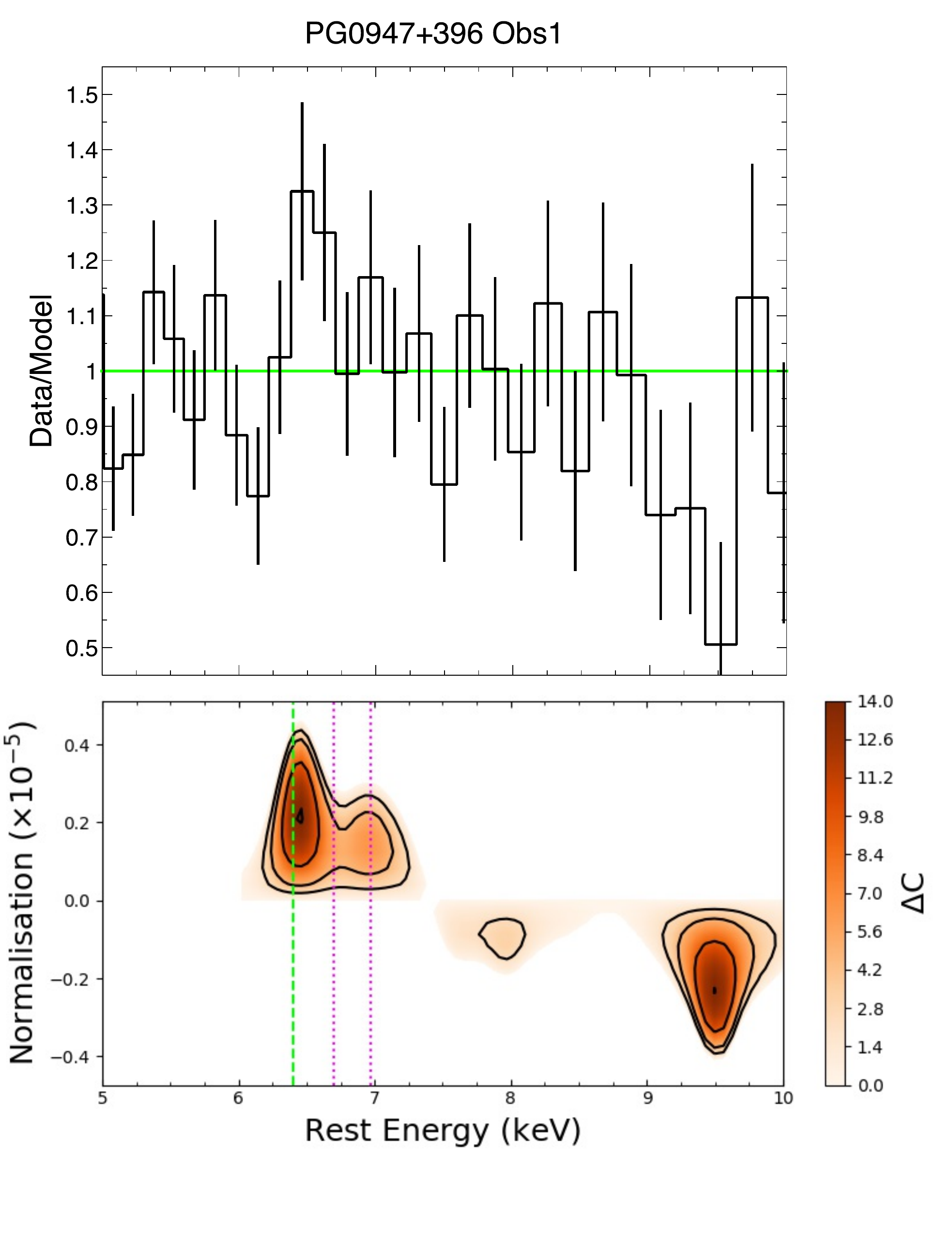}
\includegraphics[scale=0.21]{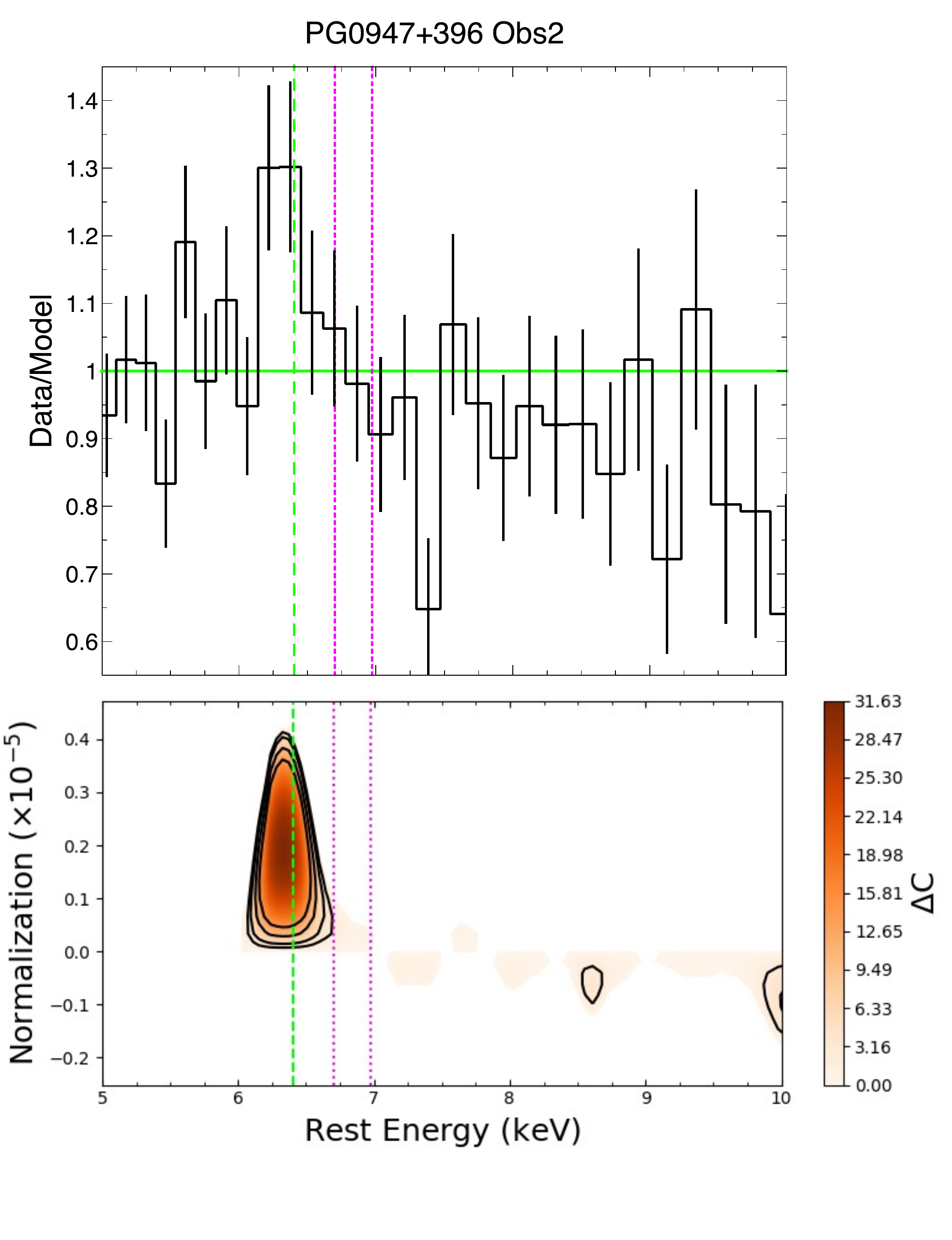}
\includegraphics[scale=0.21]{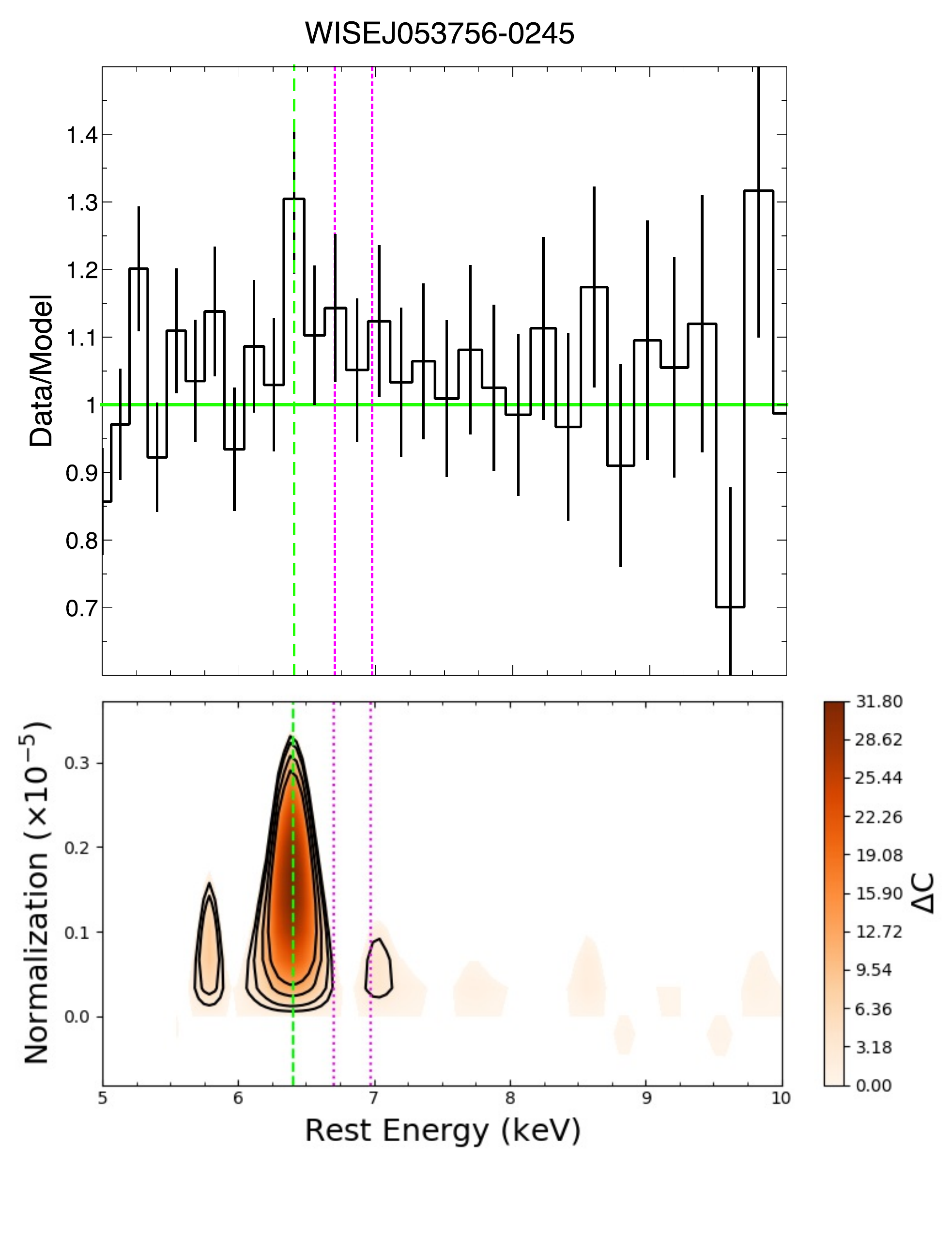}
\includegraphics[scale=0.21]{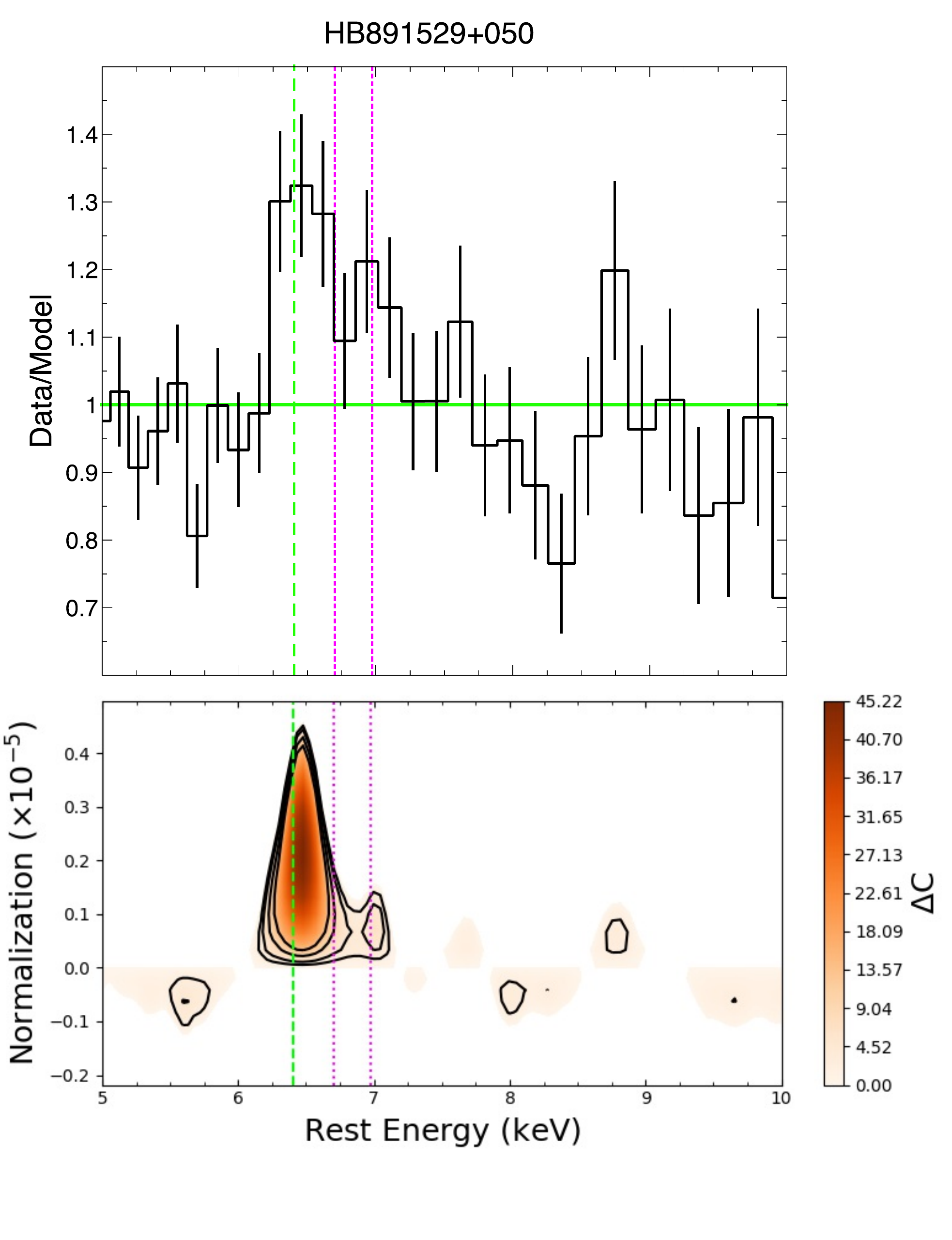}
\includegraphics[scale=0.21]{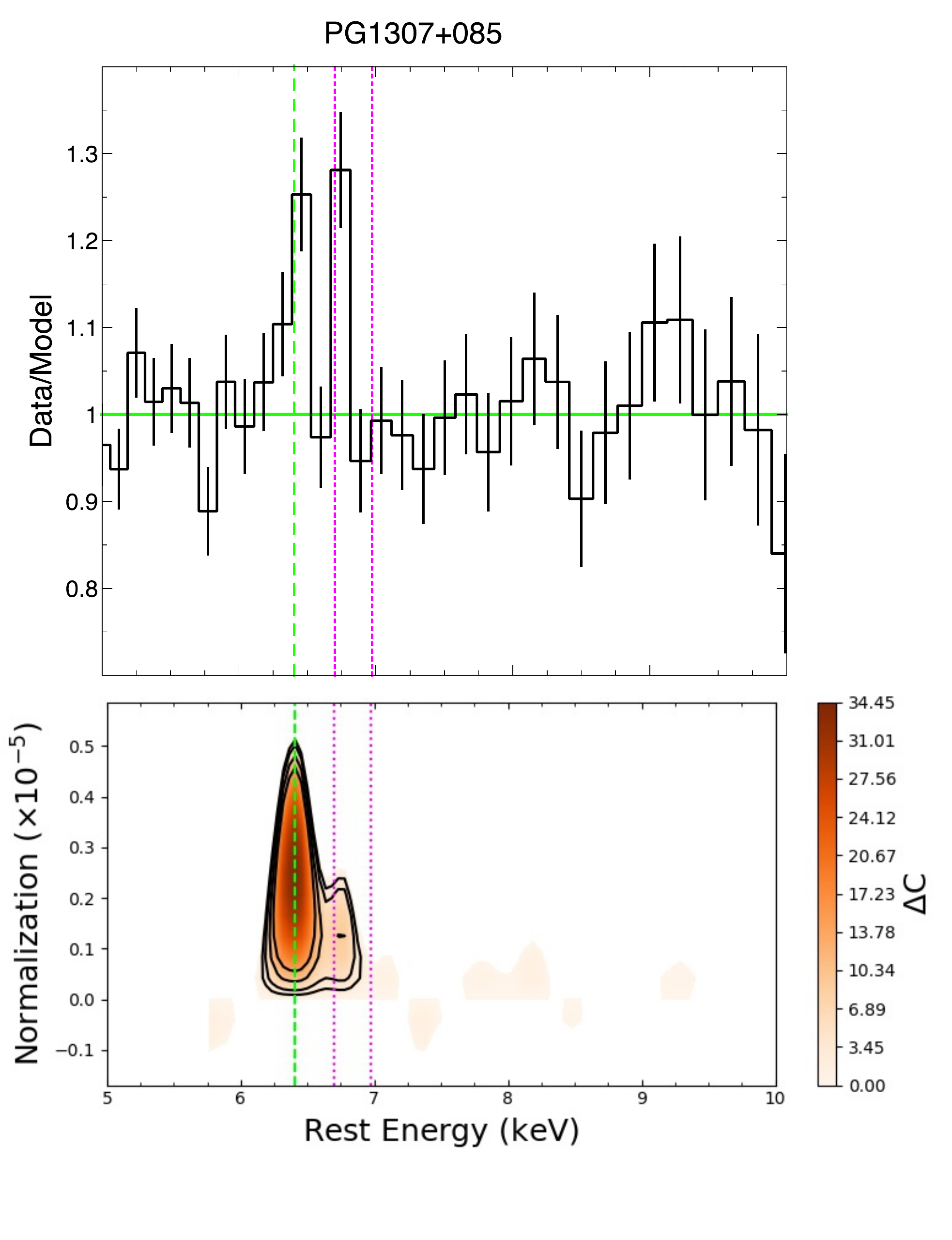}
\includegraphics[scale=0.21]{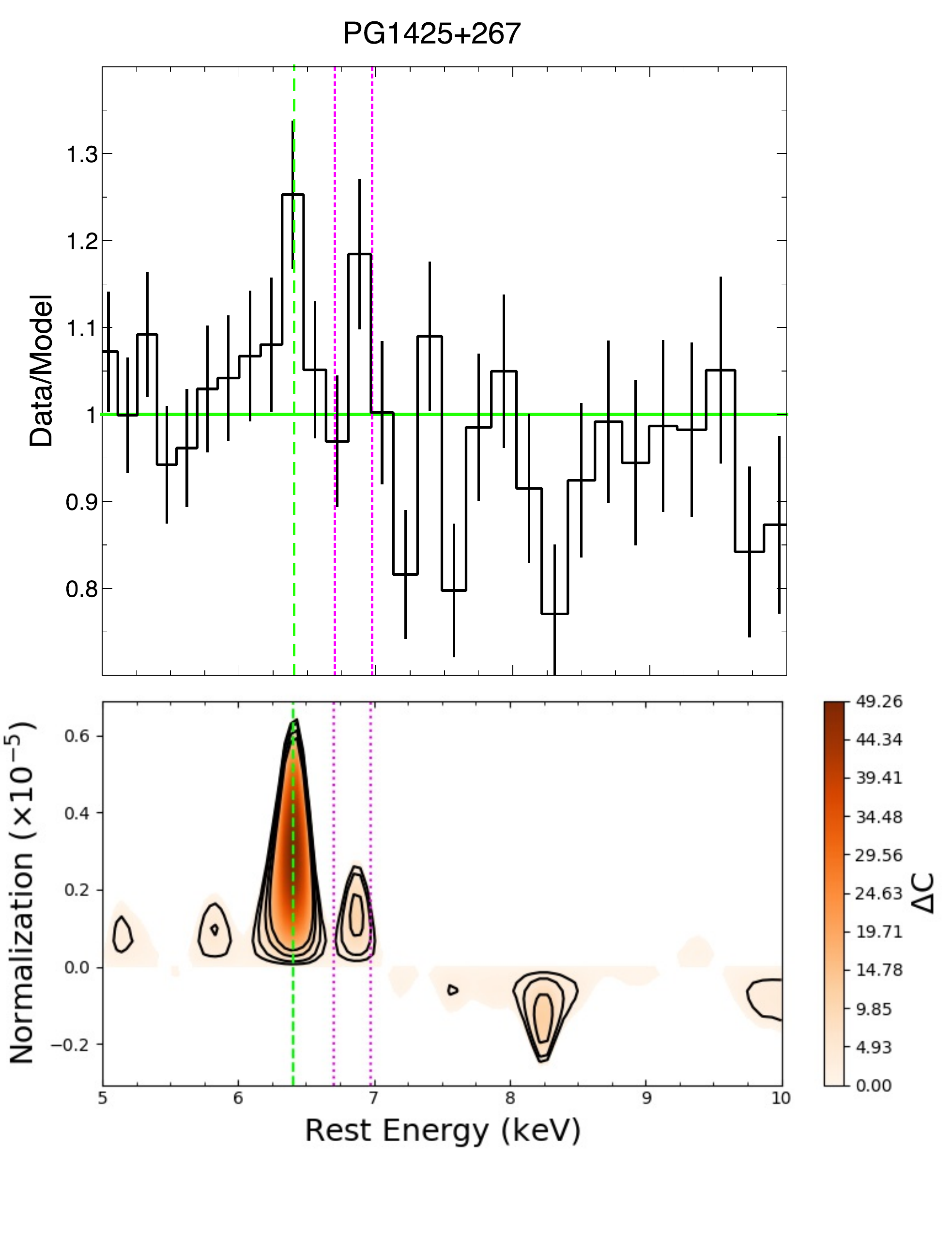}
\includegraphics[scale=0.21]{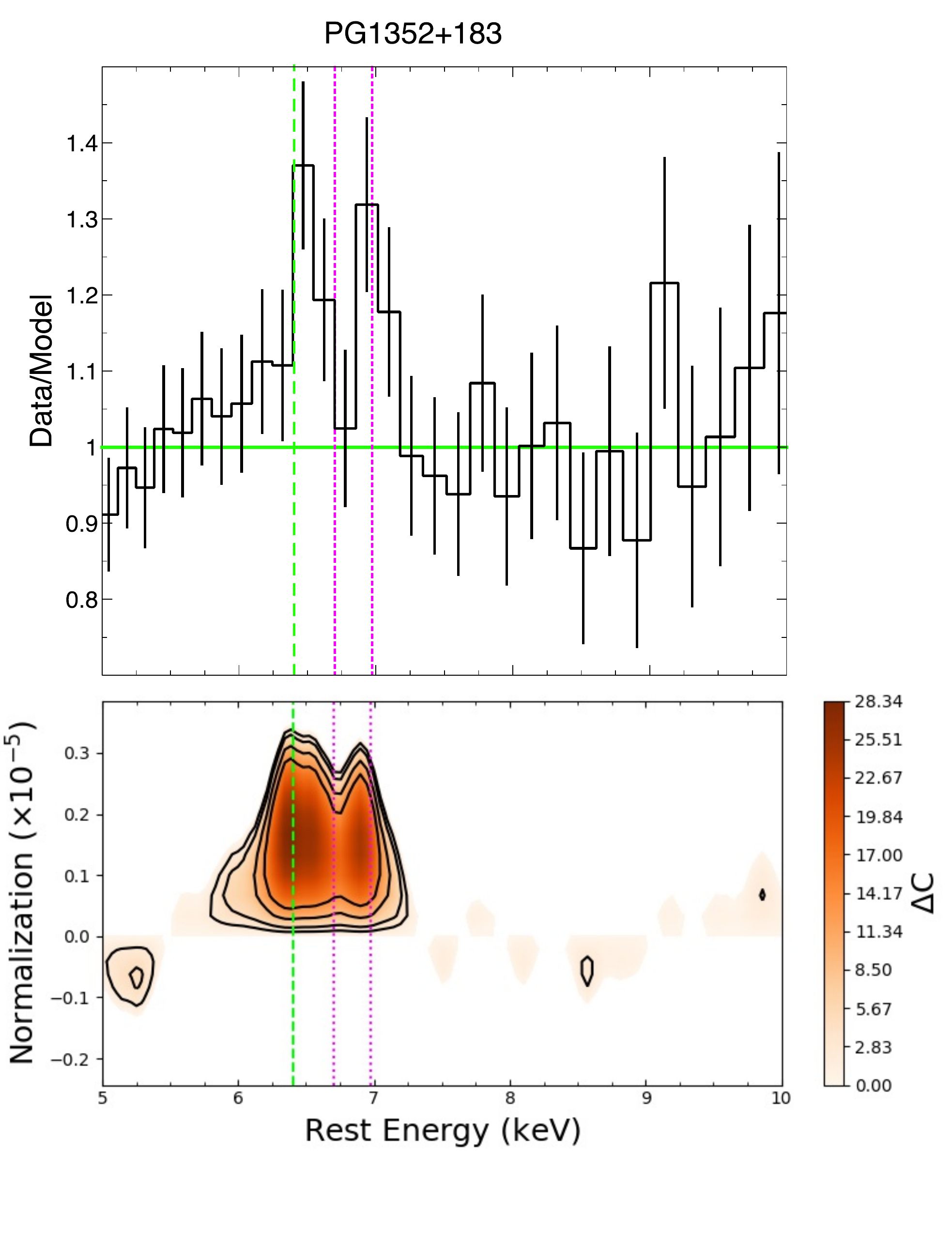}

\end{figure*}
\begin{figure*} 
\centering
\includegraphics[scale=0.21]{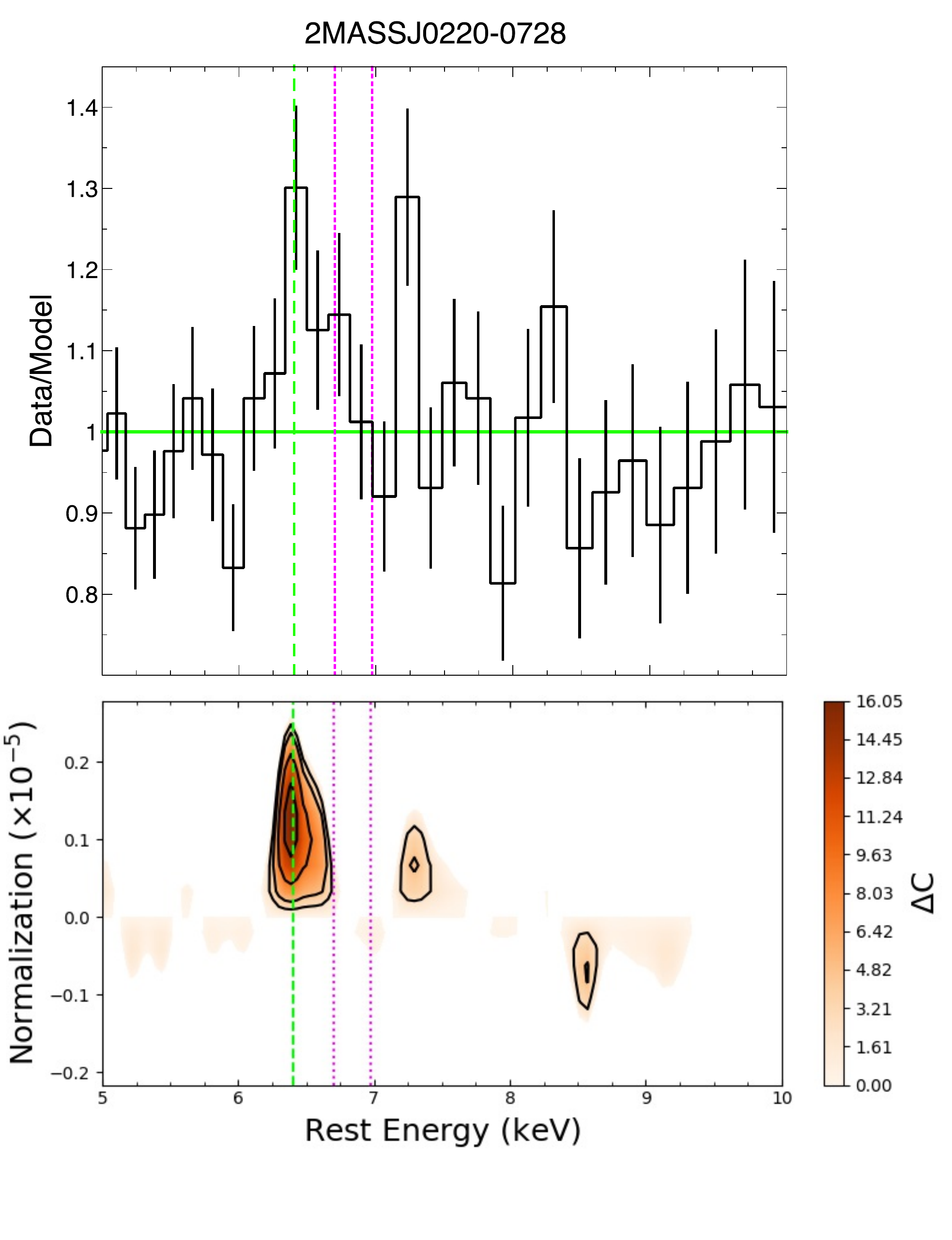}
\includegraphics[scale=0.21]{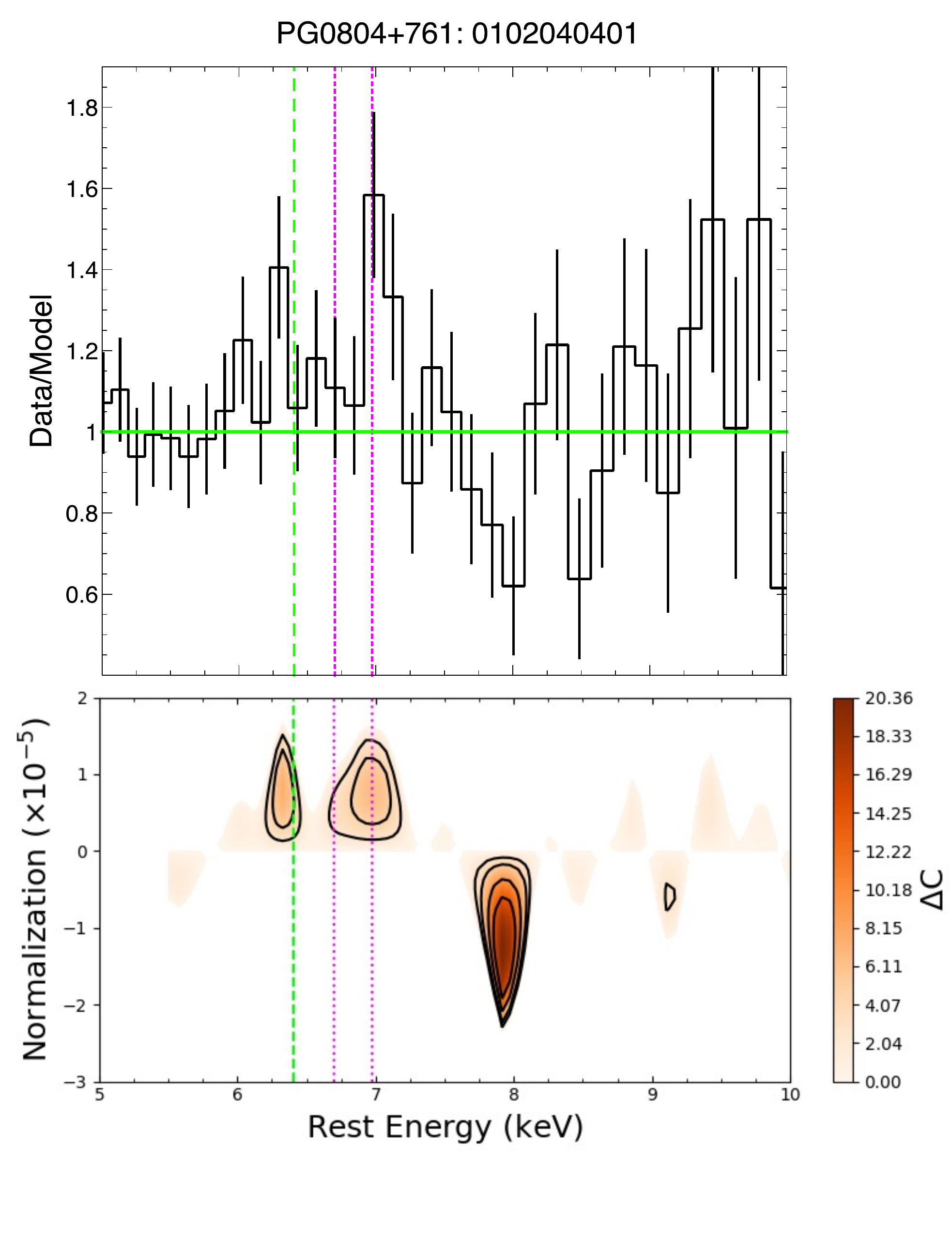}
\includegraphics[scale=0.21]{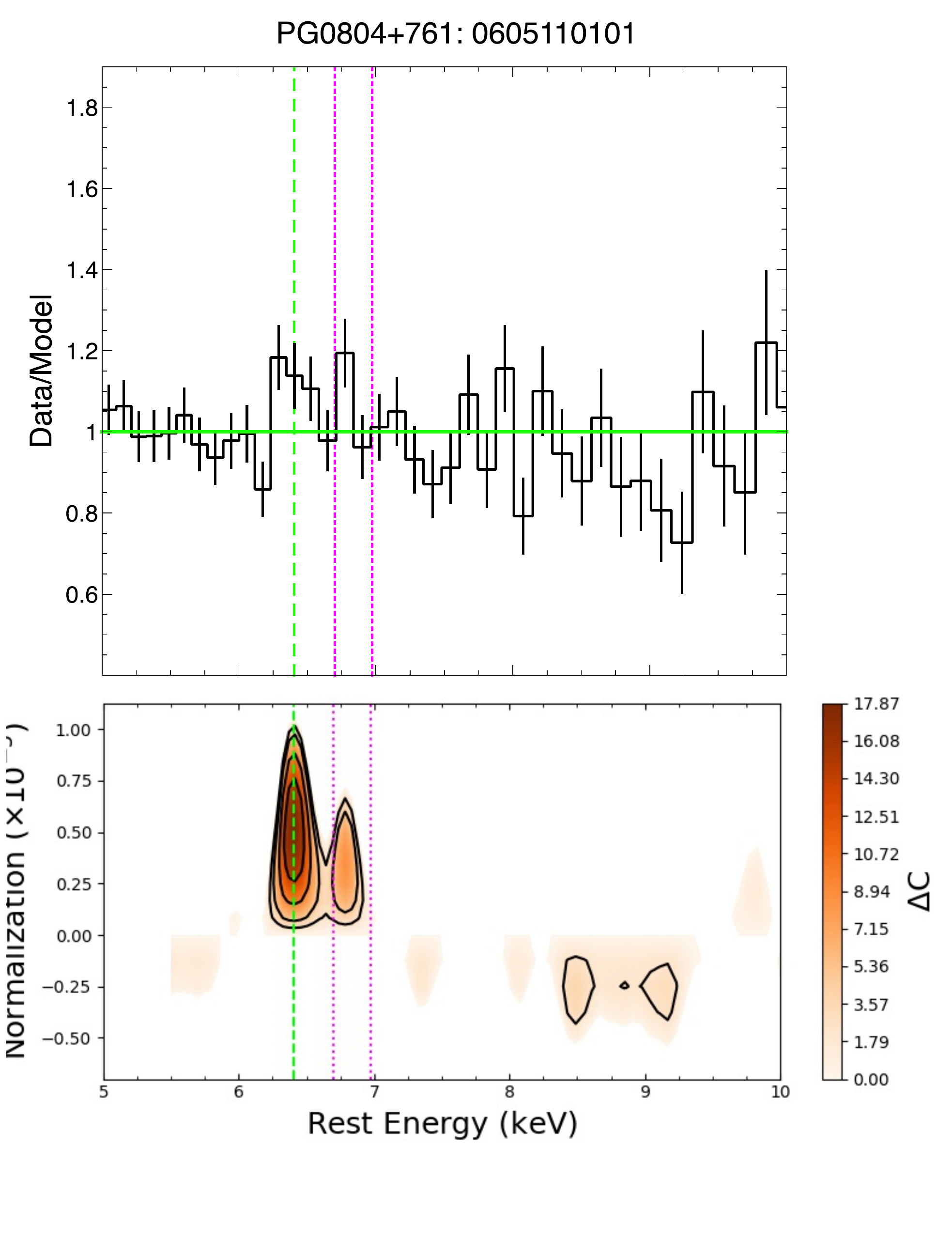}
\includegraphics[scale=0.21]{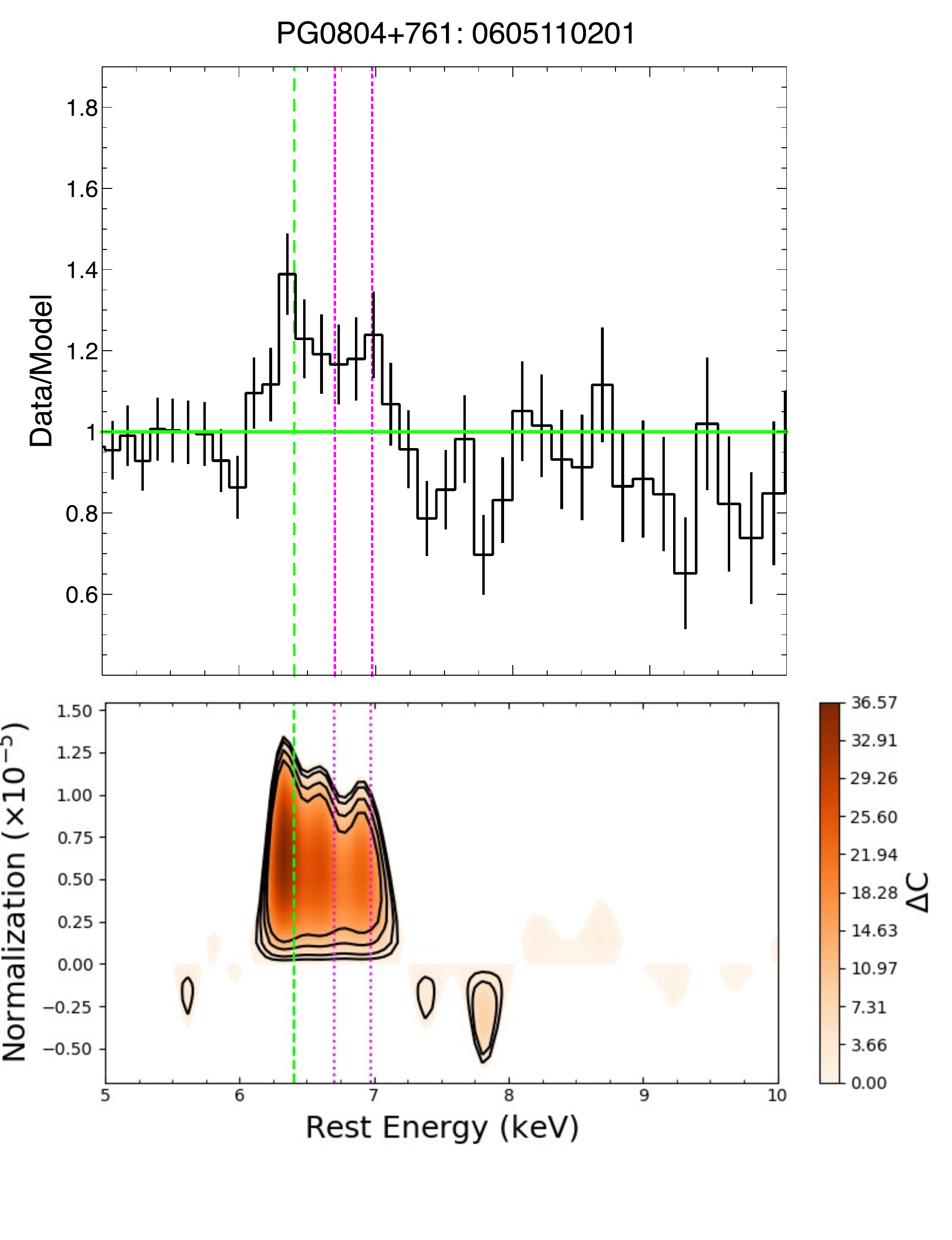}
\includegraphics[scale=0.21]{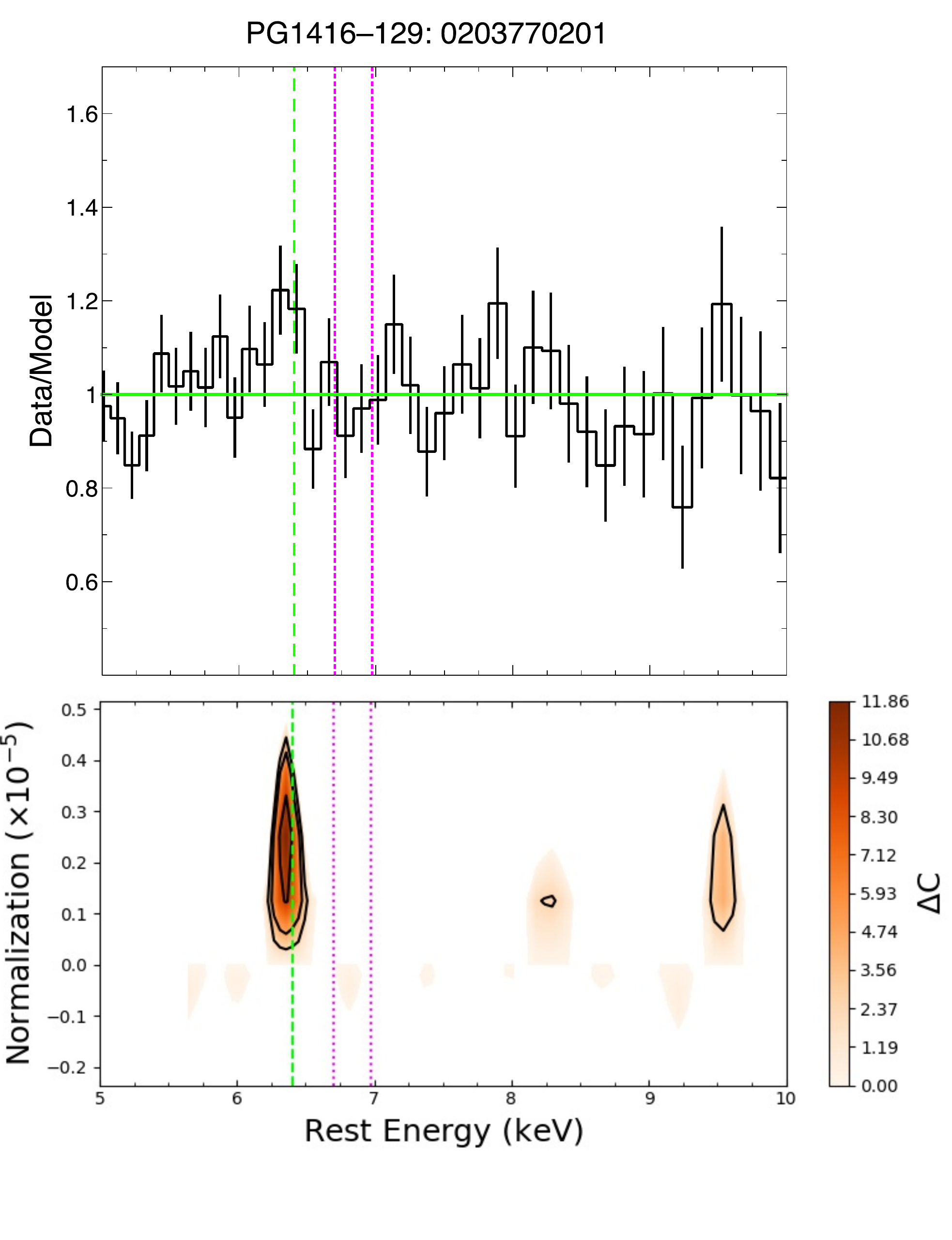}
\includegraphics[scale=0.21]{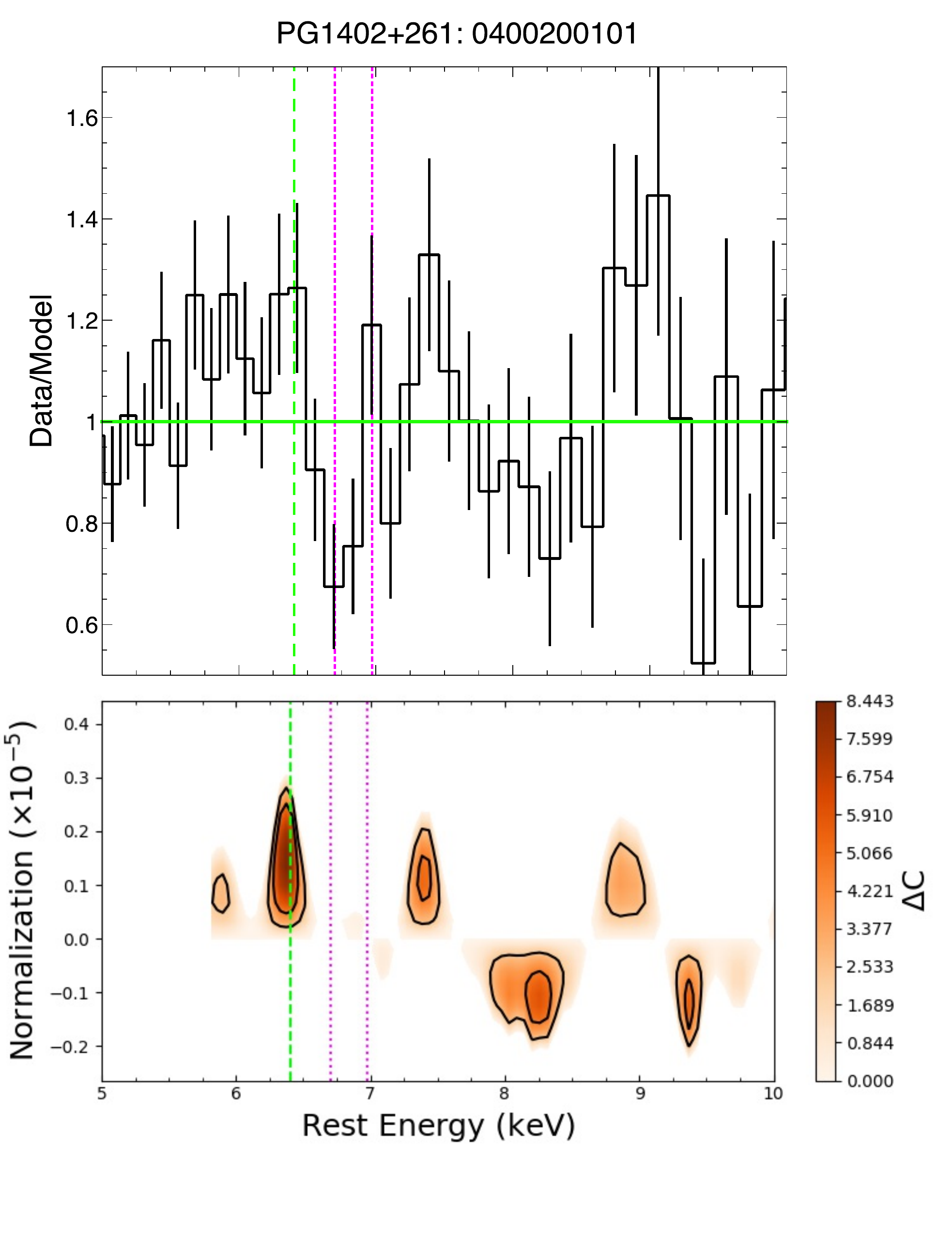}
\includegraphics[scale=0.21]{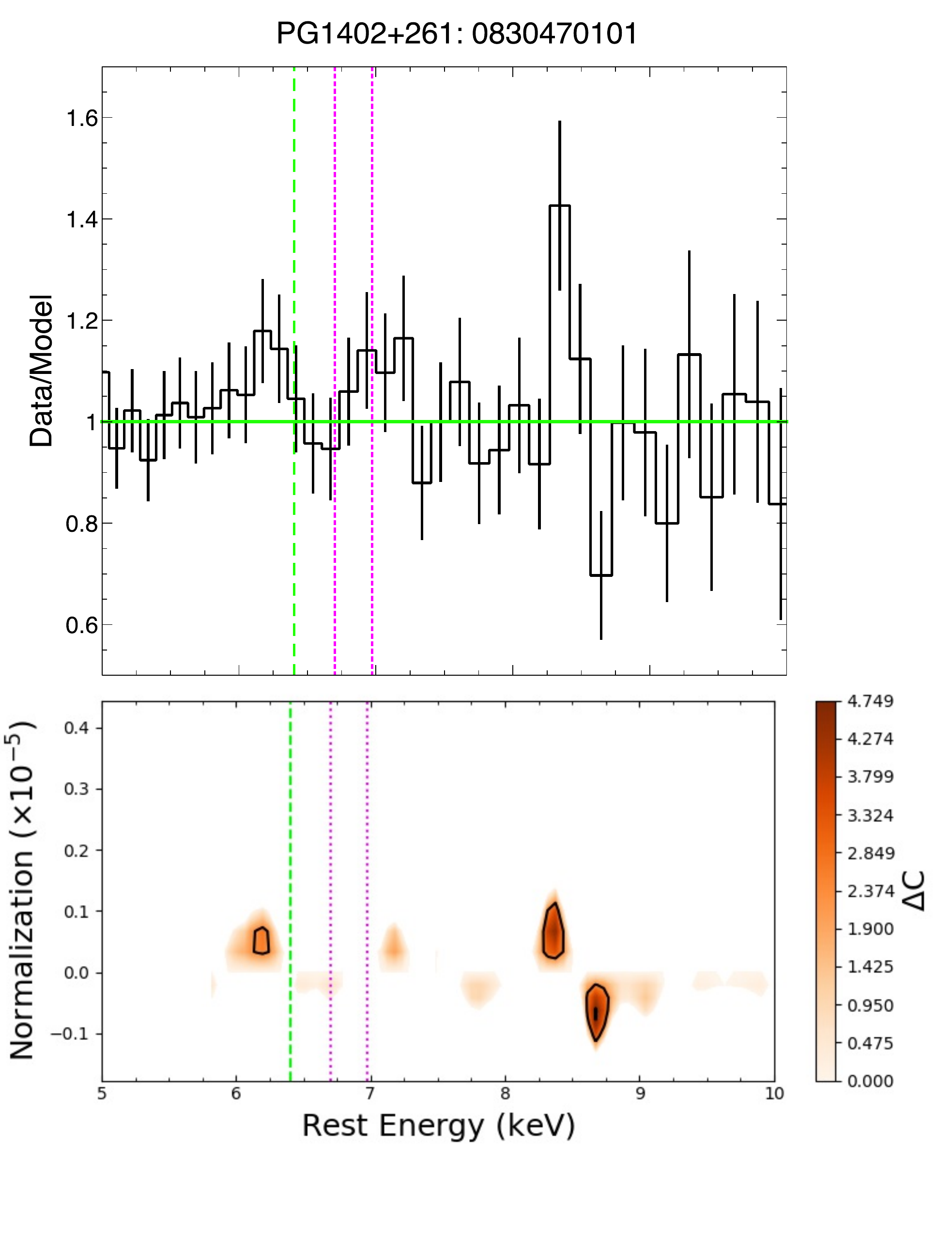}
\includegraphics[scale=0.21]{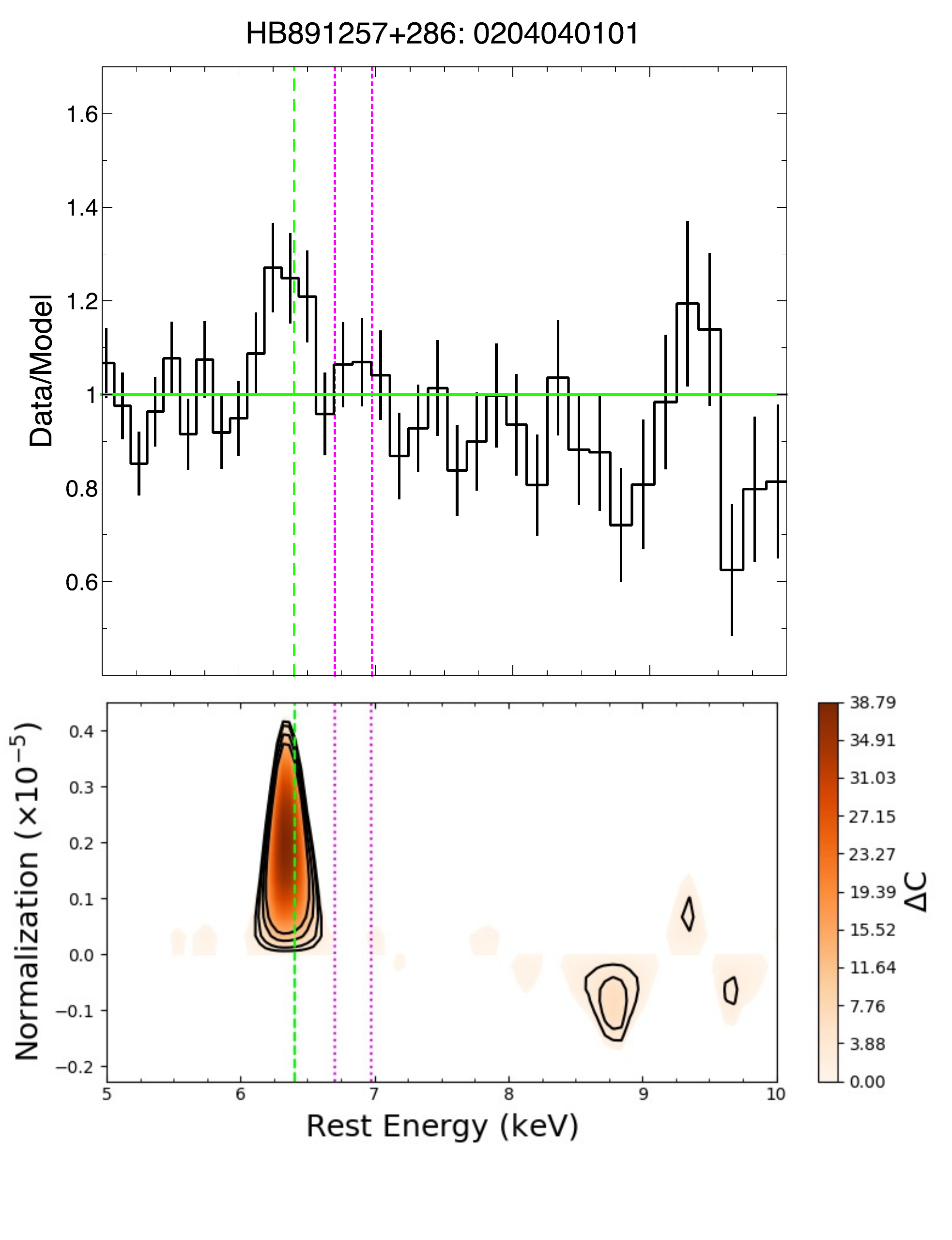}	
\includegraphics[scale=0.21]{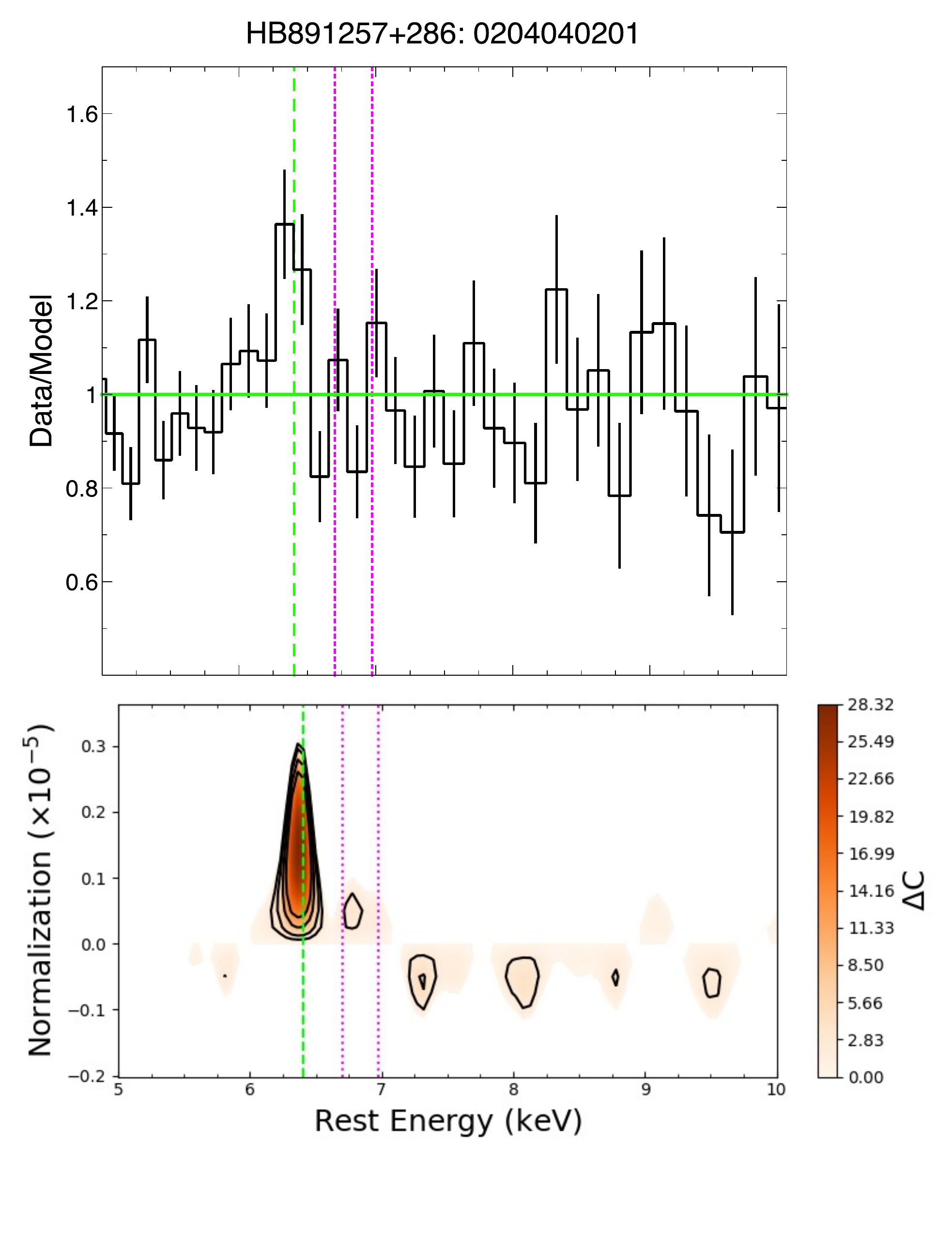}
\includegraphics[scale=0.21]{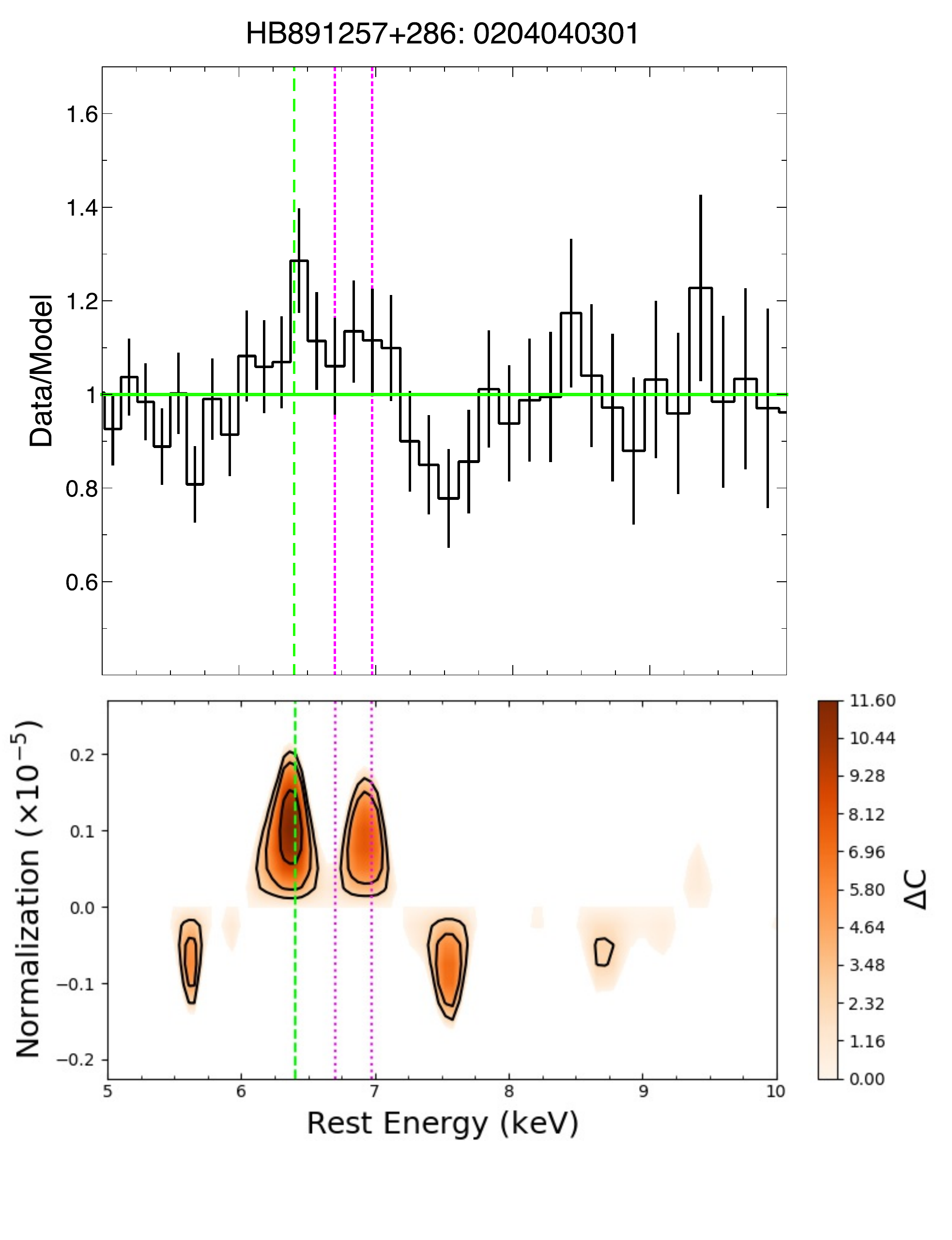}
\includegraphics[scale=0.21]{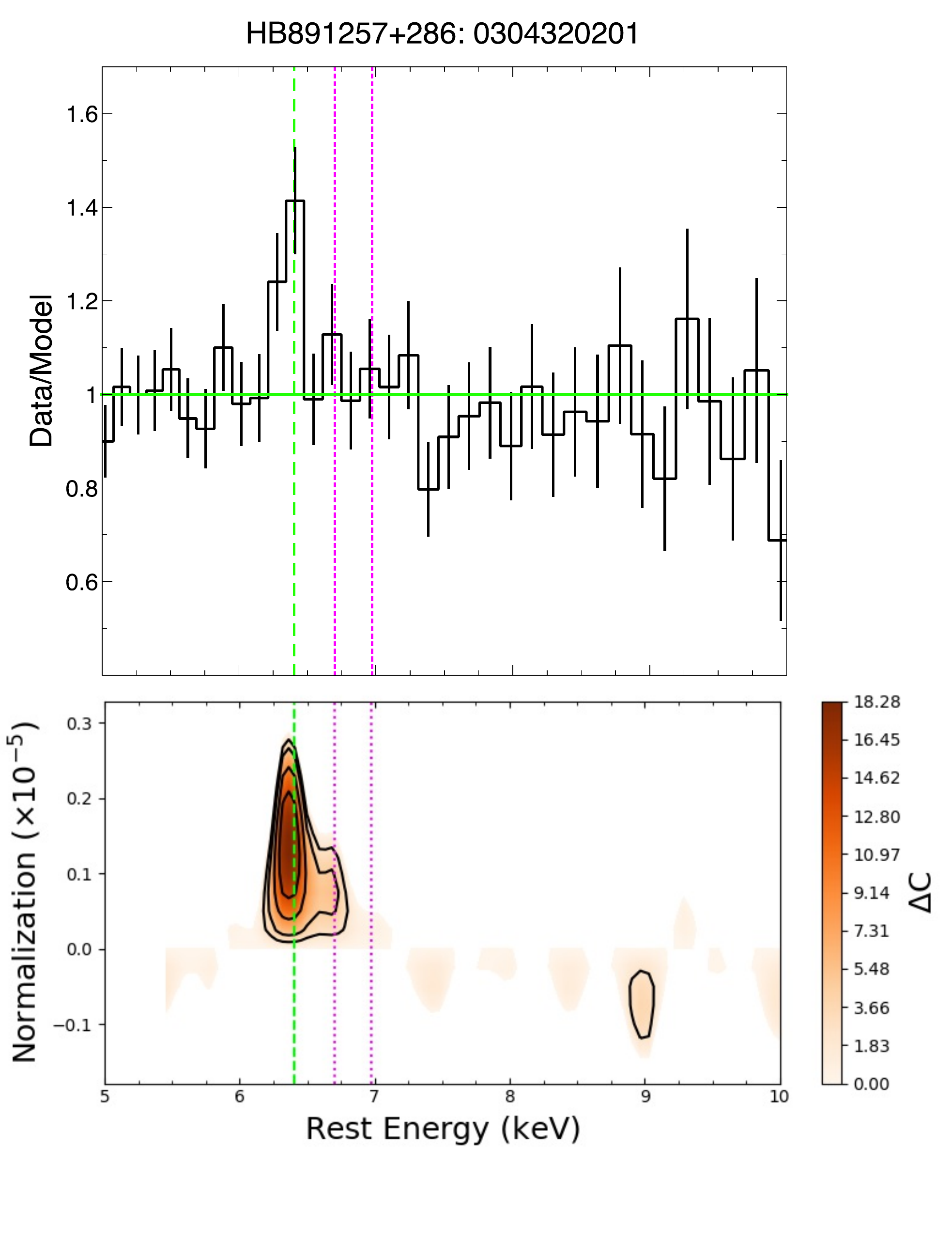}
\includegraphics[scale=0.21]{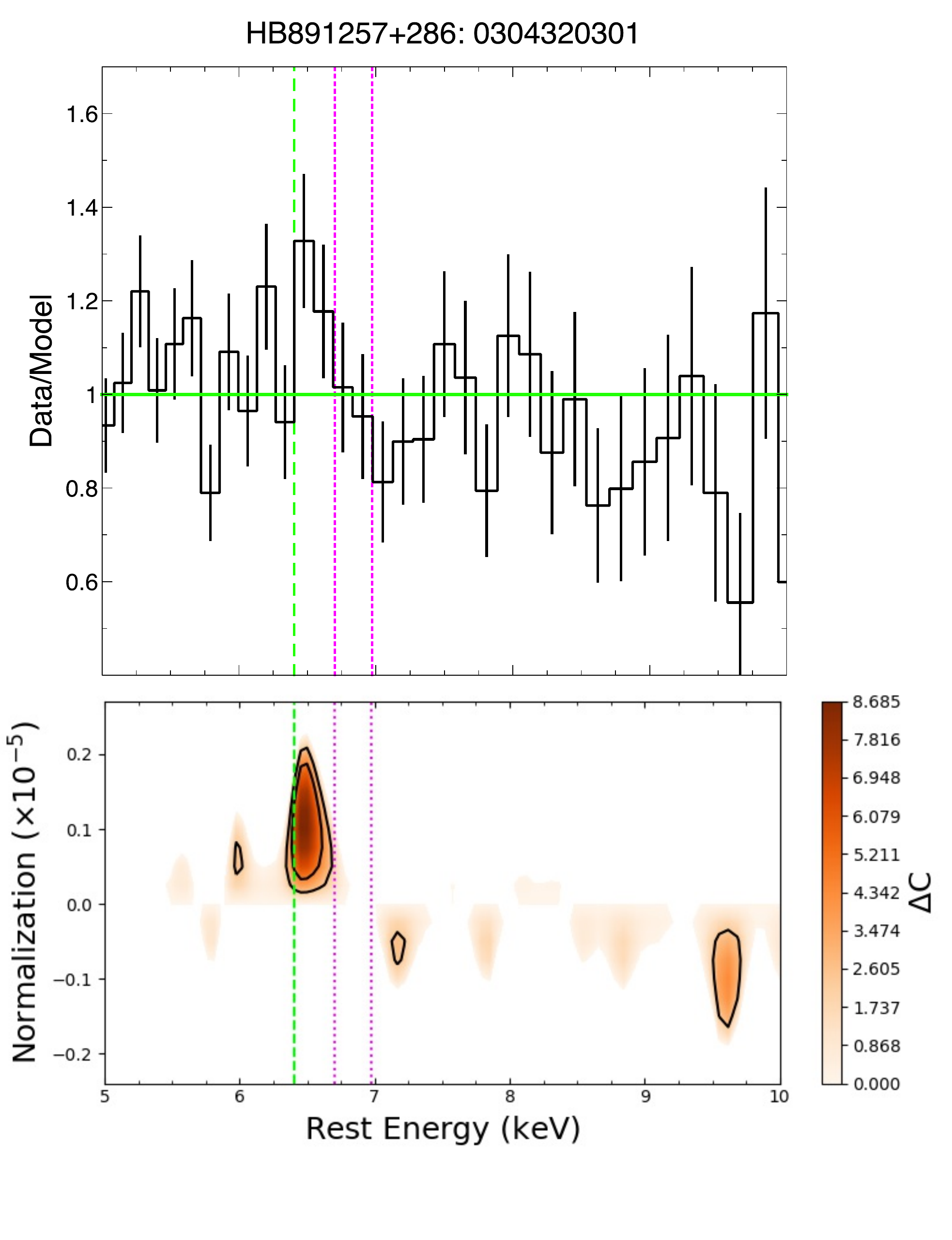}
\includegraphics[scale=0.21]{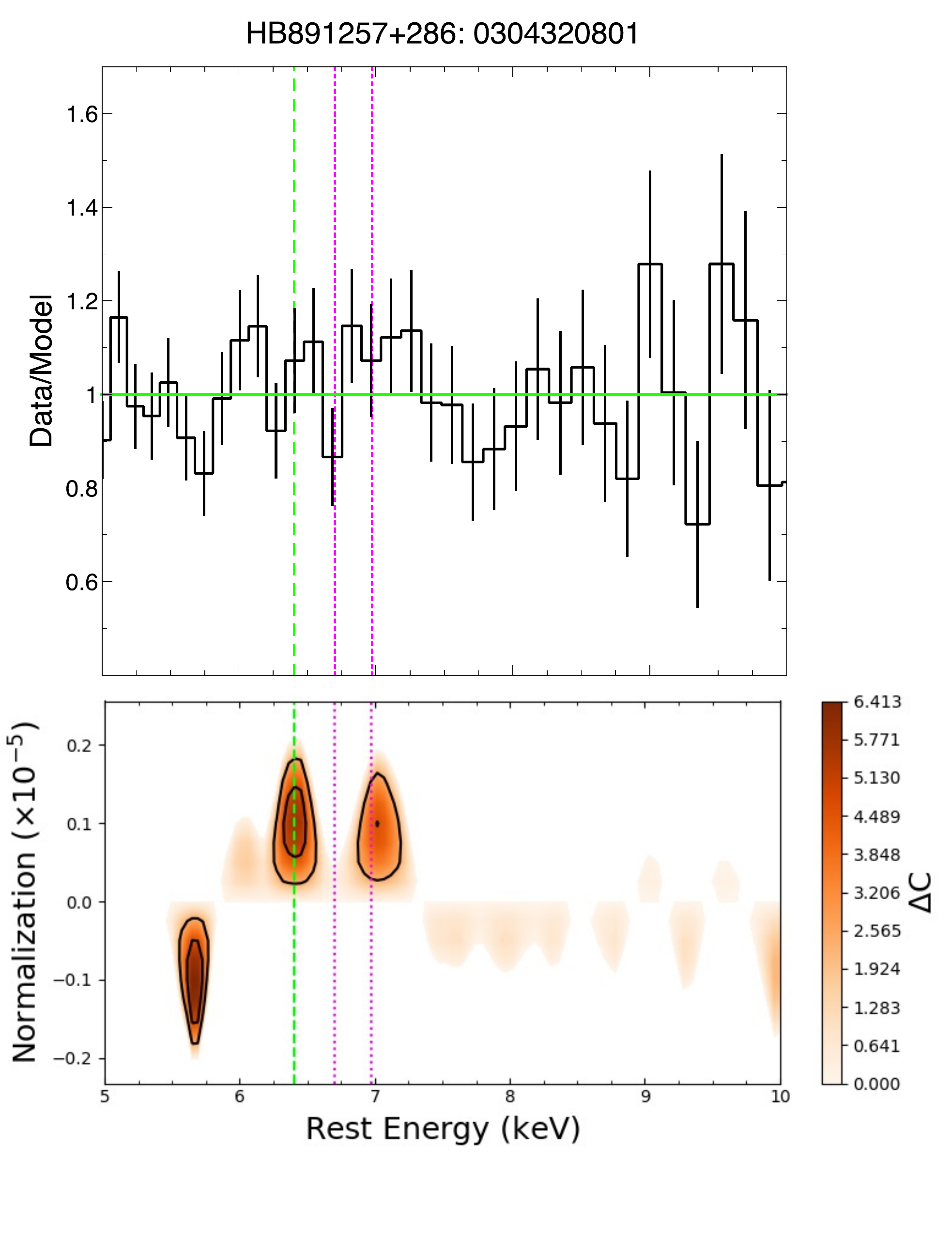}
\includegraphics[scale=0.21]{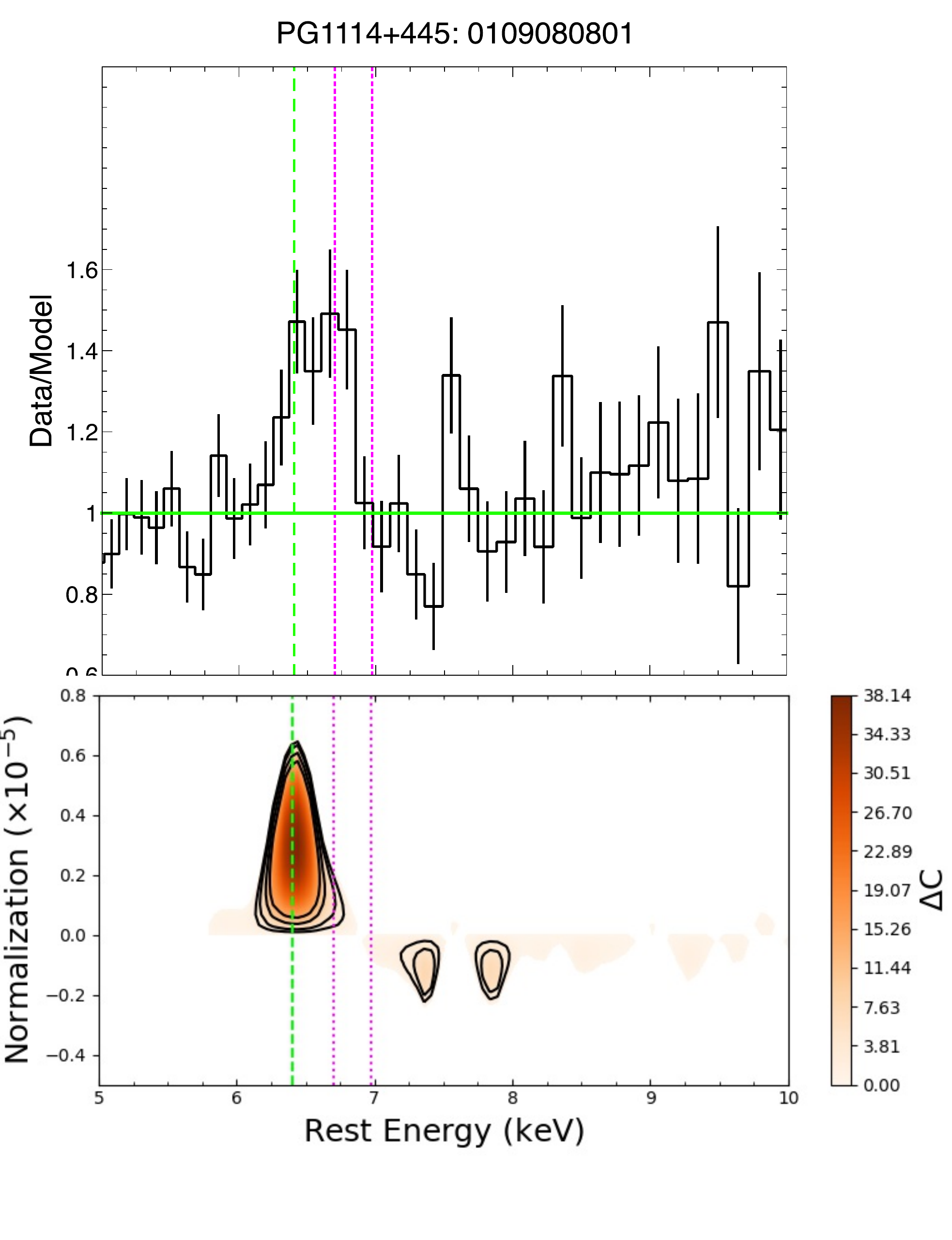}
\includegraphics[scale=0.21]{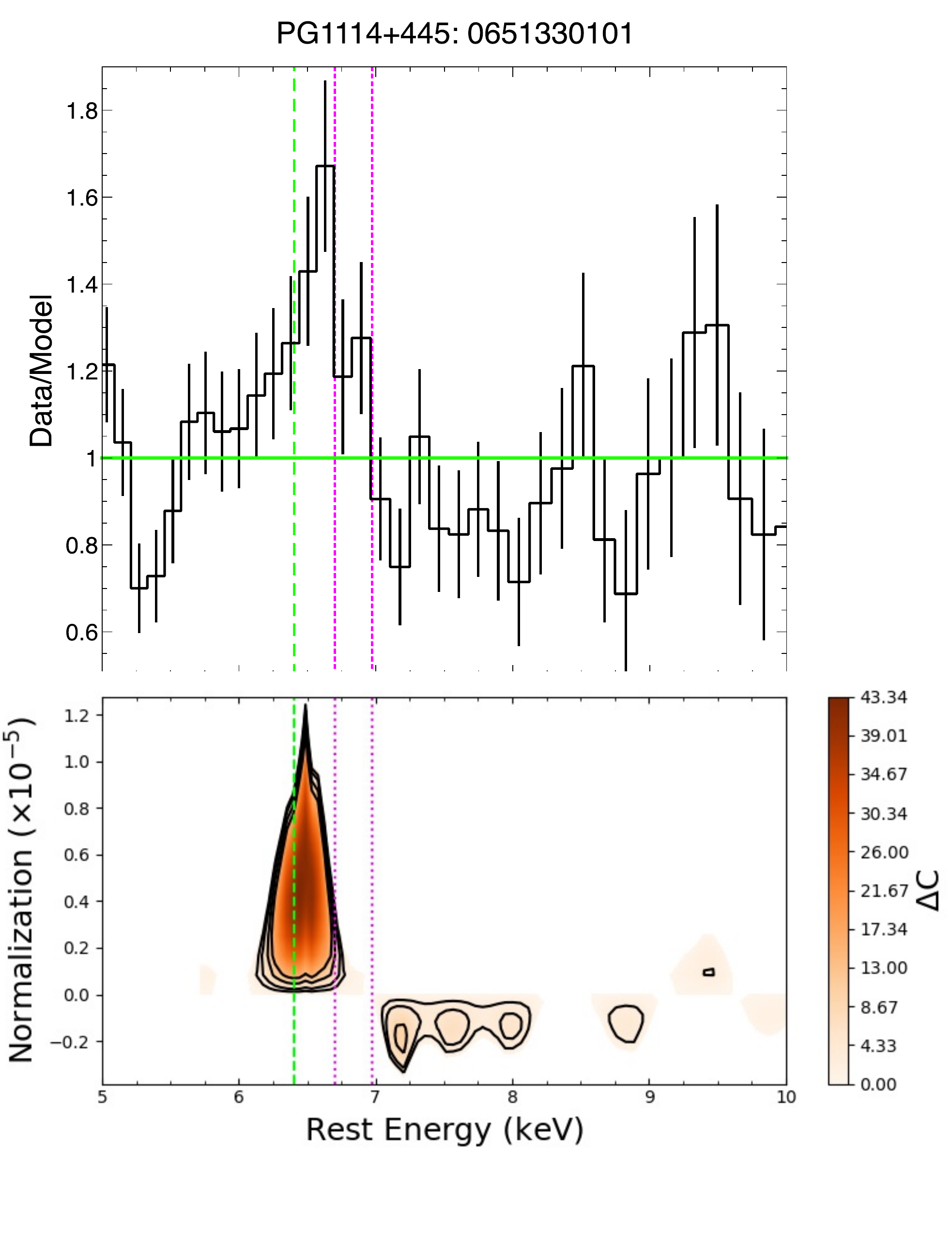}
\includegraphics[scale=0.21]{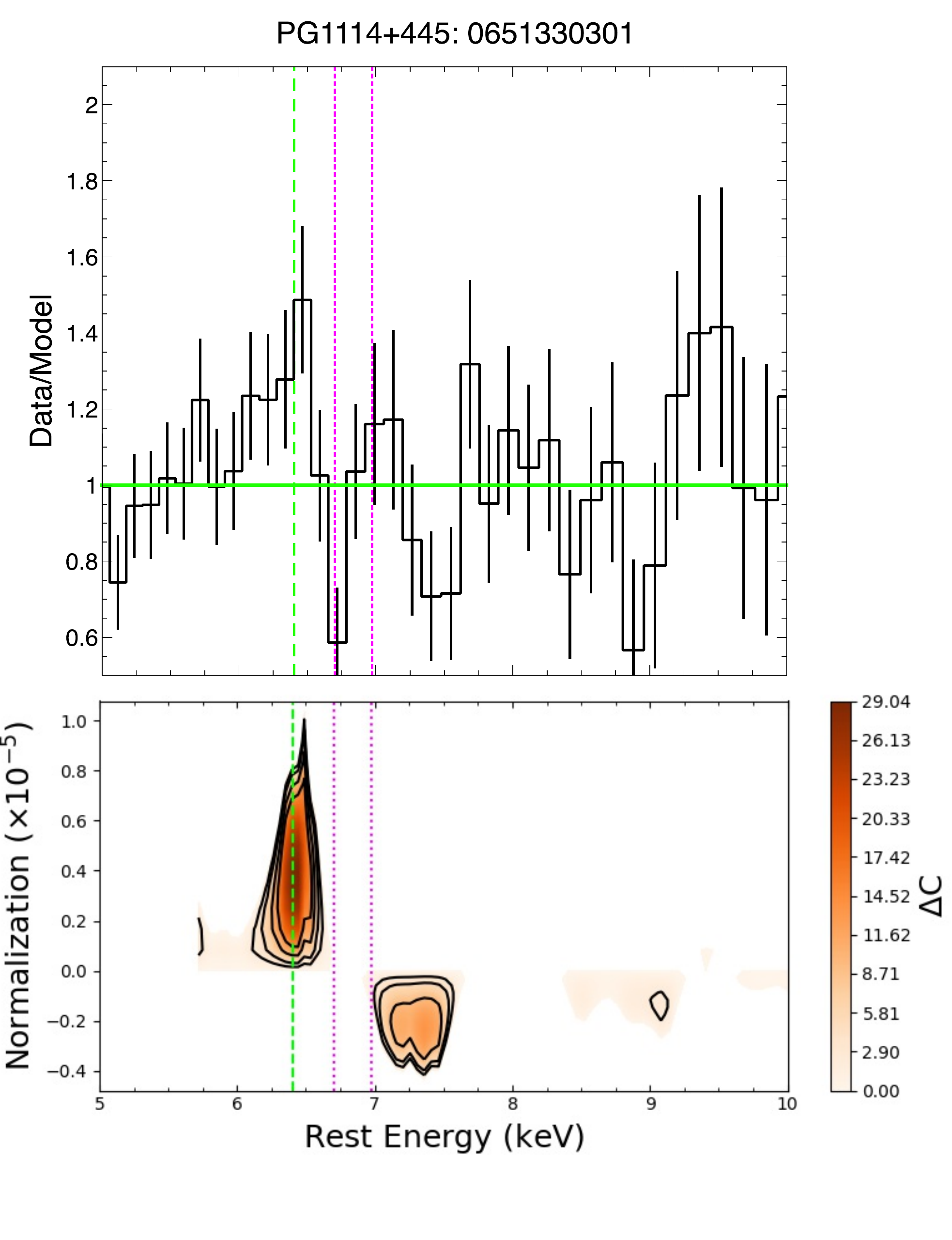}

\end{figure*}
\begin{figure*} 
\centering
\includegraphics[scale=0.21]{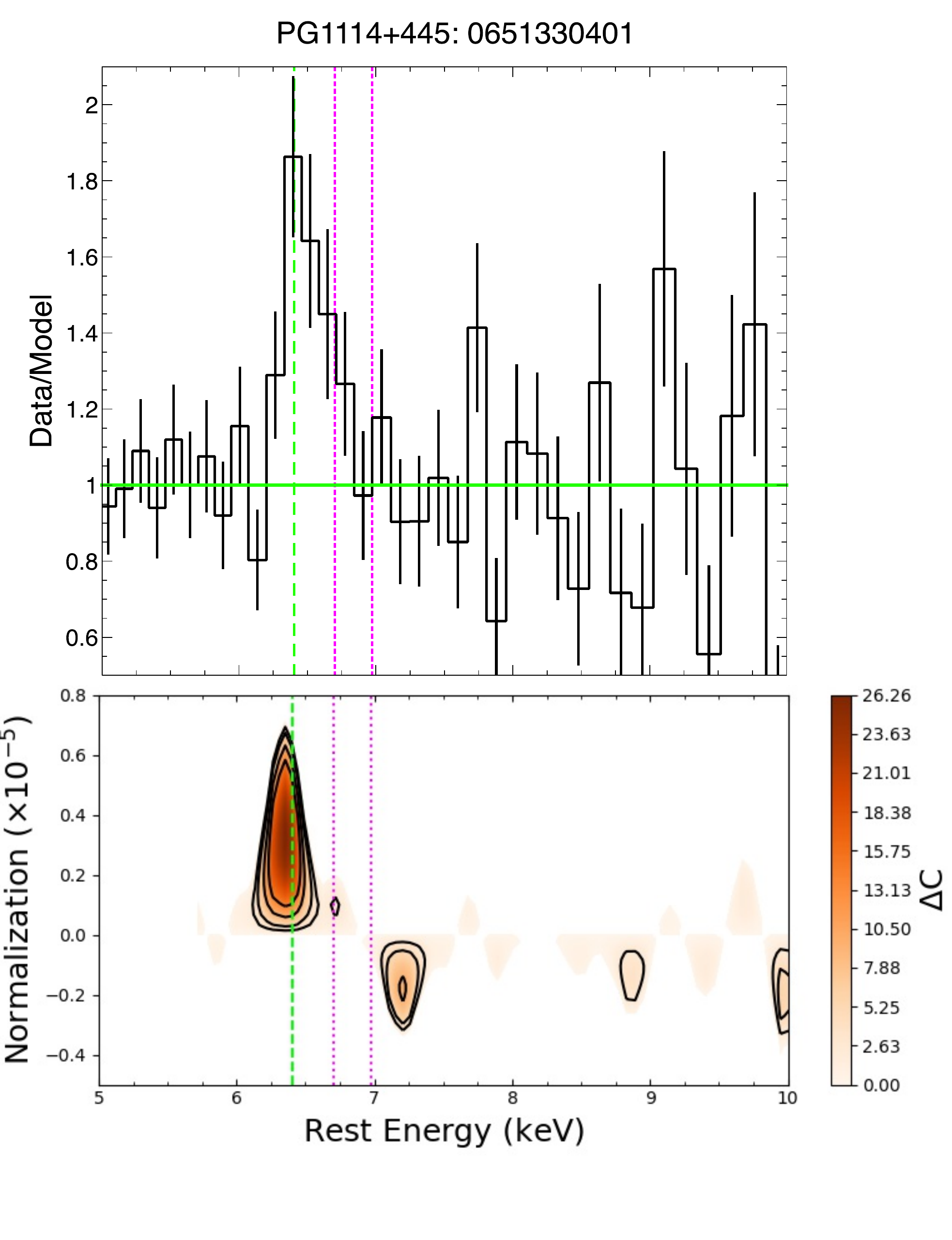}
\includegraphics[scale=0.21]{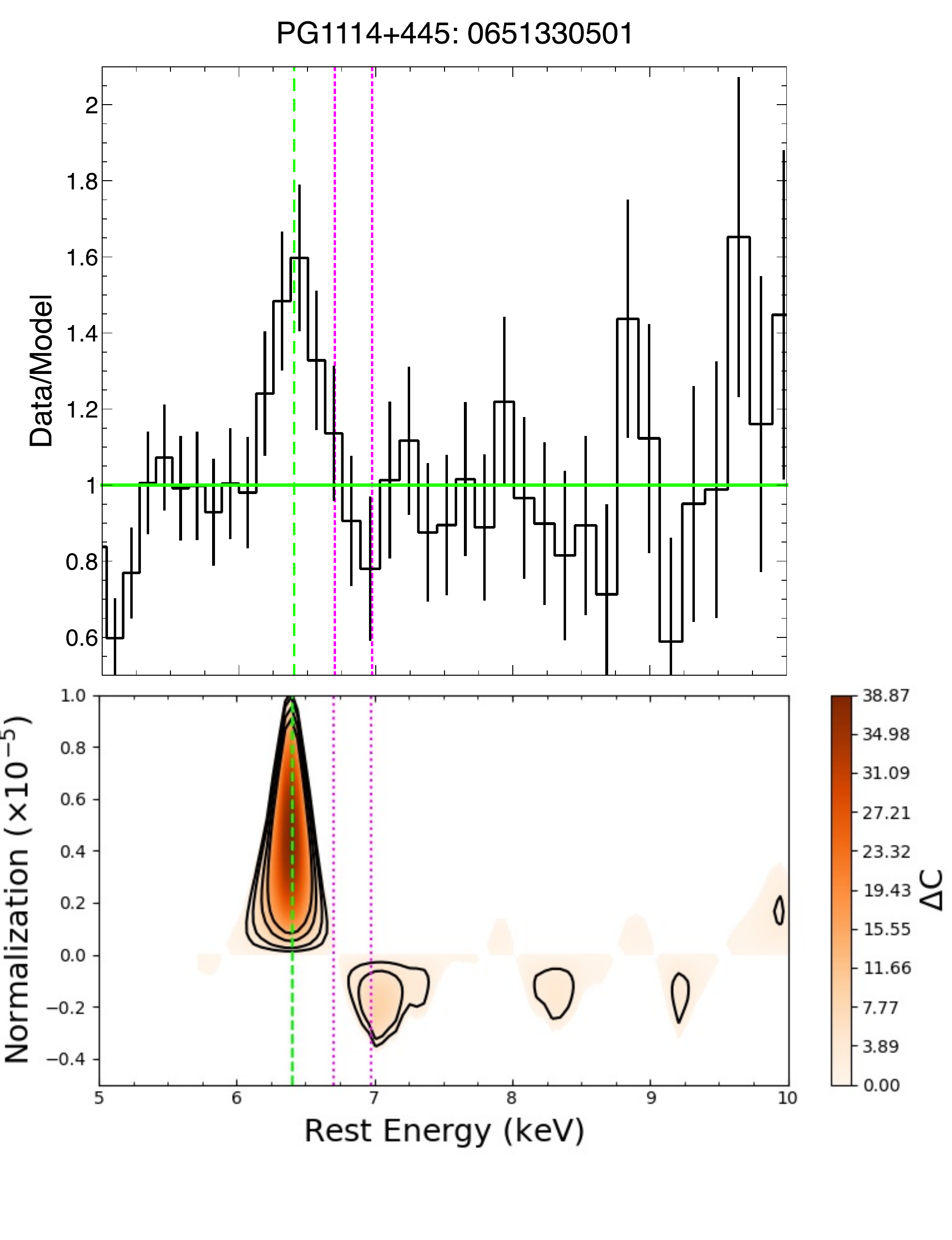}
\includegraphics[scale=0.21]{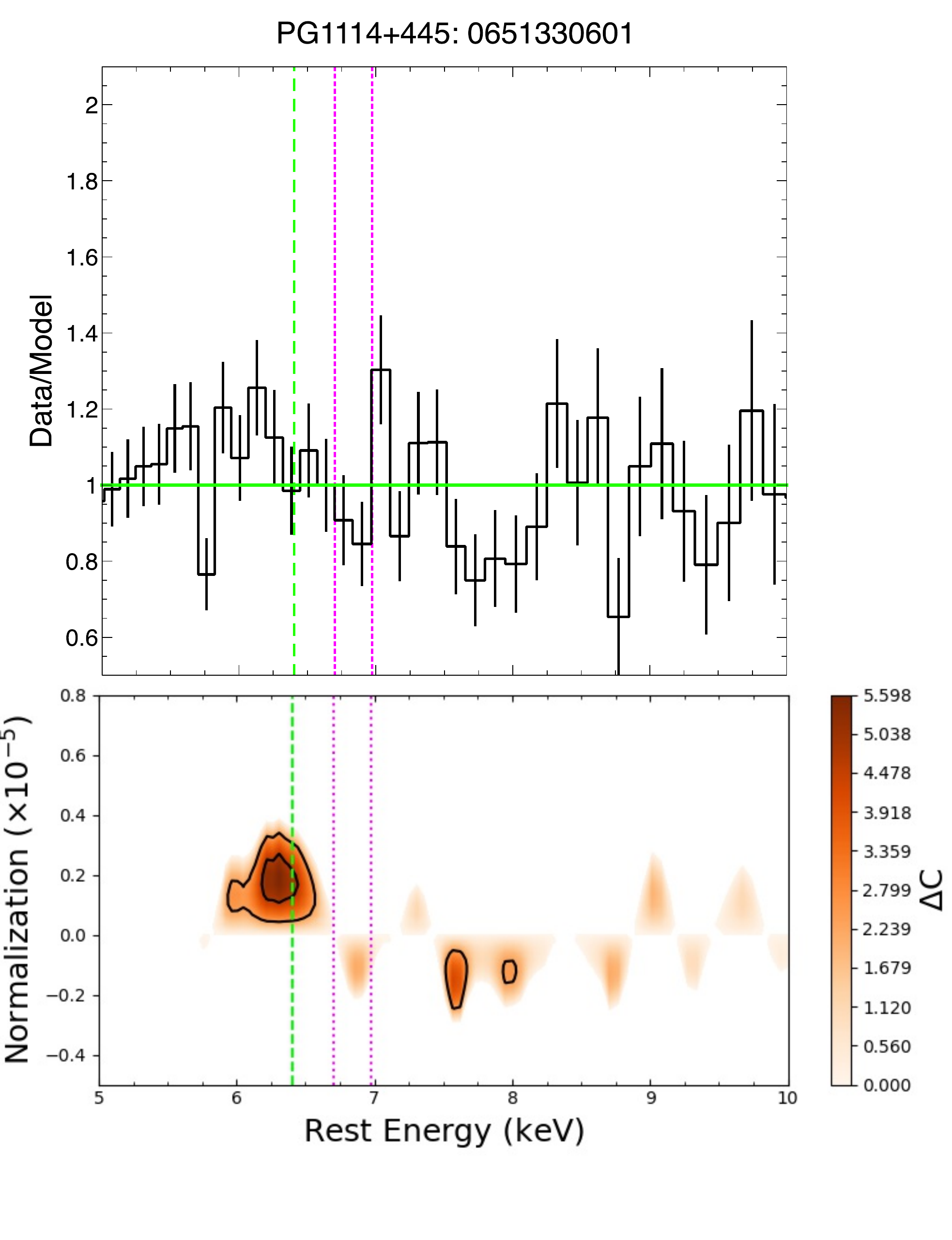}
\includegraphics[scale=0.21]{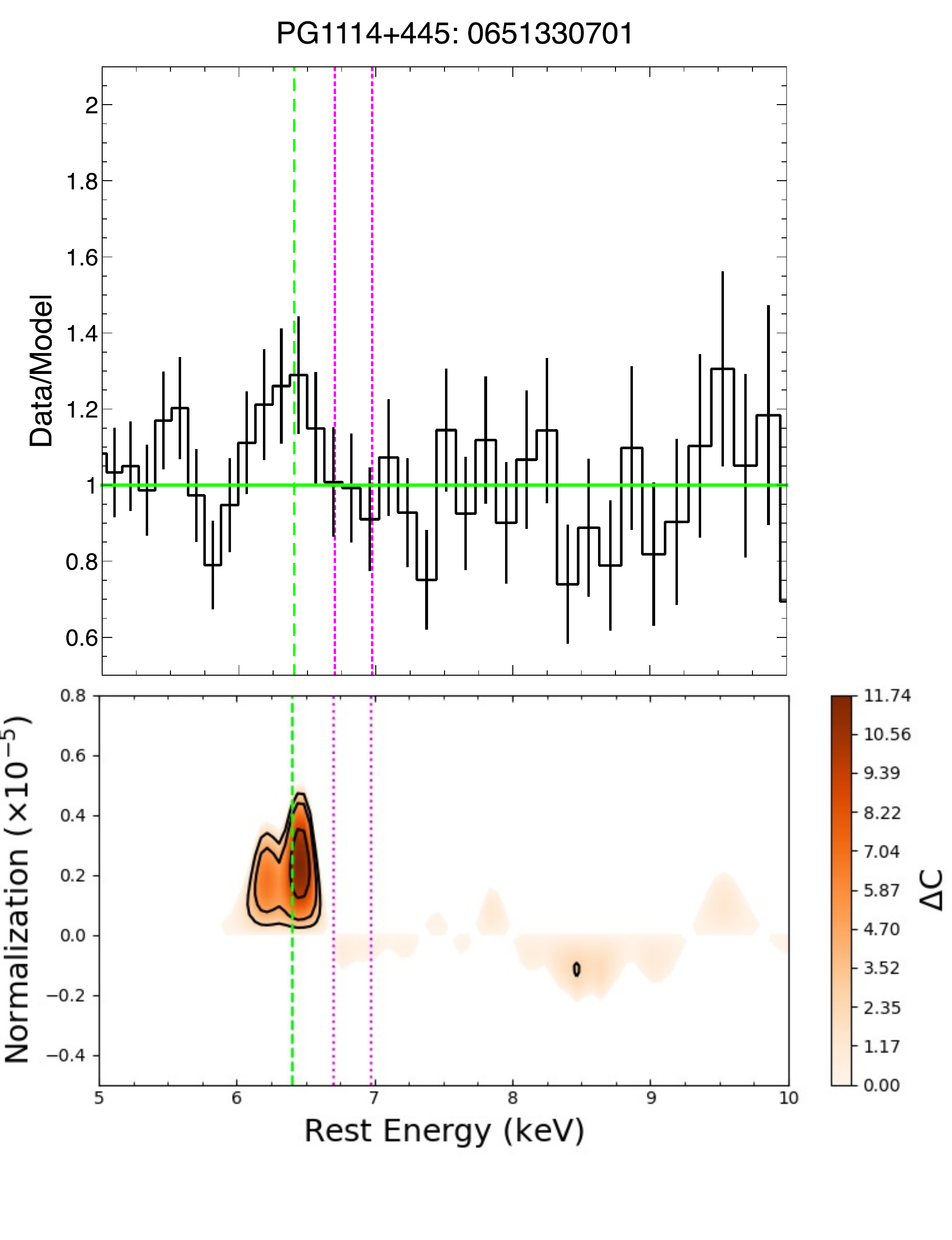}
\includegraphics[scale=0.21]{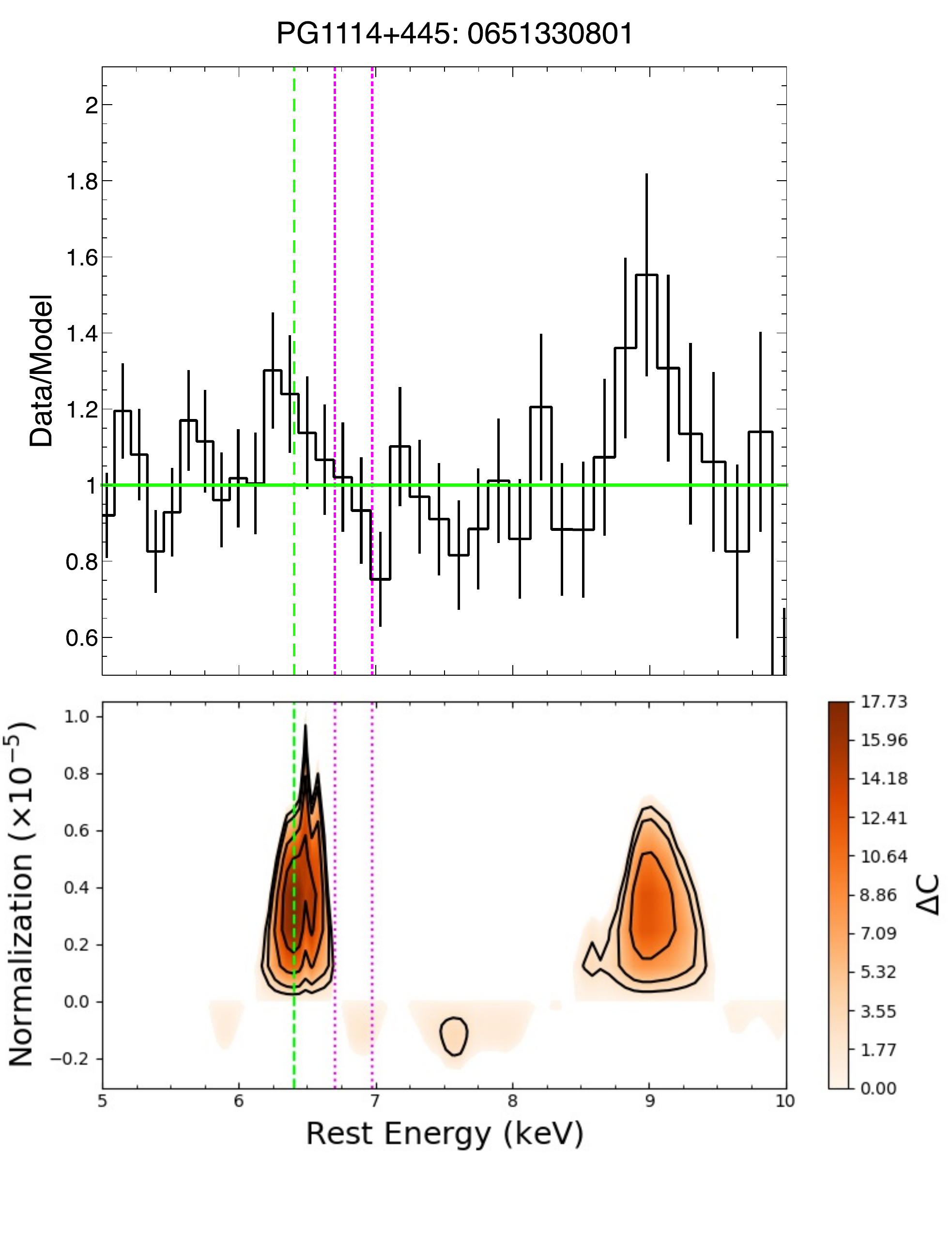}
\includegraphics[scale=0.21]{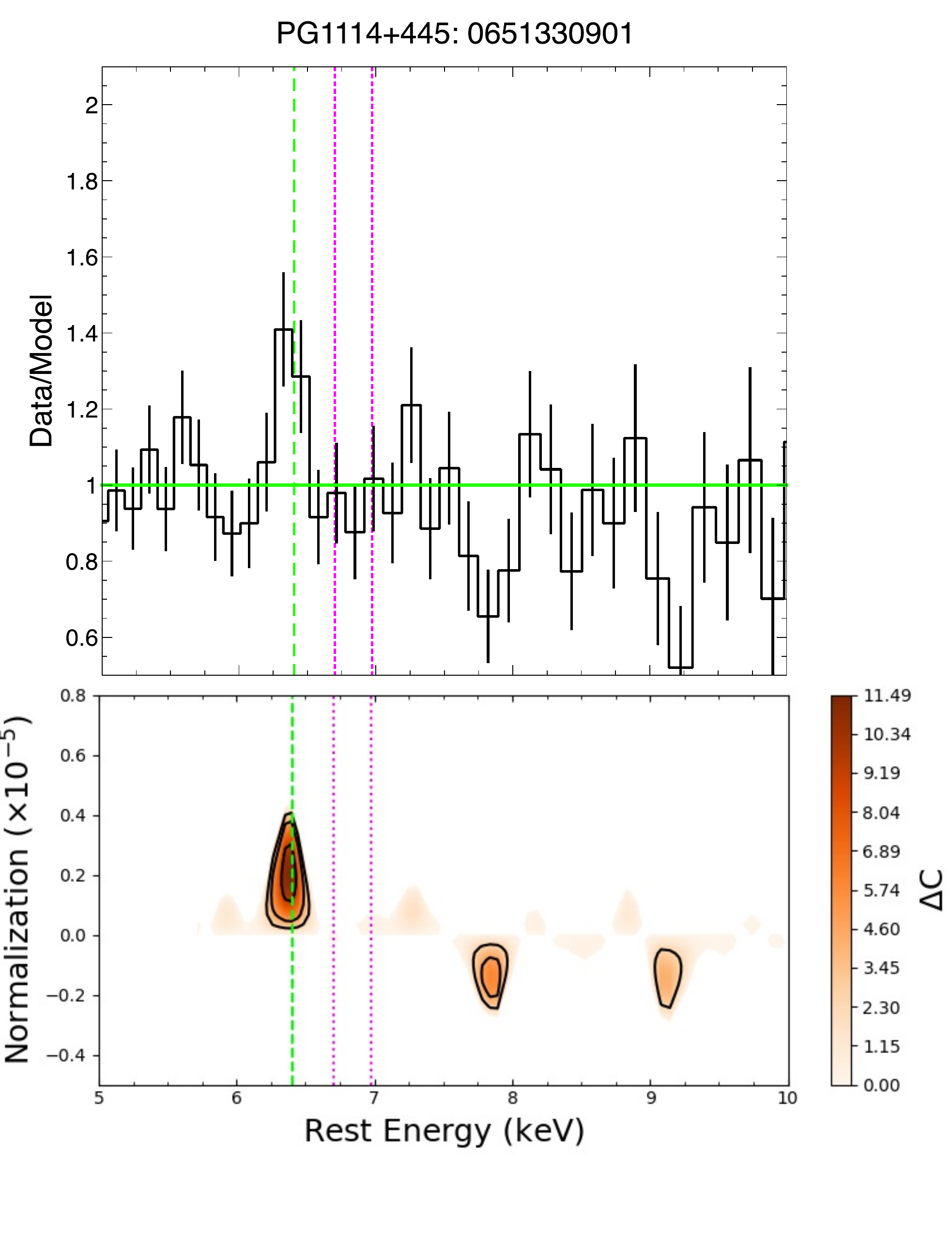}
\includegraphics[scale=0.21]{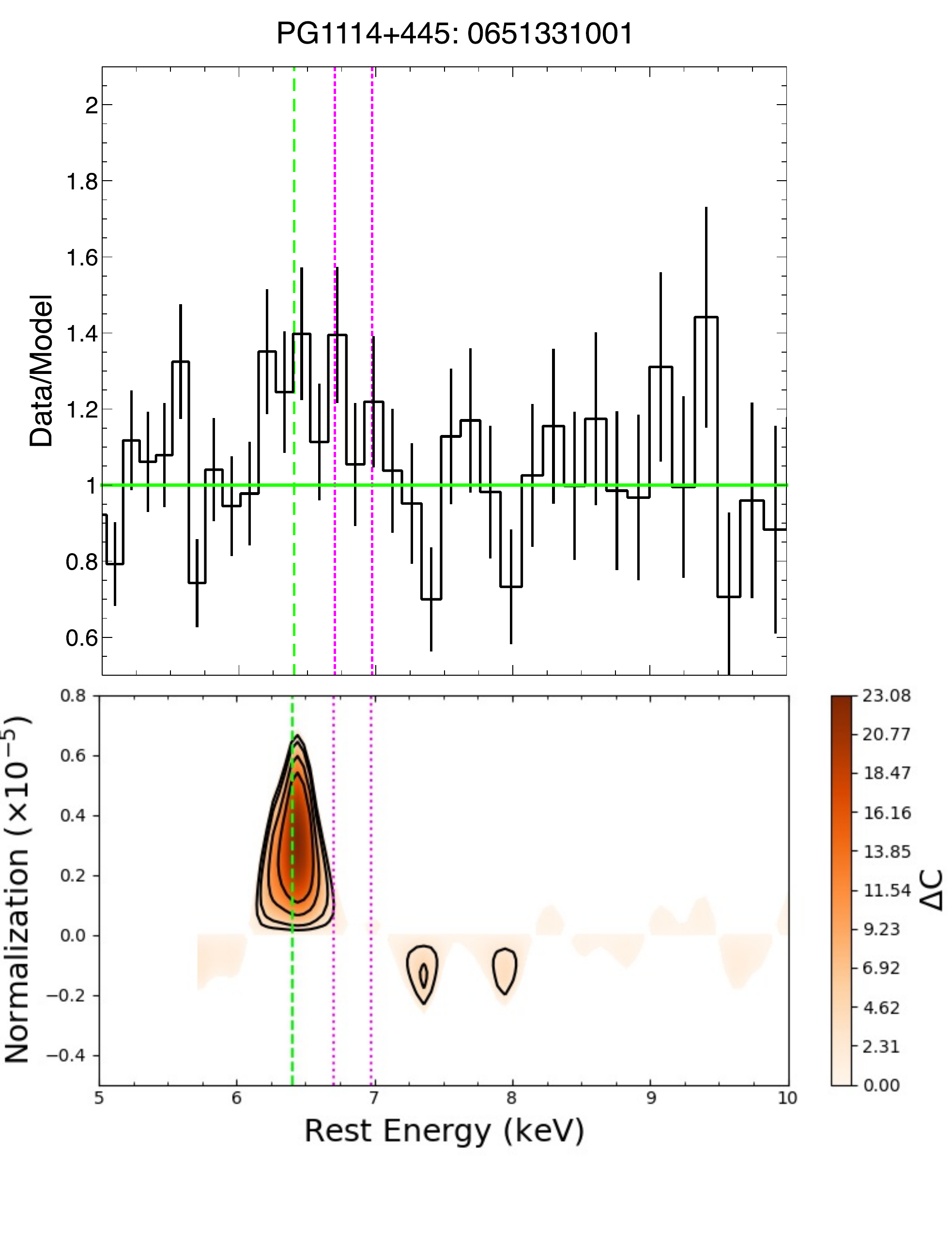}
\includegraphics[scale=0.21]{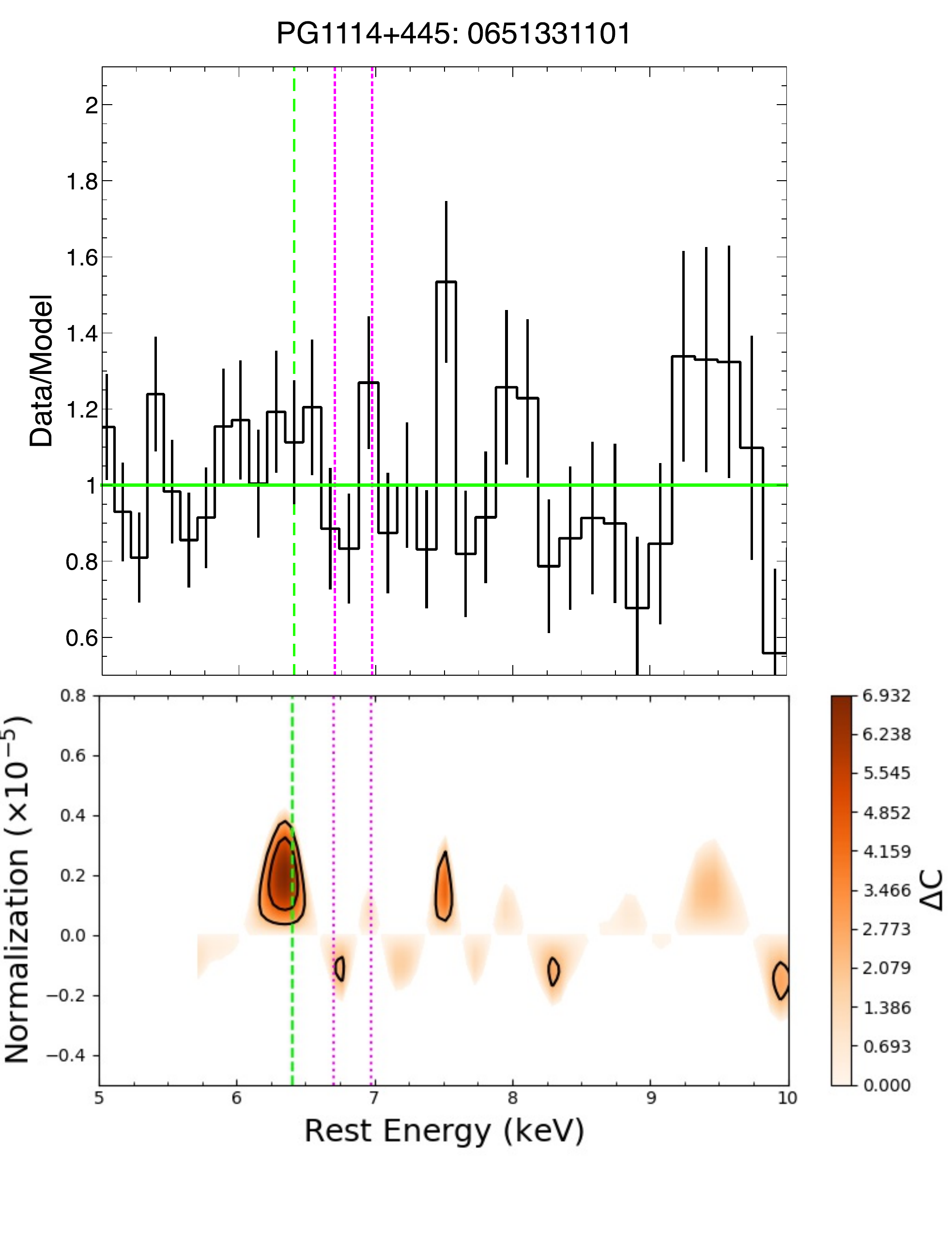}

\end{figure*}

\end{document}